\documentclass{article}
\usepackage[utf8]{inputenc}
\usepackage{graphicx}
\usepackage{float}
\usepackage{amsmath,amssymb}
\usepackage{hyperref}
\usepackage[T1]{fontenc}
\usepackage[backend=bibtex,style=numeric,natbib=true,sorting=none]{biblatex} 
\usepackage{geometry}
\usepackage{tikz}
\usetikzlibrary{positioning,arrows}
\usetikzlibrary{decorations.pathmorphing}
\usetikzlibrary{decorations.markings}
\usetikzlibrary{snakes}
\usetikzlibrary{matrix}
\usepackage{bbold}
\usepackage{environ}         
\usepackage{etoolbox}        
\usepackage{graphicx}        
\usepackage{nomencl}
\makenomenclature
\usepackage{etoolbox}
\renewcommand\nomgroup[1]{%
  \item[\bfseries
  \ifstrequal{#1}{C}{CFT symbols}{%
  \ifstrequal{#1}{H}{Heun symbols}{%
  \ifstrequal{#1}{F}{CFT symbols - semiclassics}{}}}%
]}
\numberwithin{equation}{subsection}

\newlength{\myl}
\let\origequation=\equation
\let\origendequation=\endequation

\RenewEnviron{equation}{
  \settowidth{\myl}{$\BODY$}                       
  \origequation
  \ifdimcomp{\the\linewidth}{>}{\the\myl}
  {\ensuremath{\BODY}}                             
  {\resizebox{\linewidth}{!}{\ensuremath{\BODY}}}  
  \origendequation
}
\newcommand{\FIV}[9]{\mathfrak{F}\bigg( \begin{matrix} #2 \\ #1 \end{matrix} \, #3 \, \begin{matrix} #4  \\ \, \end{matrix} \,#5 \, \begin{matrix} #6 \\ #7 \end{matrix} ; #8, #9 \bigg)}
\newcommand{\FIVhat}[9]{\widetilde{\mathfrak{F}}\bigg( \begin{matrix} #2 \\ #1 \end{matrix} \, #3 \, \begin{matrix} #4  \\ \, \end{matrix} \,#5 \, \begin{matrix} #6 \\ #7 \end{matrix} ; #8, #9 \bigg)}
\newcommand{\FIVsc}[9]{\mathcal{F}\bigg( \begin{matrix} #2 \\ #1 \end{matrix} \, #3 \, \begin{matrix} #4  \\ \, \end{matrix} \,#5 \, \begin{matrix} #6 \\ #7 \end{matrix} ; #8, #9 \bigg)}
\newcommand{\FIII}[8]{{}_1 \mathfrak{F} \left(#1 \, \, #2 \, \begin{matrix} #3 \\ \, \end{matrix} \,  #4 \, \begin{matrix} #5 \\ #6 \end{matrix} ; #7, #8 \right)}

\newcommand{\FIIIsc}[8]{{}_1 \mathcal{F} \left(#1 \, \, #2 \, \begin{matrix} #3 \\ \, \end{matrix} \,  #4 \, \begin{matrix} #5 \\ #6 \end{matrix} ; #7, #8 \right)}
\newcommand{\FIIIhat}[8]{{}_1 \widetilde{\mathfrak{F}} \left(#1 \, \, #2 \, \begin{matrix} #3 \\ \, \end{matrix} \,  #4 \, \begin{matrix} #5 \\ #6 \end{matrix} ; #7, #8 \right)}
\newcommand{\DIII}[8]{{}_1 \mathfrak{D}\bigg( #1 \,\,\begin{matrix} #2 \\ \, \end{matrix} \, \, #3 \, \, #4 \, \, \begin{matrix} #5 \\ #6 \end{matrix}; #7, #8 \bigg)}
\newcommand{\DIIIlargeLambda}[8]{{}_1 \mathfrak{D}\bigg( #1\,\, \begin{matrix} #2 \\ \, \end{matrix} \, \, #3  \, \, \begin{matrix} #4 \\ {} \end{matrix}\,\, #5\,\,#6; #7, #8 \bigg)}
\newcommand{\DIIIlargeLambdasc}[8]{{}_1 \mathcal{D}\bigg( #1\,\, \begin{matrix} #2 \\ \, \end{matrix} \, \, #3  \, \, \begin{matrix} #4 \\ {} \end{matrix}\,\, #5\,\,#6; #7, #8 \bigg)}
\newcommand{\DIIIhat}[8]{{}_1 \widetilde{\mathfrak{D}} \bigg( #1 \begin{matrix} #2 \\ \, \end{matrix} \, \, #3 \, \, #4 \, \, \begin{matrix} #5 \\ #6 \end{matrix}; #7, #8 \bigg)}
\newcommand{\DIIIsc}[8]{{}_1 \mathcal{D} \bigg( #1 \begin{matrix} #2 \\ \, \end{matrix} \, \, #3 \, \, #4 \, \, \begin{matrix} #5 \\ #6 \end{matrix}; #7, #8 \bigg)}
\newcommand{\FII}[7]{{}_1 \mathfrak{F}_1 \left( #1 \, \, #2 \, \begin{matrix} #3 \\ \, \end{matrix} #4 \, \, #5; #6, #7 \right)}
\newcommand{\FIIhat}[7]{{}_1 \widetilde{\mathfrak{F}}_1 \left( #1 \, \, #2 \, \begin{matrix} #3 \\ \, \end{matrix} #4 \, \, #5; #6, #7 \right)}

\newcommand{\FIIA}[7]{{}_{\frac{1}{2}} \mathfrak{F} \left( #1 \, \begin{matrix} #2 \\ \, \end{matrix} \,  #3 \, \begin{matrix} #4 \\ #5 \end{matrix} ; #6, #7 \right)}
\newcommand{\FIIAsc}[7]{{}_{\frac{1}{2}} \mathcal{F} \left( #1 \, \begin{matrix} #2 \\ \, \end{matrix} \,  #3 \, \begin{matrix} #4 \\ #5 \end{matrix} ; #6, #7 \right)}
\newcommand{\DII}[7]{{}_1 \mathfrak{D}_1 \bigg( #1\, \begin{matrix} #2 \\ \,  \end{matrix} \, \, #3 \, \, #4 \, \, #5; #6, #7 \bigg)}
\newcommand{\DIIhat}[7]{{}_1 \widetilde{\mathfrak{D}}_1 \bigg( #1\, \begin{matrix} #2 \\ \,  \end{matrix} \, \, #3 \, \, #4 \, \, #5; #6, #7 \bigg)}
\newcommand{\DIIsc}[7]{{}_1 \mathcal{D}_1 \bigg( #1\, \begin{matrix} #2 \\ \,  \end{matrix} \, \, #3 \, \, #4 \, \, #5; #6, #7 \bigg)}
\newcommand{\DIIA}[7]{{}_{\frac{1}{2}}\mathfrak{E}^{(#1)}\bigg( #2 \, \, #3 \, \, \begin{matrix} #4 \\ #5 \end{matrix}; #6, #7 \bigg)}
\newcommand{\DIIAsc}[7]{{}_{\frac{1}{2}}\mathcal{E}^{(#1)}\bigg( #2 \, \, #3 \, \, \begin{matrix} #4 \\ #5 \end{matrix}; #6, #7 \bigg)}
\newcommand{\DI}[6]{{}_1 \mathfrak{D}_{\frac{1}{2}} \bigg( #1\, \begin{matrix} #2 \\ \,  \end{matrix} \, \, #3 \, \, #4 \, ; #5, #6 \bigg)}
\newcommand{\DIsc}[6]{{}_1 \mathcal{D}_{\frac{1}{2}} \bigg( #1\, \begin{matrix} #2 \\ \,  \end{matrix} \, \, #3 \, \, #4 \, ; #5, #6 \bigg)}
\newcommand{\EI}[6]{{}_1 \mathfrak{E}_{\frac{1}{2}}^{(#1)}\bigg( #2\, #3\, #4\, ; #5 ,#6\bigg)}
\newcommand{\EIsc}[6]{{}_1 \mathcal{E}_{\frac{1}{2}}^{(#1)}\bigg( #2\, #3\, #4\, ; #5 ,#6\bigg)}
\newcommand{\FI}[6]{{}_1 \mathfrak{F}_{\frac{1}{2}} \left( #1 \, \, #2 \, \begin{matrix} #3 \\ \, \end{matrix} #4\, \, ; #5, #6 \right)}
\newcommand{\FIhat}[6]{{}_1 \widetilde{\mathfrak{F}}_{\frac{1}{2}} \left( #1 \, \, #2 \, \begin{matrix} #3 \\ \, \end{matrix} #4\, \, ; #5, #6 \right)}

\newcommand{\Fz}[5]{{}_\frac{1}{2} \mathfrak{F}_{\frac{1}{2}} \left( #1 \, #2 \, #3 \, ; #4, #5 \right)} 
 
\newcommand{\Ez}[5]{{}_\frac{1}{2} \mathfrak{E}_{\frac{1}{2}}^{(#1)} \left( #2 \, #3 \, ; #4, #5 \right)}
\newcommand{\Ezsc}[5]{{}_\frac{1}{2} \mathcal{E}_{\frac{1}{2}}^{(#1)} \left( #2 \, #3 \, ; #4, #5 \right)}

\title{Irregular Liouville correlators and connection formulae for \\ Heun functions}

\author{Giulio Bonelli\footnote{bonelli@sissa.it}, Cristoforo Iossa\footnote{ciossa@sissa.it}, Daniel Panea Lichtig\footnote{daniel.panea@sissa.it}, Alessandro Tanzini\footnote{tanzini@sissa.it}\\ \, \\
\normalsize International School of Advanced Studies (SISSA) via Bonomea 265, 34136 Trieste, Italy \\ \normalsize INFN, Sezione di Trieste \\
\normalsize Institute for Geometry and Physics, IGAP, via Beirut 2, 34136 Trieste, Italy}
\date{}

\begin{document}
\tikzset{
line/.style={thick, decorate, draw=black,}
 }

\maketitle

\begin{abstract} 
We perform a detailed study of a class of irregular correlators in Liouville Conformal Field Theory, of the related 
Virasoro conformal blocks with irregular singularities and of their connection formulae.
Upon considering their semi-classical limit, we provide explicit expressions of the connection matrices for the Heun function and a class of its confluences. Their calculation is reduced to concrete combinatorial formulae from conformal block expansions.
\end{abstract}
\newpage

\tableofcontents


\nomenclature[C001]{\(b\)}{Liouville coupling constant}
\nomenclature[C002]{\(Q\)}{Liouville background charge $Q = b + b^{-1}$}
\nomenclature[C003]{\(\alpha_\infty, \, \alpha_1, \, \alpha_t, \, \alpha_0\)}{Liouville momenta of non-degenerate primary insertions}
\nomenclature[C004]{\(\Delta_\infty, \, \Delta_1, \, \Delta_t, \, \Delta_0\)}{Scaling dimensions of non-degenerate primary insertions, $\Delta_i = \frac{Q^2}{4} - \alpha_i^2$}
\nomenclature[C005]{\(\alpha\)}{Intermediate momentum, with corresponding scaling dimension $\Delta$}
\nomenclature[C006]{\(\alpha_{2,1}\)}{Degenerate Liouville momentum $\alpha_{2,1} = -\frac{2b+b^{-1}}{2}$, with corresponding scaling dimension $\Delta_{2,1}$}
\nomenclature[C008]{\(\alpha_{i \theta}\)}{Momentum shifted by $- \theta \frac{b}{2}$ with $\theta = \pm 1$, namely $\alpha_{i \theta}=\alpha_i - \theta \frac{b}{2}$}
\nomenclature[C009]{\( V_{\alpha_i}(x) \)}{Primary operator of momentum $\alpha_i$ inserted at $x$}
\nomenclature[C010]{\(\vert \Delta_i \rangle\)}{Primary state, corresponding to a primary operator of dimension $\Delta_i$ inserted at zero (at infinity if $\langle \Delta_\infty \vert$)}
\nomenclature[C011]{\(\mu\)}{$L_1-$momentum of an irregular insertion of rank 1}
\nomenclature[C012]{\(\mu'\)}{Intermediate $L_1-$momentum}
\nomenclature[C013]{\(\mu_{i \theta}\)}{$L_1-$momentum shifted by $- \theta \frac{b}{2}$ with $\theta = \pm 1$}
\nomenclature[C014]{\(\vert \mu, \Lambda \rangle\)}{Irregular state of rank 1 with eigenvalues $\mu \Lambda, - \frac{\Lambda^2}{4}$ inserted at zero (at infinity if $\langle \mu, \Lambda \vert$)}
\nomenclature[C015]{\(\vert \Lambda^2 \rangle\)}{Irregular state of rank $\frac{1}{2}$ with eigenvalues $-\frac{\Lambda^2}{4}$ inserted at zero (at infinity if $\langle \Lambda^2 \vert$)}
\nomenclature[C016]{\({}_x \mathfrak{F}_y\)}{Conformal block (CB) expanded around regular insertions, with an insertion of rank\footnote{Here and in the following, if $x$ or $y$ are zero we drop the label for simplicity.} $x$ resp. $y$ at $\infty$ resp. $0$}
\nomenclature[C017]{\({}_x \mathfrak{D}_y\)}{CB with at least one variable expanded around an irreg. singularity of rank 1, with an insertion of rank $x$ resp. $y$ at $\infty$ resp. $0$}
\nomenclature[C018]{\({}_x \mathfrak{E}_y\)}{CB with at least one variable expanded around an irreg. singularity of rank $\frac{1}{2}$,  with an insertion of rank $x$ resp. $y$ at $\infty$ resp. $0$}
\nomenclature[C019]{\({}_x \widetilde{\mathfrak{F}}_y, \, {}_x \widetilde{\mathfrak{D}}_y,  \, {}_x \widetilde{\mathfrak{E}}_y\)}{CB without classical part, i.e. normalized as $1 + \dots$}
\nomenclature[C020]{\(G_{\alpha}\)}{Liouville two point function}
\nomenclature[C021]{\(C_{\alpha_1 \alpha_2 \alpha_3}\)}{Liouville three point function}
\nomenclature[C022]{\(C_{\mu \alpha}\)}{Pairing of a primary and a rank 1 irregular state}
\nomenclature[C023]{\(C_{\alpha}\)}{Pairing of a primary and a rank $\frac{1}{2}$ irregular state}
\nomenclature[C024]{\(C_{\alpha_1 \alpha_2}^{\alpha_3}\)}{OPE coefficient involving three primaries}
\nomenclature[C025]{\(B_{\mu \alpha}^{\mu'}\)}{OPE coefficient involving one primary and two irregular vertices of rank 1}
\nomenclature[C026]{\(B_{\alpha_{2,1}}\)}{OPE coefficient involving a degenerate field and two irregular vertices of rank 1/2}
\nomenclature[F001]{\(a_\infty, \, a_1, \, a_t, \, a_0\)}{Semiclassical Liouville momenta}
\nomenclature[F002]{\(a\)}{Semiclassical intermediate momentum}
\nomenclature[F003]{\(a_{i \theta}\)}{Semiclassical momentum shifted by $- \theta \frac{b^2}{2}$ with $\theta = \pm 1$}
\nomenclature[F004]{\(m\)}{Semiclassical $L_1-$momentum}
\nomenclature[F005]{\(m'\)}{Semiclassical intermediate $L_1-$momentum}
\nomenclature[F006]{\(m_\theta\)}{Semiclassical $L_1-$momentum shifted by $- \theta \frac{b^2}{2}$ with $\theta = \pm 1$}
\nomenclature[F007]{\(L\)}{Semiclassical highest eigenvalue of irregular states}
\nomenclature[F008]{\( {}_x \mathcal{F}_y \)}{Semiclassical CB expanded around regular insertions, with an insertion of rank $x$ resp. $y$ at $\infty$ resp. $0$}
\nomenclature[F009]{\( {}_x \mathcal{D}_y \)}{Semiclassical CB with at least one variable expanded around an irreg. singularity of rank 1, with an insertion of rank $x$ resp. $y$ at $\infty$ resp. $0$}
\nomenclature[F010]{\({}_x \mathcal{E}_y \)}{Semiclassical CB with at least one variable expanded around an irreg. singularity of rank $\frac{1}{2}$, with an insertion of rank $x$ resp. $y$ at $\infty$ resp. $0$}
\nomenclature[F011]{\(F\)}{Logarithm of a classical conformal block}
\nomenclature[F012]{\(W\)}{Logarithm of a semiclassical conformal block}
\nomenclature[F013]{\(u\)}{Log-derivative of a classical CB, up to constants}
\nomenclature[F014]{\({}_x \widetilde{\mathcal{F}}_y, \, {}_x \widetilde{\mathcal{D}}_y,  \, {}_x \widetilde{\mathcal{E}}_y\)}{Semiclassical CB rescaled so that they start as $1 + \dots$}
\nomenclature[H001]{\(\alpha, \, \beta, \, \gamma, \, \delta, \, \epsilon\)}{Parameters of the Heun equations}
\nomenclature[H002]{\(q\)}{Accessory parameter of the Heun equations}
\nomenclature[H003]{\(w(z)\)}{Solutions of the Heun equations in standard form}
\nomenclature[H004]{\(P_i^{-1}(z)w(z)\)}{Solutions of the Heun equations in normal form}
\nomenclature[H005]{HeunG}{General Heun function}
\nomenclature[H006]{HeunC}{Confluent Heun function expanded near a regular singularity}
\nomenclature[H007]{\(\text{HeunC}_\infty\)}{Confluent Heun function expanded near the irregular singularity}
\nomenclature[H008]{HeunRC}{Reduced confluent Heun function expanded near a regular singularity}
\nomenclature[H009]{\(\text{HeunRC}_\infty\)}{Reduced confluent Heun function expanded near the irregular singularity}
\nomenclature[H010]{HeunDC}{Doubly confluent Heun function expanded near an irregular singularity}
\nomenclature[H011]{\(\text{HeunRDC}_0\)}{Reduced doubly confluent Heun function expanded near the irregular singularity at zero}
\nomenclature[H012]{\(\text{HeunRDC}_\infty\)}{Reduced doubly confluent Heun function expanded near the irregular singularity at infinity}
\nomenclature[H013]{HeunDRDC}{Doubly reduced doubly confluent Heun function expanded near an iregular singularity}
\renewcommand{\nomname}{}
    
\section{Introduction}

In this paper we perform a detailed study of irregular correlators in Liouville Conformal Field Theory (CFT), of the related Virasoro conformal blocks with irregular singularities and of their connection formulae. Upon considering their semi-classical limit, we provide explicit expressions of the connection matrices for the Heun function and a class of its confluences. 
These result from the semi-classical limit of Virasoro conformal blocks for the five-point correlation function of four primaries and a degenerate field 
and a class of its coalescence limits to irregular conformal blocks. 
While the five-point correlator satisfies 
a linear PDE, namely the 
BPZ equation \cite{Belavin:1984vu}, its confluences satisfy a PDE obtained by an appropriate rescaling procedure. As we will discuss in detail in the paper, BPZ equations reduce in the semi-classical limit to ODEs.
For the particular five-point correlation function mentioned above, this gets identified with Heun's equation upon a suitable dictionary.
Let us also mention that the method we use can be generalised to general Fuchsian equations and their confluences upon considering the relevant
conformal blocks.

Heun’s equation \cite{heun1888theorie} is the most general second order linear differential equation with four regular singularities on the Riemann sphere. It is the next case in the Fuchsian series after the hypergeometric equation, which displays three regular singularities \cite{gauss1866carl}.
The Heun equation — along with its confluences — enters many problems
in theoretical and mathematical physics, geometry and other branches of quantitative sciences\footnote{For a huge bibliography take a look at https://theheunproject.org .} (see for example
\cite{Hortacsu:2011rr, 2015arXiv151204025F}).
For this reason, many studies appeared in the literature about it, see for example \cite{ronveaux1995heun} for a general introduction and
\cite{dekar,takemura}
for studies on the connection problem. Let us stress that the approach we follow in this paper allows to provide an explicit calculation
of the local expansions of Heun functions and their connection coefficients in terms of combinatorial formulae for convergent perturbative series, which derive from the relation with conformal block expansions.

Let us notice in particular that
Heun's equation enters the computation of surface operators\footnote{See the following coalescence diagram for a precise dictionary.}  
\cite{2010,Awata:2010bz}
in ${\cal N}=2$ $SU(2)$ supersymmetric gauge theory with $N_f\leq4$
\cite{Seiberg:1994aj}.
Moreover, 
the problem of linear perturbations of cylindrically symmetric black holes, governed by the 
Teukolsky equation \cite{Teukolsky:1972my}, is solved in terms of the confluent Heun function.
Indeed, the technique that we implement in this paper has already been developed for the confluent Heun function
for linear perturbations of Kerr black holes in
\cite{Bonelli:2021uvf} and here it is further refined and generalised.
By its very definition, Heun function solves the classical Poincar\'e uniformisation problem of a Riemann sphere with four punctures \cite{zbMATH02714348,10.1007/BF02418420}
We also remind that Heun's equation arises from the linear system whose isomonodromic deformation problem is described by the Painlev\'e VI equation
\cite{JIMBO1981306,JIMBO1981407,Jimbo1981MonodromyPD}.

Following a class of coalescences of the singularities and/or specific parameter scalings, from the configuration of four regular points
one naturally obtains a set of confluent 
irregular blocks satisfying the corresponding confluent BPZ equations. 
The Heun functions and its confluences are solutions of the 
resulting semiclassical reduced equations.

According to the Alday-Gaiotto-Tachikawa (AGT) correspondence
\cite{Alday_2010}, a precise gauge theoretical counterpart of Liouville CFT is given
by the BPS sector of four dimensional ${\cal N}=2$ $SU(2)$ gauge theory 
in the so-called $\Omega$-background \cite{Nekrasov:2003rj}. In particular the four-point conformal block 
of Liouville primary fields on the Riemann sphere gets identified with the Nekrasov partition function \cite{Nekrasov:2002qd}
of $SU(2)$ gauge theory with four fundamental hypermultiplets.
In this context, the confluence procedure is interpreted as the 
decoupling of massive hypermultiplets \cite{Gaiotto:2009ma} or the limit to strongly interacting Argyres-Douglas 
theories \cite{Bonelli_2017, Gaiotto:2012sf} in the $SU(2)$ Seiberg-Witten theory.
Degenerate field insertions in the CFT correlator correspond to 
surface operator insertions in the gauge theory
\cite{Alday_2010a}. The latter therefore satisfy BPZ equations and their confluent limits.
The importance of the AGT correspondence is that 
it maps more complicated aspects of one side to easier ones of the other, basically it
provides a proof of gauge theory dualities 
once reinterpreted as modular properties in CFT
\cite{Belavin:1984vu}. Moreover, it
provides an explicit combinatorial expression for Virasoro 
conformal blocks in terms of Nekrasov partition function.
We exploit this correspondence to provide concrete formulae for the connection matrices
for the relevant conformal blocks and their confluences. 
The semi-classical limit of CFT coincides via AGT correspondence with an asymmetric limit in the $\Omega$-background parameters
known as the Nekrasov-Shatashvili (NS) limit \cite{Nekrasov:2009rc}. This provides a quantization procedure of the classical integrable systems 
associated to the Seiberg-Witten theory \cite{Seiberg:1994aj}. From this viewpoint Heun equations can be interpreted as Schr\"odinger equations
for these quantum systems.

All in all, the connection problem for (confluent) Heun equations can be restated as
a connection problem for semi-classical conformal blocks. The latter can be computed in very explicit terms
via AGT correspondence by equivariant localisation in supersymmetric gauge theory in the NS limit.
Let us here notice that 
the classifying group of the solutions of the Heun equation \cite{2007MaCom..76..811M} is the $D_4$ Coxeter group, generated by the permutations of the four regular singular points and by the swaps of each couple of indices of the local solutions but a reference one. 
This concretely realises in the NS limit the action of the $D_4$ group on the 
vevs of surface operators in the $N_f=4$ $SU(2)$ gauge theory.

As mentioned above, the analysis of the confluences of the BPZ equations involves the appearance of {\it irregular}
conformal blocks \cite{Gaiotto:2009ma, Bonelli_2012, Gaiotto:2012sf}, which arise from the collision of regular singularities and suitable rescaling of their parameters.
In this paper we perform a detailed analysis of the irregular conformal blocks involved in the confluence process, of the related
three-point functions and of their connection matrices.


The mathematical interest of
Liouville quantum field theory has been 
highlighted by A.M. Polyakov who proposed to
interpret it as a quantum extension of the 
Poincar\'e uniformisation 
problem \cite{unpolyakov}.
A consequence of the above interpretation is that one can make use of the classical limit of Liouville theory to
obtain new exact solutions of classical uniformisation
\cite{zbMATH03989776}.
This inspired the work of several authors
\cite{Matone:1993tj,
Cantini:2001wr, takzo02, Hadasz:2006rb}
and received a renewed interest after the discovery of AGT correspondence \cite{Lit2014, Menotti:2014kra,david2015liouville, Piatek:2017fyn, hollands2017higher,saebyeok2020,lisovyy2021}.

\paragraph{Open questions:}

There is a number of open questions left for further investigation.
\begin{itemize}
    \item The generalization to $n-$point conformal blocks can be done along the same lines as the ones we have been following. 
    This produces explicit connection formulae for n-point Fuchsian systems in terms of Gamma functions and Nekrasov partition functions of linear quiver gauge theories. Via coalescence, this will provide connection formulae for higher rank singularities.
    \item In this paper we considered the class of confluences producing irregular singularities up to Poincar\'e rank one.
    This is implied by the fact that their gauge theory description can be given in a weakly coupled frame. It would be interesting to extend our analysis to higher rank singularities. These are related to Argyres-Douglas points in the gauge theory.
    \item As already mentioned, Heun functions play a relevant r\^ole in the study of linear perturbations of spinning black-holes. This topic was
    recently explored in connection to quantum Seiberg-Witten geometry in \cite{aminov2020black,Bianchi:2021xpr,Bianchi:2021mft,Hatsuda:2021gtn} and isomonodromic deformation theory \cite{Motl:2003cd,Castro:2013lba,Carneiro_da_Cunha_2016,Carneiro_da_Cunha_2020,daCunha:2021jkm,Cavalcante:2021scq,Amado:2021erf}. In our viewpoint this intriguing correspondence could be further clarified in CFT terms, as started in \cite{Bonelli:2021uvf}, and other massive gravitational sources can be studied along the same lines by making use of the results of this paper.
    For related topics, see also \cite{Casals:2021ugr,Bianchi:2021yqs,Nakajima:2021yfz,Blake:2021hjj,Pereniguez:2021xcj,Fioravanti:2021dce}.
    \item Our analysis can be extended to irregular blocks on 
    Riemann surfaces of higher genus. For example the genus one case is related to circular quiver gauge theories \cite{Bonelli:2019boe,Bonelli:2019yjd}
    \item By considering BPZ equations corresponding to higher level degenerate vertices, one can extend our analysis to higher order linear ODEs with rational coefficients.
    \item The uplift to q-difference equations can also be considered. This corresponds to consider q-Virasoro blocks
    and supersymmetric gauge theories in five dimensions
    \cite{Awata:2009ur}.
    This is related to q-Painlev\'e equations and topological strings \cite{Bershtein:2016aef,Bonelli:2017gdk}
\end{itemize}

The paper is organised as follows.
In section two, as a warm-up, we recall the relation between four-point conformal blocks with the insertion of three primary fields and
one level 2 degenerate field and hypergeometric functions and we study in detail the confluences to irregular conformal blocks and the related special functions. We obtain the connection formulae for the latter as solutions
of the constraints imposed by crossing symmetry.
In section three we systematically study the five point conformal blocks with the insertion of four primary fields and
one level 2 degenerate field. We focus on the explicit computation of the connection formulae as solutions
of the constraints imposed by crossing symmetry for the regular case and a class of its confluences. 
In each case, we also compute the semi-classical limit.
In section four we provide a dictionary between semiclassical CFT data and Heun equations in the standard form,
we apply the results of the previous section identifying the relevant semiclassical CFT blocks
with Heun functions and provide the connection formulae.
Few technical points are relegated to the Appendices. 
A final list of symbols should help the reader in following our computations.

The accompanying table collects the dictionary between (irregular) conformal blocks, supersymmetric gauge theories and the corresponding Heun functions.
\begin{center}
\begin{tabular}{ |l|l|l|l| } 
 \hline
 \multicolumn{2}{|l|}{CFT - CB} & $SU(2)$ Gauge Theory & Heun \\ 
 \hline
 $\mathfrak{F}$ & Regular &  $N_f = 4$ & HeunG \\ 
 ${}_1 \mathfrak{F}$ & Confluent & $N_f = 3$ & HeunC \\ 
 ${}_\frac{1}{2} \mathfrak{F}$ & Reduced Confluent & $N_f = 2$ asymmetric & HeunRC \\
 ${}_1 \mathfrak{D}_1$ & Doubly Confluent & $N_f = 2$ symmetric & HeunDC \\
 ${}_1 \mathfrak{E}_{\frac{1}{2}}$ & Reduced Doubly Confluent & $N_f = 1$ & HeunRDC \\
 ${}_\frac{1}{2} \mathfrak{E}_{\frac{1}{2}}$ & Doubly Reduced Doubly Confluent & $N_f = 0$ & HeunDRDC \\
 \hline
\end{tabular}
\end{center}

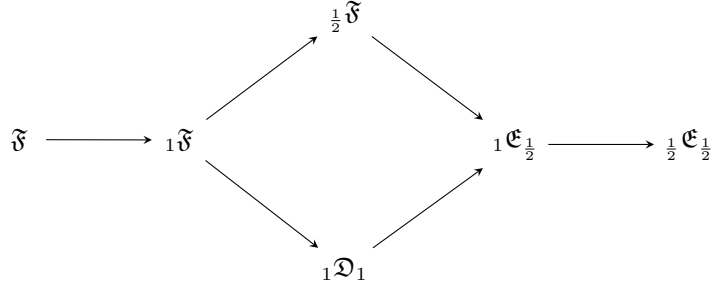
\begin{figure}[!h]\centering
\begin{tikzpicture}
  \matrix (m) [matrix of math nodes,row sep=3em,column sep=4em,minimum width=2em]
  {
  {}&{} & {}_\frac{1}{2} \mathfrak{F} & {}& {}\\
  \mathfrak{F} & {}_1 \mathfrak{F} & {} & {}_1 \mathfrak{E}_{\frac{1}{2}} & {}_\frac{1}{2} \mathfrak{E}_{\frac{1}{2}}\\
  {}& {} & {}_1 \mathfrak{D}_1 &{} &{}\\};
  \path[-stealth]
    (m-2-1) edge  (m-2-2)
    (m-2-2) edge  (m-1-3)
    (m-2-2) edge  (m-3-3)
    (m-1-3) edge  (m-2-4)
    (m-3-3) edge  (m-2-4)
    (m-2-4) edge  (m-2-5)
;
\end{tikzpicture}
\caption{Confluence diagram of conformal blocks.}
\end{figure}
\vspace{1cm}

\textbf{Acknowledgements}: We would like to thank 
S. Giusto, 
M. Hortacsu,
O. Lisovyy, 
W. Mueck,
R. Russo
for fruitful discussions. 
Some results contained in this paper have been presented by G.B. 
at the workshop "Rikkyo MathPhys 2022" and by D.P.L. at the INFN
workshop "Young Days". 
This research is partially supported by the INFN Research Projects GAST and ST\&FI, by PRIN "Geometria delle varietà algebriche" and by PRIN "Non-perturbative Aspects Of Gauge Theories And Strings".

\section{Warm-up: 4-point degenerate conformal blocks and classical special functions}

We start reviewing standard facts about four-point degenerate conformal blocks on the sphere and their confluence limits. In particular we review their relation to the hypergeometric function and its confluent limits, namely Whittaker and Bessel functions. 

The hypergeometric function is the solution to the most general second-order linear ODE with three regular singularities. On the CFT side it arises 
as the four-point conformal block on the Riemann sphere when one of the insertions is a degenerate vertex operator.

\subsection{Hypergeometric functions}\label{warmup:hypergeometric}
Consider the four-point conformal block on the sphere with one degenerate field insertion $\Phi_{2,1}$ of momentum $\alpha_{2,1}=-\frac{2b+b^{-1}}{2}$ (corresponding to $\Delta_{2,1}=-\frac{1}{2}-\frac{3b^2}{4}$):
\begin{equation}\label{corr2F1}
    \langle \Delta_\infty|V_1(1) \Phi_{2,1}(z)|\Delta_0\rangle\,.
\end{equation}
In the following we will drop the subscript $2,1$ and just denote by $\Phi(z)$ this degenerate field. The corresponding BPZ equation takes the form
\begin{equation}
    \left( b^{-2}\partial_z^2 -\left(\frac{1}{z-1}+ \frac{1}{z}\right)\partial_z + \frac{\Delta_1}{(z-1)^2}+\frac{\Delta_0}{z^2} +\frac{\Delta_\infty-\Delta_1-\Delta_{2,1}-\Delta_0}{z(z-1)} \right)\langle \Delta_\infty|V_1(1) \Phi(z)|\Delta_0\rangle=0\,.
\end{equation}
This equation has regular singularities at $0,1,\infty$. As mentioned above, the corresponding conformal blocks should therefore be expressed in terms of  hypergeometric functions. Indeed, the above differential equation by definition is solved by the conformal blocks corresponding to the correlator \eqref{corr2F1}, which in turn are given in terms of hypergeometric functions. In particular, the conformal block corresponding to the expansion $z\sim0$ is
\begin{equation}
    \mathfrak{F} \left( \begin{matrix} \alpha_1 \\ \alpha_\infty \end{matrix} \alpha_{0 \theta} \, \begin{matrix} \alpha_{2,1} \\ \alpha_0 \end{matrix} ; z \right) = z^{\frac{bQ}{2} + \theta b \alpha_0} (1-z)^{\frac{bQ}{2} + b \alpha_1} {}_2 F_1 \left( \frac{1}{2} + b \left( \theta \alpha_0+\alpha_1  - \alpha_\infty \right), \frac{1}{2} + b \left(\theta \alpha_0 +\alpha_1 + \alpha_\infty \right), 1 + 2 b \theta \alpha_0, z \right)\,,
\end{equation}
where $\theta =\pm$ and $\alpha_{0\pm} = \alpha_0 \pm \frac{-b}{2}$ are the two fusion channels allowed by the degenerate fusion rules. Similar formulae hold for the expansions around $z\sim 1$ and $\infty$. Conventionally, this conformal block is denoted diagrammatically by
\begin{equation}
\mathfrak{F} \left( \begin{matrix} \alpha_1 \\ \alpha_\infty \end{matrix} \alpha_{0 \theta} \, \begin{matrix} \alpha_{2,1} \\ \alpha_0 \end{matrix} ; z \right) =
\begin{tikzpicture}[baseline={(current bounding box.center)}, node distance=1cm and 1.5cm]
\coordinate[label=above:$\alpha_1$] (e1);
\coordinate[below=of e1] (aux1);
\coordinate[left=of aux1,label=left:$\alpha_\infty$] (e2);
\coordinate[right=1.5cm of aux1] (aux2);
\coordinate[above=of aux2,label=above:$\alpha_{2,1}$] (e3);
\coordinate[right=of aux2,label=right:$\alpha_0$] (e4);

\draw[line] (e1) -- (aux1);
\draw[line] (aux1) -- (e2);
\draw[line,dashed,red] (aux2) -- (e3);
\draw[line] (aux2) -- (e4);
\draw[line] (aux1) -- node[label=below:$\alpha_{0\theta}$] {} (aux2);
\end{tikzpicture} \,.
\end{equation}
We now want to expose the interplay between crossing symmetry, DOZZ factors and the connection formulae for the hypergeometric functions. To this end, let us expand the correlator once for $z \sim 0$ and once for $z \sim 1$:
\begin{equation}\label{eq:hypergeometric_constraint}
\begin{aligned}
        \langle \Delta_\infty|V_1(1) \Phi_{2,1}(z)|\Delta_0\rangle &= \sum_{\theta = \pm} C^{\alpha_{0\theta}}_{\alpha_{2,1} \alpha_0} C_{\alpha_\infty \alpha_1 \alpha_{0\theta}}\left|\mathfrak{F} \left( \begin{matrix} \alpha_1 \\ \alpha_\infty \end{matrix} \alpha_{0 \theta} \, \begin{matrix} \alpha_{2,1} \\ \alpha_0 \end{matrix} ; z \right)\right|^2 = \\
        &=\sum_{\theta' = \pm} C^{\alpha_{1\theta'}}_{\alpha_{2,1} \alpha_1} C_{\alpha_\infty \alpha_{1\theta'} \alpha_{0}}\left|\mathfrak{F} \left( \begin{matrix} \alpha_0 \\ \alpha_\infty \end{matrix} \alpha_{1 \theta'} \, \begin{matrix} \alpha_{2,1} \\ \alpha_1 \end{matrix} ; 1-z \right)\right|^2\,.
\end{aligned}
\end{equation}
Here $C_{\alpha \beta \gamma}$ are the DOZZ three-point functions, and $C^{\alpha}_{\beta\gamma}=G^{-1}_\alpha C_{\alpha \beta \gamma}$ are the OPE coefficients (see Appendix \ref{app:DOZZ}). Equation \eqref{eq:hypergeometric_constraint} is just the statement of crossing symmetry, due to the associativity of the OPE. The two expansions are related by the connection matrix $\mathcal{M}_{\theta \theta'}$ as follows
\begin{equation}
    \mathfrak{F} \left( \begin{matrix} \alpha_1 \\ \alpha_\infty \end{matrix} \alpha_{0 \theta} \, \begin{matrix} \alpha_{2,1} \\ \alpha_0 \end{matrix} ; z \right) = \sum_{\theta'=\pm}\mathcal{M}_{\theta \theta'}(b\alpha_0,b\alpha_1;b\alpha_\infty) \mathfrak{F} \left( \begin{matrix} \alpha_0 \\ \alpha_\infty \end{matrix} \alpha_{1 \theta'} \, \begin{matrix} \alpha_{2,1} \\ \alpha_1 \end{matrix} ; 1-z \right)\,.
\label{eq:hypconstraintsolution}
\end{equation}
Plugging the latter into \eqref{eq:hypergeometric_constraint} determines $\mathcal{M}_{\theta \theta'}$ to be
\begin{equation}
    \mathcal{M}_{\theta \theta'}(b\alpha_0,b\alpha_1;b\alpha_\infty) = \frac{\Gamma(-2\theta'b\alpha_1)\Gamma(1+2\theta b \alpha_0)}{\Gamma\left(\frac{1}{2}+\theta b \alpha_0-\theta' b \alpha_1 + b \alpha_\infty\right) \Gamma\left(\frac{1}{2}+\theta b \alpha_0-\theta' b\alpha_1 - b \alpha_\infty \right)}\,,
\label{eq:hypconnectioncoeff}
\end{equation}
which is indeed the connection matrix for hypergeometric functions.
Diagrammatically, we can express the connection formula as
\begin{equation}\label{diag:hypergeometric}
\begin{tikzpicture}[baseline={(current bounding box.center)}, node distance=1cm and 1.5cm]
\coordinate[label=above:$\alpha_1$] (e1);
\coordinate[below=of e1] (aux1);
\coordinate[left=of aux1,label=left:$\alpha_\infty$] (e2);
\coordinate[right=1.5cm of aux1] (aux2);
\coordinate[above=of aux2,label=above:$\alpha_{2,1}$] (e3);
\coordinate[right=of aux2,label=right:$\alpha_0$] (e4);

\draw[line] (e1) -- (aux1);
\draw[line] (aux1) -- (e2);
\draw[line,dashed,red] (aux2) -- (e3);
\draw[line] (aux2) -- (e4);
\draw[line] (aux1) -- node[label=below:$\alpha_{0\theta}$] {} (aux2);
\end{tikzpicture} = \sum_{\theta'=\pm}\mathcal{M}_{\theta \theta'}
\begin{tikzpicture}[baseline={(current bounding box.center)}, node distance=1cm and 1.5cm]
\coordinate[label=above:$\alpha_0$] (e1);
\coordinate[below=of e1] (aux1);
\coordinate[left=of aux1,label=left:$\alpha_\infty$] (e2);
\coordinate[right=1.5cm of aux1] (aux2);
\coordinate[above=of aux2,label=above:$\alpha_{2,1}$] (e3);
\coordinate[right=of aux2,label=right:$\alpha_1$] (e4);

\draw[line] (e1) -- (aux1);
\draw[line] (aux1) -- (e2);
\draw[line,dashed,red] (aux2) -- (e3);
\draw[line] (aux2) -- (e4);
\draw[line] (aux1) -- node[label=below:$\alpha_{1\theta'}$] {} (aux2);
\end{tikzpicture}\,.
\end{equation}

\subsection{Whittaker functions}\label{warmup:whittaker}
Colliding the singularities at 1 and $\infty$ of the hypergeometric functions we obtained the Whittaker functions, which are simply related to the confluent hypergeometric function. They have a regular singularity at $0$ and an irregular singularity of rank 1 at $\infty$. To describe the confluence of two regular singularities in CFT we introduce the rank 1 irregular state, denoted by $\langle \mu,\Lambda|$. It lives in a Whittaker module and it is defined by the following properties
\begin{equation}
    \begin{aligned}
    &\langle \mu,\Lambda|L_0  = \Lambda \partial_{\Lambda} \langle \mu,\Lambda| \\
    &\langle \mu,\Lambda|L_{-1}  = \mu \Lambda \langle \mu,\Lambda|\\
    &\langle \mu,\Lambda|L_{-2}  = -\frac{\Lambda^2}{4}\langle \mu,\Lambda|\\
    &\langle \mu,\Lambda|L_{-n}  = 0 \, , \quad n>2 \,.
    \end{aligned}
\end{equation}
Note that the action of $L_0$ is not diagonal, and hence $\langle \mu, \Lambda|$ makes no reference to any Verma module. Equivalently, one can describe this state by a confluence limit of primary operators:
\begin{equation}
    \langle \mu,\Lambda| \propto \lim_{\eta\to\infty} t^{\Delta_t-\Delta} \langle \Delta|V_t(t)\, 
\end{equation}
with\footnote{Note that this procedure mimics the decoupling of a mass in the AGT-dual gauge theory.}
\begin{equation}\label{eq:collision}
    \Delta = \frac{Q^2}{4}-\left(\frac{\mu+\eta}{2}\right)^2\,,\quad \Delta_t = \frac{Q^2}{4}-\left(\frac{\mu-\eta}{2}\right)^2\,,\quad t=\frac{\eta}{\Lambda}\,.
\end{equation}
We fix the normalization of the irregular state by giving its overlap with a primary state, namely
\begin{equation}\label{irrNormalization}
    \langle \mu,\Lambda| \Delta \rangle = |\Lambda|^{2\Delta} C_{\mu \alpha}\,,
\end{equation}
with 
\begin{equation}
    C_{\mu \alpha} = \frac{e^{-i\pi\Delta}\Upsilon_b(Q+2\alpha)}{\Upsilon_b\left(\frac{Q}{2}+\mu + \alpha\right)\Upsilon_b\left(\frac{Q}{2}+\mu-\alpha\right)}\,.
\end{equation}
The $\Lambda$-dependence is fixed by the $L_0$-action, and $C_{\mu \alpha}$ is a normalization function that only depends on $\mu$ and $\alpha$, and is calculated in Appendices \ref{app:rank1collision}, \ref{app:rank1}. The notation reflects the fact that $C$ can be interpreted as a collided three-point function \cite{Gaiotto:2012sf}. The correlator 
\begin{equation}
    \langle \mu,\Lambda|\Phi(z)|\Delta\rangle\,
\end{equation}
satisfies the BPZ equation
\begin{equation}
    \left( b^{-2}\partial_z^2 - \frac{1}{z}\partial_z +\frac{\Delta}{z^2} +\frac{\mu\Lambda}{z}-\frac{\Lambda^2}{4} \right)\langle \mu,\Lambda|\Phi(z)|\Delta\rangle=0\,,
\end{equation}
that has a rank 1 irregular singularity at $z=\infty$ and a regular singularity at $z = 0$. Correspondingly, we expect this correlator to be given in terms of confluent hypergeometric functions. Indeed, for $z\sim 0$ one finds by solving the differential equation that the corresponding \textit{confluent} (or \textit{irregular}) conformal block is given by a Whittaker function. In particular, the two solutions are $z^{\frac{b^2}{2}} M_{b\mu,\pm b\alpha}(b\Lambda z)$, where the Whittaker $M$-function has a simple expansion around $z\sim 0$:
\begin{equation}
    M_{b\mu,b\alpha}(b\Lambda z) = (b\Lambda z)^{\frac{1}{2}+b\alpha}\left(1+\mathcal{O}(b\Lambda z)\right)\,.
\end{equation}
We can compute the confluent conformal block as
\begin{equation}
    {}_1\mathfrak{F} \left( \mu \, \alpha_\theta\, \begin{matrix} \alpha_{2,1}\\ \alpha \end{matrix} ;\Lambda z\right) = \Lambda^{\Delta_\theta} (b\Lambda)^{-\frac{1}{2}-\theta b\alpha} z^{\frac{b^2}{2}} M_{b\mu,\theta b\alpha}(b\Lambda z) \,.
\end{equation}
by expanding the OPE between $\Phi(z)$ and $|\Delta\rangle$ and projecting on $\langle \mu,\Lambda|$. Comparing this with the expansion of $M$ one obtains the prefactors written above. Here the subscript $1$ indicates the presence of a rank 1 irregular singularity at infinity. We represent this block diagramatically by
\begin{equation}
    {}_1\mathfrak{F} \left( \mu \, \alpha_\theta\, \begin{matrix} \alpha_{2,1}\\ \alpha \end{matrix} ;\Lambda z\right) =
    \begin{tikzpicture}[baseline={(current bounding box.center)}, node distance=1cm and 1.5cm]
    \coordinate[circle,fill,inner sep=2pt] (aux1);
    \coordinate[left=of aux1,label=left:$\mu$] (e2);
    \coordinate[right=1.5cm of aux1] (aux2);
    \coordinate[above=of aux2,label=above:$\alpha_{2,1}$] (e3);
    \coordinate[right=of aux2,label=right:$\alpha$] (e4);

    \draw[line,double] (aux1) -- (e2);
    \draw[line,dashed,red] (aux2) -- (e3);
    \draw[line] (aux2) -- (e4);
    \draw[line] (aux1) -- node[label=below:$\alpha_{\theta}$] {} (aux2);
    \end{tikzpicture} \,.
\end{equation}
The double line denotes the rank 1 irregular state, and the fat dot the projection onto a primary state. For $z \sim \infty$ we get an intrinsically different kind of confluent conformal block since we are now expanding $z$ near an {\it irregular singularity} of rank 1, dubbed in \cite{Lisovyy:2018mnj} confluent conformal block \textit{of 2nd kind}. We denote such a conformal block by the letter $\mathfrak{D}$ and find
\begin{equation}
\begin{aligned}
     &{}_1 \mathfrak{D}\left(\mu\,\begin{matrix}\alpha_{2,1}\\ {}\end{matrix}\,\mu_+\, \alpha; \frac{1}{\Lambda z}\right)= \Lambda^{\Delta+\Delta_{2,1}}e^{-i\pi b\mu} b^{b\mu}(\Lambda z)^{\frac{b^2}{2}} W_{-b\mu,b\alpha}(e^{-i \pi} b\Lambda z)\,,\\
     &{}_1 \mathfrak{D}\left(\mu\,\begin{matrix}\alpha_{2,1}\\ {}\end{matrix}\,\mu_-\, \alpha; \frac{1}{\Lambda z}\right)= \Lambda^{\Delta+\Delta_{2,1}} b^{-b\mu}(\Lambda z)^{\frac{b^2}{2}} W_{b\mu,b\alpha}(b\Lambda z)\,,
\end{aligned}
\end{equation}
where $W$ is the Whittaker function with a simple asymptotic expansion around $z\sim \infty$. This block is obtained by doing the OPE between the irregular state and the degenerate field, which is derived in Appendix \ref{app:rank1}, and then projecting on $|\Delta\rangle$. Once again, the prefactors are fixed  by comparing with the expansion of $W$. We represent this conformal block diagramatically by
\begin{equation}
    {}_1 \mathfrak{D}\left(\mu\,\begin{matrix}\alpha_{2,1}\\ {}\end{matrix}\, \mu_\theta\, \alpha; \frac{1}{\Lambda z}\right) =
    \begin{tikzpicture}[baseline={(current bounding box.center)}, node distance=1cm and 1.5cm]
    \coordinate (aux1);
    \coordinate[left=of aux1,label=left:$\mu$] (e2);
    \coordinate[right=1.5cm of aux1,circle,fill,inner sep=2pt] (aux2);
    \coordinate[above=of aux1,label=above:$\alpha_{2,1}$] (e3);
    \coordinate[right=of aux2,label=right:$\alpha$] (e4);

    \draw[line,double] (e2) -- (aux1);
    \draw[line,dashed,red] (aux1) -- (e3);
    \draw[line] (aux2) -- (e4);
    \draw[line,double] (aux1) -- node[label=below:$\mu_{\theta}$] {} (aux2);
    \end{tikzpicture} \,.
\end{equation}
Crossing symmetry now implies
\begin{equation}\label{eq:Whittaker_constraint}
    \begin{aligned}
        \langle \mu,\Lambda| \Phi(z) |\Delta \rangle = & \sum_{\theta=\pm} C_{\alpha_{2,1},\alpha}^{\alpha_\theta}  C_{\mu \alpha_\theta} \left| {}_1\mathfrak{F} \left( \mu \, \alpha_\theta\, \begin{matrix} \alpha_{2,1}\\ \alpha \end{matrix} ;\Lambda z\right) \right|^2 =\sum_{\theta' = \pm} B_{\alpha_{2,1},\mu}^{\mu_{\theta'}}  C_{\mu_{\theta'} \alpha} \left|{}_1 \mathfrak{D}\left(\mu\,\begin{matrix}\alpha_{2,1}\\ {}\end{matrix}\, \mu_{\theta'}\, \alpha; \frac{1}{\Lambda z}\right) \right|^2\,.
    \end{aligned}
\end{equation}
Here $B$ is the irregular OPE coefficient arising from the OPE between the irregular state and the degenerate field. We calculate it in Appendices \ref{app:rank1collision},  \ref{app:rank1}, and it is given by
\begin{equation}
    B_{\alpha_{2,1},\mu}^{\mu_\pm}= e^{ i\pi\left(\frac{1}{2}\pm b\mu + \frac{ b^2}{4}\right)}\,.
\end{equation}
As for the hypergeometric function, we can make an Ansatz for the connection formula for these irregular conformal blocks of the form
\begin{equation}
    b^{\theta b \alpha}{}_1\mathfrak{F} \left( \mu \, \alpha_\theta\, \begin{matrix} \alpha_{2,1}\\ \alpha \end{matrix} ;\Lambda z\right) = \sum_{\theta'=\pm} b^{-\frac{1}{2} - \theta' b\mu}\mathcal{N}_{\theta \theta'}(b\alpha,b\mu)\,\, {}_1\mathfrak{D}\left(\mu\,\begin{matrix}\alpha_{2,1}\\ {}\end{matrix}\, \mu_{\theta'}\, \alpha; \frac{1}{\Lambda z}\right) \,.
\end{equation}
The constraints coming from crossing symmetry \eqref{eq:Whittaker_constraint} are solved by the irregular connection coefficients
\begin{equation}\label{eq:WhittakerCC}
    \mathcal{N}_{\theta \theta'}(b\alpha,b\mu) = \frac{\Gamma(1+2\theta b \alpha)}{\Gamma\left(\frac{1}{2}+\theta b \alpha- \theta' b\mu\right)} e^{i\pi \left(\frac{1-\theta'}{2}\right)\left(\frac{1}{2}-b\mu+\theta b\alpha\right)}\,.
\end{equation}
These are just the connection coefficients for Whittaker functions. In fact, in Appendix \ref{app:rank1} we argue the other way around, namely we determine the normalization function $C_{\mu\alpha}$ and the irregular OPE coefficient $B_{\alpha_{2,1},\mu}^{\mu_\pm}$ by using the known connection coefficients $\mathcal{N}_{\theta \theta'}$ for Whittaker functions. This shows the consistency of our approach. Let us emphasize for latter purposes that the functions $\mathcal{N}_{\theta \theta'}$ solve the constraint \eqref{eq:Whittaker_constraint}, which will appear later in a different context. We represent this connection formula diagrammatically by
\begin{equation}
    \begin{tikzpicture}[baseline={(current bounding box.center)}, node distance=1cm and 1.5cm]
    \coordinate[circle,fill,inner sep=2pt] (aux1);
    \coordinate[left=of aux1,label=left:$\mu$] (e2);
    \coordinate[right=1.5cm of aux1] (aux2);
    \coordinate[above=of aux2,label=above:$\alpha_{2,1}$] (e3);
    \coordinate[right=of aux2,label=right:$\alpha$] (e4);

    \draw[line,double] (aux1) -- (e2);
    \draw[line,dashed,red] (aux2) -- (e3);
    \draw[line] (aux2) -- (e4);
    \draw[line] (aux1) -- node[label=below:$\alpha_{\theta}$] {} (aux2);
    \end{tikzpicture} = \sum_{\theta'=\pm} \mathcal{N}_{\theta \theta'}
    \begin{tikzpicture}[baseline={(current bounding box.center)}, node distance=1cm and 1.5cm]
    \coordinate (aux1);
    \coordinate[left=of aux1,label=left:$\mu$] (e2);
    \coordinate[right=1.5cm of aux1,circle,fill,inner sep=2pt] (aux2);
    \coordinate[above=of aux1,label=above:$\alpha_{2,1}$] (e3);
    \coordinate[right=of aux2,label=right:$\alpha$] (e4);

    \draw[line,double] (e2) -- (aux1);
    \draw[line,dashed,red] (aux1) -- (e3);
    \draw[line] (aux2) -- (e4);
    \draw[line,double] (aux1) -- node[label=below:$\mu_{\theta'}$] {} (aux2);
    \end{tikzpicture} \,.
\end{equation}

\subsection{Bessel functions}\label{warmup:Bessel}
There is a natural limiting procedure which reduces a rank 1 irregular singularity to a rank 1/2 one. 
To describe the latter in CFT, let us introduce the rank 1/2 irregular state $\langle \Lambda^2|$ via defining properties
\begin{equation}
\begin{aligned}
    &\langle \Lambda^2|L_0 = \Lambda^2 \partial_{\Lambda^2} \langle \Lambda^2|\\
    &\langle \Lambda^2|L_{-1}= -\frac{\Lambda^2}{4}\langle \Lambda^2| \\
    &\langle \Lambda^2|L_{-n}=0 \,, \quad n>1\,.
\end{aligned}
\end{equation}
It can be obtained from the rank 1 irregular state via the limit\footnote{Note that this limit corresponds to the well known holomorphic decoupling limit of a massive hypermultiplet in the AGT dual gauge theory.}
\begin{equation}\label{rankhalfdecoupling}
    \langle \Lambda^2| = \lim_{\mu\to\infty} \langle \mu, -\frac{\Lambda^2}{4\mu}|\,.
\end{equation}
We see that reducing a rank 1 to a rank 1/2 singularity corresponds to further decoupling a mass in the AGT dual gauge theory. We normalize the rank 1/2 state as
\begin{equation}
    \langle\Lambda^2|\Delta\rangle = |\Lambda^2|^{2\Delta} C_\alpha \,, \quad C_\alpha = 2^{-4\Delta} e^{-2\pi i \Delta}\Upsilon_b(Q+2\alpha)\,.
\end{equation}
This normalization function is calculated in Appendices \ref{app:rankhalfcollision}, \ref{app:rankhalf}. Consider the following correlation function involving the rank 1/2 state:
\begin{equation}
    \langle \Lambda^2| \Phi(z) |\Delta \rangle \,.
\end{equation}
which correspondingly displays a rank 1/2 singularity at infinity. This is reflected in the BPZ equation
\begin{equation}
\left( b^{-2} \partial_z^2 - \frac{1}{z} \partial_z + \frac{\Delta}{z^2} - \frac{\Lambda^2}{4 z} \right) \langle \Lambda^2| \Phi(z) |\Delta \rangle = 0 \,.
\end{equation}
Solving this differential equation one finds that the corresponding \textit{rank 1/2 irregular} conformal block is given by a modified Bessel function $I_\nu (x)$ as
\begin{equation}
    {}_{\frac{1}{2}}\mathfrak{F} \left( \alpha_\theta \, \alpha_{2,1}\,\alpha; \, \Lambda \sqrt{z} \right) = \Gamma(1+2\theta b\alpha) \Lambda^{2\Delta_\theta} \left(\frac{b\Lambda}{2}\right)^{- 2\theta b\alpha} z^{\frac{bQ}{2}} I_{ 2\theta b \alpha}(b\Lambda \sqrt{z})\,.
\end{equation}
Here the subscript $\frac{1}{2}$ indicates the presence of a rank 1/2 singularity at infinity. This conformal block is obtained by doing the OPE between $\Phi$ and $|\Delta\rangle$ and then projecting the result on $\langle \Lambda^2|$. The prefactors are fixed by comparing this with the following expansion of the Bessel function
\begin{equation}
    I_{2\theta b\alpha}(b\Lambda\sqrt{z}) = \frac{(b\Lambda\sqrt{z}/2)^{2\theta b\alpha}}{\Gamma(1+2\theta b\alpha)}\left(1+\mathcal{O}(b\Lambda\sqrt{z})\right)\,.
\end{equation}
We represent this conformal block diagramatically by
\begin{equation}
    {}_{\frac{1}{2}}\mathfrak{F} \left( \alpha_\theta \, \alpha_{2,1}\,\alpha; \, \Lambda \sqrt{z} \right) =
    \begin{tikzpicture}[baseline={(current bounding box.center)}, node distance=1cm and 1.5cm]
    \coordinate[circle,fill,inner sep=2pt] (aux1);
    \coordinate[left=of aux1] (e2);
    \coordinate[right=1.5cm of aux1] (aux2);
    \coordinate[above=of aux2,label=above:$\alpha_{2,1}$] (e3);
    \coordinate[right=of aux2,label=right:$\alpha$] (e4);

    \draw[line,decoration=snake] (aux1) -- (e2);
    \draw[line,dashed,red] (aux2) -- (e3);
    \draw[line] (aux2) -- (e4);
    \draw[line] (aux1) -- node[label=below:$\alpha_{\theta}$] {} (aux2);
    \end{tikzpicture} \,.
\end{equation}
Here the wiggly line denotes the rank 1/2 irregular state, and the fat dot represents the pairing with a primary state. For $z\sim\infty$ we get a different kind of irregular conformal block, since we are now expanding for $z$ near an irregular singularity of rank $1/2$. We denote such a conformal block by the letter $\mathfrak{E}$  
\begin{equation}
\begin{aligned}
    &{}_{\frac{1}{2}}\mathfrak{E}^{(+)}\left(\alpha_{2,1}\, \alpha; \frac{1}{\Lambda \sqrt{z}} \right) = \sqrt{\frac{2b}{\pi}}e^{-\frac{i\pi}{2}} (\Lambda^2)^{\Delta-\frac{b^2}{4}}z^{\frac{bQ}{2}} K_{ 2 b \alpha}(e^{-i\pi} b\Lambda \sqrt{z})\,,\\
    &{}_{\frac{1}{2}}\mathfrak{E}^{(-)}\left(\alpha_{2,1}\, \alpha; \frac{1}{\Lambda \sqrt{z}} \right) = \sqrt{\frac{2b}{\pi}} (\Lambda^2)^{\Delta-\frac{b^2}{4}}z^{\frac{bQ}{2}} K_{ 2 b \alpha}( b\Lambda \sqrt{z})\,,
\end{aligned}
\end{equation}
where $K$ is the modified Bessel function of the second kind, which has a nice asymptotic expansion for $z\sim \infty$. This block is obtained from the OPE between the irregular rank 1/2 state and the degenerate field which we derived in Appendix \ref{app:rankhalf}, and then by taking the scalar product with $|\Delta\rangle$. We represent this block diagramatically by
\begin{equation}
    {}_{\frac{1}{2}}\mathfrak{E}^{(\theta)}\left(\alpha_{2,1}\, \alpha; \frac{1}{\Lambda \sqrt{z}} \right)  =
    \begin{tikzpicture}[baseline={(current bounding box.center)}, node distance=1cm and 1.5cm]
    \coordinate (aux1);
    \coordinate[left=of aux1] (e2);
    \coordinate[right=1.5cm of aux1,circle,fill,inner sep=2pt] (aux2);
    \coordinate[above=of aux1,label=above:$\alpha_{2,1}$] (e3);
    \coordinate[right=of aux2,label=right:$\alpha$] (e4);

    \draw[line,decoration=snake] (e2) -- (aux1);
    \draw[line,dashed,red] (aux1) -- (e3);
    \draw[line] (aux2) -- (e4);
    \draw[line,decoration=snake] (aux1) -- node[label=below:$\theta$] {} (aux2);
    \end{tikzpicture} \,.
\end{equation}
Crossing symmetry implies that
\begin{equation}\label{eq:Bessel_constraint}
    \langle \Lambda^2|\Phi(z)|\Delta\rangle = \sum_{\theta=\pm} C_{\alpha_{2,1},\alpha}^{\alpha_\theta}  C_{\alpha_\theta} \left|{}_{\frac{1}{2}}\mathfrak{F} \left( \alpha_\theta \, \alpha_{2,1}\,\alpha; \, \Lambda \sqrt{z} \right)\right|^2 =\sum_{\theta' = \pm} B_{\alpha_{2,1}}  C_{ \alpha} \left|{}_{\frac{1}{2}}\mathfrak{E}^{(\theta')}\left(\alpha_{2,1}\, \alpha; \frac{1}{\Lambda \sqrt{z}} \right) \right|^2\,. 
\end{equation}
Here $B_{\alpha_{2,1}}$ is the irregular OPE coefficient arising from the OPE between the irregular rank 1/2 state and the degenerate field:
\begin{equation}
\begin{aligned}
    &B_{\alpha_{2,1}} = 2^{b^2} e^{\frac{i\pi bQ}{2}}\,.
\end{aligned}
\end{equation}
These functions are derived in Appendix \ref{app:rankhalf}. We can now make an Ansatz for the connection formula for these irregular conformal blocks:
\begin{equation}
    b^{2\theta b \alpha} {}_{\frac{1}{2}}\mathfrak{F} \left( \alpha_\theta \, \alpha_{2,1}\,\alpha; \, \Lambda \sqrt{z} \right) = \sum_{\theta'=\pm}b^{-1/2} \mathcal{Q}_{\theta \theta'}(b\alpha) \, \, {}_{\frac{1}{2}}\mathfrak{E}^{(\theta')}\left(\alpha_{2,1}\, \alpha; \frac{1}{\Lambda \sqrt{z}} \right)\,.
\end{equation}
The crossing symmetry condition \eqref{eq:Bessel_constraint} gives constraints on the irregular connection coefficients, which are solved by
\begin{equation}
    \mathcal{Q}_{\theta \theta'}(b\alpha) = \frac{2^{2\theta b \alpha}}{\sqrt{2\pi }} \Gamma(1+2\theta b \alpha) e^{i\pi\left(\frac{1-\theta'}{2}\right)\left(\frac{1}{2}+2\theta b \alpha \right)}\,.
\end{equation}
These are of course nothing else than the connection coefficients for Bessel functions, including the relevant prefactors. Similar constraints of the form \eqref{eq:Bessel_constraint} will reappear later. We represent the connection formula by
\begin{equation}
    \begin{tikzpicture}[baseline={(current bounding box.center)}, node distance=1cm and 1.5cm]
    \coordinate[circle,fill,inner sep=2pt] (aux1);
    \coordinate[left=of aux1] (e2);
    \coordinate[right=1.5cm of aux1] (aux2);
    \coordinate[above=of aux2,label=above:$\alpha_{2,1}$] (e3);
    \coordinate[right=of aux2,label=right:$\alpha$] (e4);

    \draw[line,decoration=snake] (aux1) -- (e2);
    \draw[line,dashed,red] (aux2) -- (e3);
    \draw[line] (aux2) -- (e4);
    \draw[line] (aux1) -- node[label=below:$\alpha_{\theta}$] {} (aux2);
    \end{tikzpicture}  = \sum_{\theta'=\pm} \mathcal{Q}_{\theta \theta'}
    \begin{tikzpicture}[baseline={(current bounding box.center)}, node distance=1cm and 1.5cm]
    \coordinate (aux1);
    \coordinate[left=of aux1] (e2);
    \coordinate[right=1.5cm of aux1,circle,fill,inner sep=2pt] (aux2);
    \coordinate[above=of aux1,label=above:$\alpha_{2,1}$] (e3);
    \coordinate[right=of aux2,label=right:$\alpha$] (e4);

    \draw[line,decoration=snake] (e2) -- (aux1);
    \draw[line,dashed,red] (aux1) -- (e3);
    \draw[line] (aux2) -- (e4);
    \draw[line,decoration=snake] (aux1) -- node[label=below:$\theta'$] {} (aux2);
    \end{tikzpicture} \,.
\end{equation}

\section{5-point degenerate conformal blocks, confluences and connection formulae}

In this section we consider the relevant CFT correlators obeying the BPZ equations which reduce to Heun equations in the appropriate classical limit. 
Notice that for more than three 
vertix insertions BPZ equations on the sphere
are richer than 
the corresponding ODE due to
the presence of the corresponding moduli. This implies that a suitable classical limit (NS limit), engineered to decouple the moduli dynamics, is needed to recover the corresponding ODE.

We derive explicit connection formulae for the relevant conformal blocks by making use of crossing symmetry of the CFT correlators.
In the classical limit, these generate explicit solutions of the connection problem for the Heun equations.

\subsection{Regular conformal blocks}\label{regCB}
\subsubsection{General case}
The five-point function with one degenerate insertion in Liouville CFT satisfies the BPZ equation 
\begin{equation}
\begin{aligned}
    \bigg( b^{-2}\partial_z^2 + \frac{\Delta_1}{(z-1)^2} - \frac{\Delta_1 + t\partial_t + \Delta_t + z\partial_z + \Delta_{2,1} + \Delta_0 -\Delta_\infty}{z(z-1)}+ \frac{\Delta_t}{(z-t)^2}+\frac{t}{z(z-t)}\partial_t - \frac{1}{z}\partial_z+\frac{\Delta_0}{z^2} \bigg) \langle \Delta_\infty| V_1(1) V_t(t) \Phi(z) | \Delta_0 \rangle = 0\,.
\end{aligned}
\label{eq:BPZNf=4}
\end{equation}
The five-point function can be expanded in the region $z \ll t \ll 1$ as follows
\begin{equation}
\begin{aligned}
    \langle \Delta_\infty | V_1 (1) V_t (t) \Phi (z) | \Delta_0 \rangle =  \sum_{\theta = \pm} \int d \alpha \, C_{\alpha_{2,1} \alpha_0}^{\alpha_{0 \theta}} C_{\alpha_t  \alpha_{0 \theta}}^{\alpha} C_{\alpha_\infty \alpha_1 \alpha} \FIV{\alpha_\infty}{\alpha_1}{\alpha}{\alpha_t}{{\alpha_{0 \theta}}}{\alpha_{2,1}}{\alpha_0}{t}{\frac{z}{t}} \FIV{\alpha_\infty}{\alpha_1}{\alpha}{\alpha_t}{{\alpha_{0 \theta}}}{\alpha_{2,1}}{\alpha_0}{\bar{t}}{\frac{\bar{z}}{\bar{t}}}\,.
\end{aligned}
\label{eq:ssco1}
\end{equation}
As usual the conformal blocks can be computed via OPEs. The result is naturally an expansion in the variables $t$ and $z/t$. Conformal blocks are usually denoted diagrammatically as
\begin{equation}
\begin{tikzpicture}[baseline={(current bounding box.center)}, node distance=1cm and 1.5cm]
\coordinate[label=above:$\alpha_1$] (e1);
\coordinate[below=of e1] (aux1);
\coordinate[left=of aux1,label=left:$\alpha_\infty$] (e2);
\coordinate[right=1.5cm of aux1] (aux2);
\coordinate[above=of aux2,label=above:$\alpha_t$] (e3);
\coordinate[right=1.5cm of aux2] (aux3);
\coordinate[above=of aux3,label=above:$\alpha_{2,1}$] (e4);
\coordinate[right=of aux3,label=right:$\alpha_0$] (e5);

\draw[line] (e1) -- (aux1);
\draw[line] (aux1) -- (e2);
\draw[line] (e3) -- (aux2);
\draw[line,dashed,red] (aux3) -- (e4);
\draw[line] (aux3) -- (e5);
\draw[line] (aux1) -- node[label=below:$\alpha$] {} (aux2);
\draw[line] (aux2) -- node[label=below:$\alpha_{0\theta}$] {} (aux3);
\end{tikzpicture} = \, \FIV{\alpha_\infty}{\alpha_1}{\alpha}{\alpha_t}{{\alpha_{0 \theta}}}{\alpha_{2,1}}{\alpha_0}{t}{\frac{z}{t}} \,.
\label{eq:sscb1}
\end{equation}
An explicit combinatorial formula for this conformal block is given in Appendix \ref{app:NekQuiver}. The same correlator can be expanded for $z \sim t$ and small $t$ after the M\"obius transformation $x \to \frac{x-t}{1-t}$, yielding
\begin{equation}
\begin{aligned}
    &\langle \Delta_\infty | V_1 (1) V_t (t) \Phi (z) | \Delta_0 \rangle = \left|(1-t)^{\Delta_\infty - \Delta_1 - \Delta_t - \Delta_{2,1} - \Delta_0}\right|^2 \langle \Delta_\infty | V_1 (1) V_0 \left( \frac{t}{t-1} \right) \Phi \left(\frac{z-t}{1-t} \right) | \Delta_t \rangle = \\ &= \sum_{\theta = \pm} \int d \alpha \, C_{\alpha_{2,1}  \alpha_t}^{\alpha_{t \theta}} C_{\alpha_0 \alpha_{t \theta}}^{\alpha} C _{\alpha_\infty \alpha_1 \alpha} \left|(1-t)^{\Delta_\infty - \Delta_1 - \Delta_t - \Delta_{2,1} - \Delta_0} \FIV{\alpha_\infty}{\alpha_1}{\alpha}{\alpha_0}{{\alpha_{t \theta}}}{\alpha_{2,1}}{\alpha_t}{\frac{t}{t-1}}{\frac{t-z}{t}} \right|^2 \,.
\end{aligned}
\label{eq:stco1}
\end{equation}
Diagramatically, this conformal block is
\begin{equation}
\begin{tikzpicture}[baseline={(current bounding box.center)}, node distance=1cm and 1.5cm]
\coordinate[label=above:$\alpha_1$] (e1);
\coordinate[below=of e1] (aux1);
\coordinate[left=of aux1,label=left:$\alpha_\infty$] (e2);
\coordinate[right=1.5cm of aux1] (aux2);
\coordinate[above=of aux2,label=above:$\alpha_0$] (e3);
\coordinate[right=1.5cm of aux2] (aux3);
\coordinate[above=of aux3,label=above:$\alpha_{2,1}$] (e4);
\coordinate[right=of aux3,label=right:$\alpha_t$] (e5);

\draw[line] (e1) -- (aux1);
\draw[line] (aux1) -- (e2);
\draw[line] (e3) -- (aux2);
\draw[line,dashed,red] (aux3) -- (e4);
\draw[line] (aux3) -- (e5);
\draw[line] (aux1) -- node[label=below:$\alpha$] {} (aux2);
\draw[line] (aux2) -- node[label=below:$\alpha_{t\theta}$] {} (aux3);
\end{tikzpicture} = \, \FIV{\alpha_\infty}{\alpha_1}{\alpha}{\alpha_0}{{\alpha_{t \theta}}}{\alpha_{2,1}}{\alpha_t}{\frac{t}{t-1}}{\frac{t-z}{t}} \,.
\end{equation}
We notice that the diagrams
just represent the order in which the OPEs are performed, neglecting factors such as Jacobians that arise from the M\"obius transformations.
By crossing symmetry the two expansions should agree, so that
\begin{equation}
\begin{aligned}
    &\sum_{\theta = \pm} \int d \alpha \, C_{\alpha_{2,1} \alpha_0}^{\alpha_{0 \theta}} C_{\alpha_t  \alpha_{0 \theta}}^{\alpha} C_{\alpha_\infty \alpha_1 \alpha} \left| \FIV{\alpha_\infty}{\alpha_1}{\alpha}{\alpha_t}{{\alpha_{0 \theta}}}{\alpha_{2,1}}{\alpha_0}{t}{\frac{z}{t}} \right|^2 = \\ &= \sum_{\theta = \pm} \int d \alpha \, C_{\alpha_{2,1}  \alpha_t}^{\alpha_{t \theta}} C_{\alpha_0 \alpha_{t \theta}}^{\alpha} C _{\alpha_\infty \alpha_1 \alpha} \left|(1-t)^{\Delta_\infty - \Delta_1 - \Delta_t - \Delta_{2,1} - \Delta_0} \FIV{\alpha_\infty}{\alpha_1}{\alpha}{\alpha_0}{{\alpha_{t \theta}}}{\alpha_{2,1}}{\alpha_t}{\frac{t}{t-1}}{\frac{t-z}{t}} \right|^2 \,.
\end{aligned}
\end{equation}
which can be conveniently recast as
\begin{equation}\label{eq:46}
\begin{aligned}
    & \int d \alpha \, C_{\alpha_\infty \alpha_1 \alpha} \sum_{\theta = \pm} \bigg( C_{\alpha_{2,1} \alpha_0}^{\alpha_{0 \theta}} C_{\alpha_t  \alpha_{0 \theta}}^{\alpha}  \left| \FIV{\alpha_\infty}{\alpha_1}{\alpha}{\alpha_t}{{\alpha_{0 \theta}}}{\alpha_{2,1}}{\alpha_0}{t}{\frac{z}{t}} \right|^2 + \\ &- C_{\alpha_{2,1}  \alpha_t}^{\alpha_{t \theta}} C_{\alpha_0 \alpha_{t \theta}}^{\alpha} \left|(1-t)^{\Delta_\infty - \Delta_1 - \Delta_t - \Delta_{2,1} - \Delta_0} \FIV{\alpha_\infty}{\alpha_1}{\alpha}{\alpha_0}{{\alpha_{t \theta}}}{\alpha_{2,1}}{\alpha_t}{\frac{t}{t-1}}{\frac{t-z}{t}} \right|^2 \bigg) = 0 \,.
\end{aligned}
\end{equation}
By imposing the vanishing of the integrand we get a constraint analogous to \eqref{eq:hypergeometric_constraint}, which analogously to \eqref{eq:hypconstraintsolution} we solve as\footnote{The phase appearing in the RHS of equation \eqref{eq:connection0t} is fixed imposing that the overall leading powers of 
\[(1-t)^{\Delta_\infty - \Delta_1 - \Delta_t - \Delta_{2,1} - \Delta_0} \FIV{\alpha_\infty}{\alpha_1}{\alpha}{\alpha_0}{{\alpha_{t \theta}}}{\alpha_{2,1}}{\alpha_t}{\frac{t}{t-1}}{\frac{t-z}{t}} \sim e^{- i \pi (\Delta - \Delta_t - \Delta_{2,1} - \Delta_0)} t^{\Delta - \Delta_0 - \Delta_{t \theta}} (t-z)^{\Delta_{t \theta} - \Delta_{2,1} - \Delta_{2,1}} \left( 1 + \dots \right) \]
agree with the leading powers of the OPEs of the full correlator, where no explicit phase appears.}
\begin{equation}
\begin{aligned}
    & \FIV{\alpha_\infty}{\alpha_1}{\alpha}{\alpha_t}{{\alpha_{0 \theta}}}{\alpha_{2,1}}{\alpha_0}{t}{\frac{z}{t}} = \\
    = & \sum_{\theta'=\pm}\mathcal{M}_{\theta \theta'}(b\alpha_0,b\alpha_t;b\alpha) e^{i \pi (\Delta - \Delta_0 - \Delta_{2,1} - \Delta_t)}(1-t)^{\Delta_\infty - \Delta_1 - \Delta_t - \Delta_{2,1} - \Delta_0} \FIV{\alpha_\infty}{\alpha_1}{\alpha}{\alpha_0}{{\alpha_{t \theta}}}{\alpha_{2,1}}{\alpha_t}{\frac{t}{t-1}}{\frac{t-z}{t}} \,,
\end{aligned}
\label{eq:connection0t}
\end{equation}
where $\mathcal{M}_{\theta \theta'}$ are the hypergeometric connection coefficients defined in \eqref{eq:hypconnectioncoeff}.
Note indeed that in \eqref{eq:connection0t} the 
functional form of the connection coefficients
depends on the local properties of the 
conformal block in the vicinity of the 
degenerate vertex insertion as can be seen form the factorized form of \eqref{eq:46}.
Diagrammatically, the connection formula \eqref{eq:connection0t} reads
\begin{equation}
\begin{tikzpicture}[baseline={(current bounding box.center)}, node distance=1cm and 1.5cm]
\coordinate[label=above:$\alpha_1$] (e1);
\coordinate[below=of e1] (aux1);
\coordinate[left=of aux1,label=left:$\alpha_\infty$] (e2);
\coordinate[right=1.5cm of aux1] (aux2);
\coordinate[above=of aux2,label=above:$\alpha_t$] (e3);
\coordinate[right=1.5cm of aux2] (aux3);
\coordinate[above=of aux3,label=above:$\alpha_{2,1}$] (e4);
\coordinate[right=of aux3,label=right:$\alpha_0$] (e5);

\draw[line] (e1) -- (aux1);
\draw[line] (aux1) -- (e2);
\draw[line] (e3) -- (aux2);
\draw[line,dashed,red] (aux3) -- (e4);
\draw[line] (aux3) -- (e5);
\draw[line] (aux1) -- node[label=below:$\alpha$] {} (aux2);
\draw[line] (aux2) -- node[label=below:$\alpha_{0 \theta}$] {} (aux3);
\end{tikzpicture} =  
\sum_{\theta' = \pm} \mathcal{M}_{\theta \theta'} \begin{tikzpicture}[baseline={(current bounding box.center)}, node distance=1cm and 1.5cm]
\coordinate[label=above:$\alpha_1$] (e1);
\coordinate[below=of e1] (aux1);
\coordinate[left=of aux1,label=left:$\alpha_\infty$] (e2);
\coordinate[right=1.5cm of aux1] (aux2);
\coordinate[above=of aux2,label=above:$\alpha_0$] (e3);
\coordinate[right=1.5cm of aux2] (aux3);
\coordinate[above=of aux3,label=above:$\alpha_{2,1}$] (e4);
\coordinate[right=of aux3,label=right:$\alpha_t$] (e5);

\draw[line] (e1) -- (aux1);
\draw[line] (aux1) -- (e2);
\draw[line] (e3) -- (aux2);
\draw[line,dashed,red] (aux3) -- (e4);
\draw[line] (aux3) -- (e5);
\draw[line] (aux1) -- node[label=below:$\alpha$] {} (aux2);
\draw[line] (aux2) -- node[label=below:$\alpha_{t \theta'}$] {} (aux3);
\end{tikzpicture}
\,.
\label{eq:sscb15}
\end{equation}
Conformal blocks for small $z$ can also be connected to the expansion for $z \sim 1, z \sim \infty$ passing through the region $t \ll z \ll 1$. The conformal block in that region is 
\begin{equation}
\begin{tikzpicture}[baseline={(current bounding box.center)}, node distance=1cm and 1.5cm]
\coordinate[label=above:$\alpha_1$] (e1);
\coordinate[below=of e1] (aux1);
\coordinate[left=of aux1,label=left:$\alpha_\infty$] (e2);
\coordinate[right=1.5cm of aux1] (aux2);
\coordinate[above=of aux2,label=above:$\alpha_{2,1}$] (e3);
\coordinate[right=1.5cm of aux2] (aux3);
\coordinate[above=of aux3,label=above:$\alpha_t$] (e4);
\coordinate[right=of aux3,label=right:$\alpha_0$] (e5);

\draw[line] (e1) -- (aux1);
\draw[line] (aux1) -- (e2);
\draw[line,dashed,red] (e3) -- (aux2);
\draw[line] (aux3) -- (e4);
\draw[line] (aux3) -- (e5);
\draw[line] (aux1) -- node[label=below:$\alpha$] {} (aux2);
\draw[line] (aux2) -- node[label=below:$\alpha_{\theta}$] {} (aux3);
\end{tikzpicture} = \, \FIV{\alpha_\infty}{\alpha_1}{\alpha}{\alpha_{2,1}}{{\alpha_{ \theta}}}{\alpha_t}{\alpha_0}{z}{\frac{t}{z}} \,.
\end{equation}
Then, crossing symmetry relates this block to the expansion for $z \sim 0$ via
\begin{equation}\label{eq:50}
    \langle \Delta_\infty | V_1 (1) V_t (t) \Phi(z) | \Delta_0 \rangle = \langle \Delta_\infty | V_1 (1) \Phi(z) V_t (t) | \Delta_0 \rangle \,,
\end{equation}
therefore, by comparing \eqref{eq:50} with \eqref{eq:sscb1} we get
\begin{equation}
\begin{aligned}
    &\sum_{\theta = \pm} \int d \alpha \, C_{\alpha_{2,1} \alpha_0}^{\alpha_{0 \theta}} C_{\alpha_t  \alpha_{0 \theta}}^{\alpha} C_{\alpha_\infty \alpha_1 \alpha} \left| \FIV{\alpha_\infty}{\alpha_1}{\alpha}{\alpha_t}{{\alpha_{0 \theta}}}{\alpha_{2,1}}{\alpha_0}{t}{\frac{z}{t}} \right|^2 = \\ &= \sum_{\theta = \pm} \int d \alpha \, C_{\alpha_{2,1}  \alpha_\theta}^{\alpha} C_{\alpha_t \alpha_0}^{\alpha_\theta} C _{\alpha_\infty \alpha_1 \alpha} \left| \FIV{\alpha_\infty}{\alpha_1}{\alpha}{\alpha_{2,1}}{{\alpha_{ \theta}}}{\alpha_t}{\alpha_0}{z}{\frac{t}{z}} \right|^2 \,,
\end{aligned}
\end{equation}
and following the same argument as for the previous case we find 
\begin{equation}
     \FIV{\alpha_\infty}{\alpha_1}{\alpha}{\alpha_t}{{\alpha_{0 \theta}}}{\alpha_{2,1}}{\alpha_0}{t}{\frac{z}{t}} = \sum_{\theta'=\pm}\mathcal{M}_{\theta \theta'}(b\alpha_0,b\alpha;b\alpha_t) \FIV{\alpha_\infty}{\alpha_1}{\alpha}{\alpha_{2,1}}{\alpha_{\theta'}}{\alpha_t}{\alpha_0}{z}{\frac{t}{z}} \,.
\label{eq:connection0int}
\end{equation}
Now we can connect expansions in the intermediate region to expansions for $z \sim \infty$ again invoking crossing symmetry. Performing the transformation $x \to t / x$ on the LHS of \eqref{eq:50} we get
\begin{equation}
    \langle \Delta_\infty | V_1 (1) \Phi(z) V_t (t) | \Delta_0 \rangle =  \left| t^{\Delta_\infty + \Delta_1 + \Delta_{2,1} - \Delta_0 - \Delta_t} z^{- 2 \Delta_{2,1}} \right|^2 \langle \Delta_0 | V_t (1) V_1 (t) \Phi \left( \frac{t}{z} \right) | \Delta_\infty \rangle \,,
\end{equation}
that implies
\begin{equation}
\begin{aligned}
    &\sum_{\theta = \pm} \int d \alpha \, C_{\alpha_t \alpha_0 \alpha} C_{\alpha_{2,1}  \alpha_\theta}^{\alpha} C _{\alpha_\infty \alpha_1}^{\alpha_\theta} \left| \FIV{\alpha_\infty}{\alpha_1}{\alpha_\theta}{\alpha_{2,1}}{\alpha}{\alpha_t}{\alpha_0}{z}{\frac{t}{z}} \right|^2 = \\ &= \sum_{\theta = \pm} \int d \alpha \, C_{\alpha_t \alpha_0 \alpha} C_{\alpha_{2,1}  \alpha_\infty}^{\alpha_{\infty \theta}} C _{\alpha_{\infty \theta} \alpha_1}^{\alpha} \left| t^{\Delta_\infty + \Delta_1 + \Delta_{2,1} - \Delta_0 - \Delta_t} z^{- 2 \Delta_{2,1}} \FIV{\alpha_0}{\alpha_t}{\alpha}{\alpha_1}{\alpha_{\infty {\theta'}}}{\alpha_{2,1}}{\alpha_\infty}{t}{\frac{1}{z}} \right|^2 \,,
\end{aligned}
\end{equation}
and finally
\begin{equation}
     \FIV{\alpha_\infty}{\alpha_1}{\alpha_\theta}{\alpha_{2,1}}{\alpha}{\alpha_t}{\alpha_0}{z}{\frac{t}{z}}= \sum_{\theta'} \mathcal{M}_{\theta \theta'} (b\alpha, b\alpha_\infty ; b\alpha_1) t^{\Delta_\infty + \Delta_1 + \Delta_{2,1} - \Delta_0 - \Delta_t} z^{- 2 \Delta_{2,1}} \FIV{\alpha_0}{\alpha_t}{\alpha}{\alpha_1}{\alpha_{\infty {\theta'}}}{\alpha_{2,1}}{\alpha_\infty}{t}{\frac{1}{z}} \,.
\label{eq:connectionintinfinity}
\end{equation}
Combining equations \eqref{eq:connection0int} and \eqref{eq:connectionintinfinity} we can write
\begin{equation}
\begin{aligned}
    &\FIV{\alpha_\infty}{\alpha_1}{\alpha}{\alpha_t}{{\alpha_{0 \theta_1}}}{\alpha_{2,1}}{\alpha_0}{t}{\frac{z}{t}} = \\ &= \sum_{\theta_2 \theta_3} \mathcal{M}_{\theta_1 \theta_2} (b\alpha_0,b \alpha ;b \alpha_t) \mathcal{M}_{(-\theta_2) \theta_3} (b\alpha,b \alpha_\infty ; b\alpha_1) t^{\Delta_\infty + \Delta_1 + \Delta_{2,1} - \Delta_0 - \Delta_t} z^{- 2 \Delta_{2,1}} \FIV{\alpha_0}{\alpha_t}{\alpha_{\theta_2}}{\alpha_1}{\alpha_{\infty {\theta_3}}}{\alpha_{2,1}}{\alpha_\infty}{t}{\frac{1}{z}} \,.
\end{aligned}
\label{eq:connectionthoughintermediateNf4}
\end{equation}
Diagrammatically, this reads
\begin{equation}
\begin{tikzpicture}[baseline={(current bounding box.center)}, node distance=1cm and 1.5cm]
\coordinate[label=above:$\alpha_1$] (e1);
\coordinate[below=of e1] (aux1);
\coordinate[left=of aux1,label=left:$\alpha_\infty$] (e2);
\coordinate[right=1.5cm of aux1] (aux2);
\coordinate[above=of aux2,label=above:$\alpha_t$] (e3);
\coordinate[right=1.5cm of aux2] (aux3);
\coordinate[above=of aux3,label=above:$\alpha_{2,1}$] (e4);
\coordinate[right=of aux3,label=right:$\alpha_0$] (e5);

\draw[line] (e1) -- (aux1);
\draw[line] (aux1) -- (e2);
\draw[line] (e3) -- (aux2);
\draw[line,dashed,red] (aux3) -- (e4);
\draw[line] (aux3) -- (e5);
\draw[line] (aux1) -- node[label=below:$\alpha$] {} (aux2);
\draw[line] (aux2) -- node[label=below:$\alpha_{0 \theta_1}$] {} (aux3);
\end{tikzpicture} = \sum_{\theta_2 \theta_3} \mathcal{M}_{\theta_1 \theta_2} \mathcal{M}_{(-\theta_2) \theta_3} \begin{tikzpicture}[baseline={(current bounding box.center)}, node distance=1cm and 1.5cm]
\coordinate[label=above:$\alpha_t$] (e1);
\coordinate[below=of e1] (aux1);
\coordinate[left=of aux1,label=left:$\alpha_0$] (e2);
\coordinate[right=1.5cm of aux1] (aux2);
\coordinate[above=of aux2,label=above:$\alpha_1$] (e3);
\coordinate[right=1.5cm of aux2] (aux3);
\coordinate[above=of aux3,label=above:$\alpha_{2,1}$] (e4);
\coordinate[right=of aux3,label=right:$\alpha_\infty$] (e5);

\draw[line] (e1) -- (aux1);
\draw[line] (aux1) -- (e2);
\draw[line] (e3) -- (aux2);
\draw[line,dashed,red] (aux3) -- (e4);
\draw[line] (aux3) -- (e5);
\draw[line] (aux1) -- node[label=below:$\alpha_{\theta_2}$] {} (aux2);
\draw[line] (aux2) -- node[label=below:$\alpha_{\infty \theta_3}$] {} (aux3);
\end{tikzpicture}
\,.
\label{eq:sscb14}
\end{equation}
The diagrams provide a straightforward way to generalize the connection formula to an arbitrary pair of points. Indeed, writing down the diagram it is immediate to guess the correct $\mathcal{M}_{\theta \theta'}$ factors and the conformal blocks that will enter the connection formula. As an example, the connection formula for the expansions for $z \sim 1$ and $z \sim \infty$ with $t \ll 1$ are given by
\begin{equation}
\begin{tikzpicture}[baseline={(current bounding box.center)}, node distance=1cm and 1.5cm]
\coordinate[label=above:$\alpha_t$] (e1);
\coordinate[below=of e1] (aux1);
\coordinate[left=of aux1,label=left:$\alpha_0$] (e2);
\coordinate[right=1.5cm of aux1] (aux2);
\coordinate[above=of aux2,label=above:$\alpha_\infty$] (e3);
\coordinate[right=1.5cm of aux2] (aux3);
\coordinate[above=of aux3,label=above:$\alpha_{2,1}$] (e4);
\coordinate[right=of aux3,label=right:$\alpha_1$] (e5);

\draw[line] (e1) -- (aux1);
\draw[line] (aux1) -- (e2);
\draw[line] (e3) -- (aux2);
\draw[line,dashed,red] (aux3) -- (e4);
\draw[line] (aux3) -- (e5);
\draw[line] (aux1) -- node[label=below:$\alpha$] {} (aux2);
\draw[line] (aux2) -- node[label=below:$\alpha_{1 \theta}$] {} (aux3);
\end{tikzpicture} =  
\sum_{\theta'} \mathcal{M}_{\theta \theta'} \begin{tikzpicture}[baseline={(current bounding box.center)}, node distance=1cm and 1.5cm]
\coordinate[label=above:$\alpha_t$] (e1);
\coordinate[below=of e1] (aux1);
\coordinate[left=of aux1,label=left:$\alpha_0$] (e2);
\coordinate[right=1.5cm of aux1] (aux2);
\coordinate[above=of aux2,label=above:$\alpha_1$] (e3);
\coordinate[right=1.5cm of aux2] (aux3);
\coordinate[above=of aux3,label=above:$\alpha_{2,1}$] (e4);
\coordinate[right=of aux3,label=right:$\alpha_\infty$] (e5);

\draw[line] (e1) -- (aux1);
\draw[line] (aux1) -- (e2);
\draw[line] (e3) -- (aux2);
\draw[line,dashed,red] (aux3) -- (e4);
\draw[line] (aux3) -- (e5);
\draw[line] (aux1) -- node[label=below:$\alpha$] {} (aux2);
\draw[line] (aux2) -- node[label=below:$\alpha_{\infty \theta'}$] {} (aux3);
\end{tikzpicture}
\,,
\label{eq:sscb13}
\end{equation}
that is
\begin{equation}
\begin{aligned}
    &t^{\Delta_\infty + \Delta_1 + \Delta_{2,1} - \Delta_t - \Delta_0} (1-t)^{\Delta_\infty + \Delta_0 + \Delta_{2,1} - \Delta_t - \Delta_1} (z-t)^{- 2 \Delta_{2,1}} \FIV{\alpha_t}{\alpha_0}{\alpha}{\alpha_\infty}{\alpha_{1 \theta}}{\alpha_{2,1}}{\alpha_1}{t}{\frac{z-1}{z-t}} = \\ & = \sum_{\theta'} \mathcal{M}_{\theta \theta'} (b\alpha_1, b\alpha_\infty ; b\alpha) t^{\Delta_\infty + \Delta_1 + \Delta_{2,1} - \Delta_0 - \Delta_t} z^{- 2 \Delta_{2,1}} \FIV{\alpha_0}{\alpha_t}{\alpha}{\alpha_1}{\alpha_{\infty {\theta'}}}{\alpha_{2,1}}{\alpha_\infty}{t}{\frac{1}{z}} \,.
\end{aligned}
\label{eq:regfrom1toinfinity}
\end{equation}
Note that combining all the previous formulae we manage to analytically continue the expansion in $z \sim 0$ of the conformal block in all the complex plane for $t \ll 1$. It is straightforward to generalize the previous formulae for $t \sim 1, t \sim \infty$. All in all, for any value of $t$ we can connect all the possible expansions in $z$. The analytic continuation in the $t-$plane is more involved and can be done via the fusion kernel. As a concluding remark, note that there is a M\"obius tranformation in each region of expansions of the correlator, say $z \ll t \ll 1$ for reference, that only exchanges $\alpha_\infty$ and $\alpha_1$ and that does not change the region of validity of the expansion. This transformation is usually called \textit{braiding}. This gives, up to a Jacobian,
\begin{equation}
\begin{tikzpicture}[baseline={(current bounding box.center)}, node distance=1cm and 1.5cm]
\coordinate[label=above:$\alpha_\infty$] (e1);
\coordinate[below=of e1] (aux1);
\coordinate[left=of aux1,label=left:$\alpha_1$] (e2);
\coordinate[right=1.5cm of aux1] (aux2);
\coordinate[above=of aux2,label=above:$\alpha_t$] (e3);
\coordinate[right=1.5cm of aux2] (aux3);
\coordinate[above=of aux3,label=above:$\alpha_{2,1}$] (e4);
\coordinate[right=of aux3,label=right:$\alpha_0$] (e5);

\draw[line] (e1) -- (aux1);
\draw[line] (aux1) -- (e2);
\draw[line] (e3) -- (aux2);
\draw[line,dashed,red] (aux3) -- (e4);
\draw[line] (aux3) -- (e5);
\draw[line] (aux1) -- node[label=below:$\alpha$] {} (aux2);
\draw[line] (aux2) -- node[label=below:$\alpha_{0\pm}$] {} (aux3);
\end{tikzpicture} = \, \FIV{\alpha_1}{\alpha_\infty}{\alpha}{\alpha_t}{{\alpha_{0 \theta}}}{\alpha_{2,1}}{\alpha_0}{\frac{t}{t-1}}{\frac{z}{t} \frac{t-1}{z-1}} \,.
\label{eq:sscb12}
\end{equation}
Braiding changes the expansion variables in the conformal blocks according to the new positions of the insertions and as such can be used to generate other expansions and the related connection coefficients.

\subsubsection{Semiclassical limit}
Let us consider the semiclassical limit of Liouville theory, that is the double scaling limit 
\begin{equation}\label{semicl}
b \to 0, \, \alpha_i \to \infty, \, b \alpha_i = a_i \, \, {\rm finite.}
\end{equation}
In this limit the conformal blocks and the corresponding BPZ equation greatly simplify. The divergence exponentiates and the $z$ dependence becomes subleading, namely\footnote{Here and in the following we do not indicate the dependence of
$F$ and $W$ on the rescaled momenta.}
\begin{equation}
     \FIV{\alpha_\infty}{\alpha_1}{\alpha}{\alpha_t}{{\alpha_{0 \theta}}}{\alpha_{2,1}}{\alpha_0}{t}{\frac{z}{t}} = t^{\Delta - \Delta_t - \Delta_{0 \theta}} z^{\frac{b Q}{2} + \theta b \alpha_0} \exp \left[ \frac{1}{b^2} \left( F(t) + b^2 W(z/t, t) + \mathcal{O} (b^4)  \right) \right] \,.
\end{equation}
Here $F(t)$ is the classical conformal block, related to the conformal block without degenerate insertion via
\begin{equation}
    \mathfrak{F} \left( \begin{matrix} \alpha_1 \\ \alpha_\infty \end{matrix} \alpha \, \begin{matrix} \alpha_t \\ \alpha_0 \end{matrix} ; t \right) = t^{\Delta - \Delta_t - \Delta_0} e^{b^{-2} \left( F(t) + \mathcal{O}(b^2) \right)} \,.
\label{eq:defclasscbNf41}
\end{equation}
The divergences in the conformal blocks can be cured by dividing by the conformal block without the degenerate insertion. We denote the resulting finite, semiclassical conformal block by the letter $\mathcal{F}$:
\begin{equation}
    \FIVsc{a_\infty}{a_1}{a}{a_t}{{a_{0 \theta}}}{a_{2,1}}{a_0}{t}{\frac{z}{t}} = \lim_{b \to 0} \frac{\FIV{\alpha_\infty}{\alpha_1}{\alpha}{\alpha_t}{{\alpha_{0 \theta}}}{\alpha_{2,1}}{\alpha_0}{t}{\frac{z}{t}}}{\mathfrak{F} \left( \begin{matrix} \alpha_1 \\ \alpha_\infty \end{matrix} \alpha \, \begin{matrix} \alpha_t \\ \alpha_0 \end{matrix} ; t \right)} = t^{- \theta a_0} z^{\frac{1}{2} + \theta a_0} e^{-\frac{\theta}{2} \partial_{a_0} F \left(t \right)} \left( 1 + \mathcal{O}(t,z/t) \right) \,.
\label{eq:smalltsmallzsccbNf4}
\end{equation}
Note that the conformal block with the degenerate insertion and $z, t \sim 0$ contains a classical conformal block depending on $a_{0\theta}=a_0 -\theta \frac{b^2}{2}$. Dividing by the four-point function without the degenerate insertion, which depends on $a_0$, gives an incremental ratio that in the limit \eqref{semicl} becomes the derivative $\partial_{a_0} F \left(t \right)$. The BPZ equation \eqref{eq:BPZNf=4} simplifies in the semiclassical limit as well. The $t-$derivative acting on the conformal block gives
\begin{equation}
    t \partial_t \FIV{\alpha_\infty}{\alpha_1}{\alpha}{\alpha_t}{{\alpha_{0 \theta}}}{\alpha_{2,1}}{\alpha_0}{t}{\frac{z}{t}} = b^{-2}\left(-\frac{1}{4}-a^2 +a_t^2 +a_0^2 + t \partial_t F (a_i, a, t) + \mathcal{O}(b^2) \right) \FIV{\alpha_\infty}{\alpha_1}{\alpha}{\alpha_t}{{\alpha_{0 \theta}}}{\alpha_{2,1}}{\alpha_0}{t}{\frac{z}{t}} \,,
\end{equation}
therefore the $t-$derivative becomes a multiplication by a $z$-independent factor at leading order in $b^2$ and the BPZ equation becomes an ODE. Defining
\begin{equation}
    u^{(0)} = \lim_{b \to 0} b^2 t \partial_t \log \mathfrak{F} \left( \begin{matrix} \alpha_1 \\ \alpha_\infty \end{matrix} \alpha \, \begin{matrix} \alpha_t \\ \alpha_0 \end{matrix} ; t \right) \,,
\label{eq:MatoneNf40}
\end{equation}
where the superscript indicates that the block is expanded for $t \sim 0$, the BPZ equation \eqref{eq:BPZNf=4} in the semiclassical limit reads
\begin{equation}
\begin{aligned}
    \bigg( \partial_z^2 + \frac{\frac{1}{4}-a_1^2}{(z-1)^2} - \frac{\frac{1}{2}-a_1^2 -a_t^2 -a_0^2 +a_\infty^2 + u^{(0)}}{z(z-1)}+ \frac{\frac{1}{4}-a_t^2}{(z-t)^2}+\frac{u^{(0)}}{z(z-t)} +\frac{\frac{1}{4}-a_0^2}{z^2} \bigg) \FIVsc{a_\infty}{a_1}{a}{a_t}{{a_{0 \theta}}}{a_{2,1}}{a_0}{t}{\frac{z}{t}} = 0\,.
\end{aligned}
\label{eq:BPZNf=4sc}
\end{equation}
The solution of the previous ODE for $z \sim t$ is given by the semiclassical block
\begin{equation}
\begin{aligned}
    &(t-1)^{\frac{1}{2}} \FIVsc{a_\infty}{a_1}{a}{a_0}{{a_{t \theta}}}{a_{2,1}}{a_t}{\frac{t}{t-1}}{\frac{t-z}{t}} = \lim_{b \to 0} (t-1)^{- \Delta_{2,1}} \frac{\FIV{\alpha_\infty}{\alpha_1}{\alpha}{\alpha_0}{{\alpha_{t \theta}}}{\alpha_{2,1}}{\alpha_t}{\frac{t}{t-1}}{\frac{t-z}{t}}}{\mathfrak{F} \left( \begin{matrix} \alpha_1 \\ \alpha_\infty \end{matrix} \alpha \, \begin{matrix} \alpha_0 \\ \alpha_t \end{matrix} ; \frac{t}{t-1} \right)} = \\ &= \lim_{b \to 0} \frac{e^{i \pi (\Delta - \Delta_0 - \Delta_{2,1} - \Delta_t)} (1-t)^{\Delta_\infty - \Delta_1 - \Delta_t - \Delta_{2,1} - \Delta_0} \FIV{\alpha_\infty}{\alpha_1}{\alpha}{\alpha_0}{{\alpha_{t \theta}}}{\alpha_{2,1}}{\alpha_t}{\frac{t}{t-1}}{\frac{t-z}{t}}}{\mathfrak{F} \left( \begin{matrix} \alpha_1 \\ \alpha_\infty \end{matrix} \alpha \, \begin{matrix} \alpha_t \\ \alpha_0 \end{matrix} ; t \right)} \,,
\end{aligned}
\label{eq:zneartsmalltsccbNf4}
\end{equation}
therefore the connection formula \eqref{eq:connection0t} descends to the semiclassical blocks to be
\begin{equation}
    \FIVsc{a_\infty}{a_1}{a}{a_t}{{a_{0 \theta}}}{a_{,
    2,1}}{a_0}{t}{\frac{z}{t}} = \sum_{\theta'}\mathcal{M}_{\theta \theta'}(a_0,a_t;a) (t-1)^{\frac{1}{2}} \FIVsc{a_\infty}{a_1}{a}{a_0}{{a_{t \theta'}}}{a_{2,1}}{a_t}{\frac{t}{t-1}}{\frac{t-z}{t}}\,.
\label{eq:scconnection0t}
\end{equation}
Note that the intermediate momentum $a$ can be computed as a function of the parameters appearing in the semiclassical BPZ equation inverting the relation \eqref{eq:MatoneNf40}. Similarly, keeping $t \sim 0$ we can analytically continue the solution to the other singularities, that is for $z \sim 1$ and $z \sim \infty$.  In particular, we can directly connect $z \sim 0$ and $z \sim \infty$ passing though the intermediate region. The semiclassical block for $z \sim \infty$ reads
\begin{equation}
\begin{aligned}
        &t^{-\frac{1}{2}} z \FIVsc{a_0}{a_t}{a}{a_1}{{a_{\infty \theta}}}{a_{2,1}}{a_\infty}{t}{\frac{1}{z}} = \\ & = \lim_{b \to 0} \frac{t^{\Delta_\infty + \Delta_1 + \Delta_{2,1} - \Delta_0 - \Delta_t} z^{- 2 \Delta_{2,1}} \FIV{\alpha_0}{\alpha_t}{\alpha}{\alpha_1}{\alpha_{\infty {\theta'}}}{\alpha_{2,1}}{\alpha_\infty}{t}{\frac{1}{z}}}{\mathfrak{F} \left( \begin{matrix} \alpha_1 \\ \alpha_\infty \end{matrix} \alpha \, \begin{matrix} \alpha_t \\ \alpha_0 \end{matrix} ; t \right)} = \lim_{b \to 0} \frac{t^{\Delta_{2,1}} z^{- 2 \Delta_{2,1}} \FIV{\alpha_0}{\alpha_t}{\alpha}{\alpha_1}{\alpha_{\infty {\theta'}}}{\alpha_{2,1}}{\alpha_\infty}{t}{\frac{1}{z}}}{\mathfrak{F} \left( \begin{matrix} \alpha_t \\ \alpha_0 \end{matrix} \alpha \, \begin{matrix} \alpha_1 \\ \alpha_\infty \end{matrix} ; t \right)} \,.
\end{aligned}
\label{eq:zbigsmalltsccbNf4}
\end{equation}
The connection formula \eqref{eq:connectionthoughintermediateNf4} from $z \sim 0$ to $z \sim \infty$ involves a conformal block with two shifted momenta, that is
\begin{equation}
    \FIV{\alpha_0}{\alpha_t}{\alpha_{\theta'}}{\alpha_1}{\alpha_{\infty \theta}}{\alpha_{2,1}}{\alpha_\infty}{t}{\frac{1}{z}} = t^{\Delta_{\theta'} - \Delta_1 - \Delta_{\infty \theta}} \left(\frac{t}{z} \right)^{\frac{b Q}{2} + \theta b \alpha_\infty} \exp \left[ \frac{1}{b^2} F\left(a - \theta' \frac{b^2}{2}, t\right) +  W \left( a - \theta' \frac{b^2}{2}, t \right) + 
    \mathcal{O} (b^2) \right] \,.
\end{equation}
At first order in $b^2$
\begin{equation}
    F\left(a - \theta' \frac{b^2}{2}, t\right) + b^2 W \left( a- \theta' \frac{b^2}{2}, t \right) = F(a,t) - \frac{\theta' b^2 }{2} \partial_a F(a,t) + b^2 W \left( a, t \right) + \mathcal{O} (b^4) \,,
\end{equation}
therefore in the semiclassical limit
\begin{equation}\label{twiddle}
    \FIV{\alpha_0}{\alpha_t}{\alpha_{\theta'}}{\alpha_1}{\alpha_{\infty \theta}}{\alpha_{2,1}}{\alpha_\infty}{t}{\frac{1}{z}} \sim t^{- \theta' \alpha} e^{- \frac{\theta'}{2}\partial_a F(t)} \FIV{\alpha_0}{\alpha_t}{\alpha}{\alpha_1}{\alpha_{\infty \theta}}{\alpha_{2,1}}{\alpha_\infty}{t}{\frac{1}{z}}\,,\quad \mathrm{as } \, \,b\to 0 \,.
\end{equation}
This is consistent with the fact that we expect only two linearly independent $z$ behaviors. The connection formula \eqref{eq:connectionthoughintermediateNf4} simplifies to
\begin{equation}
\begin{aligned}
    &\FIVsc{a_\infty}{a_1}{a}{a_t}{{a_{0 \theta}}}{a_{2,1}}{a_0}{t}{\frac{z}{t}} =\\=& \sum_{\theta'} \left( \sum_{\sigma} \mathcal{M}_{\theta \sigma} (a_0, a ; a_t) \mathcal{M}_{(-\sigma) \theta'} (a, a_\infty ; a_1) t^{- \sigma a} e^{- \frac{\sigma}{2}\partial_a F} \right) t^{-\frac{1}{2}} z \FIVsc{a_0}{a_t}{a}{a_1}{{a_{\infty \theta'}}}{a_{2,1}}{a_\infty}{t}{\frac{1}{z}} \,.
\end{aligned}
\label{eq:sconnection0inft}
\end{equation}
Explicitly, the connection coefficients are
\begin{equation}
\begin{aligned}
     &\sum_{\sigma=\pm} \mathcal{M}_{\theta \sigma} (a_0, a ; a_t) \mathcal{M}_{(-\sigma) \theta'} (a, a_\infty ; a_1) t^{- \sigma a} e^{- \frac{\sigma}{2}\partial_a F} = \\ =& \,\sum_{\sigma=\pm}
     \frac{\Gamma(1-2\sigma a)\Gamma(-2\sigma a)\Gamma(1+2\theta a_0)\Gamma(-2\theta' a_\infty)t^{- \sigma a} e^{- \frac{\sigma}{2}\partial_a F}}{\Gamma\left(\frac{1}{2}+\theta a_0-\sigma  a +  a_t\right) \Gamma\left(\frac{1}{2}+\theta a_0-\sigma  a - a_t \right)\Gamma\left(\frac{1}{2}-\sigma a-\theta' a_\infty +  a_1\right)\Gamma\left(\frac{1}{2}-\sigma a-\theta' a_\infty - a_1\right)}\,.
\end{aligned}
\end{equation}
For future reference, the semiclassical block for small $t$ and $z \sim 1$ is given by
\begin{equation}
    \left( t(1-t) \right)^{-\frac{1}{2}} (t-z) \FIVsc{a_t}{a_0}{a}{a_\infty}{{a_{1 \theta}}}{a_{2,1}}{a_1}{t}{\frac{1-z}{t-z}} = \lim_{b \to 0} \left( t(1-t) \right)^{\Delta_{2,1}} (t-z)^{- 2 \Delta_{2,1}} \frac{\FIV{\alpha_t}{\alpha_0}{\alpha}{\alpha_\infty}{{\alpha_{1 \theta}}}{\alpha_{2,1}}{\alpha_1}{t}{\frac{1-z}{t-z}}}{\mathfrak{F} \left( \begin{matrix} \alpha_0 \\ \alpha_t \end{matrix} \alpha \, \begin{matrix} \alpha_\infty \\ \alpha_1 \end{matrix} ; t \right)} \,.
\label{eq:znear1smalltsccbNf4}
\end{equation}
Similarly one can obtain the connection coefficients for the other $t-$expansions. As an example, let us schematically consider the case $t \gg 1$. The semiclassical block for $z \sim 0$ reads
\begin{equation}
        t^{\frac{1}{2}} \FIVsc{a_\infty}{a_t}{a}{a_1}{{a_{0 \theta}}}{a_{2,1}}{a_0}{\frac{1}{t}}{z} = \lim_{b \to 0} \frac{t^{- \Delta_{2,1}} \FIV{\alpha_\infty}{\alpha_t}{\alpha}{\alpha_1}{{\alpha_{0 \theta}}}{\alpha_{2,1}}{\alpha_0}{\frac{1}{t}}{z}}{\mathfrak{F} \left( \begin{matrix} \alpha_t \\ \alpha_\infty \end{matrix} \alpha \, \begin{matrix} \alpha_1 \\ \alpha_0 \end{matrix} ; \frac{1}{t} \right)} \,.
\end{equation}
Still the $t-$derivative decouples, leaving behind
\begin{equation}
        u^{(\infty)} = \lim_{b \to 0} b^2 t \partial_t \log t^{\Delta-\Delta_t-\Delta_1-\Delta_0} \mathfrak{F} \left( \begin{matrix} \alpha_t \\ \alpha_\infty \end{matrix} \alpha \, \begin{matrix} \alpha_1 \\ \alpha_0 \end{matrix} ; \frac{1}{t} \right) \,.
\label{eq:MatoneNf4inf}
\end{equation}
Note that the semiclassical BPZ equation formally remains the same, with the substitution\footnote{From the gauge theory viewpoint this amounts to a change of frame from the electric to the monopole one.} of $u^{(0)}$ with $u^{(\infty)}$. Indeed,
the intermediate momentum $\alpha$ is now determined in terms of $u^{(\infty)}$. 
The $z \sim 1$ expansion gives
\begin{equation}
        (t-1)^{\frac{1}{2}} e^{i \theta \pi a} \FIVsc{a_\infty}{a_t}{a}{a_1}{{a_{0 \theta}}}{a_{2,1}}{a_0}{\frac{1}{t-1}}{1-z} = \lim_{b \to 0} (t-1)^{- \Delta_{2,1}} e^{i \theta \pi b \alpha} \frac{\FIV{\alpha_\infty}{\alpha_t}{\alpha}{\alpha_0}{{\alpha_{1 \theta}}}{\alpha_{2,1}}{\alpha_1}{\frac{1}{1-t}}{1-z}}{\mathfrak{F} \left( \begin{matrix} \alpha_t \\ \alpha_\infty \end{matrix} \alpha \, \begin{matrix} \alpha_0 \\ \alpha_1 \end{matrix} ; \frac{1}{1-t} \right)} \,,
\end{equation}
and the corresponding connection formula reads
\begin{equation}
    t^{\frac{1}{2}} \FIVsc{a_\infty}{a_t}{a}{a_1}{{a_{0 \theta}}}{a_{2,1}}{a_0}{\frac{1}{t}}{z} = \sum_{\theta' = \pm 1} \mathcal{M}_{\theta \theta'} \left( a_0, a_1 ; a \right) (t-1)^{\frac{1}{2}} e^{i \theta \pi a} \FIVsc{a_\infty}{a_t}{a}{a_1}{{a_{0 \theta'}}}{a_{2,1}}{a_0}{\frac{1}{t-1}}{1-z} \,.
\end{equation}
All other connection formulae at $t \gg 1$ can be obtained similarly. 
The same can be done when $t \sim 1$. Note that again the semiclassical BPZ equation looks formally as \eqref{eq:BPZNf=4sc} upon the substitution\footnote{This is the dyon frame.} of $u^{(0)}$ with
\begin{equation}
    u^{(1)} = \lim_{b \to 0} b^2 t \partial_t \log \mathfrak{F} \left( \begin{matrix} \alpha_0 \\ \alpha_\infty \end{matrix} \alpha \, \begin{matrix} \alpha_t \\ \alpha_1 \end{matrix} ; 1-t \right) \, .
\end{equation}


\subsection{Confluent conformal blocks}\label{Nf3}
\subsubsection{General case}
Consider the correlation function 
\begin{equation}
    \langle \mu, \Lambda|  V_1(1) \Phi(z)|\Delta_0\rangle\,.
\end{equation}
It solves the BPZ equation
\begin{equation}\label{BPZNf3}
\begin{aligned}
    \left(b^{-2} \partial_z^2 - \left(\frac{1}{z}+\frac{1}{z-1}\right)\partial_z + \frac{\Lambda \partial_{\Lambda} - \Delta_{2,1} - \Delta_1 - \Delta_0}{z(z-1)} + \frac{\Delta_1}{(z-1)^2} + \frac{\Delta_0}{z^2} + \frac{ \mu \Lambda}{z} -\frac{\Lambda^2}{4}\right) \langle \mu, \Lambda| \Phi(z) V_1(1)|\Delta_0\rangle=0\,,
\end{aligned}
\end{equation}
and can be decomposed into {\it confluent} conformal blocks in different ways. They are all given as collision limits of {\it regular} conformal blocks. 
\paragraph{Small $\Lambda$ blocks}
We focus first on the case where the conformal blocks are given as an expansion in $\Lambda$. The block for $z\sim 0$ is defined as\footnote{The argument $\frac{\eta+\mu}{2}$ should appear with a minus sign as in Appendix \ref{app:rank1collision}. Here and in the following we don't write it due to the symmetry of the conformal block. The reader wishing to compare with the Nekrasov partition function should take this sign into account as in Appendix \ref{app:Nek}.}
\begin{equation}\label{FIII}
     \FIII{\mu}{\alpha}{\alpha_1}{\alpha_{0 \theta}}{\alpha_{2,1}}{\alpha_0}{\Lambda}{z}=\Lambda^\Delta z^{\frac{bQ}{2}+\theta b \alpha_0}\FIIIhat{\mu}{\alpha}{\alpha_1}{\alpha_{0 \theta}}{\alpha_{2,1}}{\alpha_0}{\Lambda}{z} = \Lambda^\Delta  z^{\frac{bQ}{2}+\theta b \alpha_0} \lim_{\eta \to \infty} \FIVhat{\frac{\eta+\mu}{2}}{\frac{\eta-\mu}{2}}{\alpha}{\alpha_1}{\alpha_{0 \theta}}{\alpha_{2,1}}{\alpha_0}{\frac{\Lambda}{\eta}}{z}\,.
\end{equation}
This is nothing but the standard collision limit of $\langle \Delta_\infty|$ and $V_t(t)$ as defined in \eqref{eq:collision}. The tilde on the conformal block means it has no classical part, i.e. is normalized such that the first term is $1$. This conformal block can also be computed directly by doing the OPE of $\Phi(z)$ with $|\Delta_0\rangle$, then the OPE of $V_1(1)$ with the result which we specify to be in the Verma module $\Delta_\alpha$, and then contracting with $\langle \mu,\Lambda|$. In the diagrammatic notation introduced in section \ref{warmup:whittaker}, we represent it by
\begin{equation}
    \FIII{\mu}{\alpha}{\alpha_1}{\alpha_{0 \theta}}{\alpha_{2,1}}{\alpha_0}{\Lambda}{z} = 
    \begin{tikzpicture}[baseline={(current bounding box.center)}, node distance=1cm and 1.5cm]
    \coordinate[circle,fill,inner sep=2pt] (aux1);
    \coordinate[left=of aux1,label=left:$\mu$] (e1);
    \coordinate[right=1.5cm of aux1] (aux2);
    \coordinate[above=of aux2,label=above:$\alpha_1$] (e2);
    \coordinate[right=1.5cm of aux2] (aux3);
    \coordinate[above=of aux3,label=above:$\alpha_{2,1}$] (e3);
    \coordinate[right=of aux3,label=right:$\alpha_0$] (e4);

    \draw[line,double] (aux1) -- (e1);
    \draw[line] (aux2) -- (e2);
    \draw[line,dashed,red] (aux3) -- (e3);
    \draw[line] (aux3) -- (e4);
    \draw[line] (aux1) -- node[label=below:$\alpha$] {} (aux2);
    \draw[line] (aux2) -- node[label=below:$\alpha_{0\theta}$] {} (aux3);
    \end{tikzpicture} \,.
\end{equation}
The double line represents the rank 1 irregular state, and the dot the pairing with a primary state. For $z\sim 1$, the corresponding block can be expressed as
\begin{equation}
    e^{\mu \Lambda} \FIII{-\mu}{\alpha}{\alpha_0}{\alpha_{1 \theta}}{\alpha_{2,1}}{\alpha_1}{\Lambda}{1-z} =
    \begin{tikzpicture}[baseline={(current bounding box.center)}, node distance=1cm and 1.5cm]
    \coordinate[circle,fill,inner sep=2pt] (aux1);
    \coordinate[left=of aux1,label=left:$-\mu$] (e1);
    \coordinate[right=1.5cm of aux1] (aux2);
    \coordinate[above=of aux2,label=above:$\alpha_0$] (e2);
    \coordinate[right=1.5cm of aux2] (aux3);
    \coordinate[above=of aux3,label=above:$\alpha_{2,1}$] (e3);
    \coordinate[right=of aux3,label=right:$\alpha_1$] (e4);

    \draw[line,double] (aux1) -- (e1);
    \draw[line] (aux2) -- (e2);
    \draw[line,dashed,red] (aux3) -- (e3);
    \draw[line] (aux3) -- (e4);
    \draw[line] (aux1) -- node[label=below:$\alpha$] {} (aux2);
    \draw[line] (aux2) -- node[label=below:$\alpha_{1\theta}$] {} (aux3);
    \end{tikzpicture}\,,
\end{equation} 
where the exponential factor and the argument $-\mu$ arise from the corresponding M\"obius transformation\footnote{Actually, doing the M\"obius transformation one gets $-\Lambda$ but since the block depends only on $\mu\Lambda$ and $\Lambda^2$ except for the classical part, one can trade $-\Lambda$ for $-\mu$.}. In the intermediate region, where $z \gg 1$ but $\Lambda z \ll 1$, the corresponding block is
\begin{equation}
    z^{-\Delta_{2,1}-\Delta_1-\Delta_0} \FIII{\mu}{\alpha_{ \theta}}{\alpha_{2,1}}{\alpha}{\alpha_1}{\alpha_0}{\Lambda z}{\frac{1}{z}} = 
    \begin{tikzpicture}[baseline={(current bounding box.center)}, node distance=1cm and 1.5cm]
    \coordinate[circle,fill,inner sep=2pt] (aux1);
    \coordinate[left=of aux1,label=left:$\mu$] (e1);
    \coordinate[right=1.5cm of aux1] (aux2);
    \coordinate[above=of aux2,label=above:$\alpha_{2,1}$] (e2);
    \coordinate[right=1.5cm of aux2] (aux3);
    \coordinate[above=of aux3,label=above:$\alpha_1$] (e3);
    \coordinate[right=of aux3,label=right:$\alpha_0$] (e4);

    \draw[line,double] (aux1) -- (e1);
    \draw[line,dashed,red] (aux2) -- (e2);
    \draw[line] (aux3) -- (e3);
    \draw[line] (aux3) -- (e4);
    \draw[line] (aux1) -- node[label=below:$\alpha_\theta$] {} (aux2);
    \draw[line] (aux2) -- node[label=below:$\alpha$] {} (aux3);
    \end{tikzpicture}\,.
\end{equation}
In the deep irregular region where $z\gg1$ and $\Lambda z \gg 1$, the conformal block is given by a different collision limit, proposed in \cite{Lisovyy:2018mnj}:
\begin{equation}
\begin{aligned}
    &\DIII{\mu}{\alpha_{2,1}}{\mu_{\theta}}{\alpha}{\alpha_1}{\alpha_0}{\Lambda}{\frac{1}{\Lambda z}} = e^{\theta b\Lambda z/2} \Lambda^{\Delta_{2,1}+\Delta} \left( \Lambda z\right)^{-\theta b \mu + \frac{b^2}{2}}  \times \\ \times&\lim_{\eta \to \infty} \left( 1- \frac{\eta}{\Lambda z} \right)^{-\frac{bQ}{2} -\theta \frac{b}{2}(\mu-\eta)}  \FIVhat{\alpha_0}{\alpha_1}{\alpha}{\frac{\mu-\eta}{2}}{{\scriptstyle \frac{\mu+\eta-\theta b}{2}}}{\alpha_{2,1}}{\frac{\mu+\eta}{2}}{\frac{\Lambda}{\eta}}{\frac{\eta}{\Lambda z}} \,.
\end{aligned}
\label{eq:LLimD3}
\end{equation}
Whenever $z$ approaches an irregular singularity of rank $1$, we denote the corresponding conformal block by $\mathfrak{D}$. This conformal block can also be computed directly by doing the OPE between $\langle \mu,\Lambda|$ and $\Phi(z)$, then the OPE of the result with $V_1(1)$ and contracting with $|\Delta_0\rangle$. Diagramatically, we write
\begin{equation}
    \DIII{\mu}{\alpha_{2,1}}{\mu_{\theta}}{\alpha}{\alpha_1}{\alpha_0}{\Lambda}{\frac{1}{\Lambda z}} = 
    \begin{tikzpicture}[baseline={(current bounding box.center)}, node distance=1cm and 1.5cm]
    \coordinate (aux1);
    \coordinate[left=of aux1,label=left:$\mu$] (e1);
    \coordinate[right=1.5cm of aux1,circle,fill,inner sep=2pt] (aux2);
    \coordinate[above=of aux1,label=above:$\alpha_{2,1}$] (e2);
    \coordinate[right=1.5cm of aux2] (aux3);
    \coordinate[above=of aux3,label=above:$\alpha_1$] (e3);
    \coordinate[right=of aux3,label=right:$\alpha_0$] (e4);

    \draw[line,double] (aux1) -- (e1);
    \draw[line,dashed,red] (aux1) -- (e2);
    \draw[line] (aux3) -- (e3);
    \draw[line] (aux3) -- (e4);
    \draw[line,double] (aux1) -- node[label=below:$\mu_\theta$] {} (aux2);
    \draw[line] (aux2) -- node[label=below:$\alpha$] {} (aux3);
    \end{tikzpicture}\,.
\end{equation}
The connection problem between $0$ and $1$ is solved in the same way as for the regular conformal blocks, since we are never near the irregular singularity. The result is
\begin{equation}
    \FIII{\mu}{\alpha}{\alpha_1}{\alpha_{0 \theta}}{\alpha_{2,1}}{\alpha_0}{\Lambda}{z} = \sum_{\theta'=\pm}\mathcal{M}_{\theta \theta'}(b\alpha_0,b\alpha_1;b\alpha) e^{\mu \Lambda} \FIII{-\mu}{\alpha}{\alpha_0}{\alpha_{1 \theta'}}{\alpha_{2,1}}{\alpha_1}{\Lambda}{1-z}\,.
\end{equation}
Diagrammatically:
\begin{equation}
    \begin{tikzpicture}[baseline={(current bounding box.center)}, node distance=1cm and 1.5cm]
    \coordinate[circle,fill,inner sep=2pt] (aux1);
    \coordinate[left=of aux1,label=left:$\mu$] (e1);
    \coordinate[right=1.5cm of aux1] (aux2);
    \coordinate[above=of aux2,label=above:$\alpha_1$] (e2);
    \coordinate[right=1.5cm of aux2] (aux3);
    \coordinate[above=of aux3,label=above:$\alpha_{2,1}$] (e3);
    \coordinate[right=of aux3,label=right:$\alpha_0$] (e4);

    \draw[line,double] (aux1) -- (e1);
    \draw[line] (aux2) -- (e2);
    \draw[line,dashed,red] (aux3) -- (e3);
    \draw[line] (aux3) -- (e4);
    \draw[line] (aux1) -- node[label=below:$\alpha$] {} (aux2);
    \draw[line] (aux2) -- node[label=below:$\alpha_{0\theta}$] {} (aux3);
    \end{tikzpicture} = \sum_{\theta'=\pm}\mathcal{M}_{\theta \theta'} \begin{tikzpicture}[baseline={(current bounding box.center)}, node distance=1cm and 1.5cm]
    \coordinate[circle,fill,inner sep=2pt] (aux1);
    \coordinate[left=of aux1,label=left:$-\mu$] (e1);
    \coordinate[right=1.5cm of aux1] (aux2);
    \coordinate[above=of aux2,label=above:$\alpha_0$] (e2);
    \coordinate[right=1.5cm of aux2] (aux3);
    \coordinate[above=of aux3,label=above:$\alpha_{2,1}$] (e3);
    \coordinate[right=of aux3,label=right:$\alpha_1$] (e4);

    \draw[line,double] (aux1) -- (e1);
    \draw[line] (aux2) -- (e2);
    \draw[line,dashed,red] (aux3) -- (e3);
    \draw[line] (aux3) -- (e4);
    \draw[line] (aux1) -- node[label=below:$\alpha$] {} (aux2);
    \draw[line] (aux2) -- node[label=below:$\alpha_{1\theta'}$] {} (aux3);
    \end{tikzpicture}\,.
\end{equation}
Instead, to solve the connection problem between $1$ and $\infty$ one has to do two steps: from $1$ to the intermediate region, and then to $\infty$. At each step we decompose the correlator into conformal blocks in the different regions and then use crossing symmetry to determine the connection coefficients. The relevant formulae for the irregular state are reviewed in Appendix \ref{app:rank1}. We have
\begin{equation}
\begin{aligned}
    \langle \mu, \Lambda| \Phi(z) V_1(1)|\Delta_0\rangle &= \int d \alpha \, C_{\mu \alpha} \sum_{\theta = \pm} C_{\alpha_{2,1}\alpha_1}^{\alpha_{1\theta}} C_{\alpha_{1\theta}\alpha_0}^\alpha \left|e^{\mu \Lambda} \FIII{-\mu}{\alpha}{\alpha_0}{\alpha_{1 \theta}}{\alpha_{2,1}}{\alpha_1}{\Lambda}{1-z}\right|^2=\\
    &= \int d \alpha \,  C_{\mu \alpha}\sum_{\theta'=\pm} C_{\alpha_{2,1} \alpha_{\theta'} }^{\alpha} C_{\alpha_1\alpha_0}^{\alpha_{\theta'}} \left| z^{-\Delta_{2,1}-\Delta_1-\Delta_0} \FIII{\mu}{\alpha}{\alpha_{2,1}}{\alpha_{\theta'}}{\alpha_1}{\alpha_0}{\Lambda z}{\frac{1}{z}}\right|^2\,.
\end{aligned}
\end{equation}
We recognize this condition from the hypergeometric function \eqref{eq:hypergeometric_constraint}. Therefore we can readily solve it in terms of the hypergeometric connection coefficients $\mathcal{M}$ and the connection formula between $0$ and the intermediate region is then
\begin{equation}\label{Nf3from1tointermediate}
    e^{\mu \Lambda} \FIII{-\mu}{\alpha}{\alpha_0}{\alpha_{1 \theta}}{\alpha_{2,1}}{\alpha_1}{\Lambda}{1-z} = \sum_{\theta'=\pm} \mathcal{M}_{\theta\theta'}(b\alpha_1,b\alpha;b\alpha_0) z^{-\Delta_{2,1}-\Delta_1-\Delta_0}\FIII{\mu}{\alpha}{\alpha_{2,1}}{\alpha_{ \theta'}}{\alpha_1}{\alpha_0}{\Lambda z}{\frac{1}{z}}\,.
\end{equation}
Diagrammatically:
\begin{equation}
    \begin{tikzpicture}[baseline={(current bounding box.center)}, node distance=1cm and 1.5cm]
    \coordinate[circle,fill,inner sep=2pt] (aux1);
    \coordinate[left=of aux1,label=left:$-\mu$] (e1);
    \coordinate[right=1.5cm of aux1] (aux2);
    \coordinate[above=of aux2,label=above:$\alpha_0$] (e2);
    \coordinate[right=1.5cm of aux2] (aux3);
    \coordinate[above=of aux3,label=above:$\alpha_{2,1}$] (e3);
    \coordinate[right=of aux3,label=right:$\alpha_1$] (e4);

    \draw[line,double] (aux1) -- (e1);
    \draw[line] (aux2) -- (e2);
    \draw[line,dashed,red] (aux3) -- (e3);
    \draw[line] (aux3) -- (e4);
    \draw[line] (aux1) -- node[label=below:$\alpha$] {} (aux2);
    \draw[line] (aux2) -- node[label=below:$\alpha_{1\theta}$] {} (aux3);
    \end{tikzpicture} = \sum_{\theta'=\pm} \mathcal{M}_{\theta\theta'}
    \begin{tikzpicture}[baseline={(current bounding box.center)}, node distance=1cm and 1.5cm]
    \coordinate[circle,fill,inner sep=2pt] (aux1);
    \coordinate[left=of aux1,label=left:$\mu$] (e1);
    \coordinate[right=1.5cm of aux1] (aux2);
    \coordinate[above=of aux2,label=above:$\alpha_{2,1}$] (e2);
    \coordinate[right=1.5cm of aux2] (aux3);
    \coordinate[above=of aux3,label=above:$\alpha_1$] (e3);
    \coordinate[right=of aux3,label=right:$\alpha_0$] (e4);

    \draw[line,double] (aux1) -- (e1);
    \draw[line,dashed,red] (aux2) -- (e2);
    \draw[line] (aux3) -- (e3);
    \draw[line] (aux3) -- (e4);
    \draw[line] (aux1) -- node[label=below:$\alpha$] {} (aux2);
    \draw[line] (aux2) -- node[label=below:$\alpha_{\theta'}$] {} (aux3);
    \end{tikzpicture}\,.
\end{equation}
If one decomposes the correlator into conformal blocks in the intermediate region and near $\infty$, one obtains the crossing symmetry condition
\begin{equation}
\begin{aligned}
   \langle \mu, \Lambda| \Phi(z) V_1(1)|\Delta_0\rangle &= \int d \alpha \, C_{\alpha_1 \alpha_0}^\alpha \sum_{\theta=\pm} C_{\mu \alpha_\theta} C_{\alpha_{2,1} \alpha}^{\alpha_{\theta}}  \left|z^{-\Delta_{2,1}-\Delta_1-\Delta_0}\FIII{\mu}{\alpha_{ \theta}}{\alpha_{2,1}}{\alpha}{\alpha_1}{\alpha_0}{\Lambda z}{\frac{1}{z}}\right|^2=\\
   &= \int d \alpha \, C_{\alpha_1 \alpha_0}^\alpha \sum_{\theta'=\pm} C_{\mu_{\theta'} \alpha} B_{\alpha_{2,1} \mu}^{\mu_{\theta'}} \left|\DIII{\mu}{\alpha_{2,1}}{\mu_{\theta'}}{\alpha}{\alpha_1}{\alpha_0}{\Lambda}{\frac{1}{\Lambda z}} \right|^2\,.
\end{aligned}
\end{equation}
This condition is analogous to the one we found for the Whittaker functions \eqref{eq:Whittaker_constraint} so that the connection formula between the intermediate region and $\infty$ reads
\begin{equation}\label{Nf3fromintermediatetoinfty}
    b^{\theta b \alpha }z^{-\Delta_{2,1}-\Delta_1-\Delta_0}\FIII{\mu}{\alpha_{\theta}}{\alpha_{2,1}}{\alpha}{\alpha_1}{\alpha_0}{\Lambda z}{\frac{1}{z}} = \sum_{\theta'=\pm} b^{-\frac{1}{2}- \theta' b\mu}\mathcal{N}_{\theta \theta'}(b\alpha,b\mu) \DIII{\mu}{\alpha_{2,1}}{\mu_{\theta'}}{\alpha}{\alpha_1}{\alpha_0}{\Lambda}{\frac{1}{\Lambda z}}
\end{equation}
with irregular connection coefficients as in \eqref{connectionW}:
\begin{equation}
    \mathcal{N}_{\theta \theta'}(b\alpha, b\mu) = \frac{\Gamma(1+2\theta b \alpha)}{\Gamma\left(\frac{1}{2}+\theta b \alpha- \theta' b\mu\right)} e^{i\pi \left(\frac{1-\theta'}{2}\right)\left(\frac{1}{2}-b\mu+\theta b\alpha\right)}\,.
\end{equation}
In diagrams:
\begin{equation}
    \begin{tikzpicture}[baseline={(current bounding box.center)}, node distance=1cm and 1.5cm]
    \coordinate[circle,fill,inner sep=2pt] (aux1);
    \coordinate[left=of aux1,label=left:$\mu$] (e1);
    \coordinate[right=1.5cm of aux1] (aux2);
    \coordinate[above=of aux2,label=above:$\alpha_{2,1}$] (e2);
    \coordinate[right=1.5cm of aux2] (aux3);
    \coordinate[above=of aux3,label=above:$\alpha_1$] (e3);
    \coordinate[right=of aux3,label=right:$\alpha_0$] (e4);

    \draw[line,double] (aux1) -- (e1);
    \draw[line,dashed,red] (aux2) -- (e2);
    \draw[line] (aux3) -- (e3);
    \draw[line] (aux3) -- (e4);
    \draw[line] (aux1) -- node[label=below:$\alpha_\theta$] {} (aux2);
    \draw[line] (aux2) -- node[label=below:$\alpha$] {} (aux3);
    \end{tikzpicture} =  \sum_{\theta'=\pm} \mathcal{N}_{\theta \theta'}
    \begin{tikzpicture}[baseline={(current bounding box.center)}, node distance=1cm and 1.5cm]
    \coordinate (aux1);
    \coordinate[left=of aux1,label=left:$\mu$] (e1);
    \coordinate[right=1.5cm of aux1,circle,fill,inner sep=2pt] (aux2);
    \coordinate[above=of aux1,label=above:$\alpha_{2,1}$] (e2);
    \coordinate[right=1.5cm of aux2] (aux3);
    \coordinate[above=of aux3,label=above:$\alpha_1$] (e3);
    \coordinate[right=of aux3,label=right:$\alpha_0$] (e4);

    \draw[line,double] (aux1) -- (e1);
    \draw[line,dashed,red] (aux1) -- (e2);
    \draw[line] (aux3) -- (e3);
    \draw[line] (aux3) -- (e4);
    \draw[line,double] (aux1) -- node[label=below:$\mu_{\theta'}$] {} (aux2);
    \draw[line] (aux2) -- node[label=below:$\alpha$] {} (aux3);
    \end{tikzpicture}\,.
\end{equation}
Let us write explicitly the more interesting connection formula between $1$ and $\infty$, which is obtained by concatenating the two connection formulae above. Since the $\mathfrak{F}$ block in the intermediate region has different arguments in formula \eqref{Nf3from1tointermediate} and  \eqref{Nf3fromintermediatetoinfty}, we need to rename some of them. In the end we obtain the following connection formula from $1$ directly to $\infty$:
\begin{equation}\label{Nf3from1toinf}
\begin{aligned}
    &e^{\mu \Lambda} \FIII{-\mu}{\alpha}{\alpha_0}{\alpha_{1 \theta_1}}{\alpha_{2,1}}{\alpha_1}{\Lambda}{1-z} = \\ = 
    &\sum_{\theta_2,\theta_3=\pm} b^{-\frac{1}{2}+\theta_2 b \alpha_{\theta_2}- \theta_3 b\mu}\mathcal{M}_{\theta_1 \theta_2}(b\alpha_1,b\alpha;b\alpha_0)\mathcal{N}_{(-\theta_2) \theta_3}(b\alpha_{\theta_2}, b\mu)\DIII{\mu}{\alpha_{2,1}}{\mu_{\theta_3}}{\alpha_{\theta_2}}{\alpha_1}{\alpha_0}{\Lambda}{\frac{1}{\Lambda z}}\,.
\end{aligned}
\end{equation}
Again, in diagrams this is represented by:
\begin{equation}
    \begin{tikzpicture}[baseline={(current bounding box.center)}, node distance=1cm and 1.5cm]
    \coordinate[circle,fill,inner sep=2pt] (aux1);
    \coordinate[left=of aux1,label=left:$-\mu$] (e1);
    \coordinate[right=1.5cm of aux1] (aux2);
    \coordinate[above=of aux2,label=above:$\alpha_0$] (e2);
    \coordinate[right=1.5cm of aux2] (aux3);
    \coordinate[above=of aux3,label=above:$\alpha_{2,1}$] (e3);
    \coordinate[right=of aux3,label=right:$\alpha_1$] (e4);

    \draw[line,double] (aux1) -- (e1);
    \draw[line] (aux2) -- (e2);
    \draw[line,dashed,red] (aux3) -- (e3);
    \draw[line] (aux3) -- (e4);
    \draw[line] (aux1) -- node[label=below:$\alpha$] {} (aux2);
    \draw[line] (aux2) -- node[label=below:$\alpha_{1\theta_1}$] {} (aux3);
    \end{tikzpicture} = \sum_{\theta_2,\theta_3=\pm} \mathcal{M}_{\theta_1 \theta_2}\mathcal{N}_{(-\theta_2) \theta_3}
    \begin{tikzpicture}[baseline={(current bounding box.center)}, node distance=1cm and 1.5cm]
    \coordinate (aux1);
    \coordinate[left=of aux1,label=left:$\mu$] (e1);
    \coordinate[right=1.5cm of aux1,circle,fill,inner sep=2pt] (aux2);
    \coordinate[above=of aux1,label=above:$\alpha_{2,1}$] (e2);
    \coordinate[right=1.5cm of aux2] (aux3);
    \coordinate[above=of aux3,label=above:$\alpha_1$] (e3);
    \coordinate[right=of aux3,label=right:$\alpha_0$] (e4);

    \draw[line,double] (aux1) -- (e1);
    \draw[line,dashed,red] (aux1) -- (e2);
    \draw[line] (aux3) -- (e3);
    \draw[line] (aux3) -- (e4);
    \draw[line,double] (aux1) -- node[label=below:$\mu_{\theta_3}$] {} (aux2);
    \draw[line] (aux2) -- node[label=below:$\alpha_{\theta_2}$] {} (aux3);
    \end{tikzpicture}\,,
\end{equation}
where we have suppressed the arguments of the connection coefficients for brevity. \newline
\paragraph{Large $\Lambda$ blocks}
The conformal blocks considered up to now are expansions in $\Lambda$. One can however play the same game using expansions in $\frac{1}{\Lambda}$. For example, for large $\Lambda$ and for $z\sim 0$, we have
\begin{equation}
    \DIII{\mu}{\alpha_1}{\mu'}{\alpha_{0 \theta}}{\alpha_{2,1}}{\alpha_0}{\frac{1}{\Lambda}}{\Lambda z} = 
    \begin{tikzpicture}[baseline={(current bounding box.center)}, node distance=1cm and 1.5cm]
    \coordinate[] (aux1);
    \coordinate[left=of aux1,label=left:$\mu$] (e1);
    \coordinate[right=1.5cm of aux1,circle,fill,inner sep=2pt] (aux2);
    \coordinate[above=of aux1,label=above:$\alpha_1$] (e2);
    \coordinate[right=1.5cm of aux2] (aux3);
    \coordinate[above=of aux3,label=above:$\alpha_{2,1}$] (e3);
    \coordinate[right=of aux3,label=right:$\alpha_0$] (e4);

    \draw[line,double] (aux1) -- (e1);
    \draw[line] (aux1) -- (e2);
    \draw[line,dashed,red] (aux3) -- (e3);
    \draw[line] (aux3) -- (e4);
    \draw[line,double] (aux1) -- node[label=below:$\mu'$] {} (aux2);
    \draw[line] (aux2) -- node[label=below:$\alpha_{0\theta}$] {} (aux3);
    \end{tikzpicture} \,.
\end{equation}
One can compute it via OPE as in \eqref{irrOPE} or as a collision limit of a regular conformal block as proposed in \cite{Lisovyy:2018mnj}:
\begin{equation}
\begin{aligned}
    \DIII{\mu}{\alpha_1}{\mu'}{\alpha_{0 \theta}}{\alpha_{2,1}}{\alpha_0}{\frac{1}{\Lambda}}{\Lambda z} =& e^{-(\mu'-\mu)\Lambda} \Lambda^{\Delta_{0\theta}+2\mu'(\mu'-\mu)} z^{\frac{bQ}{2}+\theta b \alpha_0} \times \\ \times&
    \lim_{\eta\to\infty}\left(1-\frac{\eta}{\Lambda}\right)^{\Delta_1 - (\mu'-\mu)(\eta - \mu')} \FIVhat{\frac{\eta+\mu}{2}}{\alpha_1}{{\scriptstyle \frac{\eta-\mu}{2}+\mu'}}{\frac{\eta-\mu}{2}}{\alpha_{0\theta}}{\alpha_{2,1}}{\alpha_0}{\frac{\eta}{\Lambda}}{\frac{\Lambda z}{\eta}} \,.
\end{aligned}
\end{equation}
Similarly, we have a conformal block for large $\Lambda$ and $z\sim 1$, which as usual we can write in the same form as the one for $z\sim 0$ by doing a M\"obius transformation:
\begin{equation}\label{4222}
\begin{aligned}
    e^{\mu\Lambda} \DIII{-\mu}{\alpha_0}{\mu'-\mu}{\alpha_{1 \theta}}{\alpha_{2,1}}{\alpha_1}{\frac{1}{\Lambda}}{\Lambda(1-z)} &= 
    \begin{tikzpicture}[baseline={(current bounding box.center)}, node distance=1cm and 1.5cm]
    \coordinate[] (aux1);
    \coordinate[left=of aux1,label=left:$-\mu$] (e1);
    \coordinate[right=1.5cm of aux1,circle,fill,inner sep=2pt] (aux2);
    \coordinate[above=of aux1,label=above:$\alpha_0$] (e2);
    \coordinate[right=1.5cm of aux2] (aux3);
    \coordinate[above=of aux3,label=above:$\alpha_{2,1}$] (e3);
    \coordinate[right=of aux3,label=right:$\alpha_1$] (e4);

    \draw[line,double] (aux1) -- (e1);
    \draw[line] (aux1) -- (e2);
    \draw[line,dashed,red] (aux3) -- (e3);
    \draw[line] (aux3) -- (e4);
    \draw[line,double] (aux1) -- node[label=below:$\mu'-\mu$] {} (aux2);
    \draw[line] (aux2) -- node[label=below:$\alpha_{1\theta}$] {} (aux3);
    \end{tikzpicture}=\\
    &=\begin{tikzpicture}[baseline={(current bounding box.center)}, node distance=1cm and 1.5cm]
    \coordinate[] (aux1);
    \coordinate[left=of aux1,label=left:$\mu$] (e1);
    \coordinate[right=1.5cm of aux1,circle,fill,inner sep=2pt] (aux2);
    \coordinate[above=2 cm of aux1,label=above:$\alpha_1$] (e2);
    \coordinate[above=1 cm of aux1] (aux3);
    \coordinate[right=of aux3,label=right:$\alpha_{2,1}$] (e3);
    \coordinate[right=of aux2,label=right:$\alpha_0$] (e4);

    \draw[line,double] (aux1) -- (e1);
    \draw[line] (aux1) -- node[label=left:$\alpha_{1\theta}$] {} (aux3);
    \draw[line,dashed,red] (aux3) -- (e3);
    \draw[line] (aux2) -- (e4);
    \draw[line,double] (aux1) -- node[label=below:$\mu'$] {} (aux2);
    \draw[line] (aux3) --  (e2);
    \end{tikzpicture}\,.
\end{aligned}
\end{equation}
The first line of \eqref{4222} is the diagrammatic representation of the conformal block, while the second line is an equality of two a priori seemingly different conformal blocks, which can be checked by explicit computation. This is consistent with the fact that the corresponding DOZZ factors are equal:
\begin{equation}
    B_{-\mu\alpha_0}^{\mu'-\mu}C_{\mu'-\mu,\alpha_{1\theta}}=B_{\mu\alpha_{1\theta}}^{\mu'}C_{\mu' ,\alpha_0}\,,
\end{equation}
as can easily be proven by using their explicit expressions given in Appendix \ref{app:rank1collision}. The most exotic block is the one for large $\Lambda$ and large $z$, which by a slight abuse of notation we still denote by $\mathfrak{D}$:
\begin{equation}
\begin{aligned}\label{eq:largeDblock}
    \DIIIlargeLambda{\mu}{\alpha_{2,1}}{\mu_\theta }{\alpha_1}{\mu'}{\alpha_0}{\frac{1}{\Lambda}}{\frac{1}{z}} &= 
    \begin{tikzpicture}[baseline={(current bounding box.center)}, node distance=1cm and 1.5cm]
    \coordinate[] (aux1);
    \coordinate[left=of aux1,label=left:$\mu$] (e1);
    \coordinate[right=1.5cm of aux1] (aux2);
    \coordinate[above=of aux1,label=above:$\alpha_{2,1}$] (e2);
    \coordinate[right=1.5cm of aux2,circle,fill,inner sep=2pt] (aux3);
    \coordinate[above=of aux2,label=above:$\alpha_1$] (e3);
    \coordinate[right=of aux3,label=right:$\alpha_0$] (e4);

    \draw[line,double] (aux1) -- (e1);
    \draw[line,dashed,red] (aux1) -- (e2);
    \draw[line] (aux2) -- (e3);
    \draw[line] (aux3) -- (e4);
    \draw[line,double] (aux1) -- node[label=below:$\mu_\theta$] {} (aux2);
    \draw[line,double] (aux2) -- node[label=below:$\mu'$] {} (aux3);
    \end{tikzpicture} \,.
\end{aligned}
\end{equation}
This block is fully irregular in the sense that to calculate it, we have to perform two irregular OPEs as indicated by the diagram. It is more convenient to calculate it as a collision limit of a regular block:
\begin{equation}
\begin{aligned}\label{eq:bigCollision}
    &\DIIIlargeLambda{\mu}{\alpha_{2,1}}{\mu_\theta }{\alpha_1}{\mu'}{\alpha_0}{\frac{1}{\Lambda}}{\frac{1}{z}} =e^{\theta b\Lambda z/2} \Lambda^{\Delta_{2,1}} \left( \Lambda z\right)^{-\theta b \mu + \frac{b^2}{2}}  e^{-(\mu'-\mu_\theta)\Lambda} \Lambda^{\Delta_0+\Delta_1+2\mu'(\mu'-\mu_\theta)} \times \\ \times&
    \lim_{\eta\to\infty}\left(1-\frac{\eta}{\Lambda z}\right)^{\Delta_{2,1} - (\mu_\theta-\mu)(\eta - \mu_\theta)}\left(1-\frac{\eta}{\Lambda}\right)^{\Delta_1 - (\mu'-\mu_\theta)(\eta - \mu')-(\mu'-\mu_\theta)(\mu_\theta-\mu)} \FIVhat{\frac{\eta+\mu}{2}}{\alpha_{2,1}}{{\scriptstyle \frac{\eta-\mu}{2}+\mu_\theta}}{\alpha_1}{{\scriptstyle \frac{\eta-\mu}{2}+\mu'}}{\frac{\eta-\mu}{2}}{\alpha_0}{\frac{\eta}{\Lambda}}{\frac{\Lambda z}{\eta}} \,.
\end{aligned}
\end{equation}
Having defined all the necessary conformal blocks we now derive their connection formulae. Let us start by connecting $z\sim1$ with $\infty$. Expanding the correlator in these regions, we get the crossing symmetry condition
\begin{equation}
\begin{aligned}
    \langle \mu,\Lambda|\Phi(z) V_1(1)|\Delta_0\rangle &=  \int \mathrm{d}\mu' \sum_{\theta=\pm} B_{-\mu\alpha_0}^{\mu'-\mu} C_{\mu'-\mu, \alpha_{1\theta}} C_{\alpha_1 \alpha_{2,1}}^{\alpha_{1\theta}}\left|e^{\mu\Lambda}\, \DIII{-\mu}{\alpha_0}{\mu'-\mu}{\alpha_{1 \theta}}{\alpha_{2,1}}{\alpha_1}{\frac{1}{\Lambda}}{\Lambda(1-z)}\right|^2 = \\
    &= \int \mathrm{d}\mu'\sum_{\theta'=\pm}B^{\mu_{\theta'}}_{\mu\alpha_{2,1}} B^{\mu'}_{\mu_{\theta'}\alpha_1}C_{\mu',\alpha_0} \left|\DIIIlargeLambda{\mu}{\alpha_{2,1}}{\mu_{\theta'} }{\alpha_1}{\mu'}{\alpha_0}{\frac{1}{\Lambda}}{\frac{1}{z}}\right|^2\,.
\end{aligned}
\end{equation}
Using the following remarkable identity, which can easily be proven using the explicit expression of the structure functions given in Appendix \ref{app:rank1collision},
\begin{equation}\label{eq:magicidentity}
    B^{\mu_{\theta'}}_{\mu\alpha_{2,1}} B^{\mu'}_{\mu_{\theta'}\alpha_1}C_{\mu' \alpha_0}=B^{\mu'-\mu}_{-\mu \alpha_0}B^{\mu'-\mu_{\theta'}}_{\mu'-\mu,\alpha_{2,1}}C_{\mu'-\mu_{\theta'},\alpha_1}  \,,
\end{equation}
we find that the above crossing symmetry condition (after relabelling the dummy variable $\theta' \to - \theta'$) becomes:
\begin{equation}
\begin{aligned}
    \langle \mu,\Lambda|\Phi(z) V_1(1)|\Delta_0\rangle &=  \int \mathrm{d}\mu' B_{-\mu\alpha_0}^{\mu'-\mu}\sum_{\theta=\pm} C_{\mu'-\mu, \alpha_{1\theta}} C_{\alpha_1 \alpha_{2,1}}^{\alpha_{1\theta}}\left|e^{\mu\Lambda}\, \DIII{-\mu}{\alpha_0}{\mu'-\mu}{\alpha_{1 \theta}}{\alpha_{2,1}}{\alpha_1}{\frac{1}{\Lambda}}{\Lambda(1-z)}\right|^2 = \\
    &= \int \mathrm{d}\mu' B^{\mu'-\mu}_{-\mu \alpha_0}\sum_{\theta'=\pm} B^{\mu'_{\theta'}-\mu}_{\mu'-\mu,\alpha_{2,1}}C_{\mu'_{\theta'}-\mu,\alpha_1}\left|\DIIIlargeLambda{\mu}{\alpha_{2,1}}{\mu_{-\theta'} }{\alpha_1}{\mu'}{\alpha_0}{\frac{1}{\Lambda}}{\frac{1}{z}}\right|^2\,.
\end{aligned}
\end{equation}
We recognize this constraint from the Whittaker functions \eqref{eq:WhittakerCC}, and can readily write the connection formula from $1$ to $\infty$:
\begin{equation}\label{1toinftylargeLambda}
    b^{\theta b \alpha_1}e^{\mu\Lambda} \DIII{-\mu}{\alpha_0}{\mu'-\mu}{\alpha_{1 \theta}}{\alpha_{2,1}}{\alpha_1}{\frac{1}{\Lambda}}{\Lambda(1-z)} =\sum_{\theta'}b^{-\frac{1}{2}+ \theta' b(\mu'-\mu)}  \mathcal{N}_{\theta(-\theta')}(b\alpha_1,b\mu'-b\mu) \DIIIlargeLambda{\mu}{\alpha_{2,1}}{\mu_{\theta'} }{\alpha_1}{\mu'}{\alpha_0}{\frac{1}{\Lambda}}{\frac{1}{z}}\,,
\end{equation}
where $\mathcal{N}$ are the connection coefficients for the Whittaker functions \eqref{eq:WhittakerCC}.
Diagrammatically this is clear:
\begin{equation}
    \begin{tikzpicture}[baseline={(current bounding box.center)}, node distance=1cm and 1.5cm]
    \coordinate[] (aux1);
    \coordinate[left=of aux1,label=left:$\mu$] (e1);
    \coordinate[right=1.5cm of aux1,circle,fill,inner sep=2pt] (aux2);
    \coordinate[above=2 cm of aux1,label=above:$\alpha_1$] (e2);
    \coordinate[above=1 cm of aux1] (aux3);
    \coordinate[left=of aux3,label=left:$\alpha_{2,1}$] (e3);
    \coordinate[right=of aux2,label=right:$\alpha_0$] (e4);

    \draw[line,double] (aux1) -- (e1);
    \draw[line] (aux1) -- node[label=left:$\alpha_{1\theta}$] {} (aux3);
    \draw[line,dashed,red] (aux3) -- (e3);
    \draw[line] (aux2) -- (e4);
    \draw[line,double] (aux1) -- node[label=below:$\mu'$] {} (aux2);
    \draw[line] (aux3) --  (e2);
    \end{tikzpicture} = \sum_{\theta'=\pm}  \mathcal{N}_{\theta(-\theta')}\begin{tikzpicture}[baseline={(current bounding box.center)}, node distance=1cm and 1.5cm]
    \coordinate[] (aux1);
    \coordinate[left=of aux1,label=left:$\mu$] (e1);
    \coordinate[right=1.5cm of aux1] (aux2);
    \coordinate[above=of aux1,label=above:$\alpha_{2,1}$] (e2);
    \coordinate[right=1.5cm of aux2,circle,fill,inner sep=2pt] (aux3);
    \coordinate[above=of aux2,label=above:$\alpha_1$] (e3);
    \coordinate[right=of aux3,label=right:$\alpha_0$] (e4);

    \draw[line,double] (aux1) -- (e1);
    \draw[line,dashed,red] (aux1) -- (e2);
    \draw[line] (aux2) -- (e3);
    \draw[line] (aux3) -- (e4);
    \draw[line,double] (aux1) -- node[label=below:$\mu_{\theta'}$] {} (aux2);
    \draw[line,double] (aux2) -- node[label=below:$\mu'$] {} (aux3);
    \end{tikzpicture} \,.
\end{equation}
To connect $0$ and $\infty$ we expand the correlator in the relevant  regions. By crossing symmetry we have:
\begin{equation}
\begin{aligned}
    \langle \mu,\Lambda|V_1(1)\Phi(z)|\Delta_0\rangle &= \int \mathrm{d}\mu'  \sum_{\theta=\pm} B_{\mu \alpha_1}^{\mu'} C_{\mu' \alpha_{0\theta}} C_{\alpha_{2,1} \alpha_0}^{\alpha_{0\theta}} \left|\DIII{\mu}{\alpha_1}{\mu'}{\alpha_{0 \theta}}{\alpha_{2,1}}{\alpha_0}{\frac{1}{\Lambda}}{\Lambda z} \right|^2=\\
    &=  \int \mathrm{d}\mu'\sum_{\theta'=\pm}B^{\mu_{\theta'}}_{\mu\alpha_{2,1}} B^{\mu'_{\theta'}}_{\mu_{\theta'}\alpha_1}C_{\mu'_{\theta'},\alpha_0} \left|\DIIIlargeLambda{\mu}{\alpha_{2,1}}{\mu_{\theta'} }{\alpha_1}{\mu'_{\theta'}}{\alpha_0}{\frac{1}{\Lambda}}{\frac{1}{z}}\right|^2\,,
\end{aligned}
\end{equation}
for later convenience we have labelled the intermediate channel in the second line by $\mu'_{\theta'}$ instead of $\mu'$. By using an identity similar to \eqref{eq:magicidentity}:
\begin{equation}\label{magicidentity2}
    B^{\mu_{\theta'}}_{\mu\alpha_{2,1}} B^{\mu'_{\theta'}}_{\mu_{\theta'}\alpha_1}C_{\mu'_{\theta'},\alpha_0}=B_{\mu \alpha_1}^{\mu'} B_{\mu' \alpha_{2,1}}^{\mu'_{\theta'}}C_{\mu'_{\theta'}\alpha_0}\,,
\end{equation}
the above crossing symmetry equation then becomes:
\begin{equation}
\begin{aligned}
    \langle \mu,\Lambda|V_1(1)\Phi(z)|\Delta_0\rangle &= \int \mathrm{d}\mu' B_{\mu \alpha_1}^{\mu'} \sum_{\theta=\pm} C_{\mu' \alpha_{0\theta}} C_{\alpha_{2,1} \alpha_0}^{\alpha_{0\theta}} \left|\DIII{\mu}{\alpha_1}{\mu'}{\alpha_{0 \theta}}{\alpha_{2,1}}{\alpha_0}{\frac{1}{\Lambda}}{\Lambda z} \right|^2=\\
    &= \int \mathrm{d}\mu' B_{\mu \alpha_1}^{\mu'} \sum_{\theta'=\pm} B_{\mu' \alpha_{2,1}}^{\mu'_{\theta'}}C_{\mu'_{\theta'}\alpha_0}\left|\DIIIlargeLambda{\mu}{\alpha_{2,1}}{\mu_{\theta'} }{\alpha_1}{\mu'_{\theta'}}{\alpha_0}{\frac{1}{\Lambda}}{\frac{1}{z}}\right|^2\,.
\end{aligned}
\end{equation}
We recognize this constraint from the Whittaker functions \eqref{eq:Whittaker_constraint} and can readily write the connection formula from $0$ to $\infty$:
\begin{equation}\label{4234}
    b^{\theta b \alpha_0}\DIII{\mu}{\alpha_1}{\mu'}{\alpha_{0 \theta}}{\alpha_{2,1}}{\alpha_0}{\frac{1}{\Lambda}}{\Lambda z} = \sum_{\theta'=\pm} b^{-\frac{1}{2}- \theta' b\mu'}  \mathcal{N}_{\theta \theta'}(b\alpha_0,b\mu') \DIIIlargeLambda{\mu}{\alpha_{2,1}}{\mu_{\theta'} }{\alpha_1}{\mu'_{\theta'}}{\alpha_0}{\frac{1}{\Lambda}}{\frac{1}{z}}\,.
\end{equation}
Combining \eqref{4234} with the inverse of \eqref{1toinftylargeLambda} we obtain the connection formula from $0$ to $1$:
\begin{equation}\label{0to1largeLambda}
\begin{aligned}
    &b^{\theta_1 b \alpha_0}\DIII{\mu}{\alpha_1}{\mu'}{\alpha_{0 \theta_1}}{\alpha_{2,1}}{\alpha_0}{\frac{1}{\Lambda}}{\Lambda z} =\\=& \sum_{\theta_2,\theta_3=\pm} b^{-\frac{1}{2}-\theta_2 b \mu'} \mathcal{N}_{\theta_1 \theta_2}(b\alpha_0,b\mu')b^{\frac{1}{2}-\theta_2 b(\mu'_{\theta_2}-\mu)+\theta_3 b \alpha_1}\mathcal{N}^{-1}_{(-\theta_2)\theta_3}(b\mu'_{\theta_2}-b\mu,b\alpha_1)e^{\mu\Lambda}\DIII{-\mu}{\alpha_0}{\mu'_{\theta_2}-\mu}{\alpha_{1 \theta_3}}{\alpha_{2,1}}{\alpha_1}{\frac{1}{\Lambda}}{\Lambda(1-z)}\,.
\end{aligned}
\end{equation}
Diagrammatically:
\begin{equation}
    \begin{tikzpicture}[baseline={(current bounding box.center)}, node distance=1cm and 1.5cm]
    \coordinate[] (aux1);
    \coordinate[left=of aux1,label=left:$\mu$] (e1);
    \coordinate[right=1.5cm of aux1,circle,fill,inner sep=2pt] (aux2);
    \coordinate[above=of aux1,label=above:$\alpha_1$] (e2);
    \coordinate[right=1.5cm of aux2] (aux3);
    \coordinate[above=of aux3,label=above:$\alpha_{2,1}$] (e3);
    \coordinate[right=of aux3,label=right:$\alpha_0$] (e4);

    \draw[line,double] (aux1) -- (e1);
    \draw[line] (aux1) -- (e2);
    \draw[line,dashed,red] (aux3) -- (e3);
    \draw[line] (aux3) -- (e4);
    \draw[line,double] (aux1) -- node[label=below:$\mu'$] {} (aux2);
    \draw[line] (aux2) -- node[label=below:$\alpha_{0\theta_1}$] {} (aux3);
    \end{tikzpicture} = \sum_{\theta_2,\theta_3=\pm}  \mathcal{N}_{\theta_1 \theta_2}\mathcal{N}^{-1}_{(-\theta_2)\theta_3}\begin{tikzpicture}[baseline={(current bounding box.center)}, node distance=1cm and 1.5cm]
    \coordinate[] (aux1);
    \coordinate[left=of aux1,label=left:$\mu$] (e1);
    \coordinate[right=1.5cm of aux1,circle,fill,inner sep=2pt] (aux2);
    \coordinate[above=2 cm of aux1,label=above:$\alpha_1$] (e2);
    \coordinate[above=1 cm of aux1] (aux3);
    \coordinate[right=of aux3,label=right:$\alpha_{2,1}$] (e3);
    \coordinate[right=of aux2,label=right:$\alpha_0$] (e4);

    \draw[line,double] (aux1) -- (e1);
    \draw[line] (aux1) -- node[label=left:$\alpha_{1\theta}$] {} (aux3);
    \draw[line,dashed,red] (aux3) -- (e3);
    \draw[line] (aux2) -- (e4);
    \draw[line,double] (aux1) -- node[label=below:$\mu'_{\theta_2}$] {} (aux2);
    \draw[line] (aux3) --  (e2);
    \end{tikzpicture}\,.
\end{equation}
One might expect the existence of conformal blocks expanded in an intermediate region, as was the case for small $\Lambda$. Indeed, in the case of large $\Lambda$ one can define a block expanded in the intermediate region $\frac{1}{\Lambda} \ll z \ll 1$. However, by the identity \eqref{magicidentity2}, this block is actually the same as the block \eqref{eq:largeDblock} corresponding to $z\sim \infty$, in the sense that the analytic continuation between the two is trivial. Similarly, one can define another intermediate block in the region $\frac{1}{\Lambda} \ll 1-z \ll 1$ which is also the same as \eqref{eq:largeDblock} by virtue of the identity \eqref{eq:magicidentity}.

\subsubsection{Semiclassical limit}\label{Nf3SC}
In the semiclassical limit $b\to 0$ and $\alpha_i,\mu,\Lambda \to \infty$ such that $a_i=b\alpha_i,\,m=b\mu,\,L=b\Lambda$ are finite. We denote the quantities which are finite in the semiclassical limit by latin letters instead of greek ones.
\paragraph{Small $L$ blocks}
The conformal blocks in this limit are expected to exponentiate, and the $z$-dependence becomes subleading: schematically they take the form
\begin{equation}
    \mathfrak{F}(\Lambda,z) \sim e^{\frac{1}{b^2}F(L)+W(L,z)+\mathcal{O}(b^2)}\,,
\end{equation}
and they diverge in this limit. The classical conformal block $F(L)$ is related to the conformal block $\mathfrak{F}$ without the degenerate field insertion, i.e.
\begin{equation}
    {}_1\mathfrak{F} \left( \mu \, \alpha\, \begin{matrix} \alpha_1\\ \alpha_0 \end{matrix} ;\Lambda \right) = \Lambda^{\Delta} e^{\frac{1}{b^2}\left(F(L)+\mathcal{O}(b^2)\right)}\,.
\label{eq:cNf34pointsmall}
\end{equation}
Normalizing by this block, we obtain finite semiclassical conformal blocks. Consider for concreteness the block corresponding to the expansion for $z\sim0$. We define the corresponding (finite) semiclassical conformal block by
\begin{equation}
    \FIIIsc{m}{a}{a_1}{a_{0 \theta}}{a_{2,1}}{a_0}{L}{z} = \lim_{b\to 0} \frac{\FIII{\mu}{\alpha}{\alpha_1}{\alpha_{0 \theta}}{\alpha_{2,1}}{\alpha_0}{\Lambda}{z}}{{}_1\mathfrak{F} \left( \mu \, \alpha\, \begin{matrix} \alpha_1\\ \alpha_0 \end{matrix} ;\Lambda \right)} = e^{-\frac{\theta}{2}\partial_{a_0}F}z^{\frac{1}{2}+\theta a_0}(1+\mathcal{O}(L,z))  \,.
\end{equation}
The term $\exp{-\frac{\theta}{2}\partial_{a_0}F}$ on the RHS of the above equation comes from the fact that the leading behaviour of the numerator is $\exp{b^{-2}F(a_{0\theta})}$ while the denominator behaves as $\exp{b^{-2}F(a_0)}$. The fact that the $z$-dependence is subleading means that to leading order, the $\Lambda$-derivative in the BPZ equation \eqref{BPZNf3} becomes $z$-independent, since we have $\Lambda\partial_\Lambda \mathfrak{F}(\Lambda,z) \sim b^{-2} \Lambda\partial_\Lambda F(\Lambda) \mathfrak{F}(\Lambda,z)$. Then the BPZ equation in the semiclassical limit reduces to an ODE. In particular, multiplying \eqref{BPZNf3} by $b^2$, this semiclassical conformal block now satisfies the equation
\begin{equation}\label{BPZNf3NS}
\begin{aligned}
    \left(\partial_z^2 + \frac{u -\frac{1}{2}+ a_0^2 +a_1^2}{z(z-1)} + \frac{\frac{1}{4}-a_1^2}{(z-1)^2} + \frac{\frac{1}{4}-a_0^2}{z^2} + \frac{ m L}{z} -\frac{L^2}{4}\right) \FIIIsc{m}{a}{a_1}{a_{0 \theta}}{a_{2,1}}{a_0}{L}{z} =0\,.
\end{aligned}
\end{equation}
We have introduced 
\begin{equation}
    u=\lim_{b\to0} b^2 \Lambda \partial_\Lambda \log {}_1\mathfrak{F} \left( \mu \, \alpha\, \begin{matrix} \alpha_1\\ \alpha_0 \end{matrix} ;\Lambda \right) = \frac{1}{4}-a^2+ \mathcal{O}(L)
\label{eq:uelectricNf3confl}
\end{equation}
Similarly, we define the semiclassical block for $z\sim 1$ to be
\begin{equation}
\begin{aligned}
    &\FIIIsc{-m}{a}{a_0}{a_{1 \theta}}{a_{2,1}}{a_1}{L}{1-z} =\lim_{b\to0}\frac{e^{\mu \Lambda}\FIII{-\mu}{\alpha}{\alpha_0}{\alpha_{1 \theta}}{\alpha_{2,1}}{\alpha_1}{\Lambda}{1-z} }{{}_1\mathfrak{F} \left( \mu \, \alpha\, \begin{matrix} \alpha_1\\ \alpha_0 \end{matrix} ;\Lambda \right)} =\\ &=\lim_{b\to0}\frac{\FIII{-\mu}{\alpha}{\alpha_0}{\alpha_{1 \theta}}{\alpha_{2,1}}{\alpha_1}{\Lambda}{1-z} }{{}_1\mathfrak{F} \left( -\mu\, \alpha\, \begin{matrix} \alpha_0\\ \alpha_1 \end{matrix} ;\Lambda \right)}= e^{-\frac{\theta}{2}\partial_{a_1}F}(1-z)^{\frac{1}{2}+\theta a_1}(1+\mathcal{O}(L,1-z)) \,,
\end{aligned}
\end{equation}
and in the deep irregular region:
\begin{equation}
    \DIIIsc{m}{\,a_{2,1}}{m_{\theta}}{a}{a_1}{a_0}{L}{\frac{1}{L z}} =  \lim_{b\to 0} b^{-\frac{1}{2}-\theta m}\frac{\DIII{\mu}{\alpha_{2,1}}{\mu_{\theta}}{\alpha}{\alpha_1}{\alpha_0}{\Lambda}{\frac{1}{\Lambda z}}}{{}_1\mathfrak{F} \left(\mu \, \alpha\, \begin{matrix} \alpha_1\\ \alpha_0 \end{matrix} ;\Lambda \right)} = e^{-\frac{\theta}{2}\partial_{m}F} e^{\theta L z/2} L^{-\frac{1}{2}-\theta m}  z^{-\theta m}(1+\mathcal{O}(L,1/Lz))
    \,.
\end{equation}
The explicit power of $b$ is needed to combine with $\Lambda$ to form $L$. All these blocks satisfy the same equation \eqref{BPZNf3NS}. Note that in the connection formula \eqref{Nf3from1toinf} we have four different conformal blocks on the right hand side. Since in the semiclassical limit the BPZ equation becomes a second-order ODE, these four different blocks have to reduce to the two linearly independent solutions near the irregular singular point. They are given by
\begin{equation}
    \DIII{\mu}{\alpha_{2,1}}{\mu_{\theta}}{\alpha}{\alpha_1}{\alpha_0}{\Lambda}{\frac{1}{\Lambda z}} =  e^{\theta b\Lambda z/2} \Lambda^{\Delta_{2,1}+\Delta} \left( \Lambda z\right)^{-\theta b \mu + \frac{b^2}{2}} e^{\frac{1}{b^2}F(a)+W(a)+\mathcal{O}(b^2)}\,,
\end{equation}
where we have suppressed the dependence of $F$ and $W$ on the other parameters. Instead, in \eqref{Nf3from1toinf} we have
\begin{equation}
    \DIII{\mu}{\alpha_{2,1}}{\mu_{\theta}}{\alpha_{\theta'}}{\alpha_1}{\alpha_0}{\Lambda}{\frac{1}{\Lambda z}} =  e^{\theta b\Lambda z/2} \Lambda^{\Delta_{2,1}+\Delta_{\theta'}} \left( \Lambda z\right)^{-\theta b \mu + \frac{b^2}{2}} e^{\frac{1}{b^2}F(a_{\theta'})+W(a_{\theta'})+\mathcal{O}(b^2)}\,.
\end{equation}
Since we are taking the limit $b\to0$, we can safely substitute $W(a_{\theta'})\to W(a)$. This is not true for $F(a_{\theta'})$ however, since it multiplies a pole in $b^2$. Instead, in the semiclassical limit we have
\begin{equation}
     \DIII{\mu}{\alpha_{2,1}}{\mu_{\theta}}{\alpha_{\theta'}}{\alpha_1}{\alpha_0}{\Lambda}{\frac{1}{\Lambda z}} \sim \Lambda^{\theta' a} e^{-\frac{\theta'}{2}\partial_{a}F(a)} \DIII{\mu}{\alpha_{2,1}}{\mu_{\theta}}{\alpha}{\alpha_1}{\alpha_0}{\Lambda}{\frac{1}{\Lambda z}}\,,\quad \mathrm{as } \, b\to 0\,,
\end{equation}
as in \eqref{twiddle}. Therefore, we can simplify the connection formula from $1$ to $\infty$ \eqref{Nf3from1toinf} in the semiclassical limit and state it as
\begin{equation}\label{Nf31toinfSC}
    \FIIIsc{-m}{a}{a_0}{a_{1 \theta}}{a_{2,1}}{a_1}{L}{1-z} = \sum_{\theta'}\left(\sum_{\sigma=\pm} \mathcal{M}_{\theta \sigma}(a_1,a;a_0)\mathcal{N}_{(-\sigma) \theta'}(a, m)L^{\sigma a} e^{-\frac{\sigma}{2}\partial_{a}F}\right) \DIIIsc{m}{a_{2,1}}{m_{\theta'}}{a}{a_1}{a_0}{L}{\frac{1}{L z}}\,,
\end{equation}
with connection coefficients
\begin{equation}
     \sum_{\sigma=\pm}\mathcal{M}_{\theta \sigma}(a_1,a;a_0)\mathcal{N}_{(-\sigma) \theta'}(a, m)L^{\sigma a} e^{-\frac{\sigma}{2}\partial_{a}F} = \sum_{\sigma = \pm} \frac{\Gamma(1-2\sigma a)\Gamma(-2\sigma a)\Gamma(1+2\theta a_1)e^{i\pi \left(\frac{1-\theta'}{2}\right)\left(\frac{1}{2}-m-\sigma a\right)}L^{\sigma a} e^{-\frac{\sigma}{2}\partial_{a}F}}{\Gamma\left(\frac{1}{2}+\theta a_1-\sigma a+a_0\right)\Gamma\left(\frac{1}{2}+\theta a_1-\sigma a-a_0\right)\Gamma\left(\frac{1}{2}-\sigma a-\theta'm\right)}\,.
\end{equation}
Note that all the powers of $b$ appearing in \eqref{Nf3from1toinf} have been absorbed to give finite quantities.\footnote{Note also that the Gamma functions in the denominator precisely correspond to the one-loop factors of the three hypermultiplets of the corresponding AGT dual gauge theory.}\newline
The connection formula from $0$ to $1$ trivially reduces to the semiclassical one:
\begin{equation}\label{Nf3from0to1SC}
    \FIIIsc{m}{a}{a_1}{a_{0 \theta}}{a_{2,1}}{a_0}{L}{z} = \sum_{\theta'=\pm}\mathcal{M}_{\theta \theta'}(a_0,a_1;a) \FIIIsc{-m}{a}{a_0}{a_{1 \theta'}}{a_{2,1}}{a_1}{L}{1-z}\,.
\end{equation}
\paragraph{Large $L$ blocks}
For the conformal blocks valid for large $\Lambda$, the story is analogous. Taking the semiclassical limit, the conformal blocks are expected to exponentiate and the $z$-dependence becomes subleading. Schematically we have
\begin{equation}
    \mathfrak{D}(\Lambda^{-1},z) \sim e^{\frac{1}{b^2}F_D(L^{-1})+W_D(L^{-1},z)+\mathcal{O}(b^2)}\,.
\end{equation}
Here $F_D$ is the classical conformal block for large \footnote{As the notation suggests, it is nothing else but the dual prepotential of the gauge theory.} $\Lambda$ and is related to the conformal block without the degenerate field insertion, i.e.
\begin{equation}\label{FD}
    {}_1 \mathfrak{D}\left(\mu\,\begin{matrix}\alpha_1\\ {}\end{matrix}\,\mu'\, \alpha_0; \frac{1}{\Lambda}\right) = e^{-(\mu'-\mu)\Lambda}\Lambda^{\Delta_0+\Delta_1+2\mu'(\mu'-\mu)} e^{\frac{1}{b^2}(F_D(L^{-1})+\mathcal{O}(b^2))}\,.
\end{equation}
We use this block as a normalization for large $\Lambda$. For $z\sim 0$ we have
\begin{equation}\label{132}
    \DIIIsc{m}{a_1}{m'}{a_{0 \theta}}{a_{2,1}}{a_0}{\frac{1}{L}}{L z} = \lim_{b\to0} b^{\theta a_0} \frac{\DIII{\mu}{\alpha_1}{\mu'}{\alpha_{0 \theta}}{\alpha_{2,1}}{\alpha_0}{\frac{1}{\Lambda}}{\Lambda z}}{{}_1 \mathfrak{D}\left(\mu\,\begin{matrix}\alpha_1\\ {}\end{matrix}\,\mu'\, \alpha_0; \frac{1}{\Lambda}\right) } = L^{\theta a_0} e^{-\frac{\theta}{2}\partial_{a_0}F_D} z^{\frac{1}{2}+\theta a_0}(1+\mathcal{O}(L^{-1},Lz))\,.
\end{equation}
This block and all the other large-$L$ blocks defined in the following satisfy the same equation \eqref{BPZNf3NS} as the small-$L$ blocks, with the substitution 
\begin{equation}
    u \to u_D =\lim_{b\to0} b^2 \Lambda \partial_\Lambda \log {}_1 \mathfrak{D}\left(\mu\,\begin{matrix}\alpha_1\\ {}\end{matrix}\,\mu'\, \alpha_0; \frac{1}{\Lambda}\right)\,.
\end{equation} For $z\sim 1$ we have the block
\begin{equation}
\begin{aligned}\label{133}
    &\DIIIsc{-m}{a_0}{m'-m}{a_{1 \theta}}{a_{2,1}}{a_1}{\frac{1}{L}}{L(1-z)} = \lim_{b\to 0} b^{\theta a_1}\frac{e^{\mu\Lambda} \DIII{-\mu}{\alpha_0}{\mu'-\mu}{\alpha_{1 \theta}}{\alpha_{2,1}}{\alpha_1}{\frac{1}{\Lambda}}{\Lambda(1-z)}}{{}_1 \mathfrak{D}\left(\mu\,\begin{matrix}\alpha_1\\ {}\end{matrix}\,\mu'\, \alpha_0; \frac{1}{\Lambda}\right) } = \\
    &= \lim_{b\to 0} \frac{\DIII{-\mu}{\alpha_0}{\mu'-\mu}{\alpha_{1 \theta}}{\alpha_{2,1}}{\alpha_1}{\frac{1}{\Lambda}}{\Lambda(1-z)}}{{}_1 \mathfrak{D}\left(-\mu\,\begin{matrix}\alpha_0\\ {}\end{matrix}\,\mu'-\mu\, \alpha_1; \frac{1}{\Lambda}\right) }=L^{\theta a_1}e^{-\frac{\theta}{2}\partial_{a_1}F_D}(1-z)^{\frac{1}{2}+\theta a_1}(1+\mathcal{O}(L^{-1},L(1-z))) \,,
\end{aligned}
\end{equation}
and for $z\sim \infty$:
\begin{equation}\label{134}
\begin{aligned}
    \DIIIlargeLambdasc{m}{a_{2,1}}{m_\theta }{a_1}{m'}{a_0}{\frac{1}{L}}{\frac{1}{z}} &= \lim_{b\to0} b^{-\frac{1}{2}+\theta (m'-m)} \frac{\DIIIlargeLambda{\mu}{\alpha_{2,1}}{\mu_\theta }{\alpha_1}{\mu'}{\alpha_0}{\frac{1}{\Lambda}}{\frac{1}{z}}}{{}_1 \mathfrak{D}\left(\mu\,\begin{matrix}\alpha_1\\ {}\end{matrix}\,\mu'\, \alpha_0; \frac{1}{\Lambda}\right)} =\\&= e^{\theta Lz/2}e^{-\theta L/2}e^{-\frac{\theta}{2}\partial_{m}F_D}L^{-\frac{1}{2}+\theta(m'-m)} z^{-\theta m}(1+\mathcal{O}(L^{-1},z^{-1})) \,.
\end{aligned}
\end{equation}
In the connection formula from $0$ to $1$ for large $\Lambda$ \eqref{0to1largeLambda}, there appear four different conformal blocks on the right hand side. In the semiclassical limit these four reduce to two, by the same argument as for small $\Lambda$. Indeed we have
\begin{equation}
    \begin{aligned}
        &e^{\mu\Lambda}\DIII{-\mu}{\alpha_0}{\mu'_{\theta_2}-\mu}{\alpha_{1 \theta_3}}{\alpha_{2,1}}{\alpha_1}{\frac{1}{\Lambda}}{\Lambda(1-z)} = e^{-(\mu'_{\theta_2}-\mu)\Lambda} \Lambda^{\Delta_{1\theta_3}+2\mu'_{\theta_2}(\mu'_{\theta_2}-\mu)} (1-z)^{\frac{bQ}{2}+\theta b \alpha_1}e^{\frac{1}{b^2}F_D(\mu'_{\theta_2})+W_D(\mu'_{\theta_2})}\\
        &\sim e^{\theta_2 L/2} \Lambda^{-\theta_2 (2m'-m)} e^{-\frac{\theta_2}{2}\partial_{m'}F_D(m')} e^{\mu\Lambda}\DIII{-\mu}{\alpha_0}{\mu'-\mu}{\alpha_{1 \theta_3}}{\alpha_{2,1}}{\alpha_1}{\frac{1}{\Lambda}}{\Lambda(1-z)}\,,\quad \mathrm{as}\,\, b \to 0\,.
    \end{aligned}
\end{equation}
The connection formula \eqref{0to1largeLambda} from 0 to 1 in the semiclassical limit then becomes
\begin{equation}
\begin{aligned}
    &\DIIIsc{m}{a_1}{m'}{a_{0 \theta}}{a_{2,1}}{a_0}{\frac{1}{L}}{L z} = \\=&\, \sum_{\theta'=\pm}\left(\sum_{\sigma=\pm} \mathcal{N}_{\theta \sigma}(a_0,m')\mathcal{N}^{-1}_{(-\sigma)\theta'}(m'-m,a_1)e^{\frac{\sigma}{2} L}L^{-\sigma(2m'-m)}e^{-\frac{\sigma}{2}\partial_{m'}F_D(m')}\right) \DIIIsc{-m}{a_0}{m'-m}{a_{1 \theta'}}{a_{2,1}}{a_1}{\frac{1}{L}}{L(1-z)}\,,
\end{aligned}
\end{equation}
where explicitly the connection coefficients read:
\begin{equation}
\begin{aligned}
    &\sum_{\sigma=\pm} \mathcal{N}_{\theta \sigma}(a_0,m')\mathcal{N}^{-1}_{(-\sigma)\theta'}(m'-m,a_1)e^{\frac{\sigma}{2} L}L^{-\sigma(2m'-m)}e^{-\frac{\sigma}{2}\partial_{m'}F_D(m')} = \\=&\sum_{\sigma = \pm} \frac{\Gamma(1+2\theta a_0)\Gamma(-2\theta'a_1)e^{\frac{\sigma}{2} L}L^{-\sigma(2m'-m)}e^{-\frac{\sigma}{2}\partial_{m'}F_D(m')}e^{i \pi \left(\frac{1-\sigma}{2}\right)\left(\theta a_0-\theta'a_1-2m'+m\right)}}{\Gamma\left(\frac{1}{2}+\theta a_0-\sigma m'\right)\Gamma\left(\frac{1}{2}-\theta' a_1-\sigma (m'-m)\right)}\,.
\end{aligned}
\end{equation}
Again, all the spurious powers of $b$ and $\Lambda$ have beautifully recombined to give the finite combination $L$.\newline
The connection formula from $1$ to $\infty$ \eqref{1toinftylargeLambda} on the other hand becomes
\begin{equation}
    \DIIIsc{-m}{a_0}{m'-m}{a_{1 \theta}}{a_{2,1}}{a_1}{\frac{1}{L}}{L(1-z)} = \sum_{\theta'=\pm}  \mathcal{N}_{\theta(-\theta')}(a_1,m'-m) \DIIIlargeLambdasc{m}{a_{2,1}}{m_{\theta'} }{a_1}{m'}{a_0}{\frac{1}{L}}{\frac{1}{z}}\,,
\end{equation}
where $\mathcal{N}$ is:
\begin{equation}
    \mathcal{N}_{\theta(-\theta')}(a_1,m'-m,) = \frac{\Gamma(1+2\theta a_1)}{\Gamma\left(\frac{1}{2}+\theta a_1+ \theta'(m'-m) \right)}e^{i\pi \left(\frac{1+\theta'}{2}\right)\left(\frac{1}{2}-(m'-m)+\theta a_1\right)}\,.
\end{equation}

\subsection{Reduced confluent conformal blocks}\label{Nf2A}
\subsubsection{General case}
Consider the correlation function
\begin{equation}
    \langle \Lambda^2| V_1(1) \Phi(z)| \Delta_0 \rangle \,,
\end{equation}
which solves the BPZ equation
\begin{equation}
\begin{aligned}
    \left(b^{-2} \partial_z^2 - \left(\frac{1}{z}+\frac{1}{z-1}\right)\partial_z + \frac{\Lambda^2 \partial_{\Lambda^2} - \Delta_{2,1} - \Delta_1 - \Delta_0}{z(z-1)} + \frac{\Delta_1}{(z-1)^2} + \frac{\Delta_0}{z^2}  -\frac{\Lambda^2}{4z}\right) \langle \Lambda^2| \Phi(z) V_1(1)|\Delta_0\rangle=0\,.
\end{aligned}
\end{equation}
We can decompose it into irregular conformal blocks in different ways. The blocks corresponding to the expansion of $z$ around a regular singular point can be given as a further decoupling limit of the confluent conformal blocks. For the blocks corresponding to the expansion of $z$ around the irregular singular point of rank $1/2$, no closed form expression presently known to us. The block for $z\sim 0$ can be defined as
\begin{equation}
     \FIIA{\alpha}{\alpha_1}{\alpha_{0\theta}}{\alpha_{2,1}}{\alpha_0}{\Lambda^2}{z} = \lim_{\eta \to \infty} (4\eta)^\Delta\, \FIII{-\eta}{\alpha}{\alpha_1}{\alpha_{0 \theta}}{\alpha_{2,1}}{\alpha_0}{\frac{\Lambda^2}{4 \eta}}{z}\,.
\end{equation}
We multiply by the factor of $(4\eta)^\Delta$ to take care of the leading divergence in the limit. In the diagrammatic notation of section \ref{warmup:Bessel}, we represent it by
\begin{equation}
    \FIIA{\alpha}{\alpha_1}{\alpha_{0\theta}}{\alpha_{2,1}}{\alpha_0}{\Lambda^2}{z} =
    \begin{tikzpicture}[baseline={(current bounding box.center)}, node distance=1cm and 1.5cm]
    \coordinate[circle,fill,inner sep=2pt] (aux1);
    \coordinate[left=of aux1] (e1);
    \coordinate[right=1.5cm of aux1] (aux2);
    \coordinate[above=of aux2,label=above:$\alpha_1$] (e2);
    \coordinate[right=1.5cm of aux2] (aux3);
    \coordinate[above=of aux3,label=above:$\alpha_{2,1}$] (e3);
    \coordinate[right=of aux3,label=right:$\alpha_0$] (e4);

    \draw[line,decoration=snake] (aux1) -- (e1);
    \draw[line] (aux2) -- (e2);
    \draw[line,dashed,red] (aux3) -- (e3);
    \draw[line] (aux3) -- (e4);
    \draw[line] (aux1) -- node[label=below:$\alpha$] {} (aux2);
    \draw[line] (aux2) -- node[label=below:$\alpha_{0\theta}$] {} (aux3);
    \end{tikzpicture} \,.
\end{equation}
As indicated by the diagram, all OPEs are regular in this case. The wiggly line represents the rank 1/2 irregular state, and the dot the pairing with a primary. The block for $z\sim 1$ is then simply
\begin{equation}
    e^{i \pi \Delta} e^{\frac{\Lambda^2}{4}} \FIIA{\alpha}{\alpha_0}{\alpha_{1 \theta}}{\alpha_{2,1}}{\alpha_1}{e^{- i \pi} \Lambda^2}{1-z} =
    \begin{tikzpicture}[baseline={(current bounding box.center)}, node distance=1cm and 1.5cm]
    \coordinate[circle,fill,inner sep=2pt] (aux1);
    \coordinate[left=of aux1] (e1);
    \coordinate[right=1.5cm of aux1] (aux2);
    \coordinate[above=of aux2,label=above:$\alpha_0$] (e2);
    \coordinate[right=1.5cm of aux2] (aux3);
    \coordinate[above=of aux3,label=above:$\alpha_{2,1}$] (e3);
    \coordinate[right=of aux3,label=right:$\alpha_1$] (e4);

    \draw[line,decoration=snake] (aux1) -- (e1);
    \draw[line] (aux2) -- (e2);
    \draw[line,dashed,red] (aux3) -- (e3);
    \draw[line] (aux3) -- (e4);
    \draw[line] (aux1) -- node[label=below:$\alpha$] {} (aux2);
    \draw[line] (aux2) -- node[label=below:$\alpha_{1\theta}$] {} (aux3);
    \end{tikzpicture}\,.
\end{equation} 
The overall phase compensates the sign in $e^{-i\pi}\Lambda^2$ such that the classical part is still $\Lambda^{2\Delta}$. In the intermediate region where $1\ll z \ll \frac{1}{\Lambda^2}$ the corresponding block is
\begin{equation}
    z^{-\Delta_{2,1}-\Delta_1-\Delta_0}\FIIA{\alpha_\theta}{\alpha_{2,1}}{\alpha}{\alpha_1}{\alpha_0}{\Lambda^2 z}{\frac{1}{z}} = 
    \begin{tikzpicture}[baseline={(current bounding box.center)}, node distance=1cm and 1.5cm]
    \coordinate[circle,fill,inner sep=2pt] (aux1);
    \coordinate[left=of aux1] (e1);
    \coordinate[right=1.5cm of aux1] (aux2);
    \coordinate[above=of aux2,label=above:$\alpha_{2,1}$] (e2);
    \coordinate[right=1.5cm of aux2] (aux3);
    \coordinate[above=of aux3,label=above:$\alpha_1$] (e3);
    \coordinate[right=of aux3,label=right:$\alpha_0$] (e4);

    \draw[line,decoration=snake] (aux1) -- (e1);
    \draw[line,dashed,red] (aux2) -- (e2);
    \draw[line] (aux3) -- (e3);
    \draw[line] (aux3) -- (e4);
    \draw[line] (aux1) -- node[label=below:$\alpha_\theta$] {} (aux2);
    \draw[line] (aux2) -- node[label=below:$\alpha$] {} (aux3);
    \end{tikzpicture}\,.
\end{equation}
Instead, in the deep irregular region, where $z\gg \frac{1}{\Lambda^2} \gg 1$, a decoupling limit of the form \eqref{rankhalfdecoupling} does not work. Of course one can still calculate this block by solving the BPZ equation iteratively with a series Ansatz, or directly using the Ward identities determining the descendants of the OPE with the irregular state (see Appendix \ref{app:rank1}). In any case we will denote the conformal block in this region by
\begin{equation}\label{DIIA}
    \DIIA{\theta}{\alpha_{2,1}}{\alpha}{\alpha_1}{\alpha_0}{\Lambda^2}{\frac{1}{\Lambda \sqrt{z}}} \sim (\Lambda^2)^{\Delta_{2,1}+\Delta} (\Lambda \sqrt{z})^{\frac{1}{2}+b^2} e^{\theta b \Lambda \sqrt{z}} \left[1+\mathcal{O}\left(\Lambda^2,\frac{1}{\Lambda\sqrt{z}}\right)\right]\,.
\end{equation}
The $\sim$ refers to the fact that this expansion is asymptotic. In diagrams we represent this block by
\begin{equation}
    \DIIA{\theta}{\alpha_{2,1}}{\alpha}{\alpha_1}{\alpha_0}{\Lambda^2}{\frac{1}{\Lambda \sqrt{z}}} = 
    \begin{tikzpicture}[baseline={(current bounding box.center)}, node distance=1cm and 1.5cm]
    \coordinate (aux1);
    \coordinate[left=of aux1] (e1);
    \coordinate[right=1.5cm of aux1,circle,fill,inner sep=2pt] (aux2);
    \coordinate[above=of aux1,label=above:$\alpha_{2,1}$] (e2);
    \coordinate[right=1.5cm of aux2] (aux3);
    \coordinate[above=of aux3,label=above:$\alpha_1$] (e3);
    \coordinate[right=of aux3,label=right:$\alpha_0$] (e4);

    \draw[line,decoration=snake] (aux1) -- (e1);
    \draw[line,dashed,red] (aux1) -- (e2);
    \draw[line] (aux3) -- (e3);
    \draw[line] (aux3) -- (e4);
    \draw[line,decoration=snake] (aux1) -- node[label=below:$\theta$] {} (aux2);
    \draw[line] (aux2) -- node[label=below:$\alpha$] {} (aux3);
    \end{tikzpicture}\,.
\end{equation}
The solution of the connection problems goes in the same way as for the (unreduced) confluent Heun equation (section \ref{Nf3}). In particular the connection problem between 0 and 1 works in the same way as for the general Heun equation. We have
\begin{equation}\label{0to1Nf2}
    \FIIA{\alpha}{\alpha_1}{\alpha_{0\theta}}{\alpha_{2,1}}{\alpha_0}{\Lambda^2}{z} = \sum_{\theta'=\pm} \mathcal{M}_{\theta \theta'}(b\alpha_0,b\alpha_1;b\alpha) e^{i \pi \Delta} e^{\frac{\Lambda^2}{4}} \FIIA{\alpha}{\alpha_0}{\alpha_{1 \theta}}{\alpha_{2,1}}{\alpha_1}{e^{- i \pi} \Lambda^2}{1-z} \,.
\end{equation}
To solve the connection problem between $1$ and $\infty$ one has to do two steps: from $1$ to the intermediate region, and then to $\infty$. In each step we decompose the correlator into conformal blocks in the different regions and then use crossing symmetry to determine the connection coefficients. The relevant formulae for the rank 1/2 irregular state are reviewed in Appendix \ref{app:rankhalf}. We have
\begin{equation}
\begin{aligned}
    \langle \Lambda^2| \Phi(z) V_1(1)|\Delta_0\rangle &= \int d \alpha \,  C_\alpha  \sum_{\theta = \pm} C_{\alpha_{2,1}\alpha_1}^{\alpha_{1\theta}} C_{\alpha_{1\theta}\alpha_0}^\alpha \left|e^{i \pi \Delta} e^{\frac{\Lambda^2}{4}} \FIIA{\alpha}{\alpha_0}{\alpha_{1 \theta}}{\alpha_{2,1}}{\alpha_1}{e^{- i \pi} \Lambda^2}{1-z} \right|^2=\\
    &= \int d \alpha \,  C_\alpha \sum_{\theta'=\pm} C_{\alpha_{2,1} \alpha_{\theta'}}^{\alpha} C_{\alpha_1 \alpha_0}^{\alpha_{\theta'}} \left|z^{-\Delta_{2,1}-\Delta_1-\Delta_0}\FIIA{\alpha}{\alpha_{2,1}}{\alpha_{\theta'}}{\alpha_1}{\alpha_0}{\Lambda^2 z}{\frac{1}{z}}\right|^2\,.
\end{aligned}
\end{equation}
This is precisely the same condition as for the hypergeometric functions \eqref{eq:hypergeometric_constraint}. The connection formula between $1$ and the intermediate region is then
\begin{equation}\label{Nf2from1tointermediate}
    e^{i\pi\Delta} e^{\frac{\Lambda^2}{4}} \FIIA{\alpha}{\alpha_0}{\alpha_{1 \theta}}{\alpha_{2,1}}{\alpha_1}{e^{-i\pi}\Lambda^2}{1-z} = \sum_{\theta'=\pm} \mathcal{M}_{\theta\theta'}(b\alpha_1,b\alpha;b\alpha_0) z^{-\Delta_{2,1}-\Delta_1-\Delta_0}\FIIA{\alpha}{\alpha_{2,1}}{\alpha_{ \theta'}}{\alpha_1}{\alpha_0}{\Lambda^2 z}{\frac{1}{z}}\,.
\end{equation}
Diagrammatically:
\begin{equation}
    \begin{tikzpicture}[baseline={(current bounding box.center)}, node distance=1cm and 1.5cm]
    \coordinate[circle,fill,inner sep=2pt] (aux1);
    \coordinate[left=of aux1] (e1);
    \coordinate[right=1.5cm of aux1] (aux2);
    \coordinate[above=of aux2,label=above:$\alpha_0$] (e2);
    \coordinate[right=1.5cm of aux2] (aux3);
    \coordinate[above=of aux3,label=above:$\alpha_{2,1}$] (e3);
    \coordinate[right=of aux3,label=right:$\alpha_1$] (e4);

    \draw[line,decoration=snake] (aux1) -- (e1);
    \draw[line] (aux2) -- (e2);
    \draw[line,dashed,red] (aux3) -- (e3);
    \draw[line] (aux3) -- (e4);
    \draw[line] (aux1) -- node[label=below:$\alpha$] {} (aux2);
    \draw[line] (aux2) -- node[label=below:$\alpha_{1\theta}$] {} (aux3);
    \end{tikzpicture} = \sum_{\theta'=\pm} \mathcal{M}_{\theta\theta'}
    \begin{tikzpicture}[baseline={(current bounding box.center)}, node distance=1cm and 1.5cm]
    \coordinate[circle,fill,inner sep=2pt] (aux1);
    \coordinate[left=of aux1] (e1);
    \coordinate[right=1.5cm of aux1] (aux2);
    \coordinate[above=of aux2,label=above:$\alpha_{2,1}$] (e2);
    \coordinate[right=1.5cm of aux2] (aux3);
    \coordinate[above=of aux3,label=above:$\alpha_1$] (e3);
    \coordinate[right=of aux3,label=right:$\alpha_0$] (e4);

    \draw[line,decoration=snake] (aux1) -- (e1);
    \draw[line,dashed,red] (aux2) -- (e2);
    \draw[line] (aux3) -- (e3);
    \draw[line] (aux3) -- (e4);
    \draw[line] (aux1) -- node[label=below:$\alpha$] {} (aux2);
    \draw[line] (aux2) -- node[label=below:$\alpha_{\theta'}$] {} (aux3);
    \end{tikzpicture}\,.
\end{equation}
Now we decompose the correlator into conformal blocks in the intermediate region and near $\infty$, obtaining the crossing symmetry condition
\begin{equation}
\begin{aligned}
   \langle \Lambda^2| \Phi(z) V_1(1)|\Delta_0\rangle &= \int d \alpha \, C_{\alpha_1,\alpha_0}^\alpha \sum_{\theta=\pm} C_{\alpha_\theta} C_{\alpha_{2,1} \alpha}^{\alpha_{\theta}}  \left|z^{-\Delta_{2,1}-\Delta_1-\Delta_0}\FIIA{\alpha_{ \theta}}{\alpha_{2,1}}{\alpha}{\alpha_1}{\alpha_0}{\Lambda^2 z}{\frac{1}{z}}\right|^2=\\
   &= \int d \alpha \, C_{\alpha_1 \alpha_0}^\alpha \sum_{\theta'=\pm} C_\alpha B_{\alpha_{2,1}} \left|\DIIA{\theta'}{\alpha_{2,1}}{\alpha}{\alpha_1}{\alpha_0}{\Lambda^2}{\frac{1}{\Lambda \sqrt{z}}}  \right|^2\,.
\end{aligned}
\end{equation}
We recognize this condition from the Bessel functions \eqref{eq:Bessel_constraint}. We then immediately find the connection formula between the intermediate region and $\infty$:
\begin{equation}\label{Nf2fromintermediatetoinfty}
    b^{2\theta b \alpha}z^{-\Delta_{2,1}-\Delta_1-\Delta_0}\FIIA{\alpha_{\theta}}{\alpha_{2,1}}{\alpha}{\alpha_1}{\alpha_0}{\Lambda^2 z}{\frac{1}{z}} = \sum_{\theta'=\pm}b^{-\frac{1}{2}} \mathcal{Q}_{\theta \theta'}(b\alpha) \DIIA{\theta'}{\alpha_{2,1}}{\alpha}{\alpha_1}{\alpha_0}{\Lambda^2}{\frac{1}{\Lambda \sqrt{z}}} 
\end{equation}
with irregular connection coefficients as in \eqref{eq:rankhalfconnection}:
\begin{equation}
    \mathcal{Q}_{\theta \theta'}(b\alpha) = \frac{2^{2\theta b \alpha}}{\sqrt{2\pi }} \Gamma(1+2\theta b \alpha) e^{i\pi\left(\frac{1-\theta'}{2}\right)\left(\frac{1}{2}+2\theta b \alpha \right)}\,.
\end{equation}
In diagrams:
\begin{equation}
    \begin{tikzpicture}[baseline={(current bounding box.center)}, node distance=1cm and 1.5cm]
    \coordinate[circle,fill,inner sep=2pt] (aux1);
    \coordinate[left=of aux1] (e1);
    \coordinate[right=1.5cm of aux1] (aux2);
    \coordinate[above=of aux2,label=above:$\alpha_{2,1}$] (e2);
    \coordinate[right=1.5cm of aux2] (aux3);
    \coordinate[above=of aux3,label=above:$\alpha_1$] (e3);
    \coordinate[right=of aux3,label=right:$\alpha_0$] (e4);

    \draw[line,decoration=snake] (aux1) -- (e1);
    \draw[line,dashed,red] (aux2) -- (e2);
    \draw[line] (aux3) -- (e3);
    \draw[line] (aux3) -- (e4);
    \draw[line] (aux1) -- node[label=below:$\alpha_\theta$] {} (aux2);
    \draw[line] (aux2) -- node[label=below:$\alpha$] {} (aux3);
    \end{tikzpicture} = \sum_{\theta'=\pm} \mathcal{Q}_{\theta \theta'}
    \begin{tikzpicture}[baseline={(current bounding box.center)}, node distance=1cm and 1.5cm]
    \coordinate (aux1);
    \coordinate[left=of aux1] (e1);
    \coordinate[right=1.5cm of aux1,circle,fill,inner sep=2pt] (aux2);
    \coordinate[above=of aux1,label=above:$\alpha_{2,1}$] (e2);
    \coordinate[right=1.5cm of aux2] (aux3);
    \coordinate[above=of aux3,label=above:$\alpha_1$] (e3);
    \coordinate[right=of aux3,label=right:$\alpha_0$] (e4);

    \draw[line,decoration=snake] (aux1) -- (e1);
    \draw[line,dashed,red] (aux1) -- (e2);
    \draw[line] (aux3) -- (e3);
    \draw[line] (aux3) -- (e4);
    \draw[line,decoration=snake] (aux1) -- node[label=below:$\theta'$] {} (aux2);
    \draw[line] (aux2) -- node[label=below:$\alpha$] {} (aux3);
    \end{tikzpicture}\,.
\end{equation}
Let us write explicitly the more interesting connection formulae between $1$ and $\infty$, which is obtained by concatenating the two connection formulae above. Since the $\mathfrak{F}$ block in the intermediate region has different arguments in formula \eqref{Nf2from1tointermediate} and  \eqref{Nf2fromintermediatetoinfty}, we need to rename some arguments. In the end we obtain the following connection formula from $1$ directly to $\infty$:
\begin{equation}\label{1toinftyNf2}
\begin{aligned}
    &e^{i\pi\Delta} e^{\frac{\Lambda^2}{4}} \FIIA{\alpha}{\alpha_0}{\alpha_{1 \theta_1}}{\alpha_{2,1}}{\alpha_1}{e^{-i\pi}\Lambda^2}{1-z} =\\=& \sum_{\theta_2,\theta_3=\pm} \mathcal{M}_{\theta_1 \theta_2}(b\alpha_1,b\alpha;b\alpha_0)\mathcal{Q}_{(-\theta_2) \theta_3}(b\alpha_{\theta_2})b^{-\frac{1}{2}+\theta_2 b \alpha_{\theta_2}}\DIIA{{\theta_3}}{\alpha_{2,1}}{\alpha_{\theta_2}}{\alpha_1}{\alpha_0}{\Lambda^2}{\frac{1}{\Lambda \sqrt{z}}}\,.
\end{aligned}
\end{equation}
Diagrammatically we have
\begin{equation}
    \begin{tikzpicture}[baseline={(current bounding box.center)}, node distance=1cm and 1.5cm]
    \coordinate[circle,fill,inner sep=2pt] (aux1);
    \coordinate[left=of aux1] (e1);
    \coordinate[right=1.5cm of aux1] (aux2);
    \coordinate[above=of aux2,label=above:$\alpha_0$] (e2);
    \coordinate[right=1.5cm of aux2] (aux3);
    \coordinate[above=of aux3,label=above:$\alpha_{2,1}$] (e3);
    \coordinate[right=of aux3,label=right:$\alpha_1$] (e4);

    \draw[line,decoration=snake] (aux1) -- (e1);
    \draw[line] (aux2) -- (e2);
    \draw[line,dashed,red] (aux3) -- (e3);
    \draw[line] (aux3) -- (e4);
    \draw[line] (aux1) -- node[label=below:$\alpha$] {} (aux2);
    \draw[line] (aux2) -- node[label=below:$\alpha_{1\theta_1}$] {} (aux3);
    \end{tikzpicture} = \sum_{\theta_2,\theta_3=\pm} \mathcal{M}_{\theta_1 \theta_2}\mathcal{Q}_{(-\theta_2) \theta_3}
    \begin{tikzpicture}[baseline={(current bounding box.center)}, node distance=1cm and 1.5cm]
    \coordinate (aux1);
    \coordinate[left=of aux1] (e1);
    \coordinate[right=1.5cm of aux1,circle,fill,inner sep=2pt] (aux2);
    \coordinate[above=of aux1,label=above:$\alpha_{2,1}$] (e2);
    \coordinate[right=1.5cm of aux2] (aux3);
    \coordinate[above=of aux3,label=above:$\alpha_1$] (e3);
    \coordinate[right=of aux3,label=right:$\alpha_0$] (e4);

    \draw[line,decoration=snake] (aux1) -- (e1);
    \draw[line,dashed,red] (aux1) -- (e2);
    \draw[line] (aux3) -- (e3);
    \draw[line] (aux3) -- (e4);
    \draw[line,decoration=snake] (aux1) -- node[label=below:$\theta_3$] {} (aux2);
    \draw[line] (aux2) -- node[label=below:$\alpha_{\theta_2}$] {} (aux3);
    \end{tikzpicture}\,,
\end{equation}
where we have suppressed the arguments of the connection coefficients for brevity.

\subsubsection{Semiclassical limit}\label{Nf2SC}
The story works the same way here as for the confluent case. In the semiclassical limit the BPZ equation becomes
\begin{equation}\label{BPZNf2NS}
\begin{aligned}
    \left(\partial_z^2 + \frac{u -\frac{1}{2}+ a_1^2 +a_0^2}{z(z-1)} + \frac{\frac{1}{4}-a_1^2}{(z-1)^2} + \frac{\frac{1}{4}-a_0^2}{z^2} -\frac{L^2}{4z}\right) {}_{\frac{1}{2}}\mathfrak{F}(z) =0\,,
\end{aligned}
\end{equation}
for \textit{any} semiclassical block. Here $u$ is given by
\begin{equation}
    u=\lim_{b\to0} b^2 \Lambda^2 \partial_{\Lambda^2} \log {}_\frac{1}{2}\mathfrak{F} \left(\alpha\, \begin{matrix} \alpha_1\\ \alpha_0 \end{matrix} ;\Lambda^2 \right) = \frac{1}{4}-a^2+ \mathcal{O}(L^2)
\end{equation}
by the same argument as before. The finite semiclassical conformal blocks are defined by normalizing by the same block without the degenerate field insertion, i.e. the semiclassical block for $z\sim 0$ is
\begin{equation}
    \FIIAsc{a}{a_1}{a_{0\theta}}{a_{2,1}}{a_0}{L^2}{z} = \lim_{b\to0} \frac{\FIIA{\alpha}{\alpha_1}{\alpha_{0\theta}}{\alpha_{2,1}}{\alpha_0}{\Lambda^2}{z}}{{}_\frac{1}{2} \mathfrak{F} \left( \alpha\, \begin{matrix} \alpha_0\\ \alpha_1 \end{matrix} ;\Lambda^2 \right)} = e^{-\frac{\theta}{2}\partial_{a_0}F}z^{\frac{1}{2}+\theta a_0}(1+\mathcal{O}(L^2,z))  \,.
\end{equation}
Here $F =\lim_{b\to0} b^2 \log\left[ \Lambda^{-2\Delta} {}_\frac{1}{2}\mathfrak{F} \left(\alpha\, \begin{matrix} \alpha_1\\ \alpha_0 \end{matrix} ;\Lambda^2 \right)\right]$.
\begin{equation}
\begin{aligned}
    &\FIIAsc{a}{a_0}{a_{1 \theta}}{a_{2,1}}{a_1}{-L^2}{1-z} =\lim_{b\to0} \frac{e^{i\pi\Delta}e^{\frac{\Lambda^2}{4}}\FIIA{\alpha}{\alpha_0}{\alpha_{1 \theta}}{\alpha_{2,1}}{\alpha_1}{e^{-i\pi}\Lambda^2}{1-z} }{{}_{\frac{1}{2}}\mathfrak{F} \left( \alpha\, \begin{matrix} \alpha_1\\ \alpha_0 \end{matrix} ;\Lambda^2 \right)} =\\ &=\lim_{b\to0}\frac{\FIIA{\alpha}{\alpha_0}{\alpha_{1 \theta}}{\alpha_{2,1}}{\alpha_1}{e^{-i\pi}\Lambda^2}{1-z} }{{}_\frac{1}{2} \mathfrak{F} \left( \alpha\, \begin{matrix} \alpha_0\\ \alpha_1 \end{matrix} ;e^{-i\pi}\Lambda^2 \right)} = e^{-\frac{\theta}{2}\partial_{a_1}F}(1-z)^{\frac{1}{2}+\theta a_1}(1+\mathcal{O}(L^2,1-z))\,.
\end{aligned}
\end{equation}
In the deep irregular region we define the semiclassical block as
\begin{equation}
    \DIIAsc{\theta}{a_{2,1}}{a}{a_1}{a_0}{L^2}{\frac{1}{L \sqrt{z}}} =  \lim_{b\to 0} b^{-\frac{1}{2}}\frac{\DIIA{\theta}{\alpha_{2,1}}{\alpha}{\alpha_1}{\alpha_0}{\Lambda^2}{\frac{1}{\Lambda \sqrt{z}}}}{{}_\frac{1}{2}\mathfrak{F} \left(\mu \, \alpha\, \begin{matrix} \alpha_1\\ \alpha_0 \end{matrix} ;\Lambda^2 \right)} = (L\sqrt{z})^{-\frac{1}{2}} e^{\theta L \sqrt{z}} (1+\mathcal{O}(L^2,\frac{1}{L\sqrt{z}}))\,.
\end{equation}
All these blocks satisfy the same equation \eqref{BPZNf2NS}. As for the confluent case, in the connection formula between $1$ and $\infty$ we have four different $\mathfrak{E}$ blocks appearing, which should reduce to two in the semiclassical limit. Indeed, we have
\begin{equation}
     \DIIA{\theta}{\alpha_{2,1}}{\alpha_{\theta'}}{\alpha_1}{\alpha_0}{\Lambda^2}{\frac{1}{\Lambda z}} \sim (\Lambda^2)^{\theta' a} e^{-\frac{\theta'}{2}\partial_{a}F} \DIIA{\theta}{\alpha_{2,1}}{\alpha}{\alpha_1}{\alpha_0}{\Lambda^2}{\frac{1}{\Lambda z}} \,,\quad \mathrm{as } \, \,b\to 0\,,
\end{equation}
as in \eqref{twiddle}. Now that we have defined the semiclassical conformal blocks, we state the connection formulae. The connection formula from $0$ to $1$ \eqref{0to1Nf2} reduces trivially in the semiclassical limit to
\begin{equation}\label{0to1Nf2SC}
    \FIIAsc{a}{a_1}{a_{0\theta}}{a_{2,1}}{a_0}{L^2}{z} = \sum_{\theta'=\pm} \mathcal{M}_{\theta \theta'}(a_0,a_1;a) \FIIAsc{a}{a_0}{a_{1 \theta}}{a_{2,1}}{a_1}{- L^2}{1-z} \,.
\end{equation}
The connection formula from $1$ to $\infty$ \eqref{1toinftyNf2} becomes
\begin{equation}\label{1toinftyNf2SC}
    \FIIAsc{a}{a_0}{a_{1 \theta}}{a_{2,1}}{a_1}{-L^2}{1-z} = \sum_{\theta'}\left(\sum_{\sigma=\pm}\mathcal{M}_{\theta \sigma}(a_1,a;a_0)\mathcal{Q}_{(-\sigma) \theta'}(a)L^{2\sigma a}e^{-\frac{\sigma}{2}\partial_a F}\right) \DIIAsc{{\theta'}}{a_{2,1}}{a}{a_1}{a_0}{L^2}{\frac{1}{L \sqrt{z}}}\,,
\end{equation}
with connection coefficients\footnote{ Note that the Gamma functions in the denominator precisely correspond to the one-loop factors of the two hypermultiplets of the corresponding AGT dual gauge theory.
}
\begin{equation}
\begin{aligned}
    &\sum_{\sigma=\pm}\mathcal{M}_{\theta \sigma}(a_1,a;a_0)\mathcal{Q}_{(-\sigma) \theta'}(a)L^{2\sigma a}e^{-\frac{\sigma}{2}\partial_a F} =\\=& \sum_{\sigma=\pm} \frac{\Gamma(1-2\sigma a)\Gamma(-2\sigma a)\Gamma(1+2\theta a_1)2^{-2\sigma a}L^{2\sigma a}e^{-\frac{\sigma}{2}\partial_a F}e^{i\pi\left(\frac{1-\theta'}{2}\right)\left(\frac{1}{2}-2\sigma a\right)}}{\sqrt{2\pi}\Gamma\left(\frac{1}{2}+\theta a_1-\sigma a +a_0\right)\Gamma\left(\frac{1}{2}+\theta a_1-\sigma a -a_0\right)}\,.
\end{aligned}
\end{equation}

\subsection{Doubly confluent conformal blocks}
\subsubsection{General case}
Via a further collision limit we reach a correlator that solves the BPZ equation 
\begin{equation}
    \left( b^{-2} \partial_z^2 - \frac{1}{z} \partial_z + \frac{\mu_1 \Lambda_1}{z} - \frac{\Lambda_1^2}{4}  + \frac{\Lambda_2 \partial_{\Lambda_2}}{z^2}  + \frac{\mu_2 \Lambda_2}{z^3} - \frac{\Lambda_2^2}{4z^4} \right) \langle \mu_1 ,  \Lambda_1 | \Phi(z) |  \mu_2 , \Lambda_2 \rangle = 0 \,.
    \label{eq:dcHBPZ}
\end{equation}
This correlator can be expanded in the intermediate region $\Lambda_2 \ll z \ll \Lambda_1^{-1}$ and near the two irregular singularities, that is either $z \gg \Lambda_1^{-1} \gg 1$ or $z \ll \Lambda_2\ll 1$. Note that in \eqref{eq:dcHBPZ} one of the three parameters $\Lambda_1, \Lambda_2, z$ is redundant. Indeed the conformal blocks will only depend on two ratios. The conformal blocks in these regions can easily be computed as a collision limit. Explicitly, in the intermediate region $\Lambda_2 \ll z \ll \Lambda_1^{-1}$ 
\begin{equation}
    \FII{\mu_1}{\alpha_\theta}{\alpha_{2,1}}{\alpha}{\mu_2}{\Lambda_1 z}{\frac{\Lambda_2}{z}} = \Lambda_1^{\Delta_\theta} \Lambda_2^{\Delta} z^{\frac{bQ}{2}+\theta b \alpha} \lim_{\eta \to \infty} \FIIIhat{\mu_1}{\alpha_\theta}{\alpha_{2,1}}{\alpha}{\frac{\eta-\mu_2}{2}}{\frac{\eta+\mu_2}{2}}{\Lambda_1 z}{\frac{\Lambda_2}{z \eta}} \,.
\end{equation}
This conformal block is the result of the projection of the Whittaker module $| \mu_2, \Lambda_2 \rangle$ on a Verma module $\Delta$ and of $\langle \mu_1, \Lambda_1|$ on $\Delta_\theta$. We represent this block by the diagram
\begin{equation}
    \FII{\mu_1}{\alpha_\theta}{\alpha_{2,1}}{\alpha}{\mu_2}{\Lambda_1 z}{\frac{\Lambda_2}{z}} = 
    \begin{tikzpicture}[baseline={(current bounding box.center)}, node distance=1cm and 1.5cm]
    \coordinate[circle,fill,inner sep=2pt] (aux1);
    \coordinate[left=of aux1,label=left:$\mu_1$] (e1);
    \coordinate[right=1.5cm of aux1] (aux2);
    \coordinate[above=of aux2,label=above:$\alpha_{2,1}$] (e2);
    \coordinate[right=1.5cm of aux2,circle,fill,inner sep=2pt] (aux3);
    \coordinate[right=of aux3,label=right:$\mu_2$] (e4);

    \draw[line,double] (aux1) -- (e1);
    \draw[line,dashed,red] (aux2) -- (e2);
    \draw[line,double] (aux3) -- (e4);
    \draw[line] (aux1) -- node[label=below:$\alpha_\theta$] {} (aux2);
    \draw[line] (aux2) -- node[label=below:$\alpha$] {} (aux3);
    \end{tikzpicture}\,.
\end{equation}
The expansion near the irregular singularity at infinity can be  obtained by colliding in \eqref{eq:LLimD3} the insertions far from the Whittaker state in the confluent conformal block. This gives
\begin{equation}
    \DII{\mu_1}{\alpha_{2,1}}{\mu_{1 \theta}}{\alpha}{\mu_2}{\Lambda_1 \Lambda_2}{\frac{1}{\Lambda_1 z}} =  e^{\theta b\Lambda_1 z/2} \Lambda_1^{\Delta+\Delta_{2,1}} \Lambda_2^\Delta \left( \Lambda_1 z\right)^{-\theta b \mu_1 + \frac{b^2}{2}}  \lim_{\eta \to \infty} \DIIIhat{\mu_1}{\alpha_{2,1}}{\mu_{1 \theta}}{\alpha}{\frac{\eta-\mu_2}{2}}{\frac{\eta+\mu_2}{2}}{\frac{\Lambda_1 \Lambda_2}{\eta}}{\frac{1}{\Lambda_1 z}} \,.
\label{eq:D2inf}
\end{equation}
We represent this block diagrammatically by
\begin{equation}
    \DII{\mu_1}{\alpha_{2,1}}{\mu_{1 \theta}}{\alpha}{\mu_2}{\Lambda_1 \Lambda_2}{\frac{1}{\Lambda_1 z}} = 
    \begin{tikzpicture}[baseline={(current bounding box.center)}, node distance=1cm and 1.5cm]
    \coordinate (aux1);
    \coordinate[left=of aux1,label=left:$\mu_1$] (e1);
    \coordinate[right=1.5cm of aux1,circle,fill,inner sep=2pt] (aux2);
    \coordinate[above=of aux1,label=above:$\alpha_{2,1}$] (e2);
    \coordinate[right=1.5cm of aux2,circle,fill,inner sep=2pt] (aux3);
    \coordinate[right=of aux3,label=right:$\mu_2$] (e4);

    \draw[line,double] (aux1) -- (e1);
    \draw[line,dashed,red] (aux1) -- (e2);
    \draw[line,double] (aux3) -- (e4);
    \draw[line,double] (aux1) -- node[label=below:$\mu_{1\theta}$] {} (aux2);
    \draw[line] (aux2) -- node[label=below:$\alpha$] {} (aux3);
    \end{tikzpicture}\,.
\end{equation}
Finally, the expansion near the irregular singularity at zero is easily obtained from \eqref{eq:D2inf} by exchanging $\Lambda_1$ and $\Lambda_2$ and sending $z \to 1/z$, up to a Jacobian. The corresponding conformal block is 
\begin{equation}
     z^{-2 \Delta_{2,1}} \DII{\mu_2}{\alpha_{2,1}}{\mu_{2 \theta}}{\alpha}{\mu_1}{\Lambda_1 \Lambda_2}{\frac{z}{\Lambda_2}}=
    \begin{tikzpicture}[baseline={(current bounding box.center)}, node distance=1cm and 1.5cm]
    \coordinate[circle,fill,inner sep=2pt] (aux1);
    \coordinate[left=of aux1,label=left:$\mu_1$] (e1);
    \coordinate[right=1.5cm of aux1,circle,fill,inner sep=2pt] (aux2);
    \coordinate[above=of aux3,label=above:$\alpha_{2,1}$] (e2);
    \coordinate[right=1.5cm of aux2] (aux3);
    \coordinate[right=of aux3,label=right:$\mu_2$] (e4);

    \draw[line,double] (aux1) -- (e1);
    \draw[line,dashed,red] (aux3) -- (e2);
    \draw[line,double] (aux3) -- (e4);
    \draw[line] (aux1) -- node[label=below:$\alpha$] {} (aux2);
    \draw[line,double] (aux2) -- node[label=below:$\mu_{2\theta}$] {} (aux3);
    \end{tikzpicture}\,.
\end{equation}
Expanding now the correlator first near 0 and then in the intermediate region, crossing symmetry implies
\begin{equation}
\begin{aligned}
    \langle \mu_1,  \Lambda_1 | \Phi(z) |  \mu_2, \Lambda_2 \rangle &=\int d \alpha \,  G^{-1}_\alpha  C_{\mu_1 \alpha} G^{-1}_\alpha   \sum_{\theta = \pm} B_{\alpha_{2,1}, \mu_2}^{\mu_{2 \theta}}  C_{\mu_{2 \theta}  \alpha} \left| z^{-2 \Delta_{2,1}} \DII{\mu_2}{\alpha_{2,1}}{\mu_{2 \theta}}{\alpha}{\mu_1}{\Lambda_1 \Lambda_2}{\frac{z}{\Lambda_2}} \right|^2 = \\
    &=\int d\alpha\, G^{-1}_\alpha   C_{\mu_1 \alpha} \sum_{\theta'=\pm} C^{\alpha_{\theta'}}_{\alpha_{2,1} \alpha}  C_{\mu_2 \alpha_{\theta'}} \left|\FII{\mu_1}{\alpha}{\alpha_{2,1}}{\alpha_{\theta'}}{\mu_2}{\Lambda_1 z}{\frac{\Lambda_2}{z}}\right|^2\,.
\end{aligned}
\end{equation}
We recognize this condition from \eqref{eq:Whittaker_constraint}, and we can readily write down the solution to the connection problem:
\begin{equation}
    b^{\theta b \alpha}\FII{\mu_1}{\alpha}{\alpha_{2,1}}{\alpha_\theta}{\mu_2}{\Lambda_1 z}{\frac{\Lambda_2}{z}} = \sum_{\theta'=\pm}b^{-\frac{1}{2}-\theta' b \mu_2} \mathcal{N}_{\theta \theta'} (b\alpha, b\mu_2) z^{-2 \Delta_{2,1}} \DII{\mu_2}{\alpha_{2,1}}{\mu_{2 \theta'}}{\alpha}{\mu_1}{\Lambda_1 \Lambda_2}{\frac{z}{\Lambda_2}} \,.
\end{equation}
In diagrams:
\begin{equation}
    \begin{tikzpicture}[baseline={(current bounding box.center)}, node distance=1cm and 1.5cm]
    \coordinate[circle,fill,inner sep=2pt] (aux1);
    \coordinate[left=of aux1,label=left:$\mu_1$] (e1);
    \coordinate[right=1.5cm of aux1] (aux2);
    \coordinate[above=of aux2,label=above:$\alpha_{2,1}$] (e2);
    \coordinate[right=1.5cm of aux2,circle,fill,inner sep=2pt] (aux3);
    \coordinate[right=of aux3,label=right:$\mu_2$] (e4);

    \draw[line,double] (aux1) -- (e1);
    \draw[line,dashed,red] (aux2) -- (e2);
    \draw[line,double] (aux3) -- (e4);
    \draw[line] (aux1) -- node[label=below:$\alpha$] {} (aux2);
    \draw[line] (aux2) -- node[label=below:$\alpha_\theta$] {} (aux3);
    \end{tikzpicture} = \sum_{\theta'=\pm} \mathcal{N}_{\theta \theta'} 
    \begin{tikzpicture}[baseline={(current bounding box.center)}, node distance=1cm and 1.5cm]
    \coordinate[circle,fill,inner sep=2pt] (aux5);
    \coordinate[left=of aux5,label=left:$\mu_1$] (e5);
    \coordinate[right=1.5cm of aux5,circle,fill,inner sep=2pt] (aux6);
    \coordinate[right=1.5cm of aux6] (aux7);
    \coordinate[above=of aux7,label=above:$\alpha_{2,1}$] (e6);
    \coordinate[right=of aux7,label=right:$\mu_2$] (e8);

    \draw[line,double] (aux5) -- (e5);
    \draw[line,dashed,red] (aux7) -- (e6);
    \draw[line,double] (aux7) -- (e8);
    \draw[line] (aux5) -- node[label=below:$\alpha$] {} (aux6);
    \draw[line,double] (aux6) -- node[label=below:$\mu_{2\theta'}$] {} (aux7);
    \end{tikzpicture}\,.
\end{equation}
A similar argument works for the connection between the intermediate region and infinity. We obtain
\begin{equation}
    b^{\theta b \alpha}\FII{\mu_1}{\alpha_\theta}{\alpha_{2,1}}{\alpha}{\mu_2}{\Lambda_1 z}{\frac{\Lambda_2}{z}} = \sum_{\theta'=\pm}b^{-\frac{1}{2}-\theta' b \mu_1} \mathcal{N}_{\theta \theta'} (b\alpha, b\mu_1) \DII{\mu_1}{\alpha_{2,1}}{\mu_{1 \theta'}}{\alpha}{\mu_2}{\Lambda_1 \Lambda_2}{\frac{1}{\Lambda_1 z}} \,.
\end{equation}
Or, diagrammatically:
\begin{equation}
    \begin{tikzpicture}[baseline={(current bounding box.center)}, node distance=1cm and 1.5cm]
    \coordinate[circle,fill,inner sep=2pt] (aux1);
    \coordinate[left=of aux1,label=left:$\mu_1$] (e1);
    \coordinate[right=1.5cm of aux1] (aux2);
    \coordinate[above=of aux2,label=above:$\alpha_{2,1}$] (e2);
    \coordinate[right=1.5cm of aux2,circle,fill,inner sep=2pt] (aux3);
    \coordinate[right=of aux3,label=right:$\mu_2$] (e4);

    \draw[line,double] (aux1) -- (e1);
    \draw[line,dashed,red] (aux2) -- (e2);
    \draw[line,double] (aux3) -- (e4);
    \draw[line] (aux1) -- node[label=below:$\alpha_\theta$] {} (aux2);
    \draw[line] (aux2) -- node[label=below:$\alpha$] {} (aux3);
    \end{tikzpicture} = \sum_{\theta'=\pm} \mathcal{N}_{\theta \theta'}
    \begin{tikzpicture}[baseline={(current bounding box.center)}, node distance=1cm and 1.5cm]
    \coordinate (aux1);
    \coordinate[left=of aux1,label=left:$\mu_1$] (e1);
    \coordinate[right=1.5cm of aux1,circle,fill,inner sep=2pt] (aux2);
    \coordinate[above=of aux1,label=above:$\alpha_{2,1}$] (e2);
    \coordinate[right=1.5cm of aux2,circle,fill,inner sep=2pt] (aux3);
    \coordinate[right=of aux3,label=right:$\mu_2$] (e4);

    \draw[line,double] (aux1) -- (e1);
    \draw[line,dashed,red] (aux1) -- (e2);
    \draw[line,double] (aux3) -- (e4);
    \draw[line,double] (aux1) -- node[label=below:$\mu_{1\theta'}$] {} (aux2);
    \draw[line] (aux2) -- node[label=below:$\alpha$] {} (aux3);
    \end{tikzpicture}\,.
\end{equation}
Concatenating the previous connection formulae we can connect $0$ directly with $\infty$ as follows
\begin{equation}
\begin{aligned}
    &b^{-\frac{1}{2}-\theta_1 b \mu_2} z^{-2 \Delta_{2,1}} \DII{\mu_2}{\alpha_{2,1}}{\mu_{2 \theta_1}}{\alpha}{\mu_1}{\Lambda_1 \Lambda_2}{\frac{z}{\Lambda_2}} = \\ &= \sum_{\theta_2, \theta_3 = \pm } b^{\theta_2 b \alpha} \mathcal{N}^{-1}_{\theta_1 \theta_2} (b \mu_2,b \alpha)b^{-\frac{1}{2}+\theta_2 b \alpha-\theta' b \mu_1} \mathcal{N}_{(- \theta_2 ) \theta_3} (b \alpha_{\theta_2}, b \mu_1) \DII{\mu_1}{\alpha_{2,1}}{\mu_{1 \theta_3}}{\alpha_{\theta_2}}{\mu_2}{\Lambda_1 \Lambda_2}{\frac{1}{\Lambda_1 z}} \,.
\end{aligned}
\label{eq:concatenationNf2symmetric}
\end{equation}
In diagrams:
\begin{equation}
    \begin{tikzpicture}[baseline={(current bounding box.center)}, node distance=1cm and 1.5cm]
    \coordinate[circle,fill,inner sep=2pt] (aux1);
    \coordinate[left=of aux1,label=left:$\mu_1$] (e1);
    \coordinate[right=1.5cm of aux1,circle,fill,inner sep=2pt] (aux2);
    \coordinate[above=of aux3,label=above:$\alpha_{2,1}$] (e2);
    \coordinate[right=1.5cm of aux2] (aux3);
    \coordinate[right=of aux3,label=right:$\mu_2$] (e4);

    \draw[line,double] (aux1) -- (e1);
    \draw[line,dashed,red] (aux3) -- (e2);
    \draw[line,double] (aux3) -- (e4);
    \draw[line] (aux1) -- node[label=below:$\alpha$] {} (aux2);
    \draw[line,double] (aux2) -- node[label=below:$\mu_{2\theta_1}$] {} (aux3);
    \end{tikzpicture} =  \sum_{\theta_2, \theta_3 = \pm } \mathcal{N}^{-1}_{\theta_1 \theta_2} \mathcal{N}_{(- \theta_2 ) \theta_3}
    \begin{tikzpicture}[baseline={(current bounding box.center)}, node distance=1cm and 1.5cm]
    \coordinate (aux1);
    \coordinate[left=of aux1,label=left:$\mu_1$] (e1);
    \coordinate[right=1.5cm of aux1,circle,fill,inner sep=2pt] (aux2);
    \coordinate[above=of aux1,label=above:$\alpha_{2,1}$] (e2);
    \coordinate[right=1.5cm of aux2,circle,fill,inner sep=2pt] (aux3);
    \coordinate[right=of aux3,label=right:$\mu_2$] (e4);

    \draw[line,double] (aux1) -- (e1);
    \draw[line,dashed,red] (aux1) -- (e2);
    \draw[line,double] (aux3) -- (e4);
    \draw[line,double] (aux1) -- node[label=below:$\mu_{1\theta_3}$] {} (aux2);
    \draw[line] (aux2) -- node[label=below:$\alpha_{\theta_2}$] {} (aux3);
    \end{tikzpicture}\,.
\end{equation}
\subsubsection{Semiclassical limit}
Let us now consider the semiclassical limit of the doubly confluent conformal blocks. Once again, the divergence as $b \to 0$ is expected to exponentiate, that is
\begin{equation}
    z^{-2 \Delta_{2,1}} \DII{\mu_2}{\alpha_{2,1}}{\mu_{2 \theta}}{\alpha}{\mu_1}{\Lambda_1 \Lambda_2}{\frac{z}{\Lambda_2}} = z^{-2 \Delta_{2,1}} e^{\frac{\theta b \Lambda_2}{2 z}} \Lambda_2^{\Delta+\Delta_{2,1}} \Lambda_1^\Delta \left( \frac{\Lambda_2}{z} \right)^{- \theta b \mu_2 + \frac{b^2}{2}} \exp \left( b^{-2} F \left( L_1 L_2 \right) + W\left(L_1 L_2, z L_2^{-1} \right) \right) \,,
\end{equation}
where $F$ is the classical conformal block defined by
\begin{equation}
    {}_1 \mathfrak{F}_1 \left( \mu_1 \, \, \alpha \, \, \mu_2, \Lambda_1 \Lambda_2 \right) = \left( \Lambda_1 \Lambda_2 \right)^\Delta \exp \left( b^{-2} F + \mathcal{O}(b^0) \right) \,,
\end{equation}
and the ${}_1 \mathfrak{F}_1$ block is given by
\begin{equation}
    \langle \mu_1, \Lambda_1 | \mu_2, \Lambda_2 \rangle = \int d \alpha\,  C_{\mu_1 \alpha} C_{\mu_2 \alpha} \left| {}_1 \mathfrak{F}_1 \left( \mu_1 \, \, \alpha \, \, \mu_2, \Lambda_1 \Lambda_2 \right) \right|^2 \,.
\end{equation}
We define the semiclassical block near zero to be
\begin{equation}
    z \, \DIIsc{m_2}{a_{2,1}}{m_{2 \theta}}{a}{m_1}{L_1 L_2}{\frac{z}{L_2}} = \lim_{b \to 0} b^{- \frac{1}{2}- \theta b \mu_2 + \frac{b^2}{2}} z^{-2 \Delta_{2,1}} \frac{ \DII{\mu_2}{\alpha_{2,1}}{\mu_{2 \theta}}{\alpha}{\mu_1}{\Lambda_1 \Lambda_2}{\frac{z}{\Lambda_2}}}{{}_1 \mathfrak{F}_1 \left( \mu_1 \, \, \alpha \, \, \mu_2, \Lambda_1 \Lambda_2 \right)} \,,
\label{eq:nf2symsccbnear0}
\end{equation}
The semiclassical blocks satisfy the equation
\begin{equation}
    \left( \partial_z^2 + \frac{m_1 L_1}{z} - \frac{L_1^2}{4}  +  \frac{u}{z^2} + \frac{m_2 L_2}{z^3} - \frac{L_2^2}{4} \frac{1}{z^4} \right) z \DIIsc{m_2}{a_{2,1}}{m_{2 \theta}}{a}{m_1}{L_1 L_2}{\frac{z}{L_2}} = 0 \,,
\label{eq:scBPZNf2symm}
\end{equation}
with the $u$ parameter defined as usual to be the leftover of the $\Lambda_2$ derivative, that is
\begin{equation}
    u = \frac{1}{4} - a^2 + L_2 \partial_{L_2} F \left( L_1 L_2 \right) \,.
\end{equation}
Similarly, the semiclassical block near the irregular singularity at infinity is defined to be
\begin{equation}
    \DIIsc{m_1}{a_{2,1}}{m_{1 \theta}}{a}{m_2}{L_1 L_2}{\frac{1}{L_1 z}} = \lim_{b \to 0} b^{- \frac{1}{2}- \theta b \mu_1 + \frac{b^2}{2}} \frac{\DII{\mu_1}{\alpha_{2,1}}{\mu_{1 \theta}}{\alpha}{\mu_2}{\Lambda_1 \Lambda_2}{\frac{1}{\Lambda_1 z}}}{{}_1 \mathfrak{F}_1 \left( \mu_1 \, \, \alpha \, \, \mu_2, \Lambda_1 \Lambda_2 \right)} \,,
\label{eq:nf2symsccbnearinf}
\end{equation}
and satisfies the same equation \eqref{eq:scBPZNf2symm}. In equation \eqref{eq:concatenationNf2symmetric} 4 different blocks near infinity appear in the RHS. However they collapse to two of them in the semiclassical limit as in the previous cases. That is,
\begin{equation}
    \DII{\mu_1}{\alpha_{2,1}}{\mu_{1 \theta}}{\alpha_{\theta'}}{\mu_2}{\Lambda_1 \Lambda_2}{\frac{1}{\Lambda_1 z}} \sim \left( \Lambda_1 \Lambda_2 \right)^{\theta' a} e^{- \frac{\theta'}{2} \partial_a F } \DII{\mu_1}{\alpha_{2,1}}{\mu_{1 \theta}}{\alpha}{\mu_2}{\Lambda_1 \Lambda_2}{\frac{1}{\Lambda_1 z}}\,,\quad \mathrm{as } \, \,b\to 0 \,,
\end{equation}
as in \eqref{twiddle}. Finally, the connection formula \eqref{eq:concatenationNf2symmetric} in the semiclassical limit becomes
\begin{equation}
\begin{aligned}
    &z \, \DIIsc{m_2}{a_{2,1}}{m_{2 \theta}}{a}{m_1}{L_1 L_2}{\frac{z}{L_2}} = \\ &= \sum_{\theta'} \left(\sum_{\sigma=\pm} \mathcal{N}^{-1}_{\theta \sigma} (m_2,a) \mathcal{N}_{(- \sigma) \theta'} (a, m_1) \left( L_1 L_2 \right)^{\sigma a} e^{- \frac{\sigma}{2} \partial_a F } \right)\DIIsc{m_1}{a_{2,1}}{m_{1 \theta'}}{a}{m_2}{L_1 L_2}{\frac{1}{L_1 z}} \,,
\end{aligned}
\label{eq:nf2symconcsc}
\end{equation}
where explicitly the connection coefficients read
\begin{equation}
\begin{aligned}
    &\sum_{\sigma=\pm} \mathcal{N}^{-1}_{\theta \sigma} (m_2,a) \mathcal{N}_{(- \sigma) \theta'} (a, m_1) \left( L_1 L_2 \right)^{\sigma a} e^{- \frac{\sigma}{2} \partial_a F } =\\=&\, \sum_{\sigma=\pm} \frac{\Gamma(1-2\sigma a)\Gamma(-2\sigma a)\left( L_1 L_2 \right)^{\sigma a} e^{- \frac{\sigma}{2} \partial_a F }}{\Gamma\left(\frac{1}{2}+\theta m_2 - \sigma a\right)\Gamma\left(\frac{1}{2}- \theta' m_1-\sigma a\right)}  e^{i\pi \left(\frac{1+\theta}{2}\right)\left(-\frac{1}{2}- m_2 - \sigma a\right)}  e^{i\pi \left(\frac{1-\theta'}{2}\right)\left(\frac{1}{2}-m_1-\sigma a\right)}\,,
\end{aligned}
\end{equation}

\subsection{Reduced doubly confluent conformal blocks}\label{Nf1}
\subsubsection{General case}
Consider the correlation function
\begin{equation}
    \langle \mu, \Lambda_1|\Phi(z)|\Lambda_2^2\rangle\,,
\end{equation}
which solves the BPZ equation
\begin{equation}
    \left( b^{-2} \partial_z^2 - \frac{1}{z} \partial_z + \frac{\mu \Lambda_1}{z} - \frac{\Lambda_1^2}{4}  + \frac{\Lambda_2^2 \partial_{\Lambda_2^2}}{z^2}   - \frac{\Lambda_2^2}{4z^3} \right) \langle \mu ,  \Lambda_1 | \Phi(z) |  \Lambda_2^2 \rangle = 0 \,.
\end{equation}
One of the parameters among $\Lambda_1,\Lambda_2,z$ is redundant and can be set to an arbitrary value via a rescaling. We keep them all generic for convenience. We have three different conformal blocks, corresponding to the expansion of $z$ near the two irregular singular points, and for $z$ in the intermediate region. The block for $z\sim \infty$ is given by the decoupling limit of the corresponding doubly confluent block \eqref{eq:D2inf}:
\begin{equation}
     \DI{\mu}{\alpha_{2,1}}{\mu_\theta}{\alpha}{\Lambda_1 \Lambda_2^2}{\frac{1}{\Lambda_1 z}} = e^{\theta b\Lambda_1 z/2} \Lambda_1^{\Delta+\Delta_{2,1}} (\Lambda_2^2)^\Delta \left( \Lambda_1 z\right)^{-\theta b \mu + \frac{b^2}{2}} \lim_{\eta\to \infty} \DIIhat{\mu}{\alpha_{2,1}}{\mu_\theta}{\alpha}{\eta}{-\frac{\Lambda_1 \Lambda_2^2}{4\eta}}{\frac{1}{\Lambda_1 z}}\,.
\end{equation}
Equivalently, this block can be computed by doing the OPE $\langle \mu ,  \Lambda_1 | \Phi(z)$, projecting the result onto the Verma module $\Delta_\alpha$ and contracting the result with $|\Lambda_2^2\rangle$. We denote it diagrammatically by
\begin{equation}
    \DI{\mu}{\alpha_{2,1}}{\mu_\theta}{\alpha}{\Lambda_1 \Lambda_2^2}{\frac{1}{\Lambda_1 z}} = 
    \begin{tikzpicture}[baseline={(current bounding box.center)}, node distance=1cm and 1.5cm]
    \coordinate (aux1);
    \coordinate[left=of aux1,label=left:$\mu$] (e1);
    \coordinate[right=1.5cm of aux1,circle,fill,inner sep=2pt] (aux2);
    \coordinate[above=of aux1,label=above:$\alpha_{2,1}$] (e2);
    \coordinate[right=1.5cm of aux2,circle,fill,inner sep=2pt] (aux3);
    \coordinate[right=of aux3] (e4);

    \draw[line,double] (aux1) -- (e1);
    \draw[line,dashed,red] (aux1) -- (e2);
    \draw[line,decoration=snake] (aux3) -- (e4);
    \draw[line,double] (aux1) -- node[label=below:$\mu_{\theta}$] {} (aux2);
    \draw[line] (aux2) -- node[label=below:$\alpha$] {} (aux3);
    \end{tikzpicture}\,.
\end{equation}
Also for the intermediate region $\Lambda_2^2 \ll z\ll \frac{1}{\Lambda_1}$ we have a closed form expression, given by
\begin{equation}
    \FI{\mu}{\alpha_\theta}{\alpha_{2,1}\,}{\alpha}{\Lambda_1 z}{\frac{\Lambda_2^2}{z}} = \Lambda_1^{\Delta_\theta} (\Lambda_2^2)^{\Delta} z^{\frac{bQ}{2} + \theta b \alpha} \lim_{\eta\to\infty} \FIIhat{\mu}{\alpha_\theta}{\alpha_{2,1}\,}{\alpha}{\eta}{\Lambda_1 z}{-\frac{\Lambda_2^2}{4 \eta z}}\,.
\end{equation}
This conformal block can also be computed directly by projecting $|\Lambda_2^2\rangle$ onto the Verma module $\Delta$, then doing the OPE of $\Phi(z)$ term by term with the resulting expansion and then contracting with $\langle \mu,\Lambda_1|$. In diagrams
\begin{equation}
    \FI{\mu}{\alpha_\theta}{\alpha_{2,1}\,}{\alpha}{\Lambda_1 z}{\frac{\Lambda_2^2}{z}} = 
    \begin{tikzpicture}[baseline={(current bounding box.center)}, node distance=1cm and 1.5cm]
    \coordinate[circle,fill,inner sep=2pt] (aux1);
    \coordinate[left=of aux1,label=left:$\mu$] (e1);
    \coordinate[right=1.5cm of aux1] (aux2);
    \coordinate[above=of aux2,label=above:$\alpha_{2,1}$] (e2);
    \coordinate[right=1.5cm of aux2,circle,fill,inner sep=2pt] (aux3);
    \coordinate[right=of aux3] (e4);

    \draw[line,double] (aux1) -- (e1);
    \draw[line,dashed,red] (aux2) -- (e2);
    \draw[line,decoration=snake] (aux3) -- (e4);
    \draw[line] (aux1) -- node[label=below:$\alpha_\theta$] {} (aux2);
    \draw[line] (aux2) -- node[label=below:$\alpha$] {} (aux3);
    \end{tikzpicture}\,.
\end{equation}
For the expansion around the irregular singular point of half rank no explicit, closed form expression is known to us. In any case one can calculate the expansion iteratively via other methods as for \eqref{DIIA}. We denote the corresponding conformal block in this region, where $z \ll \Lambda_2^2$ and $\Lambda_1 \Lambda_2^2 \ll 1$ by
\begin{equation}
    \EI{\theta}{\mu}{\alpha}{\alpha_{2,1}}{\Lambda_1 \Lambda_2^2}{\frac{\sqrt{z}}{\Lambda_2}} \sim e^{\theta b \Lambda_2/ \sqrt{z}} \left(\frac{\sqrt{z}}{\Lambda_2}\right)^{-\frac{1}{2}-b^2} z^{-2\Delta_{2,1}} \Lambda_1^\Delta (\Lambda_2^2)^{\Delta_{2,1}+\Delta} \left[1+\mathcal{O}\left(\frac{\sqrt{z}}{\Lambda_2},\Lambda_1 \Lambda_2^2\right)\right]\,.
\end{equation}
Diagrammatically,
\begin{equation}
    \EI{\theta}{\mu}{\alpha}{\alpha_{2,1}}{\Lambda_1 \Lambda_2^2}{\frac{\sqrt{z}}{\Lambda_2}} = 
    \begin{tikzpicture}[baseline={(current bounding box.center)}, node distance=1cm and 1.5cm]
    \coordinate[circle,fill,inner sep=2pt] (aux1);
    \coordinate[left=of aux1,label=left:$\mu$] (e1);
    \coordinate[right=1.5cm of aux1,circle,fill,inner sep=2pt] (aux2);
    \coordinate[above=of aux3,label=above:$\alpha_{2,1}$] (e2);
    \coordinate[right=1.5cm of aux2] (aux3);
    \coordinate[right=of aux3] (e4);

    \draw[line,double] (aux1) -- (e1);
    \draw[line,dashed,red] (aux3) -- (e2);
    \draw[line,decoration=snake] (aux3) -- (e4);
    \draw[line] (aux1) -- node[label=below:$\alpha$] {} (aux2);
    \draw[line,decoration=snake] (aux2) -- node[label=below:$\theta$] {} (aux3);
    \end{tikzpicture}\,.
\end{equation}
To connect $0$ with the intermediate region we decompose
\begin{equation}
\begin{aligned}
    \langle \mu, \Lambda_1| \Phi(z)|\Lambda_2^2\rangle &= \int d \alpha \, C_{\mu \alpha}  G^{-1}_\alpha  \sum_{\theta=\pm}C_\alpha B_{\alpha_{2,1}} \left|\EI{\theta}{\mu}{\alpha}{\alpha_{2,1}}{\Lambda_1 \Lambda_2^2}{\frac{\sqrt{z}}{\Lambda_2}}\right|^2 = \\
    & =  \int d \alpha \, C_{\mu \alpha} G^{-1}_\alpha  \sum_{\theta'=\pm}  C_{\alpha_{\theta'}} C^{\alpha_{\theta'}}_{\alpha_{2,1},\alpha} \left|\FI{\mu}{\alpha}{\alpha_{2,1}\,}{\alpha_{\theta'}}{\Lambda_1 z}{\frac{\Lambda_2^2}{z}}\right|^2\,.
\end{aligned}
\end{equation}
We recognize this constraint from \eqref{eq:Bessel_constraint}. Its solution is
\begin{equation}
    b^{-\frac{1}{2}}\EI{\theta}{\mu}{\alpha}{\alpha_{2,1}}{\Lambda_1 \Lambda_2^2}{\frac{\sqrt{z}}{\Lambda_2}} = \sum_{\theta'=\pm} b^{2\theta' b\alpha} \mathcal{Q}^{-1}_{\theta \theta'}(b\alpha)\FI{\mu}{\alpha}{\alpha_{2,1}\,}{\alpha_{\theta'}}{\Lambda_1 z}{\frac{\Lambda_2^2}{z}}\,.
\end{equation}
In diagrams we write
\begin{equation}
    \begin{tikzpicture}[baseline={(current bounding box.center)}, node distance=1cm and 1.5cm]
    \coordinate[circle,fill,inner sep=2pt] (aux1);
    \coordinate[left=of aux1,label=left:$\mu$] (e1);
    \coordinate[right=1.5cm of aux1,circle,fill,inner sep=2pt] (aux2);
    \coordinate[above=of aux3,label=above:$\alpha_{2,1}$] (e2);
    \coordinate[right=1.5cm of aux2] (aux3);
    \coordinate[right=of aux3] (e4);

    \draw[line,double] (aux1) -- (e1);
    \draw[line,dashed,red] (aux3) -- (e2);
    \draw[line,decoration=snake] (aux3) -- (e4);
    \draw[line] (aux1) -- node[label=below:$\alpha$] {} (aux2);
    \draw[line,decoration=snake] (aux2) -- node[label=below:$\theta$] {} (aux3);
    \end{tikzpicture} = \sum_{\theta'=\pm} \mathcal{Q}^{-1}_{\theta \theta'}\begin{tikzpicture}[baseline={(current bounding box.center)}, node distance=1cm and 1.5cm]
    \coordinate[circle,fill,inner sep=2pt] (aux4);
    \coordinate[left=of aux4,label=left:$\mu$] (e5);
    \coordinate[right=1.5cm of aux4] (aux5);
    \coordinate[above=of aux5,label=above:$\alpha_{2,1}$] (e6);
    \coordinate[right=1.5cm of aux5,circle,fill,inner sep=2pt] (aux6);
    \coordinate[right=of aux6] (e7);

    \draw[line,double] (aux4) -- (e5);
    \draw[line,dashed,red] (aux5) -- (e6);
    \draw[line,decoration=snake] (aux6) -- (e7);
    \draw[line] (aux4) -- node[label=below:$\alpha$] {} (aux5);
    \draw[line] (aux5) -- node[label=below:$\alpha_{\theta'}$] {} (aux6);
    \end{tikzpicture}
    \,.
\end{equation}
Instead, to connect from the intermediate region to $\infty$ we decompose
\begin{equation}
\begin{aligned}
    \langle \mu, \Lambda_1| \Phi(z)|\Lambda_2^2\rangle
    & = \int \mathrm{d}\alpha\, C_\alpha G^{-1}_\alpha  \sum_{\theta=\pm}  C_{\mu \alpha_{\theta}} C^{\alpha_{\theta}}_{\alpha_{2,1} \alpha} \left|\FI{\mu}{\alpha_{\theta}}{\alpha_{2,1}\,}{\alpha}{\Lambda_1 z}{\frac{\Lambda_2^2}{z}}\right|^2=\\
    & = \int \mathrm{d}\alpha \, C_\alpha G^{-1}_\alpha  \sum_{\theta'=\pm} C_{\mu_{\theta'} \alpha} B^{\mu_{\theta'}}_{\alpha_{2,1} \mu} \left| \DI{\mu}{\alpha_{2,1}}{\mu_{\theta'}}{\alpha}{\Lambda_1 \Lambda_2^2}{\frac{1}{\Lambda_1 z}}\right|^2 \,.
\end{aligned}
\end{equation}
This is just the same constraint as for the Whittaker functions \eqref{eq:Whittaker_constraint}. The solution is
\begin{equation}
    b^{\theta b \alpha} \FI{\mu}{\alpha_{\theta}}{\alpha_{2,1}\,}{\alpha}{\Lambda_1 z}{\frac{\Lambda_2^2}{z}} = \sum_{\theta'=\pm} b^{-\frac{1}{2}-\theta' b \mu} \mathcal{N}_{\theta \theta'}(b\alpha, b\mu)\DI{\mu}{\alpha_{2,1}}{\mu_{\theta'}}{\alpha}{\Lambda_1 \Lambda_2^2}{\frac{1}{\Lambda_1 z}} \,.
\end{equation}
Diagrammatically
\begin{equation}
    \begin{tikzpicture}[baseline={(current bounding box.center)}, node distance=1cm and 1.5cm]
    \coordinate[circle,fill,inner sep=2pt] (aux1);
    \coordinate[left=of aux1,label=left:$\mu$] (e1);
    \coordinate[right=1.5cm of aux1] (aux2);
    \coordinate[above=of aux2,label=above:$\alpha_{2,1}$] (e2);
    \coordinate[right=1.5cm of aux2,circle,fill,inner sep=2pt] (aux3);
    \coordinate[right=of aux3] (e4);

    \draw[line,double] (aux1) -- (e1);
    \draw[line,dashed,red] (aux2) -- (e2);
    \draw[line,decoration=snake] (aux3) -- (e4);
    \draw[line] (aux1) -- node[label=below:$\alpha_{\theta}$] {} (aux2);
    \draw[line] (aux2) -- node[label=below:$\alpha$] {} (aux3);
    \end{tikzpicture} =  \sum_{\theta'=\pm} \mathcal{N}_{\theta \theta'} 
    \begin{tikzpicture}[baseline={(current bounding box.center)}, node distance=1cm and 1.5cm]
    \coordinate (aux1);
    \coordinate[left=of aux1,label=left:$\mu$] (e1);
    \coordinate[right=1.5cm of aux1,circle,fill,inner sep=2pt] (aux2);
    \coordinate[above=of aux1,label=above:$\alpha_{2,1}$] (e2);
    \coordinate[right=1.5cm of aux2,circle,fill,inner sep=2pt] (aux3);
    \coordinate[right=of aux3] (e4);

    \draw[line,double] (aux1) -- (e1);
    \draw[line,dashed,red] (aux1) -- (e2);
    \draw[line,decoration=snake] (aux3) -- (e4);
    \draw[line,double] (aux1) -- node[label=below:$\mu_{\theta'}$] {} (aux2);
    \draw[line] (aux2) -- node[label=below:$\alpha$] {} (aux3);
    \end{tikzpicture}
\end{equation}
To connect from $0$ to $\infty$ we just need to concatenate the two connection formulae above to obtain
\begin{equation}\label{193}
    b^{-\frac{1}{2}}\EI{\theta_1}{\mu}{\alpha}{\alpha_{2,1}}{\Lambda_1 \Lambda_2^2}{\frac{\sqrt{z}}{\Lambda_2}} = \sum_{\theta_2,\theta_3=\pm}b^{2\theta_2 b\alpha} \mathcal{Q}^{-1}_{\theta_1 \theta_2}(b\alpha) b^{-\frac{1}{2}+\theta_2 b \alpha - \theta_3 b \mu} \mathcal{N}_{(-\theta_2) \theta_3}(b\alpha_{\theta_2}, b\mu)\DI{\mu}{\alpha_{2,1}}{\mu_{\theta_3}}{\alpha_{\theta_2}}{\Lambda_1 \Lambda_2^2}{\frac{1}{\Lambda_1 z}} \,.
\end{equation}
In diagrams
\begin{equation}
    \begin{tikzpicture}[baseline={(current bounding box.center)}, node distance=1cm and 1.5cm]
    \coordinate[circle,fill,inner sep=2pt] (aux1);
    \coordinate[left=of aux1,label=left:$\mu$] (e1);
    \coordinate[right=1.5cm of aux1,circle,fill,inner sep=2pt] (aux2);
    \coordinate[above=of aux3,label=above:$\alpha_{2,1}$] (e2);
    \coordinate[right=1.5cm of aux2] (aux3);
    \coordinate[right=of aux3] (e4);

    \draw[line,double] (aux1) -- (e1);
    \draw[line,dashed,red] (aux3) -- (e2);
    \draw[line,decoration=snake] (aux3) -- (e4);
    \draw[line] (aux1) -- node[label=below:$\alpha$] {} (aux2);
    \draw[line,decoration=snake] (aux2) -- node[label=below:$\theta_1$] {} (aux3);
    \end{tikzpicture} = \sum_{\theta_2,\theta_3=\pm} \mathcal{Q}^{-1}_{\theta_1 \theta_2} \mathcal{N}_{(-\theta_2) \theta_3}\begin{tikzpicture}[baseline={(current bounding box.center)}, node distance=1cm and 1.5cm]
    \coordinate (aux1);
    \coordinate[left=of aux1,label=left:$\mu$] (e1);
    \coordinate[right=1.5cm of aux1,circle,fill,inner sep=2pt] (aux2);
    \coordinate[above=of aux1,label=above:$\alpha_{2,1}$] (e2);
    \coordinate[right=1.5cm of aux2,circle,fill,inner sep=2pt] (aux3);
    \coordinate[right=of aux3] (e4);

    \draw[line,double] (aux1) -- (e1);
    \draw[line,dashed,red] (aux1) -- (e2);
    \draw[line,decoration=snake] (aux3) -- (e4);
    \draw[line,double] (aux1) -- node[label=below:$\mu_{\theta_3}$] {} (aux2);
    \draw[line] (aux2) -- node[label=below:$\alpha_{\theta_2}$] {} (aux3);
    \end{tikzpicture}\,.
\end{equation}

\subsubsection{Semiclassical limit}\label{Nf1sc}
The BPZ equation in this limit becomes
\begin{equation}\label{BPZNf1SC}
    \left(\partial_z^2- \frac{L_1^2}{4}+ \frac{m L_1}{z}   + \frac{u}{z^2}   - \frac{L_2^2}{4z^3} \right) {}_1 \mathfrak{F}_{\frac{1}{2}} = 0 \,.
\end{equation}
for \textit{any} semiclassical block. Here $u$ is given by
\begin{equation}
    u=\lim_{b\to0} b^2 \Lambda_2^2 \partial_{\Lambda_2^2} \log {}_1\mathfrak{F}_\frac{1}{2} \left(\mu\,\alpha ;\Lambda_1 \Lambda_2^2 \right) = \frac{1}{4}-a^2+ \mathcal{O}(L_1 L_2^2)\,,
\end{equation}
where ${}_1\mathfrak{F}_\frac{1}{2} \left(\mu\,\alpha ;\Lambda_1 \Lambda_2^2 \right)$ is the conformal block corresponding to $\langle \mu,\Lambda_1|\Lambda_2^2\rangle$ with intermediate momentum $\alpha$. The finite semiclassical conformal blocks are defined as before by normalizing by the same block without the degenerate field insertion, i.e. for $z\sim 0$
\begin{equation}
    \EIsc{\theta}{m}{a}{a_{2,1}}{L_1 L_2^2}{\frac{\sqrt{z}}{L_2}} = \lim_{b\to0} b^{-\frac{1}{2}} \frac{\EI{\theta}{\mu}{\alpha}{\alpha_{2,1}}{\Lambda_1 \Lambda_2^2}{\frac{\sqrt{z}}{\Lambda_2}}}{{}_1\mathfrak{F}_\frac{1}{2} \left(\mu\,\alpha ;\Lambda_1 \Lambda_2^2 \right)}\sim e^{\theta L_2/\sqrt{z}} L_2^{-\frac{1}{2}} z^{\frac{3}{4}}(1+\mathcal{O}(L_1 L_2^2, \sqrt{z}/L_2))
\end{equation}
For $z\sim \infty$ instead we have
\begin{equation}\label{197}
\begin{aligned}
    \DIsc{m}{a_{2,1}}{m_\theta}{a}{L_1 L_2^2}{\frac{1}{L_1 z}}& = \lim_{b\to0} b^{-\frac{1}{2}-\theta m} \frac{\DI{\mu}{\alpha_{2,1}}{\mu_\theta}{\alpha}{\Lambda_1 \Lambda_2^2}{\frac{1}{\Lambda_1 z}}}{{}_1\mathfrak{F}_\frac{1}{2} \left(\mu\,\alpha ;\Lambda_1 \Lambda_2^2 \right)}\sim\\ &\sim \, e^{-\frac{\theta}{2}\partial_m F} e^{\theta L_1 z/2}L_1^{-\frac{1}{2}-\theta m} z^{-\theta m}(1+\mathcal{O}(L_1 L_2^2,1/L_1 z))\,.
\end{aligned}
\end{equation}
Here
\begin{equation}
    F =\lim_{b\to0} b^2 \log\left[ (\Lambda_1\Lambda_2^2)^{-\Delta} {}_1\mathfrak{F}_\frac{1}{2} \left(\mu\,\alpha ;\Lambda_1 \Lambda_2^2 \right)\right]\,.
\end{equation} 
Both these blocks satisfy the same BPZ equation \eqref{BPZNf1SC}. Analogously to the previous confluences, in the connection formula between $0$ and $\infty$ we have four different $\mathfrak{D}$ blocks appearing, which should reduce to two in the semiclassical limit. Indeed, we have
\begin{equation}
    \DI{\mu}{\alpha_{2,1}}{\mu_{\theta}}{\alpha_{\theta'}}{\Lambda_1 \Lambda_2^2}{\frac{1}{\Lambda_1 z}} \sim \left(\Lambda_1\Lambda_2^2\right)^{\theta' a} e^{-\frac{\theta'}{2}\partial_{a}F} \DI{\mu}{\alpha_{2,1}}{\mu_{\theta}}{\alpha}{\Lambda_1 \Lambda_2^2}{\frac{1}{\Lambda_1 z}}\,,\quad \mathrm{as } \, \,b\to 0\,,
\end{equation}
as in \eqref{twiddle}. Now that we have defined the semiclassical conformal blocks, we state the connection formula. \eqref{193} in the semiclassical limit becomes
\begin{equation}\label{0toinftyNf1}
    \EIsc{\theta}{m}{a}{a_{2,1}}{L_1 L_2^2}{\frac{\sqrt{z}}{L_2}} =\sum_{\theta'} \left(\sum_{\sigma=\pm} \mathcal{Q}^{-1}_{\theta \sigma}(a) \mathcal{N}_{(-\sigma) \theta'}(a, m)\left(L_1 L_2^2\right)^{\sigma a} e^{-\frac{\sigma}{2}\partial_{a}F}\right) \DIsc{m}{a_{2,1}}{m_{\theta'}}{a}{L_1 L_2^2}{\frac{1}{L_1 z}}\,.
\end{equation}
With connection coefficients\footnote{Note that the Gamma functions in the denominator precisely correspond to the one-loop factor of the single hypermultiplet of the corresponding AGT dual gauge theory.}
\begin{equation}
\begin{aligned}
    &\sum_{\sigma=\pm} \mathcal{Q}^{-1}_{\theta \sigma}(a) \mathcal{N}_{(-\sigma) \theta'}(a, m)\left(L_1 L_2^2\right)^{\sigma a} e^{-\frac{\sigma}{2}\partial_{a}F} =\\=&\,\frac{1}{\sqrt{2\pi}} \sum_{\sigma = \pm} \frac{\Gamma(1-2\sigma a)\Gamma(-2\sigma a)}{\Gamma\left(\frac{1}{2}-\theta' m - \sigma a\right)}\left(\frac{L_1 L_2^2}{4}\right)^{\sigma a} e^{-\frac{\sigma}{2}\partial_{a}F} e^{-i\pi\left(\frac{1+\theta}{2}\right)\left(\frac{1}{2}+2\sigma a\right)}e^{i\pi\left(\frac{1-\theta'}{2}\right)\left(\frac{1}{2}-m-\sigma a\right)}\,.
\end{aligned}
\end{equation}
Note that the factors of $b$ appearing in \eqref{193} precisely combine with all the factors of $\Lambda_1,\Lambda_2$ to give the finite $L_1,L_2$. 

\subsection{Doubly reduced doubly confluent conformal blocks}\label{Nf0}
\subsubsection{General case}
Decoupling the last mass we land on the last correlator of our interest, which solves the BPZ equation
\begin{equation}
    \left( b^{-2} \partial_z^2 - \frac{1}{z} \partial_z - \frac{\Lambda_1^2}{4} \frac{1}{z} +  \frac{\Lambda_2^2 \partial_{\Lambda_2^2}}{z^2}  - \frac{\Lambda_2^2}{4} \frac{1}{z^3} \right) \langle \Lambda_1^2| \Phi(z) | \Lambda_2^2 \rangle = 0 \,,
\end{equation}
Again, one of the parameters among $\Lambda_1,z,\Lambda_2$ is redundant and can be set to an arbitrary value via a rescaling. We keep them generic for convenience. We can decompose the above correlator into conformal blocks in three different regions, that is for $z \ll \Lambda_2^2 \ll 1$, $z \gg \Lambda_1^{-2} \gg 1$, or for $z$ in the intermediate region $\Lambda_2^2 \ll z \ll \Lambda_1^{-2}$. The conformal block in the intermediate region is again a block that can be expressed as a collision limit
\begin{equation}
    \Fz{\alpha_\theta}{\alpha_{2,1}}{\alpha}{\Lambda_1^2 z}{\frac{\Lambda_2^2}{z}} = (\Lambda_1^2)^{\Delta_\theta} (\Lambda_2^2)^\Delta z^{\frac{bQ}{2}+\theta b \alpha} \lim_{\eta \to \infty} \FIhat{\eta}{\alpha_\theta}{\alpha_{2,1}\,}{\alpha}{\frac{-\Lambda_1^2}{4 \eta} z}{\frac{\Lambda_2^2}{z}} \,.
\end{equation}
This conformal block can also be computed directly by projecting $|\Lambda_2^2\rangle$ onto the Verma module $\Delta$, then doing the OPE of $\Phi(z)$ term by term with the resulting expansion and then contracting with $\langle \Lambda_1^2|$. In diagrams we represent it by
\begin{equation}
    \Fz{\alpha_\theta}{\alpha_{2,1}}{\alpha}{\Lambda_1^2 z}{\frac{\Lambda_2^2}{z}} =
    \begin{tikzpicture}[baseline={(current bounding box.center)}, node distance=1cm and 1.5cm]
    \coordinate[circle,fill,inner sep=2pt] (aux1);
    \coordinate[left=of aux1] (e1);
    \coordinate[right=1.5cm of aux1] (aux2);
    \coordinate[above=of aux2,label=above:$\alpha_{2,1}$] (e2);
    \coordinate[right=1.5cm of aux2,circle,fill,inner sep=2pt] (aux3);
    \coordinate[right=of aux3] (e4);

    \draw[line,decoration=snake] (aux1) -- (e1);
    \draw[line,dashed,red] (aux2) -- (e2);
    \draw[line,decoration=snake] (aux3) -- (e4);
    \draw[line] (aux1) -- node[label=below:$\alpha_\theta$] {} (aux2);
    \draw[line] (aux2) -- node[label=below:$\alpha$] {} (aux3);
    \end{tikzpicture}\,.
\end{equation}
The block corresponding to the expansion for $z \gg \Lambda_1^{-2}$
\begin{equation}
\begin{aligned}
    \Ez{\theta}{\alpha_{2,1}}{\alpha}{\Lambda_1^2 \Lambda_2^2}{\frac{1}{\Lambda_1 \sqrt{z}}}&\sim (\Lambda_1^2)^{\Delta_{2,1}+\Delta} (\Lambda_2^2)^\Delta (\Lambda_1 \sqrt{z})^{\frac{1}{2}+b^2} e^{\theta b \Lambda_1 \sqrt{z}} \left[1+\mathcal{O}\left(\Lambda_1^2\Lambda_2^2,\frac{1}{\Lambda_1\sqrt{z}}\right)\right]\\
    &= \begin{tikzpicture}[baseline={(current bounding box.center)}, node distance=1cm and 1.5cm]
    \coordinate (aux1);
    \coordinate[left=of aux1] (e1);
    \coordinate[right=1.5cm of aux1,circle,fill,inner sep=2pt] (aux2);
    \coordinate[above=of aux1,label=above:$\alpha_{2,1}$] (e2);
    \coordinate[right=1.5cm of aux2,circle,fill,inner sep=2pt] (aux3);
    \coordinate[right=of aux3] (e4);

    \draw[line,decoration=snake] (aux1) -- (e1);
    \draw[line,dashed,red] (aux1) -- (e2);
    \draw[line,decoration=snake] (aux3) -- (e4);
    \draw[line,decoration=snake] (aux1) -- node[label=below:$\theta$] {} (aux2);
    \draw[line] (aux2) -- node[label=below:$\alpha$] {} (aux3);
    \end{tikzpicture}\,,
\end{aligned}
\end{equation}
and similarly for the expansion for $z \ll \Lambda_2^2$ 
\begin{equation}
\begin{aligned}
    z^{- 2 \Delta_{2,1}} \Ez{\theta}{\alpha}{\alpha_{2,1}}{\Lambda_1^2 \Lambda_2^2}{\frac{ \sqrt{z}}{\Lambda_2}}&\sim (\Lambda_1^2)^\Delta (\Lambda_2^2)^{\Delta_{2,1}+\Delta}z^{-2\Delta_{2,1}} \left(\frac{\sqrt{z}}{\Lambda_2}\right)^{-\frac{1}{2}-b^2}  e^{\theta b \Lambda_2/ \sqrt{z}} \left[1+\mathcal{O}\left(\Lambda_1 \Lambda_2^2,\frac{\sqrt{z}}{\Lambda_2}\right)\right] \\
    &= \begin{tikzpicture}[baseline={(current bounding box.center)}, node distance=1cm and 1.5cm]
    \coordinate[circle,fill,inner sep=2pt] (aux1);
    \coordinate[left=of aux1] (e1);
    \coordinate[right=1.5cm of aux1,circle,fill,inner sep=2pt] (aux2);
    \coordinate[above=of aux3,label=above:$\alpha_{2,1}$] (e2);
    \coordinate[right=1.5cm of aux2] (aux3);
    \coordinate[right=of aux3] (e4);

    \draw[line,decoration=snake] (aux1) -- (e1);
    \draw[line,dashed,red] (aux3) -- (e2);
    \draw[line,decoration=snake] (aux3) -- (e4);
    \draw[line] (aux1) -- node[label=below:$\alpha$] {} (aux2);
    \draw[line,decoration=snake] (aux2) -- node[label=below:$\theta$] {} (aux3);
    \end{tikzpicture}\,.
\end{aligned}
\end{equation}
To connect the intermediate region with $z \sim 0$ we decompose the correlator as
\begin{equation}
\begin{aligned}
    \langle \Lambda_1^2| \Phi(z) | \Lambda_2^2 \rangle = &\int d \alpha \, C_\alpha G^{-1}_\alpha  \sum_{\theta=\pm}  C_{\alpha_{2,1}, \alpha}^{\alpha_\theta}  C_{\alpha_\theta} \left| \Fz{\alpha}{\alpha_{2,1}}{\alpha_\theta}{\Lambda_1^2 z}{\frac{\Lambda_2^2}{z}} \right|^2 = \\ = &\int d \alpha \, C_\alpha G^{-1}_\alpha  \sum_{\theta'=\pm}
     A_{- \frac{b}{2}} C_\alpha \left| z^{- 2 \Delta_{2,1}} \Ez{\theta'}{\alpha}{\alpha_{2,1}}{\Lambda_1^2 \Lambda_2^2}{\frac{\sqrt{z}}{\Lambda_2}} \right|^2 \,.
\end{aligned}
\end{equation}
This is the same constraint as in \eqref{eq:Bessel_constraint}. Therefore the connection formula is
\begin{equation}
    b^{2\theta b \alpha}\Fz{\alpha}{\alpha_{2,1}}{\alpha_\theta}{\Lambda_1^2 z}{\frac{\Lambda_2^2}{z}} = \sum_{\theta'=\pm} b^{-\frac{1}{2}}\mathcal{Q}_{\theta \theta'} (b\alpha)  z^{- 2 \Delta_{2,1}} \Ez{\theta'}{\alpha}{\alpha_{2,1}}{\Lambda_1^2 \Lambda_2^2}{\frac{ \sqrt{z}}{\Lambda_2}} \,,
\end{equation}
Diagrammatically
\begin{equation}
    \begin{tikzpicture}[baseline={(current bounding box.center)}, node distance=1cm and 1.5cm]
    \coordinate[circle,fill,inner sep=2pt] (aux1);
    \coordinate[left=of aux1] (e1);
    \coordinate[right=1.5cm of aux1] (aux2);
    \coordinate[above=of aux2,label=above:$\alpha_{2,1}$] (e2);
    \coordinate[right=1.5cm of aux2,circle,fill,inner sep=2pt] (aux3);
    \coordinate[right=of aux3] (e4);

    \draw[line,decoration=snake] (aux1) -- (e1);
    \draw[line,dashed,red] (aux2) -- (e2);
    \draw[line,decoration=snake] (aux3) -- (e4);
    \draw[line] (aux1) -- node[label=below:$\alpha$] {} (aux2);
    \draw[line] (aux2) -- node[label=below:$\alpha_\theta$] {} (aux3);
    \end{tikzpicture} = \sum_{\theta'=\pm} \mathcal{Q}_{\theta \theta'}  \begin{tikzpicture}[baseline={(current bounding box.center)}, node distance=1cm and 1.5cm]
    \coordinate[circle,fill,inner sep=2pt] (aux1);
    \coordinate[left=of aux1] (e1);
    \coordinate[right=1.5cm of aux1,circle,fill,inner sep=2pt] (aux2);
    \coordinate[right=1.5cm of aux2] (aux3);
    \coordinate[above=of aux3,label=above:$\alpha_{2,1}$] (e2);
    \coordinate[right=of aux3] (e4);

    \draw[line,decoration=snake] (aux1) -- (e1);
    \draw[line,dashed,red] (aux3) -- (e2);
    \draw[line,decoration=snake] (aux3) -- (e4);
    \draw[line] (aux1) -- node[label=below:$\alpha$] {} (aux2);
    \draw[line,decoration=snake] (aux2) -- node[label=below:$\theta'$] {} (aux3);
    \end{tikzpicture}\,.
\end{equation}
Similarly, the connection formula between the intermediate region and $\infty$ is
\begin{equation}
    b^{2\theta b\alpha} \Fz{\alpha_\theta}{\alpha_{2,1}}{\alpha}{\Lambda_1^2 z}{\frac{\Lambda_2^2}{z}} = \sum_{\theta'=\pm}b^{-\frac{1}{2}} \mathcal{Q}_{\theta \theta'} (b\alpha) \Ez{\theta'}{\alpha_{2,1}}{\alpha}{\Lambda_1^2 \Lambda_2^2}{\frac{1}{\Lambda_1 \sqrt{z}}} \,.
\end{equation}
In diagrams:
\begin{equation}
    \begin{tikzpicture}[baseline={(current bounding box.center)}, node distance=1cm and 1.5cm]
    \coordinate[circle,fill,inner sep=2pt] (aux1);
    \coordinate[left=of aux1] (e1);
    \coordinate[right=1.5cm of aux1] (aux2);
    \coordinate[above=of aux2,label=above:$\alpha_{2,1}$] (e2);
    \coordinate[right=1.5cm of aux2,circle,fill,inner sep=2pt] (aux3);
    \coordinate[right=of aux3] (e4);

    \draw[line,decoration=snake] (aux1) -- (e1);
    \draw[line,dashed,red] (aux2) -- (e2);
    \draw[line,decoration=snake] (aux3) -- (e4);
    \draw[line] (aux1) -- node[label=below:$\alpha_\theta$] {} (aux2);
    \draw[line] (aux2) -- node[label=below:$\alpha$] {} (aux3);
    \end{tikzpicture}
    = \sum_{\theta'=\pm} \mathcal{Q}_{\theta \theta'} 
    \begin{tikzpicture}[baseline={(current bounding box.center)}, node distance=1cm and 1.5cm]
    \coordinate (aux1);
    \coordinate[left=of aux1] (e1);
    \coordinate[right=1.5cm of aux1,circle,fill,inner sep=2pt] (aux2);
    \coordinate[above=of aux1,label=above:$\alpha_{2,1}$] (e2);
    \coordinate[right=1.5cm of aux2,circle,fill,inner sep=2pt] (aux3);
    \coordinate[right=of aux3] (e4);

    \draw[line,decoration=snake] (aux1) -- (e1);
    \draw[line,dashed,red] (aux1) -- (e2);
    \draw[line,decoration=snake] (aux3) -- (e4);
    \draw[line,decoration=snake] (aux1) -- node[label=below:$\theta'$] {} (aux2);
    \draw[line] (aux2) -- node[label=below:$\alpha$] {} (aux3);
    \end{tikzpicture}\,.
\end{equation}
As in the previous cases, we can easily obtain a connection formula connecting the two irregular singularities, namely
\begin{equation}\label{212}
    b^{-\frac{1}{2}}z^{-2\Delta_{2,1}}\Ez{\theta_1}{\alpha}{\alpha_{2,1}}{\Lambda_1^2 \Lambda_2^2}{\frac{\sqrt{z}}{\Lambda_2}} = \sum_{\theta_2,\theta_3=\pm} b^{2\theta_2 b \alpha} \mathcal{Q}^{-1}_{\theta_1 \theta_2}(b\alpha) b^{-\frac{1}{2}+2\theta_2 b \alpha} \mathcal{Q}_{(-\theta_2) \theta_3}(b\alpha_{\theta_2})\Ez{\theta_3}{\alpha_{2,1}}{\alpha_{\theta_2}}{\Lambda_1^2 \Lambda_2^2}{\frac{1}{\Lambda_1 \sqrt{z}}} \,.
\end{equation}
Diagrammatically:
\begin{equation}
     \begin{tikzpicture}[baseline={(current bounding box.center)}, node distance=1cm and 1.5cm]
    \coordinate[circle,fill,inner sep=2pt] (aux1);
    \coordinate[left=of aux1] (e1);
    \coordinate[right=1.5cm of aux1,circle,fill,inner sep=2pt] (aux2);
    \coordinate[above=of aux3,label=above:$\alpha_{2,1}$] (e2);
    \coordinate[right=1.5cm of aux2] (aux3);
    \coordinate[right=of aux3] (e4);

    \draw[line,decoration=snake] (aux1) -- (e1);
    \draw[line,dashed,red] (aux3) -- (e2);
    \draw[line,decoration=snake] (aux3) -- (e4);
    \draw[line] (aux1) -- node[label=below:$\alpha$] {} (aux2);
    \draw[line,decoration=snake] (aux2) -- node[label=below:$\theta_1$] {} (aux3);
    \end{tikzpicture} = \sum_{\theta_2,\theta_3=\pm} \mathcal{Q}^{-1}_{\theta_1 \theta_2} \mathcal{Q}_{(-\theta_2) \theta_3}
    \begin{tikzpicture}[baseline={(current bounding box.center)}, node distance=1cm and 1.5cm]
    \coordinate (aux1);
    \coordinate[left=of aux1] (e1);
    \coordinate[right=1.5cm of aux1,circle,fill,inner sep=2pt] (aux2);
    \coordinate[above=of aux1,label=above:$\alpha_{2,1}$] (e2);
    \coordinate[right=1.5cm of aux2,circle,fill,inner sep=2pt] (aux3);
    \coordinate[right=of aux3] (e4);

    \draw[line,decoration=snake] (aux1) -- (e1);
    \draw[line,dashed,red] (aux1) -- (e2);
    \draw[line,decoration=snake] (aux3) -- (e4);
    \draw[line,decoration=snake] (aux1) -- node[label=below:$\theta_3$] {} (aux2);
    \draw[line] (aux2) -- node[label=below:$\alpha_{\theta_2}$] {} (aux3);
    \end{tikzpicture}\,.
\end{equation}

\subsubsection{Semiclassical limit}\label{Nf0sc}
The BPZ equation in this limit becomes
\begin{equation}\label{BPZNf0SC}
    \left(\partial_z^2- \frac{L_1^2}{4z} + \frac{u}{z^2}   - \frac{L_2^2}{4z^3} \right) {}_{\frac{1}{2}} \mathfrak{F}_{\frac{1}{2}} = 0 \,.
\end{equation}
for \textit{any} semiclassical block. Here $u$ is given by
\begin{equation}
    u=\lim_{b\to0} b^2 \Lambda_2^2 \partial_{\Lambda_2^2} \log {}_\frac{1}{2}\mathfrak{F}_\frac{1}{2} \left(\alpha ;\Lambda_1^2 \Lambda_2^2 \right) = \frac{1}{4}-a^2+ \mathcal{O}(L_1^2 L_2^2)\,,
\end{equation} where ${}_\frac{1}{2}\mathfrak{F}_\frac{1}{2} \left(\alpha ;\Lambda_1^2 \Lambda_2^2 \right)$ is the conformal block corresponding to $\langle \Lambda_1^2|\Lambda_2^2\rangle$ with intermediate momentum $\alpha$. The finite semiclassical conformal blocks are defined as before by normalizing by the same block without the degenerate field insertion, i.e. for $z\sim 0$
\begin{equation}
    z\, \Ezsc{\theta}{a}{a_{2,1}}{L_1^2 L_2^2}{\frac{\sqrt{z}}{L_2}} = \lim_{b\to0} b^{-1/2} \frac{z^{- 2 \Delta_{2,1}} \Ez{\theta}{\alpha}{\alpha_{2,1}}{\Lambda_1^2 \Lambda_2^2}{\frac{ \sqrt{z}}{\Lambda_2}}}{{}_\frac{1}{2}\mathfrak{F}_\frac{1}{2} \left(\alpha ;\Lambda_1^2 \Lambda_2^2 \right)}\sim e^{\theta L_2/\sqrt{z}} L_2^{-1/2} z^{3/4} (1+\mathcal{O}(L_1^2 L_2^2, \sqrt{z}/L_2))\,.
\end{equation}
For $z\sim \infty$ instead we have
\begin{equation}\label{216}
    \Ezsc{\theta}{a_{2,1}}{a}{L_1^2 L_2^2}{\frac{1}{L_1 \sqrt{z}}} = \lim_{b\to0} b^{-1/2} \frac{\Ez{\theta}{\alpha_{2,1}}{\alpha}{\Lambda_1^2 \Lambda_2^2}{\frac{1}{\Lambda_1 \sqrt{z}}}}{{}_\frac{1}{2}\mathfrak{F}_\frac{1}{2} \left(\alpha ;\Lambda_1^2 \Lambda_2^2 \right)}\sim e^{\theta L_1\sqrt{z}} L_1^{-1/2} z^{1/4} (1+\mathcal{O}(L_1^2 L_2^2,1/L_1 \sqrt{z}))\,.
\end{equation}
Here 
\begin{equation}
    F =\lim_{b\to0} b^2 \log\left[ (\Lambda_1^2\Lambda_2^2)^{-\Delta} {}_\frac{1}{2}\mathfrak{F}_\frac{1}{2} \left(\alpha ;\Lambda_1^2 \Lambda_2^2 \right)\right] \,.
\end{equation}
Both these blocks satisfy the same BPZ equation \eqref{BPZNf0SC}. Analogously to the previous confluences, in the connection formula between $0$ and $\infty$ we have four different $\mathfrak{E}$ blocks appearing, which should reduce to two in the semiclassical limit. Indeed, we have
\begin{equation}
    \Ez{\theta}{\alpha_{2,1}}{\alpha_{\theta'}}{\Lambda_1^2 \Lambda_2^2}{\frac{1}{\Lambda_1 \sqrt{z}}} \sim \left(\Lambda_1^2\Lambda_2^2\right)^{\theta' a}e^{-\frac{\theta'}{2}\partial_{a}F} \Ez{\theta}{\alpha_{2,1}}{\alpha}{\Lambda_1^2 \Lambda_2^2}{\frac{1}{\Lambda_1 \sqrt{z}}}\,,\quad \mathrm{as } \, \,b\to 0\,,
\end{equation}
as in \eqref{twiddle}. Now that we have defined the semiclassical conformal blocks, we state the connection formula. \eqref{212} in the semiclassical limit becomes
\begin{equation}\label{0toinftyNf0}
    z \Ezsc{\theta}{a}{a_{2,1}}{L_1^2 L_2^2}{\frac{\sqrt{z}}{L_2}}= \sum_{\theta'}\left(\sum_{\sigma=\pm} \mathcal{Q}^{-1}_{\theta \sigma}(a) \mathcal{Q}_{(-\sigma) \theta'}(a)\left(L_1 L_2\right)^{2\sigma a} e^{-\frac{\sigma}{2}\partial_a F}\right) \Ezsc{\theta'}{a_{2,1}}{a}{L_1^2 L_2^2}{\frac{1}{L_1 \sqrt{z}}}\,.
\end{equation}
With connection coefficients\footnote{ Note also that there are no Gamma functions in the denominator corresponding to the fact that we have no hypermultiplets in the corresponding AGT dual gauge theory.}
\begin{equation}
\begin{aligned}
    &\sum_{\sigma=\pm} \mathcal{Q}^{-1}_{\theta \sigma}(a) \mathcal{Q}_{(-\sigma) \theta'}(a)\left(L_1 L_2\right)^{2\sigma a} e^{-\frac{\sigma}{2}\partial_a F}= \\= &\frac{1}{2\pi} \sum_{\sigma=\pm} \Gamma(1-2\sigma a)\Gamma(-2\sigma a) \left(\frac{L_1 L_2}{4}\right)^{2\sigma a} e^{-\frac{\sigma}{2}\partial_a F} e^{-i\pi\left(\frac{1+\theta}{2}\right)\left(\frac{1}{2}+2\sigma a\right)}e^{i\pi\left(\frac{1-\theta'}{2}\right)\left(\frac{1}{2}-2\sigma a\right)} \,.
\end{aligned}
\end{equation}
Note that the factors of $b$ appearing in \eqref{212} precisely combine with all the factors of $\Lambda_1,\Lambda_2$ to give the finite $L_1,L_2$.

\section{Heun equations, confluences and connection formulae}

In this section we derive the explicit connection formulae for Heun functions and its confluences by 
identifying the semi-classical conformal blocks
with the Heun functions and using the results so far obtained.
\subsection{The Heun equation}
In the following we identify the semiclassical BPZ equation \eqref{eq:BPZNf=4sc} with Heun's equation
via a dictionary between the relevant parameters.
Moreover, we establish a precise relation between 
the Heun functions and the semiclassical \textit{regular} conformal blocks.
This is further used to obtain explicit formulae for the relevant connection coefficients. 
WLOG, we focus on the case $t \sim 0$. The connection formulae for $t \sim 1,\, t \sim \infty$ can be easily derived by matching the Heun equation and its local solutions with the corresponding semiclassical BPZ equations and the associated semiclassical conformal blocks.
\subsubsection{The dictionary}
Let us start giving the dictionary with CFT. The Heun equation reads
\begin{equation}
\begin{aligned}
    &\left( \frac{d^2 }{dz^2}+\left( \frac{\gamma}{z}+\frac{\delta}{z-1}+\frac{\epsilon}{z-t} \right)\frac{d}{dz}+\frac{\alpha \beta z - q}{z(z-1)(z-t)} \right) w(z) = 0\,,\\
    &\alpha+\beta+1=\gamma+\delta+\epsilon\,,
\end{aligned}
\label{eq:HeunGeq}
\end{equation}
where the condition $\alpha+\beta+1=\gamma+\delta+\epsilon$ ensures that the exponents of the local solutions at infinity are given by $\alpha, \beta$. Here and in the following we restrict to generic values of the parameters. Define $w(z) = P_4(z) \psi(z)$ with 
\begin{equation}
    P_4(z)=z^{-\gamma/2}(1-z)^{-\delta/2}(t-z)^{-\epsilon/2}\,.
\end{equation} 
$\psi(z)$ then satisfies the Heun equation in normal form, which is easily compared with the semiclassical BPZ equation \eqref{eq:BPZNf=4sc}. We get $2^4 = 16$ dictionaries corresponding to the $\left( \mathbb{Z}_2 \right)^4$ symmetry associated to flipping the signs of the momenta. We choose the following:
\begin{equation}
\begin{aligned}
    & a_0 = \frac{1-\gamma}{2} \,,\\
    & a_1 = \frac{1-\delta}{2} \,, \\
    & a_t = \frac{1-\epsilon}{2} \,, \\
    & a_\infty = \frac{\alpha-\beta}{2} \,, \\
    & u^{(0)} = \frac{-2q + 2t\alpha\beta + \gamma\epsilon-t(\gamma+\delta)\epsilon}{2(t-1)} \,.
\end{aligned}
\label{eq:HeundictioanaryscCFT}
\end{equation}
The inverse dictionary is
\begin{equation}
\begin{aligned}
    & \alpha = 1 - a_0 - a_1 - a_t + a_\infty \,,\\
    & \beta = 1 - a_0 - a_1 - a_t - a_\infty \,, \\
    & \gamma = 1 - 2 a_0 \,, \\
    & \delta = 1 - 2 a_1\,, \\
    & \epsilon = 1 - 2 a_t \,, \\
    & q = \frac{1}{2} + t (a_0^2 + a_t^2 + a_1^2 - a_\infty^2) - a_t - a_1 t + a_0 (2 a_t - 1 + t (2 a_1 - 1)) + (1-t) u^{(0)} \,.
\end{aligned}
\end{equation}
The two linearly independent solutions for $z \sim 0$ of \eqref{eq:BPZNf=4sc} are related by $a_0 \to - a_0$.
This corresponds to the identification of the 
two linearly independent solutions of \eqref{eq:HeunGeq} for $z \sim 0$ as
\begin{equation}\label{HeunGzero}
\begin{aligned}
    &w_-^{(0)} (z) = \text{HeunG} \left(t, q, \alpha, \beta, \gamma, \delta, z \right) \,, \\ &w_+^{(0)} (z) = z^{1-\gamma} \text{HeunG} \left(t, q - (\gamma - 1)(t \delta + \epsilon), \alpha + 1 - \gamma, \beta + 1 - \gamma, 2 -\gamma, \delta, z \right) \,,
\end{aligned}
\end{equation}
where by definition
\begin{equation}
    \text{HeunG} \left(t, q, \alpha, \beta, \gamma, \delta, z \right) = 1 + \frac{q}{t \gamma} z + \mathcal{O} (z^2) \,.
\end{equation}
The Heun function can be identified with the semiclassical conformal blocks introduced before. In particular comparing with \eqref{eq:smalltsmallzsccbNf4} we get the two solutions
\begin{equation}
\begin{aligned}
    &w_-^{(0)}(z) =  P_4(z) \,t^{\frac{1}{2} - a_t - a_0} e^{- \frac{1}{2} \partial_{a_0} F(t)} \FIVsc{a_\infty}{a_1}{a}{a_t}{{a_{0 -}}}{a_{2,1}}{a_0}{t}{\frac{z}{t}} \,, \\
    &w_+^{(0)}(z) = P_4(z) \,t^{\frac{1}{2} - a_t + a_0} e^{\frac{1}{2} \partial_{a_0} F(t)} \FIVsc{a_\infty}{a_1}{a}{a_t}{{a_{0 +}}}{a_{2,1}}{a_0}{t}{\frac{z}{t}} \,.
\end{aligned}
\end{equation}
Note that HeunG is an expansion in $z$, while the semiclassical conformal blocks are expanded both in $z$ and $t$. To match the two expansions one has to express the accessory parameter $q$ in terms of the Floquet exponent $a$ as a series in $t$. This can be done substituting the dictionary as explained in Appendix \ref{app:Nek}.\newline The solutions for $z \sim t$ are given by
\begin{equation}
\begin{aligned}
    &w_-^{(t)} (z) = \text{HeunG} \left(\frac{t}{t-1}, \frac{q-t \alpha \beta}{1-t}, \alpha, \beta, \epsilon, \delta, \frac{z-t}{1-t} \right) \,, \\ &w_+^{(t)} (z) = (t-z)^{1- \epsilon} \text{HeunG} \left( \frac{t}{t-1}, \frac{q-t \alpha \beta}{1-t} - (\epsilon - 1) \left( \frac{t}{t-1} \delta + \gamma \right), \alpha + 1 -\epsilon, \beta + 1 - \epsilon, 2 - \epsilon, \delta, \frac{z-t}{1-t} \right) \,.
\end{aligned}
\end{equation}
Comparing with the semiclassical blocks \eqref{eq:zneartsmalltsccbNf4} we get
\begin{equation}
\begin{aligned}
    & w_-^{(t)} (z) = P_4(z)\,t^{\frac{1}{2} - a_0 - a_t} (1-t)^{\frac{1}{2}-a_1} e^{- \frac{1}{2} \partial_{a_t} F(t)} \left( (t-1)^{\frac{1}{2}} \FIVsc{a_\infty}{a_1}{a}{a_0}{{a_{t -}}}{a_{2,1}}{a_t}{\frac{t}{t-1}}{\frac{t-z}{t}} \right) \,, \\
    &w_+^{(t)} (z) =P_4(z)\, t^{\frac{1}{2} - a_0 + a_t} (1-t)^{\frac{1}{2}-a_1} e^{\frac{1}{2} \partial_{a_t} F(t)} \left( (t-1)^{\frac{1}{2}} \FIVsc{a_\infty}{a_1}{a}{a_0}{{a_{t +}}}{a_{2,1}}{a_t}{\frac{t}{t-1}}{\frac{t-z}{t}} \right)  \,.
\end{aligned}
\end{equation}
The two solutions for $z \sim 1$ read
\begin{equation}
\begin{aligned}
    &w_-^{(1)} (z) = \left( \frac{z-t}{1-t} \right)^{- \alpha} \text{HeunG} \left( t, q + \alpha (\delta - \beta), \alpha, \delta + \gamma - \beta, \delta, \gamma, t \frac{1-z}{t-z} \right) \,, \\
    &w_+^{(1)} (z) = \left( \frac{z-t}{1-t} \right)^{- \alpha - 1 + \delta} (1-z)^{1- \delta} \text{HeunG} \left( t, q - \alpha (\beta + \delta - 2) + (\delta - 1) (\alpha + \beta - 1 - t \gamma), \alpha + 1 - \delta, 1 + \gamma - \beta,2 - \delta, \gamma, t \frac{1-z}{t-z} \right) \,,
\end{aligned}
\end{equation}
and matching with \eqref{eq:znear1smalltsccbNf4} gives
\begin{equation}
\begin{aligned}
    & w_-^{(1)} (z) = P_4(z) e^{\pm i \pi (a_1 + a_t)} (1-t)^{\frac{1}{2} - a_t} e^{-\frac{1}{2} \partial_{a_1} F(t)} \left( \left( t(1-t) \right)^{-\frac{1}{2}} (t-z) \FIVsc{a_t}{a_0}{a}{a_\infty}{{a_{1 -}}}{a_{2,1}}{a_1}{t}{\frac{1-z}{t-z}} \right) \\ & w_+^{(1)} (z) =P_4(z) e^{\pm i \pi (-a_1 + a_t)} (1-t)^{\frac{1}{2} - a_t } e^{\frac{1}{2} \partial_{a_1} F(t)} \left( \left( t(1-t) \right)^{-\frac{1}{2}} (t-z) \FIVsc{a_t}{a_0}{a}{a_\infty}{{a_{1 +}}}{a_{2,1}}{a_1}{t}{\frac{1-z}{t-z}} \right) \,.
\end{aligned}
\end{equation}
The $\pm$ ambuiguity in the overall phase depends on the choice of branch corresponding to
\begin{equation}
    P_4(z) \FIVsc{a_t}{a_0}{a}{a_\infty}{{a_{1 \theta}}}{a_{2,1}}{a_1}{t}{\frac{1-z}{t-z}} \propto \left( t-1 \right)^{\theta a_1 + a_t} = e^{\pm i \pi (\theta a_1 + a_t)} \left( 1 - t \right)^{\theta a_1 + a_t} \,.
\end{equation}
Finally, the two solutions near $z \sim \infty$ are given by
\begin{equation}
\begin{aligned}
    &w_+^{(\infty)} (z) = z^{- \alpha}\text{HeunG} \left( t, q - \alpha \beta (1+t) + \alpha (\delta + t \epsilon), \alpha, \alpha - \gamma + 1, \alpha - \beta + 1, \alpha + \beta + 1 - \gamma - \delta, \frac{t}{z} \right) \,, \\ &w_-^{(\infty)} (z) = z^{- \beta}\text{HeunG} \left( t, q - \alpha \beta (1+t) + \beta (\delta + t \epsilon), \beta, \beta - \gamma + 1, \beta - \alpha + 1, \alpha + \beta + 1 - \gamma - \delta, \frac{t}{z} \right) \,.
\end{aligned}
\end{equation}
Comparing with \eqref{eq:zbigsmalltsccbNf4} we get
\begin{equation}
\begin{aligned}
    & w_+^{(\infty)} (z) =P_4(z) e^{\pm i \pi (1 - a_1 - a_t)}e^{\frac{1}{2} \partial_{a_\infty} F(t)} \left( t^{-\frac{1}{2}} z \FIVsc{a_0}{a_t}{a}{a_1}{{a_{\infty +}}}{a_{2,1}}{a_\infty}{t}{\frac{1}{z}} \right) \,, \\ & w_-^{(\infty)} (z) = P_4(z) e^{\pm i \pi (1 - a_1 - a_t)} e^{ - \frac{1}{2} \partial_{a_\infty} F(t)} \left( t^{-\frac{1}{2}} z \FIVsc{a_0}{a_t}{a}{a_1}{{a_{\infty -}}}{a_{2,1}}{a_\infty}{t}{\frac{1}{z}} \right) \,,
\end{aligned}
\end{equation}
where again the $\pm$ in the phase depends on the choice of branch corresponding to
\begin{equation}
    P_4(z) = z^{-\frac{1}{2} + a_0} (1-z)^{- \frac{1}{2} + a_1} (t -z)^{-\frac{1}{2} + a_t} = e^{\mp i \pi (1 - a_1 - a_t)} z^{-\frac{1}{2} + a_0} (z-1)^{- \frac{1}{2} + a_1} (z-t)^{-\frac{1}{2} + a_t} \,.
\end{equation}

\subsubsection{Connection formulae}
Finally we are in the position to give the connection formulae for the Heun function. Let us start with $z \sim 0$ and $z \sim t$. The corresponding connection formula can be read off from \eqref{eq:scconnection0t}, which in the Heun notation reads
\begin{equation}
\begin{aligned}
    &w_-^{(0)} (z) = \frac{\Gamma(1-\epsilon) \Gamma(\gamma) e^{\frac{1}{2} \left( \partial_{a_t} - \partial_{a_0} \right) F}}{\Gamma \left( \frac{1+\gamma-\epsilon}{2} + a(q) \right) \Gamma \left( \frac{1+\gamma-\epsilon}{2} - a(q) \right)} (1-t)^{-\frac{\delta}{2}} w_-^{(t)} (z) + \frac{\Gamma(\epsilon-1) \Gamma(\gamma) e^{\frac{1}{2} \left( - \partial_{a_t} - \partial_{a_0} \right) F}}{\Gamma \left( \frac{-1+\gamma+\epsilon}{2} + a(q) \right) \Gamma \left( \frac{-1+\gamma+\epsilon}{2} - a(q) \right)} t^{\epsilon - 1} (1-t)^{-\frac{\delta}{2}} w_+^{(t)} (z) \,,
\end{aligned}
\label{eq:connectw-}
\end{equation}
for the other solution one finds
\begin{equation}
\begin{aligned}
    &w_+^{(0)} (z) = \frac{\Gamma(1-\epsilon) \Gamma(2-\gamma) e^{\frac{1}{2} \left( \partial_{a_t} + \partial_{a_0} \right) F}}{\Gamma \left( 1+\frac{1-\gamma-\epsilon}{2} + a(q) \right) \Gamma \left( 1+\frac{1-\gamma-\epsilon}{2} - a(q) \right)} t^{1-\gamma}(1-t)^{-\frac{\delta}{2}} w_-^{(t)} (z) + \frac{\Gamma(\epsilon-1) \Gamma(\gamma) e^{\frac{1}{2} \left( - \partial_{a_t} + \partial_{a_0} \right) F}}{\Gamma \left( \frac{1-\gamma+\epsilon}{2} + a(q) \right) \Gamma \left( \frac{1-\gamma+\epsilon}{2} - a(q) \right)} t^{\epsilon-\gamma} (1-t)^{-\frac{\delta}{2}} w_+^{(t)} (z) \,.
\end{aligned}
\end{equation}
Here $a(q)$ has to be computed inverting the relation \eqref{eq:MatoneNf40} and substituting the dictionary as shown explicitly in Appendix \ref{app:Nek}, formula \eqref{eq:aofq}. The result to first order is
\begin{equation}
\begin{aligned}
    &a(q) = \frac{1}{16} \sqrt{3 - 4 q + \gamma^2 + 2 \gamma (\epsilon - 1) + \epsilon (\epsilon - 2)} \times \\ &\times \left( 8 -  \frac{4(-1 + 2 q -  \epsilon (\gamma + \epsilon - 2)) (- 3 + 4 q + (\alpha - \beta) - \gamma^2 - \delta(\delta - 2) - 2 \gamma  (\epsilon - 1) - \epsilon (\epsilon - 2))}{(3 - 4 q + \gamma^2 + 2 \gamma (\epsilon - 1) + \epsilon (\epsilon - 2)) (2 - 4 q + \gamma^2 + 2 \gamma (\epsilon - 1) + \epsilon (\epsilon - 2))} t \right) + \mathcal{O} (t^2) \,.
\end{aligned}
\end{equation}
In Appendix \ref{app:Nek} we also explain how to compute the classical conformal block $F$ and its derivatives (see formula \ref{dF}). For example, to first order
\begin{equation}
    \partial_{a_t} F(t) = \frac{\left( 4 a(q)^2 - \alpha^2 + 2 \alpha \beta - \beta^2 - 2 \delta + \delta^2\right) \left(1-\epsilon \right)}{2 - 8 a(q)^2} t + \mathcal{O} (t^2) \,.
\end{equation}
The connection formula for $w_+^{(0)} (z)$ can be obtained from \eqref{eq:connectw-} by multiplying by $z^{1-\gamma}$, substituting 
\begin{equation}
    q\to q-(\gamma-1)(t \delta + \epsilon), \,\, \alpha \to \alpha+1-\gamma, \,\, \beta \to \beta +1 -\gamma, \,\, \gamma \to 2-\gamma 
\end{equation}
as in \eqref{HeunGzero}, and noting that
\begin{equation}
\begin{aligned}
    &\text{HeunG} \left( \frac{t}{t-1}, \frac{q - t \alpha \beta}{1 - t}, \alpha, \beta, \epsilon, \delta, \frac{z-t}{1-t} \right) = \\ &=  \left(\frac{z}{t}\right)^{1 - \gamma} \text{HeunG} \left( \frac{t}{t-1}, \frac{q - t (\alpha+1-\gamma)(\beta+1-\gamma)-(\gamma-1)(t \delta + \epsilon)}{1 - t}, \alpha + 1 - \gamma, \beta + 1 - \gamma, \epsilon, \delta, \frac{z-t}{1-t} \right) \,.
\end{aligned}
\end{equation}
Similarly, the connection formula from $z \sim 0$ to $z \sim \infty$ can be read off from \eqref{eq:sconnection0inft}, and gives
\begin{equation}
\begin{aligned}
     w_-^{(0)} (z) &=\left( \sum_{\sigma=\pm} \frac{\Gamma(1-2\sigma a(q))\Gamma(-2\sigma a(q))\Gamma(\gamma)\Gamma(\beta-\alpha)t^{\frac{\gamma+\epsilon-1}{2}-\sigma a(q)}e^{-\frac{1}{2}\left( \partial_{a_0} - \partial_{a_\infty} + \sigma \partial_a\right) F} e^{i \pi\left( \frac{\delta+\gamma}{2}\right)}}{\Gamma\left(\frac{\gamma-\epsilon+1}{2}-\sigma a(q)\right)\Gamma\left(\frac{\gamma+\epsilon-1}{2}-\sigma a(q)\right)\Gamma\left(1+\frac{\beta-\alpha-\delta}{2}-\sigma a(q)\right)\Gamma\left(\frac{\beta-\alpha+\delta}{2}-\sigma a(q)\right)} \right) w_+^{(\infty)}(z) + \\ &+ \left( \sum_{\sigma=\pm} \frac{\Gamma(1-2\sigma a(q))\Gamma(-2\sigma a(q))\Gamma(\gamma)\Gamma(\alpha-\beta)t^{\frac{\gamma+\epsilon-1}{2}-\sigma a(q)}e^{-\frac{1}{2}\left( \partial_{a_0} - \partial_{a_\infty} + \sigma \partial_a\right) F} e^{i \pi\left( \frac{\delta+\gamma}{2}\right)}}{\Gamma\left(\frac{\gamma-\epsilon+1}{2}-\sigma a(q)\right)\Gamma\left(\frac{\gamma+\epsilon-1}{2}-\sigma a(q)\right)\Gamma\left(1+\frac{\alpha-\beta-\delta}{2}-\sigma a(q)\right)\Gamma\left(\frac{\alpha-\beta+\delta}{2}-\sigma a(q)\right)} \right) w_-^{(\infty)}(z) \,.
\end{aligned}
\end{equation}
Let us conclude the section by giving the connection formulae from 1 to infinity. This can be derived from \eqref{eq:regfrom1toinfinity}, and gives
\begin{equation}
\begin{aligned}
    w_-^{(1)} (z) = &-(1-t)^{\frac{1}{2} - a_t} \frac{\Gamma(\beta-\alpha)\Gamma(\delta)e^{- \frac{1}{2} \left( \partial_{a_1} + \partial_{a_\infty} \right) F(t)}}{\Gamma\left(\frac{\delta-\alpha+\beta}{2}+a(q)\right) \Gamma\left(\frac{\delta-\alpha+\beta}{2}-a(q) \right)} w_+^{(\infty)} (z) +\\
    &-(1-t)^{\frac{1}{2} - a_t}  \frac{\Gamma(\alpha-\beta)\Gamma(\delta)e^{- \frac{1}{2} \left( \partial_{a_1} - \partial_{a_\infty} \right) F(t)}}{\Gamma\left(\frac{\delta+\alpha-\beta}{2}+a(q)\right) \Gamma\left(\frac{\delta+\alpha-\beta}{2}-a(q) \right)}w_-^{(\infty)} (z)\,.
\end{aligned}
\end{equation}
The connection formulae involving the other solutions can be read off from the previous ones, and the formulae involving different pairs of points can be similarly derived by considering the corresponding semiclassical conformal blocks. We conclude by stressing again that the connection formulae involving different regions in the $t-$plane are completely analogous to the previous ones, since all the singularities are regular. This will not be the case in the following.

\subsection{The confluent Heun equation}
\subsubsection{The dictionary}
Here we establish the dictionary between our results of section \ref{Nf3} on confluent conformal blocks and the confluent Heun equation (CHE) in standard notation, which reads
\begin{equation}\label{CHE_standard}
    \frac{d^2w}{dz^2}+\left(\frac{\gamma}{z}+\frac{\delta}{z-1}+\epsilon\right)\frac{dw}{dz}+\frac{\alpha z-q}{z(z-1)}w=0\,.
\end{equation}
By defining $w(z)=P_3(z)\psi(z)$ with $P_3(z)=e^{-\epsilon z/2} z^{-\gamma/2} (1-z)^{-\delta/2}$, we get rid of the first derivative and bring the equation to normal form, which can easily be compared with the semiclassical BPZ equation \eqref{BPZNf3NS}. We can read off the dictionary between the CFT parameters and the parameters of the CHE:
\begin{equation}\label{CHE_dictionary}
    \begin{aligned}
    &a_0 = \frac{1-\gamma}{2}\,,\\
    &a_1 = \frac{1-\delta}{2}\,,\\
    &m =\frac{\alpha}{\epsilon}- \frac{\gamma+\delta}{2}\,,\\
    &L = \epsilon\,,\\
    &u = \frac{1}{4}-q+\alpha - \frac{(\gamma+\delta-1)^2}{4}-\frac{\delta \epsilon}{2}\,,
    \end{aligned}
\end{equation}
where 
\begin{equation}
    u=\lim_{b\to0} b^2 \Lambda \partial_\Lambda \log {}_1\mathfrak{F} \left( \mu \, \alpha\, \begin{matrix} \alpha_1\\ \alpha_0 \end{matrix} ;\Lambda \right) = \frac{1}{4}-a^2+ \mathcal{O}(L)
\end{equation} as in \eqref{BPZNf3NS}. This relation can then be inverted to find $a$ in terms of the parameters of the CHE: we denote this by $a(q)$. We write the solutions to the CHE in standard form in the notation of Mathematica, and their relation to the conformal blocks used before. We focus first on the blocks given as an expansion for small $L$. Then, near $z=0$ we have the two linearly independent solutions
\begin{equation}
    \begin{aligned}
    & \mathrm{HeunC}(q,\alpha,\gamma,\delta,\epsilon;z) \,,\\
    & z^{1-\gamma} \mathrm{HeunC}\left(q+(1-\gamma)(\epsilon-\delta),\alpha+(1-\gamma)\epsilon, 2-\gamma,\delta,\epsilon;z \right) \,,
    \end{aligned}
\end{equation}
where the confluent Heun function has the following expansion around $z=0$:
\begin{equation}
    \mathrm{HeunC}(q,\alpha,\gamma,\delta,\epsilon;z) = 1 - \frac{q}{\gamma}z + \mathcal{O}(z^2)\,.
\end{equation}
Comparing with the semiclassical conformal blocks in \eqref{Nf3SC} we identify
\begin{equation}
    \begin{aligned}
    & \mathrm{HeunC}(q,\alpha,\gamma,\delta,\epsilon;z) = P_3(z) e^{-\frac{1}{2}\partial_{a_0}F} \FIIIsc{m}{a}{a_1}{a_{0-}}{a_{2,1}}{a_0}{L}{z}\,,\\
    & z^{1-\gamma} \mathrm{HeunC}\left(q+(1-\gamma)(\epsilon-\delta),\alpha+(1-\gamma)\epsilon, 2-\gamma,\delta,\epsilon;z \right) = P_3(z) e^{\frac{1}{2}\partial_{a_0}F} \FIIIsc{m}{a}{a_1}{a_{0+}}{a_{2,1}}{a_0}{L}{z}\,,
    \end{aligned}
\end{equation}
where 
\begin{equation}
    F = \lim_{b\to0} b^2 \log \left[\Lambda^{-\Delta}{}_1\mathfrak{F} \left( \mu \, \alpha\, \begin{matrix} \alpha_1\\ \alpha_0 \end{matrix} ;\Lambda \right)\right]\,.
\end{equation}
Doing a M\"obius transformation $z\to1-z$ we obtain solutions around $z=1$, which being a regular singularity can again be written in terms of HeunC. This amounts to sending $\gamma\to \delta,\, \delta\to\gamma,\,\epsilon\to-\epsilon,\,\alpha \to -\alpha,\,q \to q-\alpha$. The two solutions are therefore
\begin{equation}
    \begin{aligned}
    &\mathrm{HeunC}(q-\alpha,-\alpha,\delta,\gamma,-\epsilon;1-z)  \,,\\
    & (1-z)^{1-\delta} \mathrm{HeunC}\left(q-\alpha-(1-\delta)(\epsilon+\gamma),-\alpha-(1-\delta)\epsilon, 2-\delta,\gamma,-\epsilon;1-z \right)\,.
    \end{aligned}
\end{equation}
Again, comparing with the semiclassical conformal blocks in \eqref{Nf3SC}, we identify
\begin{equation}
    \begin{aligned}
    &\mathrm{HeunC}(q-\alpha,-\alpha,\delta,\gamma,-\epsilon;1-z) = P_3(z) e^{-\frac{1}{2}\partial_{a_1}F}\FIIIsc{-m}{a}{a_0}{a_{1-}}{a_{2,1}}{a_1}{L}{1-z} \,,\\
    & (1-z)^{1-\delta} \mathrm{HeunC}\left(q-\alpha-(1-\delta)(\epsilon+\gamma),-\alpha-(1-\delta)\epsilon, 2-\delta,\gamma,-\epsilon;1-z \right)=\\
    &\quad \quad=P_3(z) e^{\frac{1}{2}\partial_{a_1}F}\FIIIsc{-m}{a}{a_0}{a_{1+}}{a_{2,1}}{a_1}{L}{1-z}\,.
    \end{aligned}
\end{equation}
Around the irregular singular point $z=\infty$, we write the solutions in terms of a different function $\mathrm{HeunC}_\infty$:
\begin{equation}
    \begin{aligned}
    &z^{-\frac{\alpha}{\epsilon}}\mathrm{HeunC}_\infty(q,\alpha,\gamma,\delta,\epsilon;z^{-1})  \,\\
    &e^{-\epsilon z} z^{\frac{\alpha}{\epsilon}-\gamma-\delta} \mathrm{HeunC}_\infty(q-\gamma\epsilon,\alpha-\epsilon(\gamma+\delta),\gamma,\delta,-\epsilon;z^{-1})\,,
    \end{aligned}
\end{equation}
where the function $\mathrm{HeunC}_\infty$ has a simple asymptotic expansion around $z=\infty$:
\begin{equation}
    \mathrm{HeunC}_\infty(q,\alpha,\gamma,\delta,\epsilon;z^{-1}) \sim 1+ \frac{\alpha^2 -(\gamma+\delta-1)\alpha\epsilon+(\alpha-q)\epsilon^2}{\epsilon^3} z^{-1} + \mathcal{O}(z^{-2})\,.
\end{equation}
Comparing with the semiclassical conformal blocks we identify
\begin{equation}\label{219}
    \begin{aligned}
    &z^{-\frac{\alpha}{\epsilon}}\mathrm{HeunC}_\infty(q,\alpha,\gamma,\delta,\epsilon;z^{-1}) =e^{\mp\frac{i\pi\delta}{2}} P_3(z)e^{\frac{1}{2}\partial_{m}F} L^{\frac{1}{2}+m} \DIIIsc{m}{a_{2,1}}{m_{+}}{a}{a_1}{a_0}{L}{\frac{1}{z}} \,\\
    &e^{-\epsilon z} z^{\frac{\alpha}{\epsilon}-\gamma-\delta} \mathrm{HeunC}_\infty(q-\gamma\epsilon,\alpha-\epsilon(\gamma+\delta),\gamma,\delta,-\epsilon;z^{-1})=e^{\mp\frac{i\pi\delta}{2}} P_3(z)e^{-\frac{1}{2}\partial_{m}F}  L^{\frac{1}{2}-m} \DIIIsc{m}{a_{2,1}}{m_{-}}{a}{a_1}{a_0}{L}{\frac{1}{z}}\,.
    \end{aligned}
\end{equation}
The phase $e^{\mp \frac{i\pi\delta}{2}}$ comes from the fact that near $z=\infty$ 
\begin{equation}
    P_3(z)\sim e^{-\epsilon  z/2} z^{-\gamma/2}(-z)^{-\delta/2} =e^{\pm \frac{i\pi\delta}{2}} e^{-\epsilon  z/2} z^{-\gamma/2-\delta/2}\,.
\end{equation}
The second solution around $z=\infty$ can be found by using the manifest symmetry $(m,L)\to(-m,-L)$ of the semiclassical BPZ equation which according to the dictionary gives the symmetry $(q,\alpha,\epsilon)\to(q-\gamma\epsilon,\alpha-\epsilon(\gamma+\delta),-\epsilon)$ of the CHE in normal form. \newline
For the large-$L$ blocks the story is analogous. The dictionary \eqref{CHE_dictionary} is the same, up to the substitution
\begin{equation}
    u \to u_D=\lim_{b\to0} b^2 \Lambda \partial_\Lambda \log {}_1 \mathfrak{D}\left(\mu\,\begin{matrix}\alpha_1\\ {}\end{matrix}\,\mu'\, \alpha_0; \frac{1}{\Lambda}\right) = -(m'-m)L+\frac{1}{4}-a_0^2 + 2m'(m'-m)+\mathcal{O}(L^{-1})\,.
\end{equation} 
This relation can be inverted to find $m'$ in terms of the parameters of the CHE. We will call this $m'(q)$. With this dictionary we can identify solutions of the CHE with conformal blocks as follows: near $z=0$ we have
\begin{equation}
    \begin{aligned}
    & \mathrm{HeunC}(q,\alpha,\gamma,\delta,\epsilon;z) = P_3(z) e^{-\frac{1}{2}\partial_{a_0}F_D}\, \DIIIsc{m}{a_1}{m'}{a_{0-}}{a_{2,1}}{a_0}{\frac{1}{L}}{L z}  \,,\\
    & z^{1-\gamma} \mathrm{HeunC}\left(q+(1-\gamma)(\epsilon-\delta),\alpha+(1-\gamma)\epsilon, 2-\gamma,\delta,\epsilon;z \right) = P_3(z) e^{\frac{1}{2}\partial_{a_0}F_D}\, \DIIIsc{m}{a_1}{m'}{a_{0+}}{a_{2,1}}{a_0}{\frac{1}{L}}{L z} \,,
    \end{aligned}
\end{equation}
with $F_D$ given in \eqref{FD}. Near $z=1$ we have
\begin{equation}
    \begin{aligned}
    &\mathrm{HeunC}(q-\alpha,-\alpha,\delta,\gamma,-\epsilon;1-z) = P_3(z)e^{-\frac{1}{2}\partial_{a_1}F_D}\,\DIIIsc{-m}{a_0}{m'-m}{a_{1-}}{a_{2,1}}{a_1}{\frac{1}{L}}{L(1-z)}\,,\\
    & (1-z)^{1-\delta} \mathrm{HeunC}\left(q-\alpha-(1-\delta)(\epsilon+\gamma),-\alpha-(1-\delta)\epsilon, 2-\delta,\gamma,-\epsilon;1-z \right)=\\
    &\quad \quad=P_3(z)e^{\frac{1}{2}\partial_{a_1}F_D}\,\DIIIsc{-m}{a_0}{m'-m}{a_{1+}}{a_{2,1}}{a_1}{\frac{1}{L}}{L(1-z)}\,.
    \end{aligned}
\end{equation}
While near $z=\infty$ we have
\begin{equation}
    \begin{aligned}
    &z^{-\frac{\alpha}{\epsilon}}\mathrm{HeunC}_\infty(q,\alpha,\gamma,\delta,\epsilon;z^{-1}) =e^{\mp\frac{i\pi\delta}{2}} P_3(z)e^{L/2}e^{\frac{1}{2}\partial_{m}F_D}L^{\frac{1}{2}-(m'-m)}\DIIIlargeLambdasc{m}{a_{2,1}}{m_+ }{a_1}{m'}{a_0}{\frac{1}{L}}{\frac{1}{z}}  \,\\
    &e^{-\epsilon z} z^{\frac{\alpha}{\epsilon}-\gamma-\delta} \mathrm{HeunC}_\infty(q-\gamma\epsilon,\alpha-\epsilon(\gamma+\delta),\gamma,\delta,-\epsilon;z^{-1})=e^{\mp\frac{i\pi\delta}{2}} P_3(z)e^{-L/2}e^{-\frac{1}{2}\partial_{m}F_D}L^{\frac{1}{2}+(m'-m)}\DIIIlargeLambdasc{m}{a_{2,1}}{m_+ }{a_1}{m'}{a_0}{\frac{1}{L}}{\frac{1}{z}} \,.
    \end{aligned}
\end{equation}
As the careful reader should have noticed, we identify the small-$L$ and large-$L$ conformal blocks with the same confluent Heun functions. The only difference is in the expansion of the accessory parameter: in one case it is given in terms of the Floquet exponent $a$ as an expansion in $L$, and in the other case in terms of the parameter $m'$ as an expansion in $L^{-1}$.

\subsubsection{Connection formulae}
The connection formula between $z=0,1$ written in \eqref{Nf3from0to1SC} for the semiclassical conformal blocks can now be restated as:
\begin{equation}
    \begin{aligned}
    &\mathrm{HeunC}(q,\alpha,\gamma,\delta,\epsilon;z) = \frac{\Gamma(1-\delta)\Gamma(\gamma)e^{-\frac{1}{2}\partial_{a_0}F+\frac{1}{2}\partial_{a_1}F}}{\Gamma\left(\frac{1+\gamma-\delta}{2}+a(q)\right)\Gamma\left(\frac{1+\gamma-\delta}{2}-a(q)\right)}\mathrm{HeunC}(q-\alpha,-\alpha,\delta,\gamma,-\epsilon;1-z) + \\
    +&\, \frac{\Gamma(\delta-1)\Gamma(\gamma)e^{-\frac{1}{2}\partial_{a_0}F-\frac{1}{2}\partial_{a_1}F}}{\Gamma\left(\frac{\gamma+\delta-1}{2}+a(q)\right)\Gamma\left(\frac{\gamma+\delta-1}{2}-a(q)\right)}(1-z)^{1-\delta} \mathrm{HeunC}\left(q-\alpha-(1-\delta)(\epsilon+\gamma),-\alpha-(1-\delta)\epsilon, 2-\delta,\gamma,-\epsilon;1-z \right)\,.
    \end{aligned}
\end{equation}
The quantities $a(q)$ and $F$ can be computed as explained in Appendix \ref{app:Nek}.\newline
The connection formula between $z=1,\infty$ written in \eqref{Nf31toinfSC} reads in terms of confluent Heun functions:
\begin{equation}
    \begin{aligned}
    &\mathrm{HeunC}(q-\alpha,-\alpha,\delta,\gamma,-\epsilon;1-z) =\\=& \left( \sum_{\sigma=\pm}\frac{\Gamma(-2\sigma a(q))\Gamma(1-2\sigma a(q))\Gamma(\delta)\epsilon^{-\frac{1}{2}-\frac{\alpha}{\epsilon}+\frac{\gamma+\delta}{2}+\sigma a(q)}e^{\pm 
    \frac{i\pi\delta}{2}-\frac{1}{2}\partial_{a_1}F+\frac{1}{2}\partial_{m}F-\frac{\sigma}{2}\partial_{a} F(a)}}{\Gamma\left(\frac{1-\gamma+\delta}{2}-\sigma a(q)\right)\Gamma\left(\frac{\gamma+\delta-1}{2}-\sigma a(q)\right)\Gamma\left(\frac{1+\gamma+\delta}{2}-\frac{\alpha}{\epsilon}-\sigma a(q)\right)}\right)\times \\&\times z^{-\frac{\alpha}{\epsilon}}\mathrm{HeunC}_\infty(q,\alpha,\gamma,\delta,\epsilon;z)\,+\\+& \left(\sum_{\sigma=\pm}\frac{\Gamma(-2\sigma a(q))\Gamma(1-2\sigma a(q))\Gamma(\delta)\epsilon^{-\frac{1}{2}+\frac{\alpha}{\epsilon}-\frac{\gamma+\delta}{2}+\sigma a(q)}e^{\pm 
    \frac{i\pi\delta}{2}-\frac{1}{2}\partial_{a_1}F+\frac{1}{2}\partial_{m}F-\frac{\sigma}{2}\partial_{a} F(a)} }{\Gamma\left(\frac{1-\gamma+\delta}{2}-\sigma a(q)\right)\Gamma\left(\frac{\gamma+\delta-1}{2}-\sigma a(q)\right)\Gamma\left(\frac{1-\gamma-\delta}{2}+\frac{\alpha}{\epsilon}-\sigma a(q)\right)}\right)\times \\&\times  e^{-\epsilon z} z^{\frac{\alpha}{\epsilon}-\gamma-\delta} \mathrm{HeunC}_\infty(q-\gamma\epsilon,\alpha-\epsilon(\gamma+\delta),\gamma,\delta,-\epsilon;z)\,.
    \end{aligned}
\end{equation}
Here the phase ambiguity comes from \eqref{219}, i.e. corresponds to the choice $(-z)^{-\delta/2} = e^{\pm \frac{i\pi\delta}{2}}z^{-\delta/2}$. A similar expression can be found connecting $z=0$ and $\infty$. All connection coefficients given above are calculated in a series expansion in $L$. Therefore they are not valid for large $L$ and in that case one has to use different connection formulae, which are derived in section \ref{Nf3SC} for the large-$L$ semiclassical conformal blocks. Here we restate those results in the language of Heun functions. The connection formula from $z=0$ to $z=1$, valid for large $L$ is given by
\begin{equation}
    \begin{aligned}
    &\mathrm{HeunC}(q,\alpha,\gamma,\delta,\epsilon;z) =\\=& \left( \sum_{\sigma=\pm} \frac{\Gamma(\gamma)\Gamma(1-\delta)e^{\frac{\sigma}{2}\epsilon}\epsilon^{-\sigma\left(2m'(q)-\frac{\alpha}{\epsilon}+\frac{\gamma+\delta}{2}\right)}e^{-\frac{1}{2}\partial_{a_0}F_D+\frac{1}{2}\partial_{a_1}F_D-\frac{\sigma}{2}\partial_{m'}F_D}e^{i\pi\left(\frac{1-\sigma}{2}\right)\left(\frac{\alpha}{\epsilon}-\delta-2m'(q)\right)}}{\Gamma\left(\frac{\gamma}{2}-\sigma m'(q)\right)\Gamma\left(1-\frac{\delta}{2}-\sigma\left(m'(q)-\frac{\alpha}{\epsilon}-\frac{\gamma+\delta}{2}\right)\right)}\right)\times\\&\times\mathrm{HeunC}(q-\alpha,-\alpha,\delta,\gamma,-\epsilon;1-z) \,+ \\
    +\, &\left(\sum_{\sigma=\pm} \frac{\Gamma(\gamma)\Gamma(\delta-1)e^{\frac{\sigma}{2}\epsilon}\epsilon^{-\sigma\left(2m'(q)-\frac{\alpha}{\epsilon}+\frac{\gamma+\delta}{2}\right)}e^{-\frac{1}{2}\partial_{a_0}F_D-\frac{1}{2}\partial_{a_1}F_D-\frac{\sigma}{2}\partial_{m'}F_D}e^{i\pi\left(\frac{1-\sigma}{2}\right)\left(\frac{\alpha}{\epsilon}-2m'(q)-1\right)}}{\Gamma\left(\frac{\gamma}{2}-\sigma m'(q)\right)\Gamma\left(\frac{\delta}{2}-\sigma\left(m'(q)-\frac{\alpha}{\epsilon}-\frac{\gamma+\delta}{2}\right)\right)}\right)\times\\&\times \mathrm{HeunC}\left(q-\alpha-(1-\delta)(\epsilon+\gamma),-\alpha-(1-\delta)\epsilon, 2-\delta,\gamma,-\epsilon;1-z \right)\,,
    \end{aligned}
\end{equation}
where the quantities $m'(q)$ and $F_D$ are computed as explained in Appendix \ref{app:Nek}.\newline
The connection formula from $z=1$ to $\infty$ is simpler and reads
\begin{equation}
\begin{aligned}
    &\mathrm{HeunC}(q-\alpha,-\alpha,\delta,\gamma,-\epsilon;1-z) =\\=& e^{\pm\frac{i\pi\delta}{2}-\frac{1}{2}\partial_{a_1}F_D-\frac{1}{2}\partial_{m}F_D}\epsilon^{-\frac{1}{2}-\frac{\alpha}{\epsilon}+\frac{\gamma+\delta-\epsilon}{2}+m'(q)} \frac{\Gamma(\delta)e^{i\pi\left(\frac{\alpha}{\epsilon}-\frac{\gamma}{2}-m'(q)\right)}}{\Gamma\left(-\frac{\alpha}{\epsilon}+\frac{\gamma}{2}+\delta+m'(q)\right)} z^{-\frac{\alpha}{\epsilon}}\mathrm{HeunC}_\infty(q,\alpha,\gamma,\delta,\epsilon;z^{-1}) +\\+& e^{\pm\frac{i\pi\delta}{2}-\frac{1}{2}\partial_{a_1}F_D+\frac{1}{2}\partial_{m}F_D} \epsilon^{-\frac{1}{2}+\frac{\alpha}{\epsilon}-\frac{\gamma+\delta-\epsilon}{2}-m'(q)} \frac{\Gamma(\delta)}{\Gamma\left(\frac{\alpha}{\epsilon}-\frac{\gamma}{2}+m'(q)\right)} e^{-\epsilon z} z^{\frac{\alpha}{\epsilon}-\gamma-\delta} \mathrm{HeunC}_\infty(q-\gamma\epsilon,\alpha-\epsilon(\gamma+\delta),\gamma,\delta,-\epsilon;z^{-1})\,.
\end{aligned}
\end{equation}

\subsection{The reduced confluent Heun equation}
\subsubsection{The dictionary}
Here we establish the dictionary between our results of section \ref{Nf2A} on reduced confluent conformal blocks the reduced confluent Heun equation (RCHE) in standard notation, which reads
\begin{equation}\label{RCHE_standard}
    \frac{d^2w}{dz^2}+\left(\frac{\gamma}{z}+\frac{\delta}{z-1}\right)\frac{dw}{dz}+\frac{\beta z-q}{z(z-1)}w=0\,.
\end{equation}
This is of course just the CHE specialized to\footnote{This corresponds to the usual decoupling limit $m \to \infty, L \to 0$ such that $m L$ remains finite.} $\epsilon=0$. The interesting difference with respect to the CHE is the behaviour for $z\to \infty$, which is no longer controlled by $\epsilon$ and the degree of the singularity gets lowered to $1/2$. By defining $w(z)=P_2(z)\psi(z)$ with $P_2(z)=z^{-\gamma/2} (1-z)^{-\delta/2}$, we pass to the normal form which is easily compared with the semiclassical BPZ equation \eqref{BPZNf2NS}. The dictionary between the CFT parameters and the parameters of the RCHE reads:
\begin{equation}\label{RCHE_dictionary}
    \begin{aligned}
    &a_0 = \frac{1-\gamma}{2}\,,\\
    &a_1 = \frac{1-\delta}{2}\,,\\
    &L = 2 i \sqrt{\beta}\,,\\
    &u = \frac{1}{4}-q+\beta - \frac{(\gamma+\delta-1)^2}{4}\,,
    \end{aligned}
\end{equation}
where 
\begin{equation}
    u=\lim_{b\to0} b^2 \Lambda^2 \partial_{\Lambda^2} \log {}_{\frac{1}{2}}\mathfrak{F} \left(\alpha\, \begin{matrix} \alpha_1\\ \alpha_0 \end{matrix} ;\Lambda^2 \right) = \frac{1}{4}-a^2+ \mathcal{O}(L^2)
\end{equation}
as in \eqref{BPZNf2NS}. This relation can then be inverted to find $a$ in terms of the parameters of the RCHE: we denote this by $a(q)$. We therefore infer the relation between the solutions of the RCHE in standard form and the conformal blocks defined before. Near $z=0$ we have the following two linearly independent solutions to the RCHE in standard form \eqref{RCHE_standard}:
\begin{equation}
    \begin{aligned}
    & \mathrm{HeunRC}(q,\beta,\gamma,\delta;z) \,,\\
    & z^{1-\gamma} \mathrm{HeunRC}\left(q-(1-\gamma)\delta,\beta, 2-\gamma,\delta;z \right)\,,
    \end{aligned}
\end{equation}
where
\begin{equation}
    F =\lim_{b\to0} b^2 \log\left[ \Lambda^{-2\Delta} {}_\frac{1}{2}\mathfrak{F} \left(\alpha\, \begin{matrix} \alpha_1\\ \alpha_0 \end{matrix} ;\Lambda^2 \right)\right]\,.
\end{equation} 
Since HeunRC is nothing else than HeunC with $\epsilon=0$, it has the following expansion around $z=0$:
\begin{equation}
    \mathrm{HeunRC}(q,\beta,\gamma,\delta;z) = 1 - \frac{q}{\gamma}z + \mathcal{O}(z^2)\,.
\end{equation}
Comparing with the conformal blocks in \eqref{Nf2SC} we identify
\begin{equation}
    \begin{aligned}
    & \mathrm{HeunRC}(q,\beta,\gamma,\delta;z) = P_2(z) e^{-\frac{1}{2}\partial_{a_0}F} \FIIAsc{a}{a_1}{a_{0-}}{a_{2,1}}{a_0}{L^2}{z}\,,\\
    & z^{1-\gamma} \mathrm{HeunRC}\left(q-(1-\gamma)\delta,\beta, 2-\gamma,\delta;z \right) = P_2(z) e^{\frac{1}{2}\partial_{a_0}F} \FIIAsc{a}{a_1}{a_{0+}}{a_{2,1}}{a_0}{L^2}{z}\,,
    \end{aligned}
\end{equation}
Doing a M\"obius transformation $z\to1-z$ we obtain the solutions around $z=1$. Since this is a regular singularity the solution can again be written in terms of HeunRC. This amounts to sending $\gamma\to \delta,\, \delta\to\gamma,\,\beta \to -\beta,\,q \to q-\beta$. The two solutions are therefore
\begin{equation}
    \begin{aligned}
    &\mathrm{HeunRC}(q-\beta,-\beta,\delta,\gamma;1-z) \,,\\
    & (1-z)^{1-\delta} \mathrm{HeunRC}\left(q-\beta-(1-\delta)\gamma,-\beta, 2-\delta,\gamma;1-z \right)\,.
    \end{aligned}
\end{equation}
Comparig with the conformal blocks we identify
\begin{equation}
    \begin{aligned}
    &\mathrm{HeunRC}(q-\beta,-\beta,\delta,\gamma;1-z) = P_2(z) e^{-\frac{1}{2}\partial_{a_1}F}\FIIAsc{a}{a_0}{a_{1 -}}{a_{2,1}}{a_1}{-L^2}{1-z} \,,\\
    & (1-z)^{1-\delta} \mathrm{HeunRC}\left(q-\beta-(1-\delta)\gamma,-\beta, 2-\delta,\gamma;1-z \right)=\\
    &\quad \quad=P_2(z) e^{\frac{1}{2}\partial_{a_1}F}\FIIAsc{a}{a_0}{a_{1 +}}{a_{2,1}}{a_1}{-L^2}{1-z}\,.
    \end{aligned}
\end{equation}
The new behaviour arises for $z \to \infty$, where we write the solutions in terms of another function $\mathrm{HeunRC}_\infty$:
\begin{equation}
    \begin{aligned}
    &e^{2i\sqrt{\beta z}}z^{\frac{1}{4}-\frac{\gamma+\delta}{2}}\mathrm{HeunRC}_\infty(q,\beta,\gamma,\delta;z^{-\frac{1}{2}}) \,\\
    &e^{-2i\sqrt{\beta z}}z^{\frac{1}{4}-\frac{\gamma+\delta}{2}}\mathrm{HeunRC}_\infty(q,e^{2\pi i}\beta,\gamma,\delta;z^{-\frac{1}{2}})\,.
    \end{aligned}
\end{equation}
The function $\mathrm{HeunRC}_\infty$ has a simple asymptotic expansion around $z=\infty$:
\begin{equation}
    \mathrm{HeunRC}_\infty(q,\beta,\gamma,\delta;z^{-\frac{1}{2}}) \sim 1- \frac{q-\beta+\left(\frac{\gamma+\delta}{2}-\frac{3}{4}\right)\left(\frac{\gamma+\delta}{2}-\frac{1}{4}\right)}{i \sqrt{\beta}}z^{-\frac{1}{2}} + \mathcal{O}(z^{-1})\,.
\end{equation}
Comparing with the conformal blocks we identify
\begin{equation}\label{240}
    \begin{aligned}
    &e^{2i\sqrt{\beta z}}z^{\frac{1}{4}-\frac{\gamma+\delta}{2}}\mathrm{HeunRC}_\infty(q,\beta,\gamma,\delta;z^{-\frac{1}{2}}) =e^{\mp\frac{i\pi\delta}{2}} P_2(z) L^{\frac{1}{2}} \DIIAsc{+}{a_{2,1}}{a}{a_1}{a_0}{L^2}{\frac{1}{L \sqrt{z}}} \,\\
    &e^{-2i\sqrt{\beta z}}z^{\frac{1}{4}-\frac{\gamma+\delta}{2}}\mathrm{HeunRC}_\infty(q,e^{2\pi i}\beta,\gamma,\delta;z^{-\frac{1}{2}}) =e^{\mp\frac{i\pi\delta}{2}} P_2(z) L^{\frac{1}{2}} \DIIAsc{-}{a_{2,1}}{a}{a_1}{a_0}{L^2}{\frac{1}{L \sqrt{z}}}\,.
    \end{aligned}
\end{equation}
Note that due to the nature of the rank $1/2$ singularity at infinity, the expansion is in inverse powers of $\sqrt{z}$. The phase $e^{\mp \frac{i\pi\delta}{2}}$ comes from the fact that near $z=\infty$
\begin{equation}
    P_2(z)\sim  z^{-\gamma/2}(-z)^{-\delta/2} =e^{\pm \frac{i\pi\delta}{2}} z^{-\gamma/2-\delta/2}\,.
\end{equation} 
The second solution around $z=\infty$ can be found by using the manifest symmetry $L\to -L$ of the BPZ equation which according to the dictionary gives the symmetry $\beta \to e^{2\pi i} \beta$ of the RCHE in normal form.  

\subsubsection{Connection formulae}
The connection formula between $z=0,1$ written in \eqref{0to1Nf2SC} for the semiclassical conformal blocks can now be restated as:
\begin{equation}
    \begin{aligned}
    &\mathrm{HeunRC}(q,\beta,\gamma,\delta;z) = \frac{\Gamma(1-\delta)\Gamma(\gamma)e^{-\frac{1}{2}\partial_{a_0}F+\frac{1}{2}\partial_{a_1}F}}{\Gamma\left(\frac{1+\gamma-\delta}{2}+a(q)\right)\Gamma\left(\frac{1+\gamma-\delta}{2}-a(q)\right)}\mathrm{HeunRC}(q-\beta,-\beta,\delta,\gamma;1-z) + \\
    +&\, \frac{\Gamma(\delta-1)\Gamma(\gamma)e^{-\frac{1}{2}\partial_{a_0}F-\frac{1}{2}\partial_{a_1}F}}{\Gamma\left(\frac{\gamma+\delta-1}{2}+a(q)\right)\Gamma\left(\frac{\gamma+\delta-1}{2}-a(q)\right)}(1-z)^{1-\delta} \mathrm{HeunRC}\left(q-\beta-(1-\delta)\gamma,-\beta, 2-\delta,\gamma;1-z \right)\,,
    \end{aligned}
\end{equation}
where the quantities $a(q)$ and $F$ are computed as explained in Appendix \ref{app:Nek}.\newline
The connection formula between $z=1,\infty$ written in \eqref{1toinftyNf2SC} reads
\begin{equation}
    \begin{aligned}
    &\mathrm{HeunRC}(q-\beta,-\beta,\delta,\gamma;1-z) =\\=& \left(\sum_{\sigma=\pm}\frac{\Gamma(-2\sigma a(q))\Gamma(1-2\sigma a(q))\Gamma(\delta)\left(e^{i\pi}\beta\right)^{-\frac{1}{4}+\sigma a(q)}e^{\pm 
    \frac{i\pi\delta}{2}-\frac{1}{2}\partial_{a_1}F-\frac{\sigma}{2}\partial_{a} F}}{2\sqrt{\pi}\Gamma\left(\frac{1-\gamma+\delta}{2}-\sigma a(q)\right)\Gamma\left(\frac{\gamma+\delta-1}{2}-\sigma a(q)\right)}\right) e^{2i\sqrt{\beta z}}z^{\frac{1}{4}-\frac{\gamma+\delta}{2}}\mathrm{HeunRC}_\infty(q,\beta,\gamma,\delta;z^{-\frac{1}{2}})+\\+& \left(\sum_{\sigma=\pm}\frac{\Gamma(-2\sigma a(q))\Gamma(1-2\sigma a(q)) \Gamma(\delta)\left(e^{-i\pi}\beta\right)^{-\frac{1}{4}+\sigma a(q)}e^{\pm 
    \frac{i\pi\delta}{2}-\frac{1}{2}\partial_{a_1}F-\frac{\sigma}{2}\partial_{a} F}}{2\sqrt{\pi}\Gamma\left(\frac{1-\gamma+\delta}{2}-\sigma a(q)\right)\Gamma\left(\frac{\gamma+\delta-1}{2}-\sigma a(q)\right)}\right) e^{-2i\sqrt{\beta z}}z^{\frac{1}{4}-\frac{\gamma+\delta}{2}}\mathrm{HeunRC}_\infty(q,e^{2\pi i}\beta,\gamma,\delta;z^{-\frac{1}{2}})\,.
    \end{aligned}
\end{equation}
Here the phase ambiguity comes from \eqref{240}, i.e. corresponds to the choice $(-z)^{-\delta/2} = e^{\pm \frac{i\pi\delta}{2}}z^{-\delta/2}$. A similar expression can be found connecting $z=0$ and $\infty$. 

\subsection{The doubly confluent Heun equation}
\subsubsection{The dictionary}
The doubly confluent Heun equation (DCHE) reads
\begin{equation}
    \left( \frac{d^2}{d z^2} + \frac{\delta + \gamma z +  z^2}{z^2} \frac{d}{d z} + \frac{\alpha z - q}{z^2} \right) w(z) = 0 \,.
\end{equation}
Again putting the DCHE in its normal form via the substitution $w(z) = \Tilde{P}_2(z)\psi(z)$ with 
\begin{equation}
     \Tilde{P}_2(z) = e^{\frac{1}{2} \left( \frac{\delta}{z} -  z \right)} z^{- \frac{\gamma}{2}}
\end{equation}
we find the $2^2 = 4$ different dictionaries with \eqref{eq:scBPZNf2symm} corresponding to the $\mathbb{Z}_2^2$ symmetries $(m_i, L_i) \to (-m_i, -L_i)$ for $i = 1, 2$. For brevity we only write one of them, namely
\begin{equation}
\begin{aligned}
    &L_1 = 1 \,, \\
    &L_2 = \delta \,, \\
    &m_1 = \frac{1}{2} \left( 2 \alpha - \gamma \right) \,, \\
    &m_2 = 1 - \frac{\gamma}{2} \,, \\
    &u = \frac{1}{4} \left( - 4 q + 2 \gamma - \gamma^2 - 2 \delta \right) \,.
\end{aligned}
\end{equation}
and the inverse dictionary is
\begin{equation}
\begin{aligned}
    &\alpha = 1+m_1-m_2 \,, \\
    &\delta = L_2 \,, \\
    &\gamma = 2(1-m_2) \,, \\
    &q = - \frac{1}{2} \left( L_2 + 2 u + 2 m_2 (m_2 - 1) \right) \,, \\
    &L_1 = 1 \,.
\end{aligned}
\end{equation}
We denote the two solutions of the DCHE near the irregular singularity at zero as
\begin{equation}
\begin{aligned}
    &\text{HeunDC} \left( q, \alpha, \gamma, \delta, z \right) \,, \\  &e^{\frac{\delta}{z}} z^{2-\gamma}\text{HeunDC} \left( \delta + q + \gamma - 2, \alpha - \gamma + 2, \gamma, -\delta, z \right) \,,
\end{aligned}
\end{equation}
where $\text{HeunDC}$ has the following asymptotic expansion around $z=0$:
\begin{equation}
    \text{HeunDC} \left( q, \alpha, \gamma, \delta, z \right) \sim 1 + \frac{q}{\delta} z + \frac{q(q-\gamma) - \alpha \delta}{2 \delta^2} z^2 + \mathcal{O} (z^3)  \,.
\end{equation}
Comparing with the semiclassical block \eqref{eq:nf2symsccbnear0} we get 
\begin{equation}
\begin{aligned}
    &\text{HeunDC} \left( q, \alpha, \gamma, \delta, z \right) = \Tilde{P}_2(z) L_2^{\frac{1}{2} - m_2} e^{- \frac{1}{2} \partial_{m_2} F} \left( z\, \DIIsc{m_2}{a_{2,1}}{m_{2 -}}{a}{m_1}{ L_2}{\frac{z}{L_2}} \right) \,, \\ &\text{HeunDC} \left( \delta + q + \gamma - 2, \alpha - \gamma + 2, \gamma, -\delta, z \right) = \Tilde{P}_2(z) L_2^{\frac{1}{2} + m_2} e^{\frac{1}{2} \partial_{m_2} F} \left( z\, \DIIsc{m_2}{a_{2,1}}{m_{2 +}}{a}{m_1}{ L_2}{\frac{z}{L_2}} \right) \,.
\end{aligned}
\end{equation}
The solutions near the irregular singularity at infinity are given by 
\begin{equation}
\begin{aligned}
    &z^{- \alpha} \text{HeunDC} \left( q - \alpha (\alpha + 1 - \gamma), \alpha, 2 (\alpha + 1) - \gamma, \delta, - \frac{\delta}{z} \right) \,, \\
    &e^{-z} z^{\alpha - \gamma} \text{HeunDC} \left( q + \delta + (\gamma - \alpha) (\alpha -1), \gamma - \alpha, -2 (\alpha - 1) + \gamma, - \delta, - \frac{\delta}{z} \right) \,.
\end{aligned}
\end{equation}
Comparing with the semiclassical block \eqref{eq:nf2symsccbnearinf} we find
\begin{equation}
\begin{aligned}
    & \text{HeunDC} \left( q - \alpha (\alpha + 1 - \gamma), \alpha, 2 (\alpha + 1) - \gamma, \delta, - \frac{\delta}{z} \right) =  \Tilde{P}_2(z)e^{\frac{1}{2} \partial_{m_1} F} \DIIsc{m_1}{a_{2,1}}{m_{1+}}{a}{m_2}{ L_2}{\frac{1}{z}} \,, \\ & \text{HeunDC} \left( q + \delta + (\gamma - \alpha) (\alpha -1), \gamma - \alpha, -2 (\alpha - 1) + \gamma, - \delta, - \frac{\delta}{z} \right) = \Tilde{P}_2(z)e^{-\frac{1}{2} \partial_{m_1} F} \DIIsc{m_1}{a_{2,1}}{m_{1-}}{a}{m_2}{ L_2}{\frac{1}{ z}} \,.
\end{aligned}    
\end{equation}
\subsubsection{Connection formulae}
In this case the only connection formula is the one between zero and infinity. This can be obtained from equation \eqref{eq:nf2symconcsc} and reads
\begin{equation}
\begin{aligned}
    \text{HeunDC} \left( q, \alpha, \gamma, \delta, z \right) = &\left(\sum_{\sigma=\pm} \frac{\Gamma \left( - 2 \sigma a \right) \Gamma \left( 1 - 2 \sigma a \right) \delta^{-\frac{1}{2} +\frac{\gamma}{2} + \sigma a}}{\Gamma \left(\frac{1}{2} - \left( 1 - \frac{\gamma}{2} \right) - \sigma a \right) \Gamma \left( \frac{1}{2} - \frac{2 \alpha - \gamma}{2} - \sigma a \right)}\right) \times \\ &\times e^{\frac{1}{2} \left( - \partial_{m_1} - \partial_{m_2} - \sigma \partial_a \right) F} z^{- \alpha} \text{HeunDC} \left( q - \alpha (\alpha + 1 - \gamma), \alpha, 2 (\alpha + 1) - \gamma, \delta, - \frac{\delta}{z} \right) + \\ +& \left(\sum_{\sigma=\pm} \frac{\Gamma \left( - 2 \sigma a \right) \Gamma \left( 1 - 2 \sigma a \right) \delta^{-\frac{1}{2} +\frac{\gamma}{2} + \sigma a}e^{i\pi\left(\frac{1+\gamma}{2}-\alpha-\sigma a\right)}}{\Gamma \left(\frac{1}{2} - \left( 1 - \frac{\gamma}{2} \right) - \sigma a \right) \Gamma \left( \frac{1}{2} + \frac{2 \alpha - \gamma}{2} - \sigma a \right)} e^{\frac{1}{2} \left(\partial_{m_1} - \partial_{m_2} - \sigma \partial_a \right) F}\right) \times \\ &\times e^{-z} z^{\alpha - \gamma} \text{HeunDC} \left( q + \delta + (\gamma - \alpha) (\alpha -1), \gamma - \alpha, -2 (\alpha - 1) + \gamma, - \delta, - \frac{\delta}{z} \right) \,,
\end{aligned}
\end{equation}

\subsection{The reduced doubly confluent Heun equation}
\subsubsection{The dictionary}
Here we establish the dictionary between our results of section \ref{Nf1} on reduced doubly confluent conformal blocks and the reduced doubly confluent Heun equation (RDCHE) in the standard form, which reads
\begin{equation}\label{RDCHE_standard}
    \frac{d^2w}{dz^2}-\frac{dw}{dz}+\frac{\beta z-q+\epsilon z^{-1}}{z^2}w=0\,.
\end{equation}
By defining $w(z)=e^{z/2}\psi(z)$ we get rid of the first derivative and bring the equation to the normal form which is to be compared with the semiclassical BPZ equation \eqref{BPZNf1SC}. The resulting dictionary between the CFT parameters and the parameters of the RDCHE is
\begin{equation}\label{RDCHE_dictionary}
    \begin{aligned}
    & L_1 = 1\,,\\
    & L_2 = 2 i \sqrt{\epsilon}\,,\\
    & m = \beta\,,\\
    & u = -q\,.
    \end{aligned}
\end{equation}
The fact that $L_1=1$ is of course consistent with the fact that it is a redundant parameter. Here 
\begin{equation}
    u=\lim_{b\to0} b^2 \Lambda_2^2 \partial_{\Lambda_2^2} \log {}_1\mathfrak{F}_\frac{1}{2} \left(\mu\,\alpha ;\Lambda_1 \Lambda_2^2 \right) = \frac{1}{4}-a^2+ \mathcal{O}(L_1 L_2^2) \,
\end{equation}
as in \eqref{BPZNf1SC}. This relation can then be inverted to find $a$ in terms of the parameters of the RDCHE: we denote this by $a(q)$. We can now write the solutions to the RDCHE in standard form and their relation to the conformal blocks by comparison. Near $z=0$ we denote the two linearly independent solutions to the RDCHE in standard form \eqref{RDCHE_standard} by:
\begin{equation}
    \begin{aligned}
    &e^{2i\sqrt{\epsilon/z}}z^{3/4} \mathrm{HeunRDC}_0(q,\beta,\epsilon;\sqrt{z})\,,\\
    & e^{-2i\sqrt{\epsilon/z}}z^{3/4} \mathrm{HeunRDC}_0(q,\beta,e^{2\pi i}\epsilon;\sqrt{z})\,.
    \end{aligned}
\end{equation}
The two solutions are related by the manifest symmetry $L_2\to -L_2$ of the BPZ equation which according to the dictionary \eqref{RDCHE_dictionary} gives the symmetry $\epsilon \to e^{2\pi i} \epsilon$ of the RDCHE in normal form. The function $\mathrm{HeunRDC}_0$ has the following asymptotic expansion around $z=0$:
\begin{equation}
    \mathrm{HeunRDC}_0(q,\beta,\epsilon;\sqrt{z}) \sim 1-\frac{\frac{3}{16}+q}{i\sqrt{\epsilon}}\sqrt{z}+\mathcal{O}\left(z\right).
\end{equation}
Note again that due to the presence of a rank $1/2$ singularity, the expansion is in powers of $\sqrt{z}.$ Comparing with the semiclassical conformal blocks in \eqref{Nf1sc} we identify
\begin{equation}
    \begin{aligned}
    &e^{2i\sqrt{\epsilon/z}}z^{3/4} \mathrm{HeunRDC}_0(q,\beta,\epsilon;\sqrt{z}) = e^{z/2} L_2^{\frac{1}{2}} \EIsc{+}{m}{a}{a_{2,1}}{L_2^2}{\frac{\sqrt{z}}{L_2}}\,,\\
    & e^{-2i\sqrt{\epsilon/z}}z^{3/4} \mathrm{HeunRDC}_0(q,\beta,e^{2\pi i}\epsilon;\sqrt{z}) = e^{z/2} L_2^{\frac{1}{2}} \EIsc{-}{m}{a}{a_{2,1}}{L_2^2}{\frac{\sqrt{z}}{L_2}}\,.
    \end{aligned}
\end{equation}
For $z\sim \infty$ instead we have the two solutions
\begin{equation}
\begin{aligned}
    & z^{\beta} \mathrm{HeunRDC}_{\infty}(q,\beta,\epsilon;z^{-1})\,,\\
    & e^{z} z^{-\beta} \mathrm{HeunRDC}_{\infty}(q,-\beta,-\epsilon;-z^{-1}) \,.
\end{aligned}
\end{equation} The function $\mathrm{HeunRDC}_{\infty}(q,\beta,\epsilon;z^{-1})$ has the following asymptotic expansion around $z=\infty$:
\begin{equation}
    \mathrm{HeunRDC}_{\infty}(q,\beta,\epsilon;z^{-1}) \sim 1 + (q+\beta-\beta^2)z^{-1} + \mathcal{O}\left(z^{-2}\right)\,.
\end{equation}
Comparing with the semiclassical conformal blocks we identify
\begin{equation}
\begin{aligned}
    & z^{\beta} \mathrm{HeunRDC}_{\infty}(q,\beta,\epsilon;z^{-1}) = e^{z/2}e^{-\frac{1}{2}\partial_m F} \, \DIsc{m}{a_{2,1}}{m_-}{a}{ L_2^2}{\frac{1}{z}}\,,\\
    & e^{z} z^{-\beta} \mathrm{HeunRDC}_{\infty}(q,-\beta,-\epsilon;-z^{-1}) = e^{z/2}e^{\frac{1}{2}\partial_m F} \, \DIsc{m}{a_{2,1}}{m_+}{a}{ L_2^2}{\frac{1}{z}}\,.
\end{aligned}
\end{equation}
These solutions are related by the symmetry $(m,L_1)\to(-m,-L_1)$ of the semiclassical BPZ equation. Notice that one can rescale the BPZ equation such that it only depends on the combination $L_1 z$ and the coefficient of the cubic pole is $-L_1 L_2^2/4$. By setting $L_1=1$ according to the dictionary with the RDCHE, the above symmetry descends to the symmetry $(\beta,\epsilon,z)\to(-\beta,-\epsilon,-z)$ of the RDCHE in normal form. Furthermore, in the equation above 
\begin{equation}
    F =\lim_{b\to0} b^2 \log\left[ (\Lambda_1\Lambda_2^2)^{-\Delta} {}_1\mathfrak{F}_\frac{1}{2} \left(\mu\,\alpha ;\Lambda_1 \Lambda_2^2 \right)\right]
\end{equation}
as in \eqref{197}.

\subsubsection{Connection formulae}
The connection formula between $z=0$ and $\infty$ written in \eqref{0toinftyNf1} for the semiclassical conformal blocks can now be restated as:
\begin{equation}
\begin{aligned}
    &e^{2i\sqrt{\epsilon/z}}z^{3/4} \mathrm{HeunRDC}_0(q,\beta,\epsilon;\sqrt{z}) =\\=& \left(\sum_{\sigma = \pm} \frac{\Gamma(1-2\sigma a(q))\Gamma(-2\sigma a(q))}{\sqrt{\pi}\Gamma\left(\frac{1}{2}+\beta - \sigma a(q)\right)} \epsilon^{\frac{1}{4}+\sigma a(q)} e^{\frac{1}{2}\partial_m F-\frac{\sigma}{2}\partial_{a}F}e^{-i\pi\left(\frac{1}{4}+\sigma a(q)\right)} e^{i\pi\left(\frac{1}{2}-\beta-\sigma a(q)\right)}\right) \,z^{\beta} \mathrm{HeunRDC}_{\infty}(q,\beta,\epsilon;z^{-1})+\\+&\,  \left(\sum_{\sigma = \pm} \frac{\Gamma(1-2\sigma a(q))\Gamma(-2\sigma a(q))}{\sqrt{\pi}\Gamma\left(\frac{1}{2}-\beta - \sigma a(q)\right)} \epsilon^{\frac{1}{4}+\sigma a(q)} e^{\frac{1}{2}\partial_m F-\frac{\sigma}{2}\partial_{a}F}e^{-i\pi\left(\frac{1}{4}+\sigma a(q)\right)}\right)  e^{z} z^{-\beta} \mathrm{HeunRDC}_{\infty}(q,-\beta,-\epsilon;-z^{-1}) \,,
\end{aligned}
\end{equation}
where the quantities $a(q)$ and $F$ are computed as explained in Appendix \ref{app:Nek}.

\subsection{The doubly reduced doubly confluent Heun equation}
\subsubsection{The dictionary}
Here we establish the dictionary between our results of section \ref{Nf0} on doubly reduced doubly confluent conformal blocks and the corresponding Heun equation (DRDCHE) which reads
\begin{equation}\label{DRDCHE}
    \frac{d^2 w}{dz^2} + \frac{z-q+\epsilon z^{-1}}{z^2}w = 0\,.
\end{equation}
This already takes the normal form of the semiclassical BPZ equation \eqref{BPZNf0SC} and we immediately read off the dictionary:
\begin{equation}
\begin{aligned}
        &L_1 = 2i\,,\\
        &L_2 = 2i \sqrt{\epsilon}\,,\\
        &u = -q\,,
\end{aligned}
\end{equation}
where  
\begin{equation}
    u=\lim_{b\to0} b^2 \Lambda_2^2 \partial_{\Lambda_2^2} \log {}_\frac{1}{2}\mathfrak{F}_\frac{1}{2} \left(\alpha ;\Lambda_1^2 \Lambda_2^2 \right) = \frac{1}{4}-a^2+ \mathcal{O}(L_1^2 L_2^2)
\end{equation}
as in \eqref{BPZNf0SC}. This relation can be inverted to find $a$ in terms of the parameters of the DRDCHE: we denote this by $a(q)$. Near $z=0$ we denote the two linearly independent solutions to \eqref{DRDCHE} by
\begin{equation}
\begin{aligned}
    &e^{2i \sqrt{\epsilon/z}} z^{3/4} \mathrm{HeunDRDC}(q,\epsilon;\sqrt{z})\,,\\
    &e^{-2i \sqrt{\epsilon/z}} z^{3/4} \mathrm{HeunDRDC}(q,e^{2\pi i}\epsilon;\sqrt{z})\,.
\end{aligned}
\end{equation}
The DRDC Heun function has a simple asymptotic expansion around $z=0$:
\begin{equation}
    \mathrm{HeunDRDC}(q,\epsilon;\sqrt{z}) \sim 1-\frac{\frac{3}{16}+q}{i\sqrt{\epsilon}}\sqrt{z}+\mathcal{O}(z)\,.
\end{equation}
Note that in the expansion, $z$ appears with a square-root, and therefore mapping $z \to e^{2\pi i}z$ gives another solution. Comparing with the semiclassical conformal blocks in \eqref{Nf0sc}, we identify
\begin{equation}
\begin{aligned}
    &e^{2i \sqrt{\epsilon/z}} z^{3/4} \mathrm{HeunDRDC}(q,\epsilon;\sqrt{z}) = z L_2^{1/2} \Ezsc{+}{a}{a_{2,1}}{-4 L_2^2}{\frac{\sqrt{z}}{L_2}}\,,\\
    &e^{-2i \sqrt{\epsilon/z}} z^{3/4} \mathrm{HeunDRDC}(q,e^{2\pi i}\epsilon;\sqrt{z}) = z L_2^{1/2} \Ezsc{-}{a}{a_{2,1}}{-4 L_2^2}{\frac{\sqrt{z}}{L_2}}\,.
\end{aligned}
\end{equation}
Around $z=\infty$ we have the two linearly independent solutions
\begin{equation}
\begin{aligned}
    &e^{2i \sqrt{z}} z^{1/4} \mathrm{HeunDRDC}(q,\epsilon;\left(\epsilon z\right)^{-\frac{1}{2}}) \,,\\
    &e^{-2i \sqrt{z}} z^{1/4} \mathrm{HeunDRDC}(q,\epsilon; \left(e^{2\pi i} \epsilon z\right)^{-\frac{1}{2}})\,,
\end{aligned}
\end{equation}
which we identify with the conformal blocks
\begin{equation}
\begin{aligned}
    &e^{2i \sqrt{z}} z^{1/4} \mathrm{HeunDRDC}(q,\epsilon;\left(\epsilon z\right)^{-\frac{1}{2}}) =\sqrt{2i} \Ezsc{+}{a_{2,1}}{a}{-4 L_2^2}{\frac{1}{2i \sqrt{z}}}\,,\\
    &e^{-2i \sqrt{z}} z^{1/4} \mathrm{HeunDRDC}(q,\epsilon; \left(e^{2\pi i}\epsilon z\right)^{-\frac{1}{2}}) = \sqrt{2i}\Ezsc{-}{a_{2,1}}{a}{-4 L_2^2}{\frac{1}{2i \sqrt{z}}}\,.
\end{aligned}
\end{equation}

\subsubsection{Connection formulae}
The connection formula \eqref{0toinftyNf0} from $0$ to $\infty$ in terms of the DRDC Heun functions is
\begin{equation}
\begin{aligned}
    &e^{2i\sqrt{\epsilon/z}}z^{3/4} \mathrm{HeunDRDC}(q,\epsilon;\sqrt{z}) =\\=&\left( \frac{-i}{2\pi} \sum_{\sigma=\pm} \Gamma(1-2\sigma a(q))\Gamma(-2\sigma a(q)) \epsilon^{\frac{1}{4}+\sigma a(q)} e^{-\frac{\sigma}{2}\partial_a F}\right)  \,e^{2i\sqrt{z}}z^{1/4} \mathrm{HeunDRDC}(q,\epsilon;\left(\epsilon z\right)^{-\frac{1}{2}})+\\+&\left( \frac{1}{2\pi} \sum_{\sigma=\pm} \Gamma(1-2\sigma a(q))\Gamma(-2\sigma a(q)) \epsilon^{\frac{1}{4}+\sigma a(q)} e^{-\frac{\sigma}{2}\partial_a F}e^{-2\pi i \sigma a(q)}\right) e^{-2i \sqrt{z}} z^{1/4} \mathrm{HeunDRDC}(q, \epsilon;\left(e^{2\pi i} \epsilon z\right)^{-\frac{1}{2}}) \,,
\end{aligned}
\end{equation}
where the quantities $a(q)$ and $F$ are computed as explained in Appendix \ref{app:Nek}.

\newpage

\appendix

\section{DOZZ factors and irregular generalizations}

\subsection{Regular case}\label{app:DOZZ}
We use conventions where $\Delta=\frac{Q^2}{4}-\alpha^2$, i.e. physical range of the momentum is $\alpha \in i \mathbb{R}^+$. The formula proposed by DOZZ for the Liouville three-point function is then \cite{Dorn:1994xn,Zamolodchikov:1995aa} 
\begin{equation}
\begin{aligned}
&\langle \Delta_1|V_2(1)|\Delta_3\rangle = C_{\alpha_1 \alpha_2 \alpha_3} =\\=&\,  \frac{\Upsilon'_b(0)\Upsilon_b(Q+2\alpha_1)\Upsilon_b(Q+2\alpha_2)\Upsilon_b(Q+2\alpha_3)}{\Upsilon_b( \frac{Q}{2}+\alpha_1+\alpha_2+\alpha_3) \Upsilon_b( \frac{Q}{2}+\alpha_1+\alpha_2-\alpha_3 )\Upsilon_b( \frac{Q}{2}+\alpha_1-\alpha_2+\alpha_3 )\Upsilon_b( \frac{Q}{2}-\alpha_1+\alpha_2+\alpha_3)} \,.
\end{aligned}
\end{equation}
We neglect the dependence on the cosmological constant since its value is arbitrary and is not needed for the following discussion. We will not define the special function $\Upsilon_b$ and state all its remarkable properties, instead we refer to \cite{2001}. The most important property for us is the functional relation
\begin{equation}
    \Upsilon_b(x+b) = \gamma(bx)b^{1-2bx}\Upsilon_b(x)\,,\quad \gamma(x) = \frac{\Gamma(x)}{\Gamma(1-x)}\,.
\end{equation}
The normalization of the states is obtained from the three-point function by taking the operator in the middle to be the identity operator, i.e. with $\Delta =0$ which in our conventions means $\alpha=-\frac{Q}{2}$. One finds
\begin{equation}
    \lim_{\epsilon\to0}C_{\alpha_1,-\frac{Q}{2}+\epsilon,\alpha_2} = 2\pi \delta(\alpha_1-\alpha_2) G_{\alpha_1}\,,
\end{equation}
with the two-point function $G_\alpha$ given by
\begin{equation}
    G_\alpha = \frac{\Upsilon_b(2\alpha+Q)}{\Upsilon_b(2\alpha)}\,.
\end{equation}
We use it to raise and lower indices: For example, OPE coefficients are given by
\begin{equation}
    C^{\alpha_1}_{\alpha_2 \alpha_3} = G^{-1}_{\alpha_1}C_{\alpha_1 \alpha_2 \alpha_3}\,.
\end{equation}
We will be interested in the case where one of the fields is the degenerate field $\Phi_{2,1}$ with $\alpha_{2,1}=-\frac{2b+b^{-1}}{2}$, corresponding to $\Delta_{2,1}=-\frac{1}{2}-\frac{3b^2}{4}$. The fusion rules in this case impose that only two Verma modules appear in the OPE of this field with a primary:
\begin{equation}
    \Phi_{2,1}(z)|\Delta\rangle = \sum_{\theta=\pm}z^{\frac{bQ}{2}+\theta b \alpha} C^{\alpha_\theta}_{\alpha_{2,1},\alpha} |\Delta_{\theta}\rangle \left(1+\mathcal{O}(z)\right)\,,
\end{equation}
with 
\begin{equation}
    \alpha_\pm = \alpha \pm \left(-\frac{b}{2}\right)\,,\quad \Delta_\pm = \Delta_{\alpha_\pm} = \Delta \pm b \alpha -\frac{b^2}{4}\,.
\end{equation}
Since the degenerate field is not in the physical spectrum, i.e. $\alpha_{2,1} \not\in i \mathbb{R}^+$, the OPE coefficients $C^{\alpha_\theta}_{\alpha_{2,1},\alpha}$ have to be computed by analytic continuation of the DOZZ formula. This is tricky and is most easily performed by considering a four-point function, where the intermediate momentum is integrated over. During the analytic continuation one picks up residues of poles that cross the integration contour, and this in fact automatically imposes the fusion rules. In any case, the result is \cite{2011}:
\begin{equation}\label{degOPEcoeff}
    C^{\alpha_+}_{\alpha_{2,1},\alpha} = 1\,,\quad C^{\alpha_-}_{\alpha_{2,1},\alpha} = b^{2bQ} \frac{\gamma(2b\alpha)}{\gamma(bQ+2b\alpha)}\,.
\end{equation}

\subsection{Rank 1}\label{app:rank1collision}
In section \ref{warmup:whittaker} we introduced the rank 1 irregular state, which can be given as a confluence limit of primary operators (here we consider only the chiral half):
\begin{equation}\label{irrcollision}
    \langle \mu,\Lambda| \propto \lim_{\eta\to\infty} t^{\Delta_t-\Delta} \langle \Delta|V_t(t)\, 
\end{equation}
with
\begin{equation}\label{B22}
    \Delta = \frac{Q^2}{4}-\alpha^2\,,\quad \alpha = -\frac{\eta+\mu}{2} \,,\quad \Delta_t = \frac{Q^2}{4}-\alpha_t^2\,,\quad \alpha_t=\frac{\eta-\mu}{2}\,,\quad t=\frac{\eta}{\Lambda}\,.
\end{equation}
This reproduces the desired Ward identities for the irregular state. To determine its normalization, we perform the collision limit on a (chiral+antichiral) three-point function, keeping track of the DOZZ factors. Although irrelevant for the Ward identities, the signs of $\alpha,\alpha_t$ in \eqref{B22} are crucial now. We find
\begin{equation}\label{320}
    \lim_{\eta\to\infty} (t\bar{t})^{\Delta_t-\Delta} \langle \Delta|V_t(t,\bar{t})| \Delta_0\rangle = (\Lambda\bar{\Lambda})^{\Delta_0} \lim_{\eta\to\infty} \eta^{-2\Delta_0}  C_{-\frac{\eta+\mu}{2},\frac{\eta-\mu}{2},\alpha_0}\,.
\end{equation}
Note that consistently with the main text, we consider the chiral and antichiral parts formally as independent and distinguish them by letting the "complex conjugation" formally act only on the coordinates $t,\Lambda$ and not on the momenta $\alpha_0,\mu,\eta$. The asymptotic behaviour of the $\Upsilon_b$ function, valid for large imaginary $x$ is:
\begin{equation}\label{eq:UpsilonAsymptotics}
    \log \Upsilon_b\left(\frac{Q}{2}+x\right) = -\frac{1}{2}\Delta_x\log\Delta_x +\frac{1+Q^2}{12}\log\Delta_x+\frac{3}{2}\Delta_x+\mathcal{O}(x^0)\,.
\end{equation}
We therefore find the following asymptotic behaviour of the DOZZ factor:
\begin{equation}
    C_{-\frac{\eta+\mu}{2},\frac{\eta-\mu}{2},\alpha_0} \sim (-\eta^2)^{\Delta_0 - \mu(Q-\mu)} \frac{\Upsilon_b(Q+2\alpha_0)}{\Upsilon_b( \frac{Q}{2}+\mu+\alpha_0) \Upsilon_b( \frac{Q}{2}+\mu-\alpha_0 )} \,.
\end{equation}
This suggests that we get a finite limit in (320) if we substract the factor of $(-\eta^2)^{ - \mu(Q-\mu)}$ by hand. This can also be achieved by changing the power of $t$ that we substract in the definition \eqref{irrcollision}, but this would change the $L_0$-action on the irregular state, which we avoid. It is however precisely what is done in \cite{Gaiotto:2012sf}. In any case, we find the following normalization of the irregular state:
\begin{equation}\label{B26}
    \langle \mu,\Lambda|\Delta_0\rangle = \lim_{\eta\to\infty} (-\eta^2)^{\mu(Q-\mu)} |t|^{2\Delta_t-2\Delta} \langle \Delta|V_t(t,\bar{t})| \Delta_0\rangle = |\Lambda|^{2\Delta_0} C_{\mu \alpha_0}\,,
\end{equation}
with normalization function
\begin{equation}
    C_{\mu \alpha} = \frac{e^{-i\pi \Delta}\Upsilon_b(Q+2\alpha)}{\Upsilon_b( \frac{Q}{2}+\mu+\alpha) \Upsilon_b( \frac{Q}{2}+\mu-\alpha)} \,.
\end{equation}
The choice of the branch for the phase is consistent with the result found in \ref{app:rank1}.\newline
In the text we also consider a different kind of collision limit, which reproduces the OPE between a primary operator and the irregular state. Performing this collision limit while keeping track of the DOZZ factors, we can extract the corresponding irregular OPE coefficient. In particular, consider the following correlation function, which we expand for large $\Lambda$:
\begin{equation}\label{B28}
    \begin{aligned}
    &\langle \mu,\Lambda|V_1(1)|\Delta_0\rangle = \int d\mu' B_{\mu \alpha_1}^{\mu'} C_{\mu' \alpha_0} \left| {}_1 \mathfrak{D}\left(\mu\, \begin{matrix}\alpha_1\\{}\end{matrix}\,\mu'\,\alpha_0\, ;\, \frac{1}{\Lambda} \right) \right|^2\,.
    \end{aligned}
\end{equation}
Here $B_{\mu \alpha_1}^{\mu'}$ is the OPE coefficient corresponding to the OPE between the irregular state and $V_1$, $C_{\mu' \alpha_0}$ is the normalization function defined above and ${}_1 \mathfrak{D}$ is just the corresponding conformal block. Following \cite{Lisovyy:2018mnj}, we can express an irregular three-point function equivalently as a limit of a regular four-point function:
\begin{equation}\label{B29}
    \begin{aligned}
    &\langle \mu,\Lambda|V_1(1)|\Delta_0\rangle = \lim_{\eta \to \infty} (-\eta^2)^{\mu(Q-\mu)} \int d\mu'  C^{\alpha(\eta)}_{\alpha_\infty(\eta),\alpha_1}C_{\alpha(\eta),\alpha_t(\eta),\alpha_0}  \times \\\times& \left|e^{-(\mu'-\mu)\Lambda}\left(-\frac{\Lambda}{\eta}\right)^{\Delta_1 - (\mu'-\mu)(\eta - \mu')} \left(\frac{\Lambda}{\eta}\right)^{\Delta_\infty(\eta)-\Delta_t(\eta)} \left(1-\frac{\eta}{\Lambda}\right)^{\Delta_1 - (\mu'-\mu)(\eta - \mu')} \mathfrak{F} \left( \begin{matrix} \alpha_1\\\alpha_\infty (\eta) \, \end{matrix} \alpha(\eta) \, \begin{matrix} \alpha_t(\eta)\\ \alpha_0 \end{matrix} ; \frac{\eta}{\Lambda} \right)\right|^2\,,
    \end{aligned}
\end{equation}
with
\begin{equation}
    \alpha_\infty (\eta)=-\frac{\eta+\mu}{2}\,,\quad \alpha_t(\eta) = \frac{\eta-\mu}{2}\,,\quad \alpha(\eta)=-\frac{\eta-\mu}{2}-\mu'\,.
\end{equation}
Several comments are in order: First, notice that in line with the definition of the irregular state we have multiplied by the same factors of $(-\eta^2)^{\mu(Q-\mu)}$ and $\left(\Lambda\bar{\Lambda}/\eta^2\right)^{\Delta_\infty(\eta)-\Delta_t(\eta)}$ as in \eqref{B26}. Second, the remaining factors which we have put by hand are equal to $1$ in the limit:
\begin{equation}
    \lim_{\eta\to\infty} e^{-(\mu'-\mu)\Lambda}\left(-\frac{\Lambda}{\eta}\right)^{\Delta_1 - (\mu'-\mu)(\eta - \mu')}  \left(1-\frac{\eta}{\Lambda}\right)^{\Delta_1 - (\mu'-\mu)(\eta - \mu')} = \lim_{\eta\to\infty} e^{-(\mu'-\mu)\Lambda} \left(1-\frac{\Lambda}{\eta}\right)^{\Delta_1 - (\mu'-\mu)(\eta - \mu')}=1\,.
\end{equation}
Therefore all the factors that we put by hand are the same as if we had computed \eqref{B29} by doing the OPE between $V_1$ and $|\Delta_0\rangle$ instead of between $\langle \mu,\Lambda|$ and $V_1$. This ensures crossing symmetry of the irregular three-point function. Furthermore, the factors inside the modulus square in the limit give the irregular conformal block up to an overall divergence, i.e.:
\begin{equation}
    \begin{aligned}
    &e^{-(\mu'-\mu)\Lambda}\left(-\frac{\Lambda}{\eta}\right)^{\Delta_1 - (\mu'-\mu)(\eta - \mu')} \left(\frac{\Lambda}{\eta}\right)^{\Delta_\infty(\eta)-\Delta_t(\eta)} \left(1-\frac{\eta}{\Lambda}\right)^{\Delta_1 - (\mu'-\mu)(\eta - \mu')} \mathfrak{F} \left( \begin{matrix} \alpha_1\\\alpha_\infty (\eta) \, \end{matrix} \alpha(\eta) \, \begin{matrix} \alpha_t(\eta)\\ \alpha_0 \end{matrix} ; \frac{\eta}{\Lambda} \right) \longrightarrow\\
    &\longrightarrow \eta^{-\Delta_0-\Delta_1-2\mu'(\mu'-\mu)} {}_1 \mathfrak{D}\left(\mu\, \begin{matrix}\alpha_1\\{}\end{matrix}\,\mu'\,\alpha_0\, ;\, \frac{1}{\Lambda} \right)\,,\quad \mathrm{as}\, \eta \to \infty\,.
    \end{aligned}
\end{equation}
This leaves us with
\begin{equation}
\begin{aligned}
    &\lim_{\eta \to \infty} (-\eta^2)^{\mu(Q-\mu)}(\eta^2)^{-\Delta_0-\Delta_1-2\mu'(\mu'-\mu)}C^{\alpha(\eta)}_{\alpha_\infty(\eta),\alpha_1}C_{\alpha(\eta),\alpha_t(\eta),\alpha_0} = \\=&\, \frac{e^{-i\pi(\Delta_1+2\mu'(\mu'-\mu))}\Upsilon_b(Q+2\alpha_1)}{\Upsilon_b( \frac{Q}{2}+\mu'-\mu-\alpha_1) \Upsilon_b( \frac{Q}{2}+\mu'-\mu+\alpha_1 )}\frac{e^{-i\pi\Delta_0}\Upsilon_b(Q+2\alpha_0)}{\Upsilon_b( \frac{Q}{2}+\mu'+\alpha_0) \Upsilon_b( \frac{Q}{2}+\mu'-\alpha_0 )}\,,
\end{aligned}
\end{equation}
which remarkably has a finite limit. We recognize $C_{\mu' \alpha_0}$ and therefore we can identify
\begin{equation}
    B^{\mu'}_{\mu \alpha_1} = \frac{e^{-i\pi(\Delta_1+2\mu'(\mu'-\mu))}\Upsilon_b(Q+2\alpha_1)}{\Upsilon_b( \frac{Q}{2}+\mu'-\mu-\alpha_1) \Upsilon_b( \frac{Q}{2}+\mu'-\mu+\alpha_1 )}\,.
\end{equation}
Specializing this formula to the case when $V_1$ is a degenerate field is again tricky and involves analytic continuation. It is simpler to perform the collision limit again. The fusion rules now imply that $\alpha(\eta) = \alpha_\infty(\eta) \pm (-b/2)$, i.e. $\mu' =\mu_\pm= \mu \pm (-b/2)$. Performing the collision limit using the degenerate OPE coefficients \ref{degOPEcoeff} one finds
\begin{equation}
    B^{\mu_\theta}_{\mu \alpha_{2,1}} = e^{i\pi\left(\frac{1}{2}+\theta b \mu +\frac{b^2}{4}\right)}\,,
\end{equation}
in agreement with the result \eqref{irrOPEcoeffdeg}.

\subsection{Rank 1/2}\label{app:rankhalfcollision}
Unfortunately, for the rank 1/2 state the situation is not as nice. It is clear that if we decouple another mass, the normalization function $C_{\mu\alpha}$ will diverge badly, since there are no $\Upsilon_b$-functions in the numerator to compensate the divergence of the denominator. Indeed, it behaves as
\begin{equation}
    C_{\mu \alpha} = \frac{e^{-i\pi \Delta}\Upsilon_b(Q+2\alpha)}{\Upsilon_b( \frac{Q}{2}+\mu+\alpha) \Upsilon_b( \frac{Q}{2}+\mu-\alpha)} \to \mathrm{const.}\,\times  e^{3\mu^2} (-\mu^2)^{-\frac{1+Q^2}{6}-\mu^2 +\Delta} e^{-i\pi \Delta}\Upsilon_b(Q+2\alpha)\,,\quad \mathrm{as}\,\, \mu\to \infty\,.
\end{equation}
The constant comes from the $\mathcal{O}(x^0)$ term in the expansion of the $\Upsilon_b$-function \eqref{eq:UpsilonAsymptotics}. We neglect it in the following/consider it substracted by hand. This suggests we define
\begin{equation}
    \langle \Lambda^2|\Delta\rangle = |\Lambda^2|^{2\Delta} C_\alpha = \lim_{\mu\to\infty} e^{-3\mu^2} (-\mu^2)^{\frac{1+Q^2}{6}+\mu^2} \langle -\frac{\Lambda^2}{4\mu} |\Delta\rangle = |\Lambda^2|^{2\Delta} 2^{-4\Delta} e^{-2\pi i \Delta}\Upsilon_b(Q+2\alpha)\,,
\end{equation}
where the factor of $-\frac{1}{4}$ is needed to reproduce the Ward identity $\langle \Lambda^2|L_1 = -\frac{\Lambda^2}{4}\langle \Lambda^2|$. This gives the normalization function for the rank 1/2 state as
\begin{equation}
    C_\alpha = 2^{-4\Delta} e^{-2\pi i \Delta}\Upsilon_b(Q+2\alpha)\,,
\end{equation}
in agreement with the result \eqref{rankhalfnormalization}. Since no collision limit is known that reproduces the OPE between a primary and the rank 1/2 state, we cannot determine the corresponding OPE coefficient in the way we did in the previous section for the rank 1 state. For the case of a degenerate field however, we determine the OPE coefficient in Appendix \ref{app:rankhalf}.

\section{Irregular OPEs}

\subsection{Rank 1}\label{app:rank1}
The form of the (chiral) OPE of a general vertex operator with the irregular state introduced in section \ref{warmup:whittaker} is fixed by the Ward identities to be:
\begin{equation}\label{irrOPE}
    \langle \mu,\Lambda| V^{\Delta}_{\mu,\mu'}(z) = \sum_{k=0}^\infty z^{2 \mu'(\mu'-\mu)-k} \Lambda^{\Delta+2\mu'(\mu'-\mu)} e^{-(\mu'-\mu)\Lambda z} \langle \mu', \Lambda ; k|\,.
\end{equation}
Here $V^{\Delta}_{\mu,\mu'}(z)$ is a vertex operator of weight $\Delta$ which maps from the Whittaker module specified by $(\mu,\Lambda)$, to the module specified by $(\mu',\Lambda)$. Furthermore $\langle \mu', \Lambda ; k|$ are the ("generalized") descendants of the irregular state. They take the form
\begin{equation}
    \langle \mu', \Lambda ; k| = \sum c_{ijY}\Lambda^{-i} \partial_\Lambda^{j} \langle \mu', \Lambda|L_{Y}\,,
\end{equation}
where $c_{ijY}$ are coefficients fixed by the Ward identities and the sum runs over $i,j\geq 0$ and all Young tableaux $Y$ such that $i+j+|Y|=k$.
Furthermore we normalize $\langle \mu', \Lambda ; 0| \equiv \langle \mu', \Lambda|$. We then write the full (chiral+antichiral) OPE between the irregular state and a degenerate field as
\begin{equation}\label{degirrOPE}
    \langle \mu, \Lambda|\Phi(z) = \sum_{\theta=\pm} B_{\mu,\alpha_{2,1}}^{\mu_\theta} \left|\sum_{k=0}^\infty e^{\theta b\Lambda z/2} \Lambda^{-\theta b\mu+\Delta_{2,1}+\frac{b^2}{2}}  z^{-\theta b\mu+\frac{b^2}{2}-k} \right|^2 \langle \mu_\theta , \Lambda;k,\bar{k}|\,,
\end{equation}
where $B_{\mu,\alpha_{2,1}}^{\mu_\theta}$ are the corresponding irregular OPE coefficients. We have anticipated the fact that for the OPE with the degenerate field $\mu'=m_\pm = \mu \pm \frac{-b}{2}$ as will be shown later from the BPZ equation. Furthermore we now have both chiral and antichiral descendants which we label by $k$ and $\bar{k}$, respectively.\newline
We want to determine the irregular OPE coefficients $B$ and the normalization function $C$ introduced in \eqref{irrNormalization}. To this end consider the correlation function
\begin{equation}
    \langle \mu,\Lambda| \Phi(z) |\Delta \rangle \,.
\end{equation}
We can decompose it into irregular conformal blocks doing the OPE left or right as
\begin{equation}\label{crossingrank1}
    \begin{aligned}
        \langle \mu,\Lambda| \Phi(z) |\Delta \rangle = & \sum_{\theta=\pm} C_{\alpha_{2,1},\alpha}^{\alpha_\theta}  C_{\mu \alpha_\theta} \left| {}_1\mathfrak{F} \left( \mu \, \alpha_\theta\, \begin{matrix} \alpha_{2,1}\\ \alpha \end{matrix} ;\Lambda z\right) \right|^2 =\sum_{\theta' = \pm} B_{\alpha_{2,1},\mu}^{\mu_{\theta'}}  C_{\mu_{\theta'} \alpha} \left|{}_1 \mathfrak{D}\left(\mu\,\begin{matrix}\alpha_{2,1}\\ {}\end{matrix}\, \mu_{\theta'}\, \alpha; \frac{1}{\Lambda z}\right) \right|^2\,.
    \end{aligned}
\end{equation}
Here $C_{\alpha_{2,1},\alpha}^{\alpha_\theta}$ is just the usual (regular) OPE coefficient given in terms of the DOZZ formula, $B$ is the irregular OPE coefficient to be determined, and $C_{\mu \alpha}$ is the normalization function of the irregular state, to be determined also. It is defined by
\begin{equation}
    \langle \mu,\Lambda| \Delta \rangle = |\Lambda|^{2\Delta} C_{\mu \alpha}\,.
\end{equation}
To determine $B$ and $C$ we use the BPZ equation
\begin{equation}
\left( b^{-2} \partial_z^2 - \frac{1}{z} \partial_z + \frac{\Delta}{z^2} + \frac{\mu \Lambda}{z} - \frac{\Lambda^2}{4} \right) \langle \mu,\Lambda| \Phi(z) |\Delta \rangle = 0 \,.
\end{equation}
This equation can be solved exactly and has the two solutions $z^{\frac{b^2}{2}} M_{b\mu,\pm b\alpha}(b\Lambda z)$, where $M$ denotes the Whittaker function. It has a simple expansion around $z\sim 0$:
\begin{equation}
    M_{b\mu,b\alpha}(b\Lambda z) = (b\Lambda z)^{\frac{1}{2}+b\alpha}\left(1+\mathcal{O}(b\Lambda z)\right)\,.
\end{equation}
Comparing this expansion with the leading term in the OPE between $\Phi(z)$ and $|\Delta\rangle$ we can identify
\begin{equation}
    {}_1 \mathfrak{F} \left( \mu \, \alpha_\theta\, \begin{matrix} \alpha_{2,1}\\ \alpha \end{matrix} ;\Lambda z\right) = \Lambda^{\Delta_\theta}  z^{\frac{b^2}{2}} (b\Lambda)^{-\frac{1}{2}-\theta b\alpha} M_{b\mu,\theta b\alpha}(b\Lambda z) \,.
\end{equation}
On the other hand, there exist two other solutions to the BPZ equation which have a simple expansion around $z\sim \infty$, namely the Whittaker $W$ functions $W_{\pm b\mu,b\alpha}(\pm b\Lambda z)$. They have an asymptotic expansion at $\infty$ given by
\begin{equation}\label{eq:Wasymptotics}
    W_{b\mu,b\alpha}(b\Lambda z) \sim e^{-b\Lambda z/2} (b\Lambda z)^{b\mu} \left( 1+ \mathcal{O}((b\Lambda z)^{-1})\right)\,,
\end{equation}
valid in the Stokes sector $|\mathrm{arg}(b\Lambda z)| < \frac{3\pi}{2}$. An important fact is that this function is invariant under $\alpha \to -\alpha$. We see that the expansion of the Whittaker $W$ function (times the factor $z^{b^2/2}$) has exactly the form of the OPE between the irregular state and the degenerate field, with 
\begin{equation}
    \mu' = \mu_\pm = \mu \pm \left(-\frac{b}{2}\right)\,.
\end{equation}
(Note that with this convention, $\mu_\pm$ corresponds to $W_{\mp b\mu,b\alpha}(\mp b\Lambda z)$. This may seem confusing but we like to keep the expression $\mu_\pm$ analogous to the fusion rules with a regular state which give $\alpha_\pm = \alpha \pm \frac{-b}{2}$). \newline
Comparing the expansion of the $W$ function with the irregular OPE \eqref{degirrOPE}, we can identify
\begin{equation}
\begin{aligned}
     &{}_1 \mathfrak{D}\left(\mu\,\begin{matrix}\alpha_{2,1}\\ {}\end{matrix}\,\mu_+\, \alpha; \frac{1}{\Lambda z}\right)= \Lambda^{\Delta+\Delta_{2,1}}e^{-i\pi b\mu} b^{b\mu}(\Lambda z)^{\frac{b^2}{2}} W_{-b\mu,b\alpha}(e^{-i \pi} b\Lambda z)\,,\\
     &{}_1 \mathfrak{D}\left(\mu\,\begin{matrix}\alpha_{2,1}\\ {}\end{matrix}\,\mu_-\, \alpha; \frac{1}{\Lambda z}\right)= \Lambda^{\Delta+\Delta_{2,1}} b^{-b\mu}(\Lambda z)^{\frac{b^2}{2}} W_{b\mu,b\alpha}(b\Lambda z)\,.
\end{aligned}
\end{equation}
For simplicity we focus on the branch specified by $-\Lambda=e^{-i\pi}\Lambda$ and use the asymptotic expansion (\ref{eq:Wasymptotics}) for both $b\Lambda z$ and $e^{-i\pi} b\Lambda z \to \infty$. This is valid for $-\frac{\pi}{2}< \mathrm{arg}(b\Lambda z) < \frac{3\pi}{2}$. The modulus squared has to be understood as acting by sending $\Lambda z \to \bar{\Lambda}\bar{z}$ and correspondingly $e^{-i\pi} \Lambda z \to e^{+i\pi}\bar{\Lambda}\bar{z}$. Since we have assumed $-\frac{\pi}{2}< \mathrm{arg}(b\Lambda z) < \frac{3\pi}{2}$, we also have $-\frac{\pi}{2}< \mathrm{arg}(e^{i\pi} b\bar{\Lambda} \bar{z}) < \frac{3\pi}{2}$, so all the asymptotic expansions are in their domain of validity. Similar expressions hold in the other Stokes sectors. \newline
We can now restate the crossing symmetry condition \eqref{crossingrank1} in terms of Whittaker functions and use the known connection formulae for them (see https://dlmf.nist.gov/13.14) to determine the normalization function $C$ and the OPE coefficient $B$. We have
\begin{equation}
    M_{\kappa,\mu}(z) = \frac{\Gamma(1+2\mu)}{\Gamma\left(\frac{1}{2}+\kappa+\mu\right)}e^{i\pi\left(\frac{1}{2}-\kappa+\mu\right)}W_{\kappa,\mu}(z)+\frac{\Gamma(1+2\mu)}{\Gamma\left(\frac{1}{2}-\kappa+\mu\right)}e^{-i\pi\kappa}W_{-\kappa,\mu}(e^{-i\pi}z)\,.
\end{equation}
Plugging this into \eqref{crossingrank1} using the identifications of the conformal blocks with the Whittaker functions we obtain the condition
\begin{equation}
\begin{aligned}\label{eq:crossingrelation}
    \langle \mu,\Lambda| \Phi(z) |\Delta \rangle = & \,|\Lambda|^{2\Delta+2\Delta_{2,1}+b^2} \sum_{\theta=\pm} b^{-1-2\theta b\alpha} C_{\alpha_{2,1},\alpha}^{\alpha_\theta}  C_{\mu \alpha_\theta}\Gamma(1+2\theta b\alpha)^2 \times\\
    \times&\, \left| \frac{e^{i\pi\left(\frac{1}{2}-b\mu+\theta b\alpha\right)}}{\Gamma\left(\frac{1}{2}+b\mu+\theta b \alpha\right)}z^{\frac{b^2}{2}}W_{b\mu,b \alpha}(b\Lambda z)+\frac{e^{-i\pi b \mu}}{\Gamma\left(\frac{1}{2}-b\mu+\theta b \alpha\right)}z^{\frac{b^2}{2}}W_{-b\mu,b \alpha}(e^{-i\pi}b\Lambda z) \right|^2=\\
    =&\, |\Lambda|^{2\Delta+2\Delta_{2,1}+b^2} B_{\alpha_{2,1},\mu}^{\mu_+}  C_{\mu_+ \alpha}\left|e^{-i\pi b\mu} b^{b\mu} z^{\frac{b^2}{2}} W_{-b\mu,b\alpha}(e^{-i\pi} b\Lambda z) \right|^2+\\
    +&\,|\Lambda|^{2\Delta+2\Delta_{2,1}+b^2}B_{\alpha_{2,1},\mu}^{\mu_-}  C_{\mu_-,\alpha} \left|b^{-b\mu} z^{\frac{b^2}{2}} W_{b\mu,b\alpha}(b\Lambda z) \right|^2    \,,
\end{aligned}
\end{equation}
where we have used the fact that $W_{\kappa,-\mu}(z)=W_{\kappa,\mu}(z)$. Using the expression \eqref{degOPEcoeff} for the coefficients $C^{\alpha_\theta}_{\alpha_{2,1},\alpha}$, the cancellation of the cross-terms in the modulus squared gives the following functional equation for $C_{\mu\alpha}$:
\begin{equation}
    \frac{C_{\mu \alpha_+}}{C_{\mu \alpha_-}}  = e^{-2\pi i b \alpha}b^{2bQ+4b\alpha} \frac{\gamma(-2b\alpha)\gamma\left(\frac{1}{2}+b\mu+b\alpha\right)}{\gamma(bQ+2b\alpha)\gamma\left(\frac{1}{2}+b\mu-b\alpha\right)}\,,
\end{equation}
which is solved in terms of the usual $\Upsilon_b$-function:
\begin{equation}
    C_{\mu \alpha} = \frac{e^{-i\pi\Delta}\Upsilon_b(Q+2\alpha)}{\Upsilon_b\left(\frac{Q}{2}+\mu + \alpha\right)\Upsilon_b\left(\frac{Q}{2}+\mu-\alpha\right)}\,,
\end{equation}
up to normalization and a periodic function of $\alpha$ with period $b$. We see however that the minimal choice is consistent with the result obtained by the collision limit in \ref{app:rank1collision}. Once we know the expression for $C_{\mu\alpha}$,we can compute the irregular OPE coefficients $B_{\alpha_{2,1},\mu}^{\mu_\pm}$ from the diagonal terms in (\ref{eq:crossingrelation}). The result is
\begin{equation}\label{irrOPEcoeffdeg}
    B_{\alpha_{2,1},\mu}^{\mu_\pm}= e^{i\pi\left(\frac{1}{2}\pm  b\mu + \frac{b^2}{4}\right)}\,.
\end{equation}
Again, we find that this is in agreement with the result found by the collision limit in \ref{app:rank1collision}. For completeness, let us write the connection formula for the conformal blocks $\mathfrak{F}$ and $\mathfrak{D}$, which solves the crossing symmetry constraint \eqref{crossingrank1}. Using the identification of the conformal blocks with the Whittaker functions with the correct prefactors we find
\begin{equation}\label{connectionW}
    b^{\theta b \alpha} {}_1\mathfrak{F} \left( \mu \, \alpha_\theta\, \begin{matrix} \alpha_{2,1}\\ \alpha \end{matrix} ;\Lambda z\right) = \sum_{\theta'=\pm}b^{-\frac{1}{2}- \theta' b\mu} \mathcal{N}_{\theta \theta'}(b\alpha,b\mu) {}_1\mathfrak{D}\left(\mu\,\begin{matrix}\alpha_{2,1}\\ {}\end{matrix}\, \mu_{\theta'}\, \alpha; \frac{1}{\Lambda z}\right) \,,
\end{equation}
with irregular connection coefficients
\begin{equation}
    \mathcal{N}_{\theta \theta'}(b\alpha,b\mu) = \frac{\Gamma(1+2\theta b \alpha)}{\Gamma\left(\frac{1}{2}+\theta b \alpha- \theta' b\mu\right)} e^{i\pi \left(\frac{1-\theta'}{2}\right)\left(\frac{1}{2}-b\mu+\theta b\alpha\right)}\,.
\end{equation}
The inverse relation is
\begin{equation}\label{connectionWinverse}
     b^{-\frac{1}{2}-\theta b \mu }{}_1\mathfrak{D}\left(\mu\,\begin{matrix}\alpha_{2,1}\\ {}\end{matrix}\, \mu_{\theta}\, \alpha; \frac{1}{\Lambda z}\right)= \sum_{\theta'=\pm} b^{\theta'b\alpha}\mathcal{N}^{-1}_{\theta \theta'}(b\mu,b\alpha) {}_1\mathfrak{F} \left( \mu \, \alpha_{\theta'}\, \begin{matrix} \alpha_{2,1}\\ \alpha \end{matrix} ;\Lambda z\right) \,,
\end{equation}
with
\begin{equation}
    \mathcal{N}^{-1}_{\theta \theta'}(b\mu,b\alpha) = \frac{\Gamma(-2\theta'b \alpha)}{\Gamma\left(\frac{1}{2}+\theta b \mu - \theta' b \alpha\right)}  e^{i\pi \left(\frac{1+\theta}{2}\right)\left(-\frac{1}{2}- b \mu - \theta' b \alpha\right)}\,.
\end{equation}
As a final remark, note that the Whittaker $W$-functions have a non-trivial monodromy around $\infty$. However, since for the correlator we considered, the monodromy around $0$ and $\infty$ is the same, and by construction we have no monodromy around $0$, the combination of $W$-functions appearing in the correlator expanded for large $\Lambda z$ is precisely such that the monodromy cancels. This can be checked also purely locally by carefully using the asymptotic expansions of the $W$-functions and its Stokes sectors. In particular, any other correlator involving this irregular state will have the same asymptotic behaviour and thus the normalization function $C_{\mu\alpha}$ ensures also the absence of monodromies for any other correlator.

\subsection{Rank 1/2}\label{app:rankhalf}
Let us repeat the same arguments for the rank 1/2 irregular state introduced in section \ref{warmup:Bessel}. The (chiral) OPE between the irregular state and the degenerate field is fixed by the Ward identities to be:
\begin{equation}\label{irrOPEhalf}
    \langle \Lambda^2| \Phi_{\Lambda,\pm}(z)= \sum_{k=0}^\infty (\Lambda^2)^{-\frac{1}{4}-\frac{b^2}{4}} z^{\frac{1}{4}+\frac{b^2}{2}-\frac{k}{2}}e^{\pm b \Lambda \sqrt{z}} \langle \Lambda^2;\frac{k}{2}|\,.
\end{equation}
Here $\langle \Lambda^2;\frac{k}{2}|$ are the ("generalized") descendants of the irregular state. They take the form
\begin{equation}
    \langle \Lambda^2;\frac{k}{2}| = \sum c_{ijY}\Lambda^{-i} \partial_\Lambda^{j} \langle \Lambda^2|L_{Y}\,,
\end{equation}
where $c_{ijY}$ are coefficients fixed by the Ward identities and the sum runs over $i,j\geq 0$ and all Young tableaux $Y$ such that $i+j+2|Y|=k$. In particular, note that only the integer descendants (i.e. $k \in 2\mathbb{Z}$) can contain Virasoro generators $L_Y$.
Furthermore we normalize $\langle \Lambda^2;0| \equiv \langle \Lambda^2|$. Since both $z$-behaviours in \eqref{irrOPEhalf} given by $\pm$ live in the same Bessel module specified by $\Lambda$, there is no canonical way of choosing a basis of solutions, in contrast to the rank 1 case. This ambiguity does not affect the physical correlator, since we have to sum over both solutions with the corresponding OPE coefficients. Changing the basis of conformal blocks changes the OPE coefficients in a way that the physical correlator is invariant. Consider the following correlation function involving the rank 1/2 state:
\begin{equation}
    \langle \Lambda^2| \Phi(z) |\Delta \rangle \,.
\end{equation}
We can decompose it into conformal blocks by doing the OPE left and right:
\begin{equation}\label{eq:crossingBessel}
    \langle \Lambda^2|\Phi(z)|\Delta\rangle = \sum_{\theta=\pm} C_{\alpha_{2,1},\alpha}^{\alpha_\theta}  C_{\alpha_\theta} \left|{}_{\frac{1}{2}}\mathfrak{F} \left( \alpha_\theta \, \alpha_{2,1}\,\alpha; \, \Lambda \sqrt{z} \right)\right|^2 =\sum_{\theta' = \pm} B_{\alpha_{2,1}}  C_{ \alpha} \left|{}_{\frac{1}{2}}\mathfrak{E}^{(\theta')}\left(\alpha_{2,1}\, \alpha; \frac{1}{\Lambda \sqrt{z}} \right) \right|^2\,. 
\end{equation}
Here $C_\alpha$ is the normalization function of the irregular state, defined by
\begin{equation}
    \langle\Lambda^2|\Delta\rangle = |\Lambda^2|^{2\Delta} C_\alpha\,,
\end{equation}
which is to be determined. We also want to determine the irregular OPE coefficient $B_{\alpha_{2,1}}$. To do so, consider the BPZ equation that the correlator obeys:
\begin{equation}
\left( b^{-2} \partial_z^2 - \frac{1}{z} \partial_z + \frac{\Delta}{z^2} - \frac{\Lambda^2}{4 z} \right) \langle \Lambda^2| \Phi(z) |\Delta \rangle = 0 \,.
\end{equation}
Solving this differential equation one identifies the conformal block corresponding to the expansion near 0 with a modified Bessel function:
\begin{equation}
    {}_{\frac{1}{2}}\mathfrak{F} \left( \alpha_\theta \, \alpha_{2,1}\,\alpha; \, \Lambda \sqrt{z} \right) = \Gamma(1+2\theta b\alpha) \Lambda^{2\Delta_\theta} \left(\frac{b\Lambda}{2}\right)^{- 2\theta b\alpha} z^{\frac{bQ}{2}} I_{ 2\theta b \alpha}(b\Lambda \sqrt{z})\,.
\end{equation}
The prefactors are fixed by looking at the OPE between $\Phi$ and $|\Delta\rangle$ and using the expansion of the Bessel function:
\begin{equation}
    I_{2\theta b\alpha}(b\Lambda\sqrt{z}) = \frac{(b\Lambda\sqrt{z}/2)^{2\theta b\alpha}}{\Gamma(1+2\theta b\alpha)}\left(1+\mathcal{O}(b\Lambda\sqrt{z})\right)\,.
\end{equation}
On the other hand there are two other solutions to the BPZ equation given by the modified Bessel functions of the second kind $K_{2b\alpha}(\pm b\Lambda\sqrt{z})$. They have a nice behaviour at $\infty$, given by the asymptotic formula
\begin{equation}
    K_{2b\alpha}(b\Lambda\sqrt{z}) \sim \sqrt{\frac{\pi}{2b\Lambda\sqrt{z}}}e^{-b\Lambda\sqrt{z}}(1+\mathcal{O}((b\Lambda\sqrt{z})^{-1}))\,.
\end{equation}
Furthermore $K_{2b\alpha}(b\Lambda\sqrt{z})=K_{-2b\alpha}(b\Lambda\sqrt{z})$. This expansion has precisely the form of the OPE between the irregular state and the degenerate field \eqref{irrOPEhalf}. We can therefore identify the necessary prefactors and defi1ne the irregular conformal blocks for $z \sim \infty$:
\begin{equation}
\begin{aligned}
    &{}_{\frac{1}{2}}\mathfrak{E}^{(+)}\left(\alpha_{2,1}\, \alpha; \frac{1}{\Lambda \sqrt{z}} \right) = \sqrt{\frac{2b}{\pi}}e^{-\frac{i\pi}{2}} (\Lambda^2)^{\Delta-\frac{b^2}{4}}z^{\frac{bQ}{2}} K_{ 2 b \alpha}(e^{-i\pi} b\Lambda \sqrt{z})\,,\\
    &{}_{\frac{1}{2}}\mathfrak{E}^{(-)}\left(\alpha_{2,1}\, \alpha; \frac{1}{\Lambda \sqrt{z}} \right) = \sqrt{\frac{2b}{\pi}} (\Lambda^2)^{\Delta-\frac{b^2}{4}}z^{\frac{bQ}{2}} K_{ 2 b \alpha}( b\Lambda \sqrt{z})\,.
\end{aligned}
\end{equation}
We can now restate the crossing symmetry condition \eqref{eq:crossingBessel} in terms of Bessel functions and use the known connection formulae for them (see e.g. dlmf.nist.gov/10.27) to determine the normalization function $C$ and the OPE coefficient $B_{\alpha_{2,1}}$. We have
\begin{equation}
    I_{\nu}(z)= \frac{i}{\pi} e^{i\pi \nu} K_\nu(z) - \frac{i}{\pi} K_\nu(e^{-i\pi}z) \,.
\end{equation}
Plugging this formula into \eqref{eq:crossingBessel} using the identifications between the conformal blocks and Bessel functions, one finds that the vanishing of the cross-terms gives the condition
\begin{equation}
    \frac{C_{\alpha_+}}{C_{\alpha_-}}=2^{-8b\alpha}b^{2bQ+8b\alpha} e^{-4\pi i b \alpha}\frac{\gamma(-2b\alpha)}{\gamma(bQ+2b\alpha)}\,.
\end{equation}
We take the simplest solution, namely 
\begin{equation}\label{rankhalfnormalization}
    C_{\alpha} = 2^{-4\Delta} e^{-2\pi i \Delta} \Upsilon_b(Q+2\alpha) \,.
\end{equation} 
This is in agreement with the result found in \ref{app:rankhalfcollision}. Once we have the expression for $C_\alpha$, we can compute the irregular OPE coefficients from the diagonal terms of the crossing symmetry condition. The result is
\begin{equation}
\begin{aligned}\label{halfrankOPEcoeff}
    &B_{\alpha_{2,1}} = 2^{b^2}e^{\frac{i\pi bQ}{2}}\,.
\end{aligned}
\end{equation}
We see that the OPE coefficients are independent of $\pm$, which is a reflection of the fact that we have a symmetry rotating the basis of conformal blocks into each other and leaving the physical correlator invariant.\newline
For completeness, let us write also the connection formula for the irregular conformal blocks:
\begin{equation}\label{eq:rankhalfconnection}
    b^{2\theta b \alpha}{}_{\frac{1}{2}}\mathfrak{F} \left( \alpha_\theta \, \alpha_{2,1}\,\alpha; \, \Lambda \sqrt{z} \right) = \sum_{\theta'=\pm}b^{-\frac{1}{2}} \mathcal{Q}_{\theta \theta'}(b\alpha) {}_{\frac{1}{2}}\mathfrak{E}^{(\theta')}\left(\alpha_{2,1}\, \alpha; \frac{1}{\Lambda \sqrt{z}} \right)\,.
\end{equation}
with irregular connection coefficients
\begin{equation}
    \mathcal{Q}_{\theta \theta'}(b\alpha) = \frac{2^{2\theta b \alpha}}{\sqrt{2\pi }} \Gamma(1+2\theta b \alpha) e^{i\pi\left(\frac{1-\theta'}{2}\right)\left(\frac{1}{2}+2\theta b \alpha \right)}\,.
\end{equation}
The inverse relation is 
\begin{equation}
    b^{-\frac{1}{2}} {}_{\frac{1}{2}}\mathfrak{E}^{(\theta)}\left(\alpha_{2,1}\, \alpha; \frac{1}{\Lambda \sqrt{z}} \right)= \sum_{\theta'=\pm}b^{2 \theta' b \alpha} \mathcal{Q}^{-1}_{\theta \theta'}(b\alpha) {}_{\frac{1}{2}}\mathfrak{F} \left( \alpha_{\theta'} \, \alpha_{2,1}\,\alpha; \, \Lambda \sqrt{z} \right) \,.
\end{equation}
with irregular connection coefficients
\begin{equation}
    \mathcal{Q}^{-1}_{\theta \theta'}(b\alpha) = \frac{2^{-2 \theta' b \alpha}}{\sqrt{2\pi}}\Gamma(-2\theta'b\alpha)  e^{-i\pi\left(\frac{1+\theta}{2}\right)\left(\frac{1}{2}+2\theta' b \alpha \right)}\,.
\end{equation}

\section{Classical conformal blocks and accessory parameters}\label{app:Nek}
In this Appendix we give explicit combinatorial expressions for the classical conformal blocks used in the main text. 
\subsection{The regular case}
Let us start with the case of regular conformal blocks. Via the AGT correspondence \cite{Alday_2010} the four-point regular conformal block is given by 
\begin{equation}
\begin{aligned}
    &\mathfrak{F} \left( \begin{matrix} \alpha_1 \\ \alpha_\infty \end{matrix} \alpha \, \begin{matrix} \alpha_t \\ \alpha_0 \end{matrix} ; t \right) = t^{\Delta - \Delta_t - \Delta_0} (1-t)^{-2(\frac{Q}{2}+ \alpha_1)(\frac{Q}{2}+\alpha_t)} \times \\ &\times \sum_{\vec{Y}} t^{| \vec{Y} |} z_{\text{vec}} \left( \vec{\alpha}, \vec{Y} \right) \prod_{\theta = \pm} z_{\text{hyp}} \left( \vec{\alpha}, \vec{Y}, \alpha_t + \theta \alpha_0 \right) z_{\text{hyp}} \left( \vec{\alpha}, \vec{Y}, \alpha_1 + \theta \alpha_\infty \right) \,,
\end{aligned}
\label{eq:reg4ptasnek}
\end{equation}
where the sum runs over all pairs of Young tableaux $\left( Y_1, Y_2 \right)$. We denote the size of the pair $| \vec{Y} | = | Y_1 | + | Y_2 |$, and
\cite{Flume,Bruzzo_2003}
\begin{equation}\label{zdef}
\begin{aligned}
    &z_{\text{hyp}} \left( \vec{\alpha}, \vec{Y}, \mu \right) = \prod_{k= 1,2} \prod_{(i,j) \in Y_k} \left( \alpha_k + \mu + b^{-1} \left( i - \frac{1}{2} \right) + b \left( j - \frac{1}{2} \right) \right) \,, \\
    &z_{\text{vec}} \left( \vec{\alpha}, \vec{Y} \right) = \prod_{k,l = 1,2} \prod_{(i,j) \in Y_k}  E^{-1} \left( \alpha_k - \alpha_l, Y_k, Y_l, (i, j) \right) \prod_{(i',j') \in Y_l} \left( Q - E \left( \alpha_l - \alpha_k, Y_l, Y_k, (i', j') \right) \right)^{-1}\,, \\
    &E \left( \alpha, Y_1, Y_2, (i,j) \right) = \alpha - b^{-1} L_{Y_2} ((i,j)) + b \left( A_{Y_1} ((i,j)) + 1 \right) \,.
\end{aligned}
\end{equation}
Here $L_Y ((i,j)), A_Y((i,j))$ denote respectively the leg-length and the arm-length of the box at the site $(i,j)$ of the tableau $Y$. If we denote a Young tableau as $Y = ( \nu_1' \ge \nu_2' \ge \dots)$ and its transpose as $Y^T = ( \nu_1 \ge \nu_2 \ge \dots)$, then $L_Y$ and $A_Y$ read
\begin{equation}
    A_Y (i, j) = \nu_i' -  j \,, \, \, L_Y (i, j) = \nu_j - i \,.
\end{equation}
\begin{figure}[t!]\centering
\includegraphics[scale=0.5]{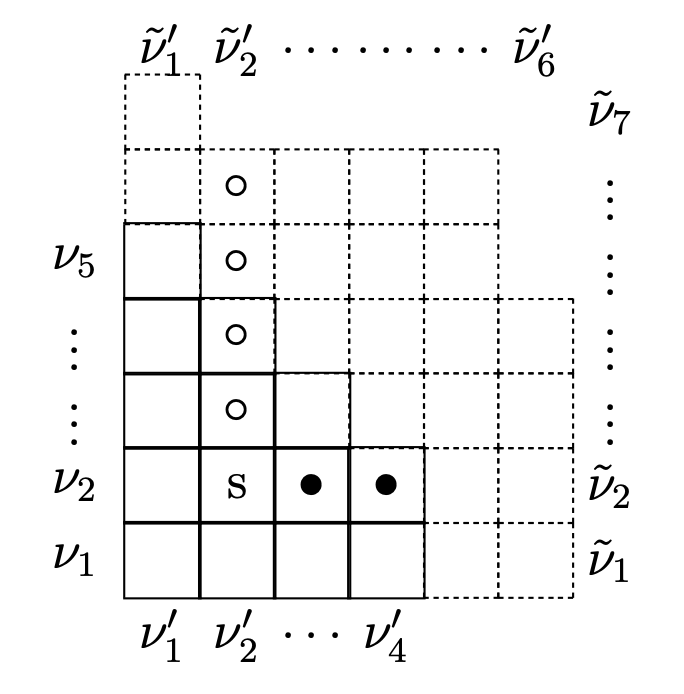}
\caption{Arm length $A_{\tilde{Y}} (s)=4$ (white circles) and leg length $L_Y(s)=2$ (black dots) of a box at the site $s = (2,2)$ for the pair of superimposed diagrams $Y$ (solid lines) and $\tilde{Y}$ (dotted lines).}
\end{figure}
Note that they can be negative if the box $(i,j)$ are the coordinates of a box outside the tableau. Also, the previous formulae has to be evaluated at $\vec{\alpha} = (\alpha_1, \alpha_2) = (\alpha, - \alpha)$. Comparing \eqref{eq:reg4ptasnek} with \eqref{eq:defclasscbNf41} we find the explicit expression for the classical conformal block $F$:
\begin{equation}
    F(t) = \lim_{b \to 0} b^2 \log \left[ (1-t)^{-2(\frac{Q}{2}+ \alpha_1)(\frac{Q}{2}+\alpha_t)} \sum_{\vec{Y}} t^{| \vec{Y} |} z_{\text{vec}} \left( \vec{\alpha}, \vec{Y} \right) \prod_{\theta = \pm} z_{\text{hyp}} \left( \vec{\alpha}, \vec{Y}, \alpha_t + \theta \alpha_0 \right) z_{\text{hyp}} \left( \vec{\alpha}, \vec{Y}, \alpha_1 + \theta \alpha_\infty \right) \right] \,.
\label{eq:expressforbigFF}
\end{equation} 
This turns into a combinatorial expression of the $u$ parameter defined as
\begin{equation}
    u^{(0)} = \lim_{b \to 0} b^2 t \partial_t \log \mathfrak{F} \left( \begin{matrix} \alpha_1 \\ \alpha_\infty \end{matrix} \alpha \, \begin{matrix} \alpha_t \\ \alpha_0 \end{matrix} ; t \right) = - \frac{1}{4} - a^2 + a_t^2 + a_0^2 + t \partial_t F(t)
\end{equation}
in terms of the intermediate momentum $\alpha$. After substituting the dictionary with the Heun equation this gives a combinatorial expression of the accessory parameter $q$ in terms of the Floquet exponent $a = b \alpha$. Inverting this relation order by order in $t$ allows us to compute the connection coefficients in terms of the accessory parameter. Let us carry out explicitly a first order computation for the sake of clarity. At one instanton the relevant pairs of Young tableaux are $\vec{Y} = ((1),(0))$ and $\vec{Y} = ((0),(1))$. The various contributions give
\begin{equation}
\begin{aligned}
    &z_{\text{hyp}} \left( \vec{\alpha}, ((1),(0)), \mu \right) = \frac{Q}{2} + \alpha + \mu \,, \\
    &z_{\text{hyp}} \left( \vec{\alpha}, ((0),(1)), \mu \right) = \frac{Q}{2} - \alpha + \mu \,, 
\end{aligned}
\end{equation}
and since $A_{(0)} (i=1,j=1) = L_{(0)} (i=1,j=1) = -1$ and $A_{(1)} (i=1,j=1) = L_{(1)} (i=1,j=1) = 0$,
\begin{equation}
\begin{aligned}
    &E \left(0, (1), (1), (i=1,j=1) \right) = b \,, \\
    &E \left(2 \alpha, (1), (0), (i=1,j=1) \right) = Q + 2 \alpha \,,
\end{aligned}
\end{equation}
therefore
\begin{equation}
\begin{aligned}
    z_{\text{vec}} \left( \vec{\alpha}, ((1),(0)) \right) &= \prod_{l = 1, 2} E^{-1} \left( \alpha - \alpha_l, (1), Y_l, (i = 1, j = 1) \right) \prod_{k=1,2} \left( Q - E \left( \alpha - \alpha_k, (1), Y_k, (i' = 1, j' = 1) \right) \right)^{-1} \\ &= \frac{1}{- 2 \alpha \left( Q + 2 \alpha \right)} \,, \\ z_{\text{vec}} \left( \vec{\alpha}, ((0),(1)) \right) &= \prod_{l = 1, 2} E^{-1} \left( -\alpha - \alpha_l, (1), Y_l, (i = 1, j = 1) \right) \prod_{k=1,2} \left( Q - E \left( -\alpha - \alpha_k, (1), Y_k, (i' = 1, j' = 1) \right) \right)^{-1} \\ &= \frac{1}{2 \alpha \left( Q - 2 \alpha \right)} \,.
\end{aligned}
\end{equation}
Note that  and that every time $(i,j)$ have to run into an empty tableau, the corresponding term contributes with $1$. Finally, substituting the previous results in \eqref{eq:expressforbigFF} we get
\begin{equation}
    F(t) = \frac{\left( \frac{1}{4} - a^2 - a_1^2 + a_\infty^2 \right) \left(  \frac{1}{4} - a^2 - a_t^2 + a_0^2 \right)}{\frac{1}{2} - 2 a^2} t + \mathcal{O} (t^2) \,.
\end{equation}
In the main text we will need the derivatives of $F$ expressed in terms of Heun parameters. For example,
\begin{equation}\label{dF}
    \partial_{a_t} F(t) = \frac{\left( 4 a^2 - \alpha^2 + 2 \alpha \beta - \beta^2 - 2 \delta + \delta^2\right) \left(1-\epsilon \right)}{2 - 8 a^2} t + \mathcal{O} (t^2) \,.
\end{equation}
Moreover,
\begin{equation}
    u^{(0)} = -\frac{1}{4} - a^2 + a_t^2 + a_0^2 + \frac{\left( \frac{1}{4} - a^2 - a_1^2 + a_\infty^2 \right) \left(  \frac{1}{4} - a^2 - a_t^2 + a_0^2 \right)}{\frac{1}{2} - 2 a^2} t + \mathcal{O} (t^2) \,.
\end{equation}
Note that the relation between $u^{(0)}$ and $a$ is quadratic at $t = 0$, therefore we will have two solutions for $a(u^{(0)})$:
\begin{equation}
    a = \pm  \sqrt{-\frac{1}{4} - u^{(0)} + a_t^2 + a_0^2}\left( 1- \frac{\left(-1+2a_0^2+2a_1^2-2a_\infty^2+2a_t^2-2u^{(0)}\right) \left(-1+4a_t^2-2u^{(0)}\right)}{2\left(-1+4a_0^2+4a_t^2-4 u^{(0)}\right) \left(-1+2a_0^2+2a_t^2-2u^{(0)}\right)} t +\mathcal{O}(t^2)\right) \,.
\end{equation}
Substituting the dictionary \eqref{eq:HeundictioanaryscCFT} we obtain
\begin{equation}
\begin{aligned}
    & a = \pm\frac{1}{2} \sqrt{(\alpha +\beta -\delta )^2-4 q}\mp\frac{t \left(\delta  (q (\alpha +\beta +1)-\gamma  (\alpha  \beta +q))+(q-\alpha  \beta ) (2 q-\gamma  (\alpha +\beta -1))+\delta ^2 (-q)\right)}{\sqrt{(\alpha +\beta -\delta )^2-4 q} (4 q-(\alpha +\beta -\delta -1) (\alpha +\beta -\delta +1))}+O\left(t^2\right)\,.
\end{aligned}
\label{eq:aofq}
\end{equation}
Note that that all the connection formulae near the various singularity are all symmetric under $a \to - a$. The sign has to be carefully chosen only when connecting to the intermediate region. Finally, we are in the position to expand the connection coefficients. For example, one would have, choosing the lower sign in $a$,
\begin{equation}
\begin{aligned}
    &\Gamma \left( \frac{1 + \gamma - \epsilon}{2} + a \right) \simeq \Gamma\left( \frac{1+\gamma-\epsilon- \sqrt{-4q+(\alpha+\beta-\delta)^2}}{2} \right) \times \\ &\times \left(1 + \frac{t \left(\delta  (q (\alpha +\beta +1)-\gamma  (\alpha  \beta +q))+(q-\alpha  \beta ) (2 q-\gamma  (\alpha +\beta -1))+\delta ^2 (-q)\right) \psi_0\left( \frac{1+\gamma-\epsilon- \sqrt{-4q+(\alpha+\beta-\delta)^2}}{2}\right)}{\sqrt{(\alpha +\beta -\delta )^2-4 q} (4 q-(\alpha +\beta -\delta -1) (\alpha +\beta -\delta +1))} \right) \,,
\end{aligned}
\end{equation}
where $\psi_0$ is the Digamma function. 
\subsection{The confluent case}
In order to discuss the confluent classical conformal block, let us write the four-point conformal block appearing in \eqref{eq:MatoneNf4inf}, that is
\begin{equation}
\begin{aligned}
    &\mathfrak{F} \left( \begin{matrix} \alpha_t \\ \alpha_\infty \end{matrix} \alpha \, \begin{matrix} \alpha_1 \\ \alpha_0 \end{matrix} ; \frac{1}{t} \right) = t^{-\Delta + \Delta_1 + \Delta_0} (1-t^{-1})^{-2(\frac{Q}{2}+ \alpha_1)(\frac{Q}{2}+\alpha_t)} \times \\ &\times \sum_{\vec{Y}} t^{-| \vec{Y} |} z_{\text{vec}} \left( \vec{\alpha}, \vec{Y} \right) \prod_{\theta = \pm} z_{\text{hyp}} \left( \vec{\alpha}, \vec{Y}, \alpha_t + \theta \alpha_\infty \right) z_{\text{hyp}} \left( \vec{\alpha}, \vec{Y}, \alpha_1 + \theta \alpha_0 \right) \,.
\end{aligned}
\label{eq:eq:appsmNf3}
\end{equation}
Note that in the decoupling limit \eqref{FIII}, that is 
\begin{equation}
    \alpha_t + \alpha_\infty = - \mu \,, \, \, \alpha_t - \alpha_\infty = \eta \,, \, \, t = \frac{\Lambda}{\eta} \,,
\end{equation}
where then $\eta \to \infty$, 
\begin{equation}
\begin{aligned}
    &z_{\text{hyp}} \left( \vec{\alpha}, \vec{Y}, \alpha_t - \alpha_\infty \right) \sim (\alpha_t - \alpha_\infty)^{2| \vec{Y} |} \sim \left( \frac{\Lambda}{t} \right)^{2| \vec{Y} |} \,, \\
    &z_{\text{hyp}} \left( \vec{\alpha}, \vec{Y}, \alpha_t + \alpha_\infty \right) = z_{\text{hyp}} \left( \vec{\alpha}, \vec{Y}, - \mu \right) \,, \\
    &(1-t^{-1})^{2(\frac{Q}{2}+ \alpha_1)(\frac{Q}{2}+\alpha_t)} \sim e^{- \left( \frac{Q}{2} + \alpha_1 \right) \Lambda} \,.
\end{aligned}
\end{equation}
Therefore the confluent 3-point function \eqref{eq:cNf34pointsmall} has the following combinatorial expression
\begin{equation}
    {}_1\mathfrak{F} \left( \mu \, \alpha\, \begin{matrix} \alpha_1\\ \alpha_0 \end{matrix} ;\Lambda \right) = \Lambda^\Delta e^{\left( \frac{Q}{2} + \alpha_1 \right) \Lambda}  \sum_{\vec{Y}} \Lambda^{| \vec{Y} |} z_{\text{vec}} \left( \vec{\alpha}, \vec{Y} \right) z_{\text{hyp}} \left( \vec{\alpha}, \vec{Y}, - \mu \right) \prod_{\theta = \pm} z_{\text{hyp}} \left( \vec{\alpha}, \vec{Y}, \alpha_1 + \theta \alpha_0 \right) \,.
\end{equation}
As for the previous case, this turns into a combinatorial expression of the $u$ parameter defined in equation \ref{eq:uelectricNf3confl} in terms of the intermediate momentum $a$, that after substituting the dictionary with the CHE gives a combinatorial expression for the accessory parameter in terms of the Floquet exponent. Again, inverting this relation is useful for computing the explicit connection coefficients. Similarly we can give an explicit expression of the classical conformal block for big $\Lambda$ appearing in \eqref{FD}, that is 
\begin{equation}
\begin{aligned}
    &{}_1 \mathfrak{D}\left(\mu\,\begin{matrix}\alpha_1\\ {}\end{matrix}\,\mu'\, \alpha_0; \frac{1}{\Lambda}\right) = \lim_{\eta \to \infty} \Lambda^{\Delta_0 + \Delta_1 + 2 \mu' (\mu'-\mu)} e^{-(\mu'-\mu)\Lambda} \left(1-\frac{\eta}{\Lambda}\right)^{\Delta_1 - (\mu'-\mu)(\eta - \mu') - \left( \frac{Q}{2} + \alpha_1\right) \left( Q + \eta - \mu \right)} \times \\ &\times \sum_{\vec{Y}} \left( \frac{\eta}{\Lambda} \right)^{|\vec{Y}|} z_{\text{vec}} \left( \vec{\alpha} (\eta), \vec{Y} \right) \prod_{\theta = \pm} z_{\text{hyp}} \left(\vec{\alpha} (\eta), \vec{Y}, \frac{\eta - \mu}{2} + \theta \alpha_0 \right) z_{\text{hyp}} \left(\vec{\alpha} (\eta), \vec{Y}, \alpha_1 + \theta \frac{- \eta - \mu}{2} \right) \,,
\end{aligned}
\end{equation}
where
\begin{equation}
    \vec{\alpha} (\eta) = \left( - \frac{\eta - \mu}{2} - \mu', \frac{\eta - \mu}{2} + \mu' \right)\,.
\end{equation}
Again, this gives an explicit expression of the classical conformal block $F_D (L^{-1})$ recalling that
\begin{equation}
    {}_1 \mathfrak{D}\left(\mu\,\begin{matrix}\alpha_1\\ {}\end{matrix}\,\mu'\, \alpha_0; \frac{1}{\Lambda}\right) = e^{-(\mu'-\mu)\Lambda}\Lambda^{\Delta_0+\Delta_1+2\mu'(\mu'-\mu)} e^{\frac{1}{b^2}(F_D(L^{-1})+\mathcal{O}(b^2))}\,.
\end{equation}
\subsection{The reduced confluent case}
To obtain the reduced confluent classical block we decouple the momentum $\mu$ starting from \eqref{eq:eq:appsmNf3} as follows
\begin{equation}
    \Lambda = - \frac{\Lambda_1 \Lambda_2}{4 \mu} \,, \,\, \text{as} \,\,  \mu \to \infty \,.
\end{equation}
This gives 
\begin{equation}
    {}_\frac{1}{2}\mathfrak{F} \left(\alpha\, \begin{matrix} \alpha_1\\ \alpha_0 \end{matrix} ;\Lambda^2 \right) = \Lambda^{2 \Delta} \sum_{\vec{Y}} \left(\frac{\Lambda^2}{4} \right)^{| \vec{Y} |} z_{\text{vec}} \left( \vec{\alpha}, \vec{Y} \right) \prod_{\theta = \pm} z_{\text{hyp}} \left( \vec{\alpha}, \vec{Y}, \alpha_1 + \theta \alpha_0 \right) \,.
\end{equation}
This gives for the classical conformal blocks
\begin{equation}
    F (L^2) = \lim_{b \to 0} b^2 \log \sum_{\vec{Y}} \left(\frac{\Lambda^2}{4} \right)^{| \vec{Y} |} z_{\text{vec}} \left( \vec{\alpha}, \vec{Y} \right) \prod_{\theta = \pm} z_{\text{hyp}} \left( \vec{\alpha}, \vec{Y}, \alpha_1 + \theta \alpha_0 \right) \,.
\end{equation}
\subsection{The doubly confluent case}
Let us consider the following decoupling limit of \eqref{eq:eq:appsmNf3}:
\begin{equation}
    \alpha_1 + \alpha_0 = - \mu_2 \,, \,\, \alpha_1 - \alpha_0 = \eta \,, \,\, \Lambda \to \frac{\Lambda_1 \Lambda_2}{\eta}\,, \,\, \text{as} \,\, \eta \to \infty \,.
\end{equation}
This gives 
\begin{equation}
    {}_1 \mathfrak{F}_1 \left( \mu_1 \, \, \alpha \, \, \mu_2, \Lambda_1 \Lambda_2 \right) = \left( \Lambda_1 \Lambda_2 \right)^{\Delta} e^{\frac{\Lambda_1 \Lambda_2}{2}} \sum_{\vec{Y}} \left( \Lambda_1 \Lambda_2 \right)^{| \vec{Y} |} z_{\text{vec}} \left( \vec{\alpha}, \vec{Y} \right) z_{\text{hyp}} \left( \vec{\alpha}, \vec{Y}, - \mu_1 \right) z_{\text{hyp}} \left( \vec{\alpha}, \vec{Y}, - \mu_2 \right) \,,
\label{eq:1F1nek}
\end{equation}
and
\begin{equation}
    F(L_1 L_2) = \lim_{b \to 0} b^2 \log \left[ e^{\frac{\Lambda_1 \Lambda_2}{2}} \sum_{\vec{Y}} \left( \Lambda_1 \Lambda_2 \right)^{| \vec{Y} |} z_{\text{vec}} \left( \vec{\alpha}, \vec{Y} \right) z_{\text{hyp}} \left( \vec{\alpha}, \vec{Y}, - \mu_1 \right) z_{\text{hyp}} \left( \vec{\alpha}, \vec{Y}, - \mu_2 \right) \right] \,.
\end{equation}
\subsection{The reduced doubly confluent case}
We now decouple $\mu_2$ in \eqref{eq:1F1nek} as follows
\begin{equation}
    \Lambda_2 \to -\frac{\Lambda_2^2}{4 \mu_2} \,, \,\, \text{as} \,\, \mu_2 \to \infty \,.
\end{equation}
Again,
\begin{equation}
    {}_1\mathfrak{F}_\frac{1}{2} \left(\mu\,\alpha ;\Lambda_1 \frac{\Lambda_2^2}{4} \right) = \left( \Lambda_1 \Lambda_2^2 \right)^{\Delta} \sum_{\vec{Y}} \left( \Lambda_1 \Lambda_2^2 \right)^{| \vec{Y} |} z_{\text{vec}} \left( \vec{\alpha}, \vec{Y} \right) z_{\text{hyp}} \left( \vec{\alpha}, \vec{Y}, - \mu \right) \,.
\label{eq:1Fhalfnek}
\end{equation}
Therefore the corresponding classical conformal block gives
\begin{equation}
    F(L_1 L_2^2) = \lim_{b \to 0} b^2 \log \sum_{\vec{Y}} \left( \Lambda_1 \frac{\Lambda_2^2}{4} \right)^{| \vec{Y} |} z_{\text{vec}} \left( \vec{\alpha}, \vec{Y} \right) z_{\text{hyp}} \left( \vec{\alpha}, \vec{Y}, - \mu \right) \,.
\end{equation}
\subsection{The doubly reduced doubly confluent case}
Decoupling the last momentum $\mu$ in \eqref{eq:1Fhalfnek} by setting
\begin{equation}
    \Lambda_1 \to -\frac{\Lambda_1^2}{4 \mu_1} \,, \,\, \text{as} \,\, \mu \to \infty
\end{equation}
gives
\begin{equation}
    {}_\frac{1}{2}\mathfrak{F}_\frac{1}{2} \left(\alpha ;\Lambda_1^2 \Lambda_2^2 \right) = \left( \Lambda_1^2 \Lambda_2^2 \right)^{\Delta} \sum_{\vec{Y}} \left(\frac{\Lambda_1^2 \Lambda_2^2}{16} \right)^{| \vec{Y} |} z_{\text{vec}} \left( \vec{\alpha}, \vec{Y} \right) \,.
\end{equation}
The corresponding classical conformal block gives
\begin{equation}
    F(L_1^2 L_2^2) = \lim_{b \to 0} b^2 \log \left[ \sum_{\vec{Y}} \left( \frac{\Lambda_1^2 \Lambda_2^2}{16} \right)^{| \vec{Y} |} z_{\text{vec}} \left( \vec{\alpha}, \vec{Y} \right) \right] \,.
\end{equation}

\section{Combinatorial formula for the degenerate 5-point block}\label{app:NekQuiver}
As for the four-point blocks in the previous Appendix, we give an explicit combinatorial expression for the degenerate 5-point conformal block introduced in \ref{regCB} via the AGT correspondence. It can be computed as the partition function of $\mathcal{N}=2$ gauge theory with four flavours and a surface operator, or equivalently as a quiver gauge theory with specific masses fixed by the fusion rules of the degenerate field. Using the representation as a quiver gauge theory we find
\begin{equation}\label{NekQuiver}
\begin{aligned}
    &\FIV{\alpha_\infty}{\alpha_1}{\alpha}{\alpha_t}{{\alpha_{0 \theta}}}{\alpha_{2,1}}{\alpha_0}{t}{\frac{z}{t}} = t^{\Delta - \Delta_t - \Delta_{0\theta}} z^{\frac{bQ}{2}+\theta b \alpha_0} (1-t)^{-2(\frac{Q}{2}+ \alpha_1)(\frac{Q}{2}-\alpha_t)}\left(1-\frac{z}{t}\right)^{-2(\frac{Q}{2}+ \alpha_t)(\frac{Q}{2}+\alpha_{2,1})}(1-z)^{-2(\frac{Q}{2}+ \alpha_1)(\frac{Q}{2}+\alpha_{2,1})} \times \\ &\times \sum_{\vec{Y},\vec{W}} t^{| \vec{Y} |} \left(\frac{z}{t}\right)^{|\vec{W}|} z_{\text{vec}} \left( \vec{\alpha}, \vec{Y} \right)z_{\text{vec}} \left( \vec{\alpha_{0\theta}}, \vec{W} \right) \prod_{\sigma = \pm}  z_{\text{hyp}} \left( \vec{\alpha}, \vec{Y}, \alpha_1 + \sigma \alpha_\infty \right) z_{\text{hyp}} \left( \vec{\alpha_{0\theta}}, \vec{W}, \alpha_{2,1} + \sigma \alpha_0 \right) z_{\text{bifund}} \left( \vec{\alpha},\vec{Y},\vec{\alpha_{0\theta}},\vec{W};\alpha_t \right) \,,
\end{aligned}
\end{equation}
where the sum runs over two pairs of Young tableaux $\vec{Y}=(Y_1,Y_2)$ and $\vec{W}=(W_1,W_2)$. $\vec{\alpha_{0\theta}}$ has to be understood as $(\alpha_{0\theta},-\alpha_{0\theta})$ and we recall that $\alpha_{2,1} = -\frac{2b+b^{-1}}{2}$. Furthermore $z_{\text{vec}}$ and $z_{\text{hyp}}$ are defined as in \eqref{zdef}. The new ingredient is the contribution of a bifundamental, defined as
\begin{equation}
\begin{aligned}
    &z_{\text{bifund}}\left( \vec{\alpha},\vec{Y},\vec{\beta},\vec{W};\alpha_t \right)=\\=&\prod_{k,l = 1,2} \prod_{(i,j) \in Y_k}  \left[E \left( \alpha_k - \beta_l, Y_k, W_l, (i, j) \right)-\left(\frac{Q}{2}+\alpha_t\right)  \right] \prod_{(i',j') \in W_l} \left[ Q - E \left( \beta_l - \alpha_k, W_l, Y_k, (i', j') \right)-\left(\frac{Q}{2}+\alpha_t\right) \right]\,,
\end{aligned}
\end{equation}
with $E$ as in \eqref{zdef}.\newline
Since all other conformal blocks are defined in terms of this degenerate 5-point block, the expression \eqref{NekQuiver} can be used to compute any other block. In particular one can verify explicitly that the various confluence limits are finite.
\newpage
\section{List of Symbols}
\printnomenclature
\newpage
\printbibliography

@article{Hatsuda:2021gtn,
    author = "Hatsuda, Yasuyuki and Kimura, Masashi",
    title = "{Spectral Problems for Quasinormal Modes of Black Holes}",
    eprint = "2111.15197",
    archivePrefix = "arXiv",
    primaryClass = "gr-qc",
    reportNumber = "RUP-21-22",
    month = "11",
    year = "2021"
}

@article{Blake:2021hjj,
    author = "Blake, Mike and Davison, Richard A.",
    title = "{Chaos and pole-skipping in rotating black holes}",
    eprint = "2111.11093",
    archivePrefix = "arXiv",
    primaryClass = "hep-th",
    month = "11",
    year = "2021"
}

@article{Nakajima:2021yfz,
    author = "Nakajima, Hiroaki and Lin, Wenbin",
    title = "{New Chandrasekhar transformation in Kerr spacetime}",
    eprint = "2111.05857",
    archivePrefix = "arXiv",
    primaryClass = "gr-qc",
    month = "11",
    year = "2021"
}

@article{Bianchi:2021yqs,
    author = "Bianchi, Massimo and Di Russo, Giorgio",
    title = "{Turning black-holes and D-branes inside out their photon-spheres}",
    eprint = "2110.09579",
    archivePrefix = "arXiv",
    primaryClass = "hep-th",
    month = "10",
    year = "2021"
}

@article{Amado:2021erf,
    author = "Amado, Juli\'an Barrag\'an and da Cunha, Bruno Carneiro and Pallante, Elisabetta",
    title = {{QNMs of scalar fields on small Reissner-Nordstr\"om-AdS$\mathbf{_5}$ black holes}},
    eprint = "2110.08349",
    archivePrefix = "arXiv",
    primaryClass = "hep-th",
    month = "10",
    year = "2021"
}

@article{Bianchi:2021mft,
    author = "Bianchi, Massimo and Consoli, Dario and Grillo, Alfredo and Morales, Jose Francisco",
    title = "{More on the SW-QNM correspondence}",
    eprint = "2109.09804",
    archivePrefix = "arXiv",
    primaryClass = "hep-th",
    month = "9",
    year = "2021"
}

@article{Cavalcante:2021scq,
    author = "Cavalcante, Jo\~ao Paulo and da Cunha, Bruno Carneiro",
    title = {{Scalar and Dirac perturbations of the Reissner-Nordstr\"om black hole and Painlev\'e transcendents}},
    eprint = "2109.06929",
    archivePrefix = "arXiv",
    primaryClass = "gr-qc",
    doi = "10.1103/PhysRevD.104.124040",
    journal = "Phys. Rev. D",
    volume = "104",
    number = "12",
    pages = "124040",
    year = "2021"
}

@article{Casals:2021ugr,
    author = "Casals, Marc and da Costa, Rita Teixeira",
    title = "{Hidden spectral symmetries and mode stability of subextremal Kerr(-dS) black holes}",
    eprint = "2105.13329",
    archivePrefix = "arXiv",
    primaryClass = "gr-qc",
    month = "5",
    year = "2021"
}

@article{daCunha:2021jkm,
    author = "da Cunha, Bruno Carneiro and Cavalcante, Jo\~ao Paulo",
    title = "{Teukolsky master equation and Painlev\'e transcendents: Numerics and extremal limit}",
    eprint = "2105.08790",
    archivePrefix = "arXiv",
    primaryClass = "hep-th",
    doi = "10.1103/PhysRevD.104.084051",
    journal = "Phys. Rev. D",
    volume = "104",
    number = "8",
    pages = "084051",
    year = "2021"
}

@article{Bianchi:2021xpr,
    author = "Bianchi, Massimo and Consoli, Dario and Grillo, Alfredo and Morales, Jos\`e Francisco",
    title = "{QNMs of branes, BHs and fuzzballs from quantum SW geometries}",
    eprint = "2105.04245",
    archivePrefix = "arXiv",
    primaryClass = "hep-th",
    doi = "10.1016/j.physletb.2021.136837",
    journal = "Phys. Lett. B",
    volume = "824",
    pages = "136837",
    year = "2022"
}

@article{Alday_2010,
   title={Liouville Correlation Functions from Four-Dimensional Gauge Theories},
   volume={91},
   ISSN={1573-0530},
   url={http://dx.doi.org/10.1007/s11005-010-0369-5},
   DOI={10.1007/s11005-010-0369-5},
   number={2},
   journal={Letters in Mathematical Physics},
   publisher={Springer Science and Business Media LLC},
   author={Alday, Luis F. and Gaiotto, Davide and Tachikawa, Yuji},
   year={2010},
   month={Jan},
   pages={167–197}
}

@article{2010,
   title={Loop and surface operators in $ \mathcal{N} = 2 $ gauge theory and Liouville modular geometry},
   volume={2010},
   ISSN={1029-8479},
   url={http://dx.doi.org/10.1007/JHEP01(2010)113},
   DOI={10.1007/jhep01(2010)113},
   number={1},
   journal={Journal of High Energy Physics},
   publisher={Springer Science and Business Media LLC},
   author={Alday, Luis F. and Gaiotto, Davide and Gukov, Sergei and Tachikawa, Yuji and Verlinde, Herman},
   year={2010},
   month={Jan}
}

@article{10.1007/BF02418420,
author = {H. Poincaré},
title = {{Sur les groupes des équations linéaires}},
volume = {4},
journal = {Acta Mathematica},
number = {none},
publisher = {Institut Mittag-Leffler},
pages = {201 -- 312},
year = {1900},
doi = {10.1007/BF02418420},
URL = {https://doi.org/10.1007/BF02418420}
}

@article{JIMBO1981407,
title = {Monodromy perserving deformation of linear ordinary differential equations with rational coefficients. II},
journal = {Physica D: Nonlinear Phenomena},
volume = {2},
number = {3},
pages = {407-448},
year = {1981},
issn = {0167-2789},
doi = {https://doi.org/10.1016/0167-2789(81)90021-X},
url = {https://www.sciencedirect.com/science/article/pii/016727898190021X},
author = {Michio Jimbo and Tetsuji Miwa}
}

@article{Jimbo1981MonodromyPD,
  title={Monodromy preserving deformation of linear ordinary differential equations with rational coefficients. III},
  author={Michio Jimbo and Tetsuji Miwa},
  journal={Physica D: Nonlinear Phenomena},
  year={1981},
  volume={4},
  pages={26-46}
}

@article{JIMBO1981306,
title = {Monodromy preserving deformation of linear ordinary differential equations with rational coefficients: I. General theory and $\tau$-function},
journal = {Physica D: Nonlinear Phenomena},
volume = {2},
number = {2},
pages = {306-352},
year = {1981},
issn = {0167-2789},
doi = {https://doi.org/10.1016/0167-2789(81)90013-0},
url = {https://www.sciencedirect.com/science/article/pii/0167278981900130},
author = {Michio Jimbo and Tetsuji Miwa and Kimio Ueno}
}

@Article{zbMATH02714348,
 Author = {F. {Klein}},
 Title = {{\"Uber eine neue Art von Riemann'schen Fl\"achen. (Zweite Mittheilung).}},
 FJournal = {{Mathematische Annalen}},
 Journal = {{Math. Ann.}},
 ISSN = {0025-5831},
 Volume = {10},
 Pages = {398--417},
 Year = {1876},
 Publisher = {Springer, Berlin/Heidelberg},
 Language = {German},
 DOI = {10.1007/BF01442321},
 MSC2010 = {30F20},
 Zbl = {08.0439.02}
}

@book{gauss1866carl,
  title={Carl Friedrich Gauss Werke: Bd. Analysis (various texts, in Latin and German, orig. publ. between 1799-1851, or found in the "Nachlass"; annotated by E.J. Schering). 1866 [i.e. 1868},
  author={Gauss, C.F. and Schering, E. and Brendel, M. and Schlesinger, L. and Kaestner, W.F. and Teubner, B.G. and Springer, J. and Dedekind, R. and Perthes, F.A. and Gesellschaft der Wissenschaften zu G{\"o}ttingen},
  series={Carl Friedrich Gauss Werke},
  url={https://books.google.it/books?id=uDMAAAAAQAAJ},
  year={1866},
  publisher={Gedruckt in der Dieterichschen Universit{\"a}ts-Druckerei W. Fr. Kaestner}
}

@article{heun1888theorie,
  title={Zur Theorie der Riemann'schen Functionen zweiter Ordnung mit vier Verzweigungspunkten},
  author={Heun, Karl},
  journal={Mathematische Annalen},
  volume={33},
  number={2},
  pages={161--179},
  year={1888},
  publisher={Springer}
}

@article{Hortacsu:2011rr,
    author = "Hortacsu, M.",
    editor = "Camci, Ugur and Semiz, Ibrahim",
    title = "{Heun Functions and Some of Their Applications in Physics}",
    eprint = "1101.0471",
    archivePrefix = "arXiv",
    primaryClass = "math-ph",
    doi = "10.1142/9789814417532_0002",
    pages = "23--39",
    year = "2012"
}

@ARTICLE{2015arXiv151204025F,
       author = {{Fiziev}, P P},
        title = "{The Heun functions as a modern powerful tool for research in different scientific domains}",
      journal = {arXiv e-prints},
         year = 2015,
        month = dec,
          eid = {arXiv:1512.04025},
        pages = {arXiv:1512.04025},
archivePrefix = {arXiv},
       eprint = {1512.04025},
 primaryClass = {math-ph},
       adsurl = {https://ui.adsabs.harvard.edu/abs/2015arXiv151204025F}
}

@ARTICLE{2007MaCom..76..811M,
       author = {{Maier}, Robert S.},
        title = "{The 192 solutions of the Heun equation}",
      journal = {Mathematics of Computation},
         year = 2007,
        month = jun,
       volume = {76},
       number = {258},
        pages = {811-843},
          doi = {10.1090/S0025-5718-06-01939-9},
archivePrefix = {arXiv},
       eprint = {math/0408317},
 primaryClass = {math.CA},
       adsurl = {https://ui.adsabs.harvard.edu/abs/2007MaCom..76..811M}
}

@article{Awata:2010bz,
    author = "Awata, Hidetoshi and Fuji, Hiroyuki and Kanno, Hiroaki and Manabe, Masahide and Yamada, Yasuhiko",
    title = "{Localization with a Surface Operator, Irregular Conformal Blocks and Open Topological String}",
    eprint = "1008.0574",
    archivePrefix = "arXiv",
    primaryClass = "hep-th",
    doi = "10.4310/ATMP.2012.v16.n3.a1",
    journal = "Adv. Theor. Math. Phys.",
    volume = "16",
    number = "3",
    pages = "725--804",
    year = "2012"
}

@article{Seiberg:1994aj,
    author = "Seiberg, N. and Witten, Edward",
    title = "{Monopoles, duality and chiral symmetry breaking in N=2 supersymmetric QCD}",
    eprint = "hep-th/9408099",
    archivePrefix = "arXiv",
    reportNumber = "RU-94-60, IASSNS-HEP-94-55",
    doi = "10.1016/0550-3213(94)90214-3",
    journal = "Nucl. Phys. B",
    volume = "431",
    pages = "484--550",
    year = "1994"
}

@article{Bonelli:2021uvf,
    author = "Bonelli, Giulio and Iossa, Cristoforo and Lichtig, Daniel Panea and Tanzini, Alessandro",
    title = "{Exact solution of Kerr black hole perturbations via CFT$_2$ and instanton counting. Greybody factor, Quasinormal modes and Love numbers}",
    eprint = "2105.04483",
    archivePrefix = "arXiv",
    primaryClass = "hep-th",
    month = "5",
    year = "2021"
}

@inproceedings{Nekrasov:2009rc,
    author = "Nekrasov, Nikita A. and Shatashvili, Samson L.",
    title = "{Quantization of Integrable Systems and Four Dimensional Gauge Theories}",
    booktitle = "{16th International Congress on Mathematical Physics}",
    eprint = "0908.4052",
    archivePrefix = "arXiv",
    primaryClass = "hep-th",
    reportNumber = "TCD-MATH-09-19, HMI-09-09, IHES-P-09-38",
    doi = "10.1142/9789814304634_0015",
    pages = "265--289",
    month = "8",
    year = "2009"
}

@article{Alday_2010a,
   title={Loop and surface operators in $ \mathcal{N} = 2 $ gauge theory and Liouville modular geometry},
   volume={2010},
   ISSN={1029-8479},
   url={http://dx.doi.org/10.1007/JHEP01(2010)113},
   DOI={10.1007/jhep01(2010)113},
   number={1},
   journal={Journal of High Energy Physics},
   publisher={Springer Science and Business Media LLC},
   author={Alday, Luis F. and Gaiotto, Davide and Gukov, Sergei and Tachikawa, Yuji and Verlinde, Herman},
   year={2010},
   month={Jan}
}

@article{Dorn:1994xn,
    author = "Dorn, Harald and Otto, H. J.",
    title = "{Two and three point functions in Liouville theory}",
    eprint = "hep-th/9403141",
    archivePrefix = "arXiv",
    reportNumber = "HU-BERLIN-IEP-94-2, DESY-94-066",
    doi = "10.1016/0550-3213(94)00352-1",
    journal = "Nucl. Phys. B",
    volume = "429",
    pages = "375--388",
    year = "1994"
}

@article{Zamolodchikov:1995aa,
    author = "Zamolodchikov, Alexander B. and Zamolodchikov, Alexei B.",
    title = "{Structure constants and conformal bootstrap in Liouville field theory}",
    eprint = "hep-th/9506136",
    archivePrefix = "arXiv",
    reportNumber = "LPM-95-24, RU-95-39",
    doi = "10.1016/0550-3213(96)00351-3",
    journal = "Nucl. Phys. B",
    volume = "477",
    pages = "577--605",
    year = "1996"
}

@article{Pereniguez:2021xcj,
    author = "Pere\~niguez, David and Cardoso, Vitor",
    title = "{Love numbers and magnetic susceptibility of charged black holes}",
    eprint = "2112.08400",
    archivePrefix = "arXiv",
    primaryClass = "gr-qc",
    month = "12",
    year = "2021"
}

@article{Castro:2013lba,
    author = "Castro, Alejandra and Lapan, Joshua M. and Maloney, Alexander and Rodriguez, Maria J.",
    title = "{Black Hole Scattering from Monodromy}",
    eprint = "1304.3781",
    archivePrefix = "arXiv",
    primaryClass = "hep-th",
    doi = "10.1088/0264-9381/30/16/165005",
    journal = "Class. Quant. Grav.",
    volume = "30",
    pages = "165005",
    year = "2013"
}

@article{Motl:2003cd,
    author = "Motl, Lubos and Neitzke, Andrew",
    title = "{Asymptotic black hole quasinormal frequencies}",
    eprint = "hep-th/0301173",
    archivePrefix = "arXiv",
    reportNumber = "HEP-UK-0017, HUTP-03-A005",
    doi = "10.4310/ATMP.2003.v7.n2.a4",
    journal = "Adv. Theor. Math. Phys.",
    volume = "7",
    number = "2",
    pages = "307--330",
    year = "2003"
}

@misc{aminov2020black,
      title={Black Hole Quasinormal Modes and Seiberg-Witten Theory}, 
      author={Gleb Aminov and Alba Grassi and Yasuyuki Hatsuda},
      year={2020},
      eprint={2006.06111},
      archivePrefix={arXiv},
      primaryClass={hep-th}
}

@article{Bruzzo_2003,
   title={Multi-instanton calculus and equivariant cohomology},
   volume={2003},
   ISSN={1029-8479},
   url={http://dx.doi.org/10.1088/1126-6708/2003/05/054},
   DOI={10.1088/1126-6708/2003/05/054},
   number={05},
   journal={Journal of High Energy Physics},
   publisher={Springer Science and Business Media LLC},
   author={Bruzzo, Ugo and Fucito, Francesco and Morales, José F and Tanzini, Alessandro},
   year={2003},
   month={May},
   pages={054–054}
}

@misc{david2015liouville,
      title={Liouville Quantum Gravity on the Riemann sphere}, 
      author={François David and Antti Kupiainen and Rémi Rhodes and Vincent Vargas},
      year={2015},
      eprint={1410.7318},
      archivePrefix={arXiv},
      primaryClass={math.PR}
}

@article{lisovyy2021,
   title={Accessory parameters in confluent Heun equations and classical irregular conformal blocks},
   volume={111},
   ISSN={1573-0530},
   url={http://dx.doi.org/10.1007/s11005-021-01400-6},
   DOI={10.1007/s11005-021-01400-6},
   number={6},
   journal={Letters in Mathematical Physics},
   publisher={Springer Science and Business Media LLC},
   author={Lisovyy, O. and Naidiuk, A.},
   year={2021},
   month={Nov}
}

@article{Belavin:1984vu,
    author = "Belavin, A. A. and Polyakov, Alexander M. and Zamolodchikov, A. B.",
    editor = "Khalatnikov, I. M. and Mineev, V. P.",
    title = "{Infinite Conformal Symmetry in Two-Dimensional Quantum Field Theory}",
    reportNumber = "CERN-TH-3827",
    doi = "10.1016/0550-3213(84)90052-X",
    journal = "Nucl. Phys. B",
    volume = "241",
    pages = "333--380",
    year = "1984"
}

@article{Gaiotto:2009ma,
    author = "Gaiotto, Davide",
    editor = "Das, Sumit R. and Shapere, Alfred D.",
    title = "{Asymptotically free $\mathcal{N} = 2$ theories and irregular conformal blocks}",
    eprint = "0908.0307",
    archivePrefix = "arXiv",
    primaryClass = "hep-th",
    doi = "10.1088/1742-6596/462/1/012014",
    journal = "J. Phys. Conf. Ser.",
    volume = "462",
    number = "1",
    pages = "012014",
    year = "2013"
}

@article{Gaiotto:2012sf,
    author = "Gaiotto, Davide and Teschner, Joerg",
    title = "{Irregular singularities in Liouville theory and Argyres-Douglas type gauge theories, I}",
    eprint = "1203.1052",
    archivePrefix = "arXiv",
    primaryClass = "hep-th",
    reportNumber = "DESY-12-046",
    doi = "10.1007/JHEP12(2012)050",
    journal = "JHEP",
    volume = "12",
    pages = "050",
    year = "2012"
}

@article{Nekrasov:2002qd,
    author = "Nekrasov, Nikita A.",
    title = "{Seiberg-Witten prepotential from instanton counting}",
    eprint = "hep-th/0206161",
    archivePrefix = "arXiv",
    reportNumber = "ITEP-TH-22-02, IHES-P-04-22",
    doi = "10.4310/ATMP.2003.v7.n5.a4",
    journal = "Adv. Theor. Math. Phys.",
    volume = "7",
    number = "5",
    pages = "831--864",
    year = "2003"
}

@article{Nekrasov:2003rj,
    author = "Nekrasov, Nikita and Okounkov, Andrei",
    title = "{Seiberg-Witten theory and random partitions}",
    eprint = "hep-th/0306238",
    archivePrefix = "arXiv",
    reportNumber = "ITEP-TH-36-03, PUDM-2003, IHES-P-03-43",
    doi = "10.1007/0-8176-4467-9_15",
    journal = "Prog. Math.",
    volume = "244",
    pages = "525--596",
    year = "2006"
}

@article{Teukolsky:1972my,
    author = "Teukolsky, S. A.",
    title = "{Rotating black holes - separable wave equations for gravitational and electromagnetic perturbations}",
    reportNumber = "OAP-291",
    doi = "10.1103/PhysRevLett.29.1114",
    journal = "Phys. Rev. Lett.",
    volume = "29",
    pages = "1114--1118",
    year = "1972"
}

@article{Hadasz:2006rb,
    author = "Hadasz, Leszek and Jaskolski, Zbigniew",
    title = "{Liouville theory and uniformization of four-punctured sphere}",
    eprint = "hep-th/0604187",
    archivePrefix = "arXiv",
    reportNumber = "BONN-TH-2006-004, IFT-UWR-0104-006",
    doi = "10.1063/1.2234272",
    journal = "J. Math. Phys.",
    volume = "47",
    pages = "082304",
    year = "2006"
}

@article{Matone:1993tj,
    author = "Matone, Marco",
    title = "{Uniformization theory and 2-D gravity. 1. Liouville action and intersection numbers}",
    eprint = "hep-th/9306150",
    archivePrefix = "arXiv",
    reportNumber = "IC-MATH-8-92, DFPD-TH-92-41",
    doi = "10.1142/S0217751X95000139",
    journal = "Int. J. Mod. Phys. A",
    volume = "10",
    pages = "289--336",
    year = "1995"
}

@article{Cantini:2001wr,
    author = "Cantini, Luigi and Menotti, Pietro and Seminara, Domenico",
    title = "{Proof of Polyakov conjecture for general elliptic singularities}",
    eprint = "hep-th/0105081",
    archivePrefix = "arXiv",
    reportNumber = "IFUP-TH-15-2001, DFF-01-05-2001",
    doi = "10.1016/S0370-2693(01)00998-4",
    journal = "Phys. Lett. B",
    volume = "517",
    pages = "203--209",
    year = "2001"
}

@article{Menotti:2014kra,
    author = "Menotti, Pietro",
    title = "{On the monodromy problem for the four-punctured sphere}",
    eprint = "1401.2409",
    archivePrefix = "arXiv",
    primaryClass = "hep-th",
    reportNumber = "IFUP-TH-2014-1",
    doi = "10.1088/1751-8113/47/41/415201",
    journal = "J. Phys. A",
    volume = "47",
    number = "41",
    pages = "415201",
    year = "2014"
}

@article{Awata:2009ur,
    author = "Awata, Hidetoshi and Yamada, Yasuhiko",
    title = "{Five-dimensional AGT Conjecture and the Deformed Virasoro Algebra}",
    eprint = "0910.4431",
    archivePrefix = "arXiv",
    primaryClass = "hep-th",
    doi = "10.1007/JHEP01(2010)125",
    journal = "JHEP",
    volume = "01",
    pages = "125",
    year = "2010"
}

@article{Bershtein:2016aef,
    author = "Bershtein, M. A. and Shchechkin, A. I.",
    title = "{q-deformed Painlev\'e $\tau$ function and q-deformed conformal blocks}",
    eprint = "1608.02566",
    archivePrefix = "arXiv",
    primaryClass = "math-ph",
    doi = "10.1088/1751-8121/aa5572",
    journal = "J. Phys. A",
    volume = "50",
    number = "8",
    pages = "085202",
    year = "2017"
}

@article{Bonelli:2017gdk,
    author = "Bonelli, Giulio and Grassi, Alba and Tanzini, Alessandro",
    title = "{Quantum curves and $q$-deformed Painlev\'e equations}",
    eprint = "1710.11603",
    archivePrefix = "arXiv",
    primaryClass = "hep-th",
    doi = "10.1007/s11005-019-01174-y",
    journal = "Lett. Math. Phys.",
    volume = "109",
    number = "9",
    pages = "1961--2001",
    year = "2019"
}

@article{Bonelli:2019boe,
    author = "Bonelli, Giulio and Del Monte, Fabrizio and Gavrylenko, Pavlo and Tanzini, Alessandro",
    title = "{${\mathcal {N}}$ = $2^*$ Gauge Theory, Free Fermions on the Torus and Painlev\'e VI}",
    eprint = "1901.10497",
    archivePrefix = "arXiv",
    primaryClass = "hep-th",
    doi = "10.1007/s00220-020-03743-y",
    journal = "Commun. Math. Phys.",
    volume = "377",
    number = "2",
    pages = "1381--1419",
    year = "2020"
}

@article{Bonelli:2019yjd,
    author = "Bonelli, Giulio and Del Monte, Fabrizio and Gavrylenko, Pavlo and Tanzini, Alessandro",
    title = "{Circular quiver gauge theories, isomonodromic deformations and $W_N$ fermions on the torus}",
    eprint = "1909.07990",
    archivePrefix = "arXiv",
    primaryClass = "hep-th",
    doi = "10.1007/s11005-020-01343-4",
    month = "9",
    year = "2019"
}

@article{Carneiro_da_Cunha_2016,
   title={Kerr–de Sitter greybody factors via isomonodromy},
   volume={93},
   ISSN={2470-0029},
   url={http://dx.doi.org/10.1103/PhysRevD.93.024045},
   DOI={10.1103/physrevd.93.024045},
   number={2},
   journal={Physical Review D},
   publisher={American Physical Society (APS)},
   author={Carneiro da Cunha, Bruno and Novaes, Fábio},
   year={2016},
   month={Jan}
}

@article{Carneiro_da_Cunha_2020,
   title={Confluent conformal blocks and the Teukolsky master equation},
   volume={102},
   ISSN={2470-0029},
   url={http://dx.doi.org/10.1103/PhysRevD.102.105013},
   DOI={10.1103/physrevd.102.105013},
   number={10},
   journal={Physical Review D},
   publisher={American Physical Society (APS)},
   author={Carneiro da Cunha, Bruno and Cavalcante, João Paulo},
   year={2020},
   month={Nov}
}

@article{Bonelli_2012,
   title={Wild quiver gauge theories},
   volume={2012},
   ISSN={1029-8479},
   url={http://dx.doi.org/10.1007/JHEP02(2012)031},
   DOI={10.1007/jhep02(2012)031},
   number={2},
   journal={Journal of High Energy Physics},
   publisher={Springer Science and Business Media LLC},
   author={Bonelli, Giulio and Maruyoshi, Kazunobu and Tanzini, Alessandro},
   year={2012},
   month={Feb}
}

@article{Bonelli_2017,
   title={On Painlevé/gauge theory correspondence},
   volume={107},
   ISSN={1573-0530},
   url={http://dx.doi.org/10.1007/s11005-017-0983-6},
   DOI={10.1007/s11005-017-0983-6},
   number={12},
   journal={Letters in Mathematical Physics},
   publisher={Springer Science and Business Media LLC},
   author={Bonelli, Giulio and Lisovyy, Oleg and Maruyoshi, Kazunobu and Sciarappa, Antonio and Tanzini, Alessandro},
   year={2017},
   month={Sep},
   pages={2359–2413}
}

@book{ronveaux1995heun,
  title={Heun's Differential Equations},
  author={Ronveaux, P.A. and Ronveaux, A. and Arscott, F.M. and S, S. and Schmidt, D. and Wolf, G. and Maroni, P. and Duval, A.},
  isbn={9780198596950},
  lccn={95194538},
  series={Oxford science publications},
  url={https://books.google.es/books?id=5p65FD8caCgC},
  year={1995},
  publisher={Oxford University Press}
}

@article{Lisovyy:2018mnj,
    author = "Lisovyy, O. and Nagoya, H. and Roussillon, J.",
    title = "{Irregular conformal blocks and connection formulae for Painlev\'e V functions}",
    eprint = "1806.08344",
    archivePrefix = "arXiv",
    primaryClass = "math-ph",
    doi = "10.1063/1.5031841",
    journal = "J. Math. Phys.",
    volume = "59",
    number = "9",
    pages = "091409",
    year = "2018"
}

@article{Piatek:2017fyn,
    author = "Pi\k{a}tek, Marcin and Pietrykowski, Artur R.",
    title = "{Solving Heun's equation using conformal blocks}",
    eprint = "1708.06135",
    archivePrefix = "arXiv",
    primaryClass = "hep-th",
    doi = "10.1016/j.nuclphysb.2018.11.021",
    journal = "Nucl. Phys. B",
    volume = "938",
    pages = "543--570",
    year = "2019"
}

@article{2001,
   title={Liouville theory revisited},
   volume={18},
   ISSN={1361-6382},
   url={http://dx.doi.org/10.1088/0264-9381/18/23/201},
   DOI={10.1088/0264-9381/18/23/201},
   number={23},
   journal={Classical and Quantum Gravity},
   publisher={IOP Publishing},
   author={Teschner, J},
   year={2001},
   month={Nov},
   pages={R153–R222}
}

@article{2011,
   title={Analytic continuation of Liouville theory},
   volume={2011},
   ISSN={1029-8479},
   url={http://dx.doi.org/10.1007/JHEP12(2011)071},
   DOI={10.1007/jhep12(2011)071},
   number={12},
   journal={Journal of High Energy Physics},
   publisher={Springer Science and Business Media LLC},
   author={Harlow, Daniel and Maltz, Jonathan and Witten, Edward},
   year={2011},
   month={Dec}
}

@Article{zbMATH03989776,
 Author = {P. G. {Zograf} and L. A. {Takhtadzhyan}},
 Title = {{Action of the Liouville equation is a generating function for the accessory parameters and the potential of the Weil-Petersson metric on the Teichm\"uller space}},
 FJournal = {{Functional Analysis and its Applications}},
 Journal = {{Funct. Anal. Appl.}},
 ISSN = {0016-2663},
 Volume = {19},
 Pages = {219--220},
 Year = {1985},
 Publisher = {Springer US, New York, NY},
 Language = {English},
 DOI = {10.1007/BF01076626},
 MSC2010 = {32G15 31C15},
 Zbl = {0612.32018}
}

@article{Lit2014,
   title={Classical conformal blocks and Painlevé VI},
   volume={2014},
   ISSN={1029-8479},
   url={http://dx.doi.org/10.1007/JHEP07(2014)144},
   DOI={10.1007/jhep07(2014)144},
   number={7},
   journal={Journal of High Energy Physics},
   publisher={Springer Science and Business Media LLC},
   author={Litvinov, Alexey and Lukyanov, Sergei and Nekrasov, Nikita and Zamolodchikov, Alexander},
   year={2014},
   month={Jul}
}

@article{takzo02,
author = {Takhtajan, Leon and Zograf, Peter},
year = {2002},
month = {01},
pages = {},
title = {Hyperbolic 2-spheres with conical singularities, accessory parameters and Kähler metrics on $\mathcal{M}_{0,n}$},
volume = {355},
journal = {Transactions of the American Mathematical Society},
doi = {10.2307/1194984}
}

@misc{hollands2017higher,
      title={Higher length-twist coordinates, generalized Heun's opers, and twisted superpotentials}, 
      author={Lotte Hollands and Omar Kidwai},
      year={2017},
      eprint={1710.04438},
      archivePrefix={arXiv},
      primaryClass={hep-th}
}

@article{unpolyakov,
    title={Lecture at Steklov institute in Leningrad},
    author={A. M. Polyakov},
    year={1982},
    journal={unpublished}
}

@article{Fioravanti:2021dce,
    author = "Fioravanti, Davide and Gregori, Daniele",
    title = "{A new method for exact results on Quasinormal Modes of Black Holes}",
    eprint = "2112.11434",
    archivePrefix = "arXiv",
    primaryClass = "hep-th",
    month = "12",
    year = "2021"
}

@article{saebyeok2020,
   title={Opers, surface defects, and Yang-Yang functional},
   volume={24},
   ISSN={1095-0753},
   url={http://dx.doi.org/10.4310/ATMP.2020.v24.n7.a4},
   DOI={10.4310/atmp.2020.v24.n7.a4},
   number={7},
   journal={Advances in Theoretical and Mathematical Physics},
   publisher={International Press of Boston},
   author={Jeong, Saebyeok and Nekrasov, Nikita},
   year={2020},
   pages={1789–1916}
}

@article{dekar,
	author = {Dekar,Li{\`e}s and Chetouani,Lyazid and Hammann,Th{\'e}ophile F.},
	doi = {10.1063/1.532407},
	eprint = {https://doi.org/10.1063/1.532407},
	journal = {Journal of Mathematical Physics},
	number = {5},
	pages = {2551-2563},
	title = {An exactly soluble Schr{\"o}dinger equation with smooth position-dependent mass},
	url = {https://doi.org/10.1063/1.532407},
	volume = {39},
	year = {1998}
}

@article{takemura,
 ISSN = {1364503X},
 URL = {http://www.jstor.org/stable/25190740},
 author = {Kouichi Takemura},
 journal = {Philosophical Transactions: Mathematical, Physical and Engineering Sciences},
 number = {1867},
 pages = {1179--1201},
 publisher = {The Royal Society},
 title = {On the Heun Equation},
 volume = {366},
 year = {2008}
}

@article{Flume,
    author = "Flume, R. and Poghossian, R.",
    title = "{An Algorithm for the microscopic evaluation of the coefficients of the Seiberg-Witten prepotential}",
    eprint = "hep-th/0208176",
    archivePrefix = "arXiv",
    doi = "10.1142/S0217751X03013685",
    journal = "Int. J. Mod. Phys. A",
    volume = "18",
    pages = "2541",
    year = "2003"
}
\end{document}